\documentclass[10pt,a4paper,twoside]{report}

\usepackage[utf8]{inputenc}
\usepackage[T1]{fontenc}
\usepackage[amsthm,libertine]{newtxmath}
\usepackage[sfdefault]{biolinum}
\usepackage[scale=.9]{noto}
\usepackage[scale=.95]{inconsolata}

\usepackage{amsmath}
\usepackage{bm}
\usepackage{stmaryrd}

\usepackage{anyfontsize}

\usepackage{xcolor}

\usepackage[margin=1.5in]{geometry}

\usepackage[sf,bf]{titlesec}

\definecolor{Prune}{cmyk}{0.24,1,0.17,0.60}
\definecolor{PruneLight}{cmyk}{0, 1, 0.56, 0.19}
\definecolor{PruneAbricot}{cmyk}{0, 1, 0.5, 0}

\definecolor{PruneMustard}{cmyk}{0.2, 0, 0.91, 0}

\definecolor{PruneBlue}{cmyk}{1, 0.37, 0.44, 0.10}
\definecolor{PruneBlueGreen}{cmyk}{0.85, 0.12, 0.46, 0.26}
\definecolor{PruneGreen}{cmyk}{1, 0, 0.66, 0.09}

\titleformat{\chapter}[display]
  {\huge\bfseries\sffamily\color{Prune}}{\chaptertitlename\ \thechapter}{20pt}{\Huge}
\titleformat*{\section}{\Large\bfseries\sffamily\color{Prune}}
\titleformat*{\subsection}{\large\bfseries\sffamily\color{PruneLight}}
\titleformat*{\subsubsection}{\large\bfseries\sffamily\color{PruneLight}}

\titleformat*{\paragraph}{\bfseries\sffamily\color{PruneLight}}

\titlespacing{\paragraph}{%
  0pt}{
  1.25ex plus .5ex minus .2ex}{
  1em}

\makeatletter
\let\c@table\c@figure
\let\ftype@table\ftype@figure
\makeatother

\usepackage{tikz}
\usetikzlibrary{arrows.meta,patterns,ipe,calc} 

\usepackage[absolute]{textpos} 
\usepackage{colortbl}
\usepackage{array}

\usepackage[all]{xy}
\usepackage{fancyvrb}
\usepackage{amsthm}
\usepackage{caption,subcaption}
\usepackage{graphicx}
\usepackage{xspace}
\usepackage{proof}
\usepackage{braket}
\usepackage{tabularx}

\usepackage{mdframed}
\usepackage{multirow} 
\usepackage{multicol} 
\usepackage{scrextend} 

\usepackage{pdfpages}

\usepackage[hyphens]{url}
\usepackage[hyperindex=true]{hyperref}
\hypersetup{
  pdftitle={On Quantum Programming Languages},
  pdfauthor={Benoît Valiron},
  pdfsubject={Habilitation à Diriger des Recherches},
  colorlinks=true,
  linkcolor=PruneBlue,
  filecolor=PruneBlueGreen,
  urlcolor=PruneBlueGreen,
  citecolor=PruneGreen
}
\usepackage{fancyhdr}

\usepackage[frozencache,cachedir=minted]{minted}
\usepackage[many]{tcolorbox}
\tcbuselibrary{minted} 

\newenvironment{mylife}{\begin{tcolorbox}[standard jigsaw, breakable,boxrule=.1pt,leftrule=5pt,arc=0pt,colframe=Prune,colback=Prune,opacityback=0.05]}{\end{tcolorbox}}
\newenvironment{mylife*}{\begin{tcolorbox}[standard jigsaw, boxrule=.1pt,leftrule=5pt,arc=0pt,colframe=Prune,colback=white,opacityback=0.1]}{\end{tcolorbox}}

\newtcblisting{haskell}{
  colframe=PruneMustard!50,
  colback=PruneMustard!5,
  listing only,
  listing engine=minted,
  minted language=haskell,
  minted options={fontsize=\small,linenos,numbersep=3mm},
}

\usepackage{etoc}

\usepackage{paralist}

\newcommand{\fullcite}[1]{}

\usepackage[english]{babel}
\usepackage{csquotes}

\usepackage{kotex}

\usepackage{imakeidx}
\makeindex

\def\myfont{\sffamily\color{Prune}}

\theoremstyle{plain}
\newtheorem{theorem}{\myfont Theorem}[chapter]

\newtheorem{proposition}[theorem]{\myfont Proposition}

\theoremstyle{definition}

\newtheorem{example}[theorem]{\myfont Example}
\newtheorem{notation}[theorem]{\myfont Notation}

\theoremstyle{remark}
\newtheorem{remark}[theorem]{\myfont Remark}

\newcommand{\itempaper}[1]{\item[\cite{#1}]\fullcite{#1}}
\newcommand{\cincludegraphics}[2]{\raisebox{-0.5\height}{\includegraphics[#1]{#2}}}

\newcommand{\simpleidx}[1]{{#1}\index{#1}\xspace}
\newcommand{\emphidx}[1]{\emph{#1}\index{#1}\xspace}
\newcommand{\emphidxalt}[2]{\emph{#1}\index{#2}\xspace}
\newcommand{\etal}{\textit{et al.}\xspace}
\newcommand{\etc}{\textit{etc}\xspace}

\newcommand{\Cpp}{{C{\!+}{\!+}}\xspace}
\newcommand{\silq}{\textsc{Silq}\xspace}
\newcommand{\quipper}{\textsc{Quipper}\xspace}
\newcommand{\protoquipper}{\textsc{ProtoQuipper}\xspace}
\newcommand{\protoquipperx}[1]{\textsc{ProtoQuipper#1}\xspace}
\newcommand{\qbrick}{\textsc{Qbrick}\xspace}
\newcommand{\qiskit}{\textsc{Qiskit}\xspace}
\newcommand{\QIO}{{QIO}\xspace}
\newcommand{\qasm}{\textsc{Qasm}\xspace}
\newcommand{\lov}{\textsc{LOv}\xspace}

\newcommand{\parr}{\mathrel{\rotatebox[origin=c]{180}{\&}}}
\newcommand{\tensor}{\otimes}
\newcommand{\loli}{\mathrel{\multimap}}
\newcommand{\tunit}{{\mathbf{1}}}
\newcommand{\punit}{{\mathbf{0}}}
\newcommand{\bor}{\bigm|}

\newcommand{\tto}{\Rightarrow}
\newcommand{\bit}{\ensuremath{\texttt{bit}}}
\newcommand{\bool}{\ensuremath{\texttt{bool}}}
\newcommand{\qbit}{\ensuremath{\texttt{qbit}}}

\newcommand{\meas}{\ensuremath{\texttt{meas}}}
\newcommand{\qinit}{\ensuremath{\texttt{qinit}}}
\newcommand{\ttrue}{\ensuremath{\texttt{t\!\!t}}}
\newcommand{\ffalse}{\ensuremath{\texttt{f\!\!f}}}
\newcommand{\lettermx}[3]{{\texttt{let}\,{#1}={#2}\,\texttt{in}\,{#3}}}
\newcommand{\letrectermx}[3]{{\texttt{let}\,\texttt{rec}\,{#1}={#2}\,\texttt{in}\,{#3}}}
\newcommand{\tuple}[1]{{\braket{#1}}}

\newcommand{\unitterm}{\tuple{}}
\newcommand{\iftermx}[3]{{\texttt{if}\,{#1}\,\texttt{then}\,{#2}\,\texttt{else}\,{#3}}}
\newcommand{\iftermq}[3]{{\texttt{if}^\circ\,{#1}\,\texttt{then}\,{#2}\,\texttt{else}\,{#3}}}
\newcommand{\unittype}{\tunit}

\newcommand{\inl}{\texttt{inl}}
\newcommand{\inr}{\texttt{inr}}
\newcommand{\iso}{\leftrightarrow}

\newcommand{\FV}{\text{FV}}
\newcommand{\tfix}{\texttt{fix}}
\newcommand{\tnil}{\texttt{nil}}
\newcommand{\tfold}{\texttt{fold}}
\newcommand{\tcons}{\mathbin{{:}\!\!{:}}}

\newcommand{\mumall}{μMALL}

\newcommand{\hilb}[1]{\mathcal{#1}}
\newcommand{\cat}[1]{\mathcal{#1}}

\newcommand{\denot}[1]{{\llbracket{#1}\rrbracket}}

\newcommand{\qH}{\ensuremath{\mathcal{Q}}}
\newcommand{\Cx}{\ensuremath{\mathbb{C}}}
\newcommand{\Nx}{\ensuremath{\mathbb{N}}}
\newcommand{\CPM}{\ensuremath{\mathbf{CPM}}\xspace}
\newcommand{\CPMfinbip}{\ensuremath{\mathbf{CPM}^{\oplus}_{\text{fin}}}\xspace}
\newcommand{\CPMd}{\ensuremath{\mathbf{CPM}_D}\xspace}
\newcommand{\CPMbip}{\ensuremath{\mathbf{CPM}^{\oplus}_D}\xspace}

\newcommand{\NOT}{\textit{NOT}}
\newcommand{\CNOT}{\textit{CNOT}}
\newcommand{\SWAP}{\textit{SWAP}}

\newcommand{\trace}{\mathsf{Tr}}

\newcommand{\utffontx}{\includegraphics[width=1.5ex]{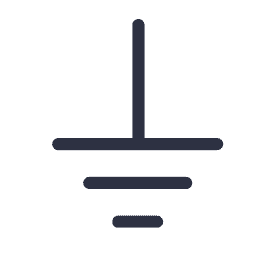}}
\newcommand{\ground}{\mbox{\utffontx}}
\newcommand{\subground}{\mbox{\utffontx}}
\newcommand{\ZXg}{ZX${}_{\subground}$\xspace}

\newcommand{\Toylang}{\Lambda_{\text{bool}}}

\newcommand{\tbox}{\texttt{box}\xspace}
\newcommand{\tunbox}{\texttt{unbox}\xspace}
\newcommand{\tcirc}{\texttt{circ}\xspace}

\newcommand{\qswitch}{\textsc{Switch}}

\newcommand{\RED}[1]{\textsc{RED}_{#1}}

\newcommand{\myciteallouchereuse}{
C. Allouche, M. Baboulin, T. Goubault de Brugière, and B. Valiron. ``Reuse method for quantum circuit
synthesis''. In: \emph{Recent Advances in Mathematical and Statistical Methods, post-proceedings of the IV
AMMCS International Conference on Applied Mathematics, Modeling and Computational Science,
Waterloo, Canada, August 20 -- 25, 2017}. Ed. by D. Marc Kilgour, Herb Kunze, Roman Makarov,
Roderick Melnik, and Xu Wang. Springer International Publishing, 2018, pp. 3--12. \textsc{isbn}:
978-3-319-99719-3. \textsc{doi}: \href {https://doi.org/10.1007/978-3-319-99719-3_1} {\nolinkurl
{10.1007/978-3-319-99719-3_1}}. \textsc{hal}: \href {https://hal.archives-ouvertes.fr/hal-01711378} {\nolinkurl
{hal-01711378}}.}
\newcommand{\mycitearrighiaddressable}{
Pablo Arrighi, Christopher Cedzich, Marin Costes, Ulysse Rémond, and Benoît Valiron. ``Addressable
quantum gates''. In: \emph{ACM Transactions on Quantum Computing} 4.3 (2023), pp. 1--41. \textsc{doi}:
\href {https://doi.org/10.1145/3581760} {\nolinkurl {10.1145/3581760}}. \textsc{hal}: \href
{https://hal.archives-ouvertes.fr/hal-03936367} {\nolinkurl {hal-03936367}}. \textsc{arXiv}: \href
{https://www.arxiv.org/abs/2109.08050} {\nolinkurl {2109.08050}}.}
\newcommand{\myciteassafcall}{
Ali Assaf, Alejandro Díaz-Caro, Simon Perdrix, Christine Tasson, and Benoît Valiron. ``Call-by-value,
call-by-name and the vectorial behaviour of the algebraic lambda-calculus''. In: \emph{Logical Methods in
Computer Science} 10.4 (2014). \textsc{doi}: \href {https://doi.org/10.2168/LMCS-10(4:8)2014} {\nolinkurl
{10.2168/LMCS-10(4:8)2014}}. \textsc{arXiv}: \href {https://www.arxiv.org/abs/1005.2897v7} {\nolinkurl
{1005.2897v7}}.}
\newcommand{\mycitearrighisubject}{
Pablo Arrighi, Alejandro Díaz-Caro, and Benoît Valiron. ``Subject reduction in a curry-style polymorphic
type system with a vectorial structure''. In: \emph{Proceedings of the 7th International Workshop on
Developments of Computational Methods, DCM 2011} (Zurich, Switzerland, July 3, 2021). Ed. by
Elham Kashefi, Jean Krivine, and Femke van Raamsdonk. Vol. 88. Electronic Proceedings in Theoretical
Computer Science. Preliminary work to the journal paper \cite{arrighi2017vectorial}. 2011, pp. 1--15. \textsc{doi}: \href
{https://doi.org/10.4204/EPTCS.88.1} {\nolinkurl {10.4204/EPTCS.88.1}}. \textsc{hal}: \href
{https://hal.archives-ouvertes.fr/hal-00924926} {\nolinkurl {hal-00924926}}. \textsc{arXiv}: \href
{https://www.arxiv.org/abs/1012.4032} {\nolinkurl {1012.4032}}.}
\newcommand{\mycitearrighivectorial}{
Pablo Arrighi, Alejandro Díaz-Caro, and Benoît Valiron. ``The vectorial lambda-calculus''. In:
\emph{Information and Computation} 254 (2017), pp. 105--139. \textsc{doi}: \href
{https://doi.org/10.1016/j.ic.2017.04.001} {\nolinkurl {10.1016/j.ic.2017.04.001}}. \textsc{hal}: \href
{https://hal.archives-ouvertes.fr/hal-00921087} {\nolinkurl {hal-00921087}}.}
\newcommand{\mycitebrugieresynthesizing}{
Timothée Goubault de Brugière, Marc Baboulin, Benoît Valiron, and Cyril Allouche. ``Synthesizing quantum
circuits via numerical optimization''. In: \emph{Proceedings of the 19th International Conference on
Computational Science, ICCS 2019, Part II} (Faro, Portugal, June 12--14, 2019). Ed. by João M. F. Rodrigues,
Pedro J. S. Cardoso, Jânio M. Monteiro, Roberto Lam, Valeria V. Krzhizhanovskaya, Michael Harold Lees,
Jack J. Dongarra, and Peter M. A. Sloot. Vol. 11537. Lecture Notes in Computer Science. Springer, 2019,
pp. 3--16. \textsc{isbn}: 978-3-030-22740-1. \textsc{doi}: \href {https://doi.org/10.1007/978-3-030-22741-8_1}
{\nolinkurl {10.1007/978-3-030-22741-8_1}}. \textsc{hal}: \href {https://hal.archives-ouvertes.fr/hal-02174967}
{\nolinkurl {hal-02174967}}. \textsc{arXiv}: \href {https://www.arxiv.org/abs/2004.07714} {\nolinkurl
{2004.07714}}.}
\newcommand{\mycitebrugierehouseholder}{
Timothée Goubault de Brugière, Marc Baboulin, Benoît Valiron, and Cyril Allouche. ``Quantum circuits
synthesis using Householder transformations''. In: \emph{Computer Physics Communications} 248 (2020),
p. 107001. \textsc{doi}: \href {https://doi.org/10.1016/j.cpc.2019.107001} {\nolinkurl
{10.1016/j.cpc.2019.107001}}. \textsc{hal}: \href {https://hal.archives-ouvertes.fr/hal-02545123} {\nolinkurl
{hal-02545123}}. \textsc{arXiv}: \href {https://www.arxiv.org/abs/2004.07710} {\nolinkurl {2004.07710}}.}
\newcommand{\mycitebrugierequantum}{
Timothée Goubault de Brugière, Marc Baboulin, Benoît Valiron, Simon Martiel, and Cyril Allouche.
``Quantum CNOT circuits synthesis for NISQ architectures using the syndrome decoding problem''. In:
[RC20], pp. 189--205. \textsc{doi}: \href {https://doi.org/10.1007/978-3-030-52482-1_11} {\nolinkurl
{10.1007/978-3-030-52482-1_11}}.}
\newcommand{\mycitebrugierereducing}{
Timothee Goubault De Brugiere, Marc Baboulin, Benoît Valiron, Simon Martiel, and Cyril Allouche.
``Reducing the depth of linear reversible quantum circuits''. In: \emph{IEEE Transactions on Quantum
Engineering} 2 (2021), p. 3102422. \textsc{doi}: \href {https://doi.org/10.1109/TQE.2021.3091648} {\nolinkurl
{10.1109/TQE.2021.3091648}}. \textsc{hal}: \href {https://hal.archives-ouvertes.fr/hal-03553916} {\nolinkurl
{hal-03553916}}.}
\newcommand{\mycitebrugieredecoding}{
Timothée Goubault de Brugière, Marc Baboulin, Benoît Valiron, Simon Martiel, and Cyril Allouche.
``Decoding techniques applied to the compilation of CNOT circuits for NISQ architectures''. In:
\emph{Science of Computer Programming} 214 (2022), p. 102726. \textsc{doi}: \href
{https://doi.org/10.1016/J.SCICO.2021.102726} {\nolinkurl {10.1016/J.SCICO.2021.102726}}. \textsc{hal}: \href
{https://hal.archives-ouvertes.fr/hal-03547113} {\nolinkurl {hal-03547113}}.
}
\newcommand{\myciteborgnahybrid}{
Agustín Borgna, Simon Perdrix, and Benoît Valiron. ``Hybrid quantum-classical circuit simplification with
the ZX-calculus''. In: \emph{Proceedings of the 19th Asian Symposium on Programming Languages and
Systems, APLAS 2021} (Chicago, IL, USA (Online Conference), Oct. 17--18, 2021). Ed. by Hakjoo Oh.
Vol. 13008. Lecture Notes in Computer Science. Springer, 2021, pp. 121--139. \textsc{doi}: \href
{https://doi.org/10.1007/978-3-030-89051-3_8} {\nolinkurl {10.1007/978-3-030-89051-3_8}}. \textsc{arXiv}: \href
{https://www.arxiv.org/abs/2109.06071} {\nolinkurl {2109.06071}}.
}
\newcommand{\mycitecharetonautomated}{
Christophe Chareton, Sébastien Bardin, François Bobot, Valentin Perrelle, and Benoît Valiron. ``An
automated deductive verification framework for circuit-building quantum programs''. In: \emph{Proceedings
of the 30th European Symposium on Programming Languages and Systems, ESOP 2021} (Luxembourg City,
Luxembourg, Mar. 27--Apr. 1, 2021). Ed. by Nobuko Yoshida. Vol. 12648. Lecture Notes in Computer Science.
Springer, 2021, pp. 148--177. \textsc{isbn}: 978-3-030-72018-6. \textsc{doi}: \href
{https://doi.org/10.1007/978-3-030-72019-3_6} {\nolinkurl {10.1007/978-3-030-72019-3_6}}. \textsc{arXiv}: \href
{https://www.arxiv.org/abs/2003.05841} {\nolinkurl {2003.05841}}.}
\newcommand{\mycitecharetonformal}{
Christophe Chareton, Sébastien Bardin, Dongho Lee, Benoît Valiron, Renaud Vilmart, and Zhaowei Xu.
``Formal Methods for Quantum Programs: A Survey''. Draft, to appear as a book chapter. 2021. \textsc{arXiv}:
\href {https://www.arxiv.org/abs/2109.06493} {\nolinkurl {2109.06493}}.
}
\newcommand{\mycitechiribellaquantum}{
G. Chiribella, G. M. D'Ariano, P. Perinotti, and B. Valiron. ``Quantum computations without definite causal
structure''. In: \emph{Physical Review A} 88 (2013), p. 022318. \textsc{doi}: \href
{https://doi.org/10.1103/PhysRevA.88.022318} {\nolinkurl {10.1103/PhysRevA.88.022318}}. \textsc{arXiv}: \href
{https://www.arxiv.org/abs/0912.0195} {\nolinkurl {0912.0195}}.
}
\newcommand{\myciteclementlov}{
Alexandre Clément, Nicolas Heurtel, Shane Mansfield, Simon Perdrix, and Benoît Valiron. ``LOv-calculus: a
graphical language for linear optical quantum circuits''. In: \emph{47th International Symposium on
Mathematical Foundations of Computer Science, MFCS 2022, August 22-26, 2022, Vienna, Austria}. Ed. by
Stefan Szeider, Robert Ganian, and Alexandra Silva. Vol. 241. LIPIcs. 2022, 35:1--35:16. \textsc{doi}: \href
{https://doi.org/10.4230/LIPICS.MFCS.2022.35} {\nolinkurl {10.4230/LIPICS.MFCS.2022.35}}. \textsc{url}: \url
{https://doi.org/10.4230/LIPIcs.MFCS.2022.35}.}
\newcommand{\myciteclementcomplete}{
Alexandre Clément, Nicolas Heurtel, Shane Mansfield, Simon Perdrix, and Benoît Valiron. ``A complete
equational theory for quantum circuits''. In: \emph{38th Annual ACM/IEEE Symposium on Logic in
Computer Science, LICS 2023, Boston, MA, USA, June 26-29, 2023}. IEEE, 2023, pp. 1--13. \textsc{doi}: \href
{https://doi.org/10.1109/LICS56636.2023.10175801} {\nolinkurl {10.1109/LICS56636.2023.10175801}}.
\textsc{hal}: \href {https://hal.archives-ouvertes.fr/hal-03926757} {\nolinkurl {hal-03926757}}.}
\newcommand{\myciteclementcompletetqc}{
Alexandre Clément, Nicolas Heurtel, Shane Mansfield, Simon Perdrix, and Benoît Valiron. ``A Complete
Equational Theory for Quantum Circuits''. Presentation accepted at the 18th Conference on the Theory of
Quantum Computation, Communication and Cryptography (TQC 2023), in Aveiro, Portugal. 2023.
\textsc{hal}: \href {https://hal.archives-ouvertes.fr/hal-04318291v1} {\nolinkurl {hal-04318291v1}}.}
\newcommand{\mycitelemonniercategorical}{
Kostia Chardonnet, Louis Lemonnier, and Benoît Valiron. ``Categorical semantics of reversible
pattern-matching''. In: \cite{mfps2021}, pp. 18--33. \textsc{doi}: \href {https://doi.org/10.4204/EPTCS.351.2}
{\nolinkurl {10.4204/EPTCS.351.2}}.}
\newcommand{\mycitechardonnetcurryrc}{
Kostia Chardonnet, Alexis Saurin, and Benoît Valiron. ``Toward a Curry-Howard equivalence for linear,
reversible computation - work-in-progress''. In: \cite{rc2020}, pp. 144--152. \textsc{doi}: \href
{https://doi.org/10.1007/978-3-030-52482-1_8} {\nolinkurl {10.1007/978-3-030-52482-1_8}}. \textsc{hal}: \href
{https://hal.archives-ouvertes.fr/hal-03103455} {\nolinkurl {hal-03103455}}.}
\newcommand{\mycitechardonnetcurry}{
Kostia Chardonnet, Alexis Saurin, and Benoît Valiron. ``Towards a Curry-Howard correspondence for linear,
reversible computation''. In: \emph{Proceedings of the 5th International Workshop on Trends in Linear Logic
and Applications (TLLA 2021)} (Rome (virtual), Italy). 2021. \textsc{hal}: \href
{https://hal.archives-ouvertes.fr/lirmm-03271484} {\nolinkurl {lirmm-03271484}}.
}
\newcommand{\mycitechapuispagerank}{
Théodore Chapuis-Chkaiban, Zeno Toffano, and Benoît Valiron. ``On new pagerank computation methods
using quantum computing''. In: \emph{Quantum Information Processing} 22.3 (2023), p. 138. \textsc{doi}:
\href {https://doi.org/10.1007/S11128-023-03856-Y} {\nolinkurl {10.1007/S11128-023-03856-Y}}. \textsc{hal}:
\href {https://hal.archives-ouvertes.fr/hal-04056045} {\nolinkurl {hal-04056045}}.
}
\newcommand{\mycitechardonnetgeometry}{
Kostia Chardonnet, Benoît Valiron, and Renaud Vilmart. ``Geometry of interaction for ZX diagrams''. In:
\emph{Proceedings of the 46th International Symposium on Mathematical Foundations of Computer
Science, MFCS 2021} (Tallinn, Estonia). Ed. by Filippo Bonchi and Simon J. Puglisi. Vol. 202. LIPIcs. Schloss
Dagstuhl - Leibniz-Zentrum fuer Informatik, 2021, 30:1--30:16. \textsc{isbn}: 978-3-95977-201-3. \textsc{doi}:
\href {https://doi.org/10.4230/LIPIcs.MFCS.2021.30} {\nolinkurl {10.4230/LIPIcs.MFCS.2021.30}}.}
\newcommand{\mycitechardonnetmanyworlds}{
Kostia Chardonnet, Marc de Visme, Benoît Valiron, and Renaud Vilmart. ``The Many-Worlds Calculus:
Representing Quantum Control''. 2022.
}
\newcommand{\mycitediazcarorealizability}{
Alejandro Díaz-Caro, Mauricio Guillermo, Alexandre Miquel, and Benoît Valiron. ``Realizability in the unitary
sphere''. In: \cite{lics2019}, pp. 1--13. \textsc{doi}: \href {https://doi.org/10.1109/LICS.2019.8785834} {\nolinkurl
{10.1109/LICS.2019.8785834}}. \textsc{hal}: \href {https://hal.archives-ouvertes.fr/hal-02175168} {\nolinkurl
{hal-02175168}}. \textsc{arXiv}: \href {https://www.arxiv.org/abs/1904.08785} {\nolinkurl {1904.08785}}.
}
\newcommand{\mycitegoubaultgaussian}{
Timothée Goubault de Brugière, Marc Baboulin, Benoît Valiron, Simon Martiel, and Cyril Allouche.
``Gaussian elimination versus greedy methods for the synthesis of linear reversible circuits''. In: \emph{ACM
Transactions on Quantum Computing} 2.3 (2021), p. 11. \textsc{doi}: \href {https://doi.org/10.1145/3474226}
{\nolinkurl {10.1145/3474226}}. \textsc{hal}: \href {https://hal.archives-ouvertes.fr/hal-03547117} {\nolinkurl
{hal-03547117}}.
}
\newcommand{\mycitegreenintroduction}{
Alexander S. Green, Peter LeFanu Lumsdaine, Neil J. Ross, Peter Selinger, and Benoît Valiron. ``An
introduction to quantum programming in quipper''. In: \cite{rc2013}, pp. 110--124. \textsc{doi}: \href
{https://doi.org/10.1007/978-3-642-38986-3_10} {\nolinkurl {10.1007/978-3-642-38986-3_10}}. \textsc{arXiv}:
\href {https://www.arxiv.org/abs/1304.5485} {\nolinkurl {1304.5485}}.}
\newcommand{\mycitegreenquipper}{
Alexander S. Green, Peter LeFanu Lumsdaine, Neil J. Ross, Peter Selinger, and Benoît Valiron. ``Quipper: a
scalable quantum programming language''. In: \emph{Proceedings of the ACM SIGPLAN Conference on
Programming Language Design and Implementation, PLDI'13} (Seattle, WA, USA). Ed. by
Hans-Juergen Boehm and Cormac Flanagan. ACM, 2013, pp. 333--342. \textsc{isbn}: 978-1-4503-2014-6.
\textsc{doi}: \href {https://doi.org/10.1145/2491956.2462177} {\nolinkurl {10.1145/2491956.2462177}}.
\textsc{arXiv}: \href {https://www.arxiv.org/abs/1304.3390} {\nolinkurl {1304.3390}}.
}
\newcommand{\myciteheurtelperceval}{
Nicolas Heurtel, Andreas Fyrillas, Grégoire de Gliniasty, Raphaël Le Bihan, Sébastien Malherbe,
Marceau Pailhas, Eric Bertasi, Boris Bourdoncle, Pierre-Emmanuel Emeriau, Rawad Mezher, Luka Music,
Nadia Belabas, Benoît Valiron, Pascale Senellart, Shane Mansfield, and Jean Senellart. ``Perceval: a software
platform for discrete variable photonic quantum computing''. In: \emph{Quantum} 7 (2023), p. 931.
\textsc{doi}: \href {https://doi.org/10.22331/Q-2023-02-21-931} {\nolinkurl {10.22331/Q-2023-02-21-931}}.
\textsc{hal}: \href {https://hal.archives-ouvertes.fr/hal-03874624} {\nolinkurl {hal-03874624}}.
}
\newcommand{\myciteheurtelstrong}{
Nicolas Heurtel, Shane Mansfield, Jean Senellart, and Benoît Valiron. ``Strong simulation of linear optical
processes''. In: \emph{Computer Physics Communications} 291 (2023), p. 108848. \textsc{doi}: \href
{https://doi.org/10.1016/J.CPC.2023.108848} {\nolinkurl {10.1016/J.CPC.2023.108848}}. \textsc{hal}: \href
{https://hal.archives-ouvertes.fr/hal-03874624v1} {\nolinkurl {hal-03874624v1}}.}
\newcommand{\mycitelagoparallelism}{
Ugo Dal Lago, Claudia Faggian, Benoît Valiron, and Akira Yoshimizu. ``Parallelism and synchronization in an
infinitary context''. In: \cite{lics2015}, pp. 559--572. \textsc{doi}: \href {https://doi.org/10.1109/LICS.2015.58}
{\nolinkurl {10.1109/LICS.2015.58}}. \textsc{hal}: \href {https://hal.archives-ouvertes.fr/hal-01231831}
{\nolinkurl {hal-01231831}}. \textsc{arXiv}: \href {https://www.arxiv.org/abs/1505.03635} {\nolinkurl
{1505.03635}}.}
\newcommand{\mycitelagogeometry}{
Ugo Dal Lago, Claudia Faggian, Benoît Valiron, and Akira Yoshimizu. ``The geometry of parallelism:
classical, probabilistic, and quantum effects''. In: \cite{popl2017}, pp. 833--845. \textsc{doi}: \href
{https://doi.org/10.1145/3009837.3009859} {\nolinkurl {10.1145/3009837.3009859}}. \textsc{hal}: \href
{https://hal.archives-ouvertes.fr/hal-01474620} {\nolinkurl {hal-01474620}}. \textsc{arXiv}: \href
{https://www.arxiv.org/abs/1610.09629} {\nolinkurl {1610.09629}}.
}
\newcommand{\myciteleeconcrete}{
Dongho Lee, Valentin Perrelle, Benoît Valiron, and Zhaowei Xu. ``Concrete categorical model of a quantum
circuit description language with measurement''. In: \emph{Proceedings of the 41st IARCS Annual
Conference on Foundations of Software Technology and Theoretical Computer Science, FSTTCS 2021}.
Ed. by Mikolaj Bojanczyk and Chandra Chekuri. Vol. 213. LIPIcs. 2021, 51:1--51:20. \textsc{doi}: \href
{https://doi.org/10.4230/LIPIcs.FSTTCS.2021.51} {\nolinkurl {10.4230/LIPIcs.FSTTCS.2021.51}}.
}
\newcommand{\mycitepaganiapplying}{
Michele Pagani, Peter Selinger, and Benoît Valiron. ``Applying quantitative semantics to higher-order
quantum computing''. In: \cite{popl2014}, pp. 647--658. \textsc{doi}: \href {https://doi.org/10.1145/2535838.2535879}
{\nolinkurl {10.1145/2535838.2535879}}. \textsc{arXiv}: \href {https://www.arxiv.org/abs/1311.2290} {\nolinkurl
{1311.2290}}.
}
\newcommand{\myciteqpl}{
Benoît Valiron, Shane Mansfield, Pablo Arrighi, and Prakash Panangaden, eds. \emph{Proceedings 17th
International Conference on Quantum Physics and Logic, QPL 2020} (Online (due to Covid), June 2--6, 2020).
Vol. 340. EPTCS. 2020.}
\newcommand{\myciteqplbis}{
Shane Mansfield, Benoît Valiron, and Vladimir Zamdzhiev, eds. \emph{Proceedings of the Twentieth
International Conference on Quantum Physics and Logic, QPL 2023} (Paris, France, July 17--21, 2023).
Vol. 384. EPTCS. 2023. \textsc{doi}: \href {https://doi.org/10.4204/EPTCS.384} {\nolinkurl {10.4204/EPTCS.384}}.}
\newcommand{\mycitesmithquipper}{
Jonathan M. Smith, Neil J. Ross, Peter Selinger, and Benoît Valiron. ``Quipper: concrete resource estimation
in quantum algorithms''. In: \emph{Informal Proceedings of QAPL'14, Grenoble, France}. 2014. \textsc{arXiv}:
\href {https://www.arxiv.org/abs/1412.0625} {\nolinkurl {1412.0625}}.}
\newcommand{\myciteselingerquantum}{
Peter Selinger and Benoît Valiron. ``Quantum lambda-calculus''. In: \cite{gay2009semantic}. Chap. 4, pp. 135--172.}
\newcommand{\myciteschererconcrete}{
Artur Scherer, Benoît Valiron, Siun-Chuon Mau, D. Scott Alexander, Eric van den Berg, and
Thomas E. Chapuran. ``Concrete resource analysis of the quantum linear-system algorithm used to compute
the electromagnetic scattering cross section of a 2D target''. In: \emph{Quantum Information Processing}
16.3 (2017), p. 60. \textsc{doi}: \href {https://doi.org/10.1007/s11128-016-1495-5} {\nolinkurl
{10.1007/s11128-016-1495-5}}. \textsc{hal}: \href {https://hal.archives-ouvertes.fr/hal-01474610} {\nolinkurl
{hal-01474610}}. \textsc{arXiv}: \href {https://www.arxiv.org/abs/1505.06552} {\nolinkurl {1505.06552}}.}
\newcommand{\mycitesabrysymmetric}{
Amr Sabry, Benoît Valiron, and Juliana Kaizer Vizzotto. ``From symmetric pattern-matching to quantum
control''. In: \emph{Proceedings of the 21st International Conference on Foundations of Software Science
and Computation Structures, FoSSaCS 2018} (Thessaloniki, Greece). Ed. by Christel Baier and Ugo Dal Lago.
Vol. 10803. Lecture Notes in Computer Science. Springer, 2018, pp. 348--364. \textsc{doi}: \href
{https://doi.org/10.1007/978-3-319-89366-2_19} {\nolinkurl {10.1007/978-3-319-89366-2_19}}. \textsc{hal}: \href
{https://hal.archives-ouvertes.fr/hal-01763568} {\nolinkurl {hal-01763568}}. \textsc{arXiv}: \href
{https://www.arxiv.org/abs/1804.00952} {\nolinkurl {1804.00952}}.
}
\newcommand{\mycitevalironorthogonality}{
Benoît Valiron. ``Orthogonality and algebraic lambda-calculus''. In: \emph{Proceedings of the 7th
International QPL Workshop Quantum Physics and Logic, QPL'10} (Oxford, UK). Ed. by Bob Coecke,
Prakash Panangaden, and Peter Selinger. 2010, pp. 169--175. \textsc{url}: \url
{http://www.cs.ox.ac.uk/people/bob.coecke/QPL_proceedings.html}.}
\newcommand{\mycitevalironsemantics}{
Benoît Valiron. ``Semantics of a typed algebraic lambda-calculus''. In: \emph{Proceedings of the Sixth
Workshop on Developments in Computational Models: Causality, Computation, and Physics, DCM 2010}
(Edinburgh, Scotland, July 9--10, 2010). Ed. by S. Barry Cooper, Prakash Panangaden, and Elham Kashefi.
Vol. 26. Electronic Proceedings in Theoretical Computer Science. Preliminary work to the journal
paper \cite{valiron2013typed}. 2010, pp. 147--158. \textsc{doi}: \href {https://doi.org/10.4204/EPTCS.26.14} {\nolinkurl
{10.4204/EPTCS.26.14}}.}
\newcommand{\mycitevalironquantumtuto}{
Benoît Valiron. ``Quantum computation: a tutorial''. In: \emph{New Generation Computing} 30.4 (2012),
pp. 271--296. \textsc{doi}: \href {https://doi.org/10.1007/s00354-012-0401-7} {\nolinkurl
{10.1007/s00354-012-0401-7}}.}
\newcommand{\mycitevalirontyped}{
Benoît Valiron. ``A typed, algebraic, computational lambda-calculus''. In: \emph{Mathematical Structures in
Computer Science} 23.2 (2013). Journal, extended version of \cite{valiron2010semantics}., pp. 504--554. \textsc{doi}: \href
{https://doi.org/10.1017/S0960129512000205} {\nolinkurl {10.1017/S0960129512000205}}.}
\newcommand{\mycitevalironquantumtutobis}{
Benoît Valiron. ``Quantum computation: from a programmer's perspective''. In: \emph{New Generation
Computing} 31.1 (2013), pp. 1--26. \textsc{doi}: \href {https://doi.org/10.1007/s00354-012-0120-0} {\nolinkurl
{10.1007/s00354-012-0120-0}}.}
\newcommand{\mycitevalirongenerating}{
Benoît Valiron. ``Generating reversible circuits from higher-order functional programs''. In:
\emph{Proceedings of the 8th International Conference on Reversible Computation, RC'16} (Bologna, Italy).
Ed. by Simon J. Devitt and Ivan Lanese. Vol. 9720. Lecture Notes in Computer Science. Springer, 2016,
pp. 289--306. \textsc{doi}: \href {https://doi.org/10.1007/978-3-319-40578-0_21} {\nolinkurl
{10.1007/978-3-319-40578-0_21}}. \textsc{hal}: \href {https://hal.archives-ouvertes.fr/hal-01474621} {\nolinkurl
{hal-01474621}}.}
\newcommand{\mycitevalironprogrammer}{
Benoît Valiron. \emph{Programmer un ordinateur quantique}. Column in MathsInfos Hors-Série Numéro 3,
published by Fondation Mathématique de Paris. 2017. \textsc{hal}: \href
{https://hal.archives-ouvertes.fr/hal-01763585} {\nolinkurl {hal-01763585}}.}
\newcommand{\mycitevalironformal}{
Benoît Valiron. ``A formal analysis of quantum algorithms''. In: \emph{ERCIM News} 112 (Jan. 2018),
pp. 23--24. \textsc{hal}: \href {https://hal.archives-ouvertes.fr/hal-01763602} {\nolinkurl {hal-01763602}}.}
\newcommand{\mycitevalironsemanticsjlamp}{
Benoît Valiron. ``Semantics of quantum programming languages: classical control, quantum control''. In:
\emph{Journal of Logical and Algebraic Methods in Programming} 128 (2022), p. 100790. \textsc{doi}: \href
{https://doi.org/10.1016/J.JLAMP.2022.100790} {\nolinkurl {10.1016/J.JLAMP.2022.100790}}. \textsc{hal}: \href
{https://hal.archives-ouvertes.fr/hal-04038653} {\nolinkurl {hal-04038653}}.
}
\newcommand{\mycitevalironprogramming}{
Benoît Valiron, Neil J. Ross, Peter Selinger, Dana Scott Alexander, and Jonathan M. Smith. ``Programming
the quantum future''. In: \emph{Communications of the ACM} 58.8 (2015), pp. 52--61. \textsc{doi}: \href
{https://doi.org/10.1145/2699415} {\nolinkurl {10.1145/2699415}}. \textsc{url}: \url
{http://doi.acm.org/10.1145/2699415}. \textsc{hal}: \href {https://hal.archives-ouvertes.fr/hal-01194416}
{\nolinkurl {hal-01194416}}.}
\newcommand{\mycitevalironfinite}{
Benoît Valiron and Steve Zdancewic. ``Finite vector spaces as model of simply-typed lambda-calculi''. In:
\emph{Proceedings of the 11th International Colloquium on Theoretical Aspects of Computing, ICTAC 2014}
(Bucharest, Romania, Sept. 17--19, 2014). Ed. by Gabriel Ciobanu and Dominique Méry. Vol. 8687. Lecture
Notes in Computer Science. See \cite{valiron2014modeling} for the long version. Springer, 2014, pp. 442--459. \textsc{doi}: \href
{https://doi.org/10.1007/978-3-319-10882-7_26} {\nolinkurl {10.1007/978-3-319-10882-7_26}}.}
\newcommand{\mycitevalironmodeling}{
Benoît Valiron and Steve Zdancewic. ``Modeling simply-typed lambda calculi in the category of finite vector
spaces''. In: \emph{Scientific Annals of Computer Science} 24.2 (2014), pp. 325--368. \textsc{doi}: \href
{https://doi.org/10.7561/SACS.2014.2.325} {\nolinkurl {10.7561/SACS.2014.2.325}}.
}
\newcommand{\mycitexureasoning}{
Zhaowei Xu, Benoît Valiron, and Mingsheng Ying. ``Reasoning about Recursive Quantum Programs''. Draft,
to appear in ACM TOCL. 2021. \textsc{arXiv}: \href {https://www.arxiv.org/abs/2107.11679} {\nolinkurl
{2107.11679}}.}

\title{On Quantum Programming Languages\\[1ex]\large Habilitation à Diriger des Recherches}

\author{Benoît Valiron}
\date{\today}

\begin{document}

\renewcommand{\thechapter}{\Alph{chapter}}

\etocsetlevel{part}{-1}
\etocsetlevel{chapter}{0}
\etocsetlevel{section}{1}
\etocsetlevel{subsection}{2}
\etocsetlevel{subsubsection}{3}
\etocsetlevel{paragraph}{4}
\etocsetlevel{subparagraph}{5}

\setcounter{tocdepth}{2}
\setcounter{secnumdepth}{3}


\begin{titlepage}
\newgeometry{left=6cm,bottom=2cm, top=1cm, right=1cm}
\tikz[remember picture,overlay]%
\node[opacity=1,inner sep=0pt] at (-13mm,-135mm){\includegraphics{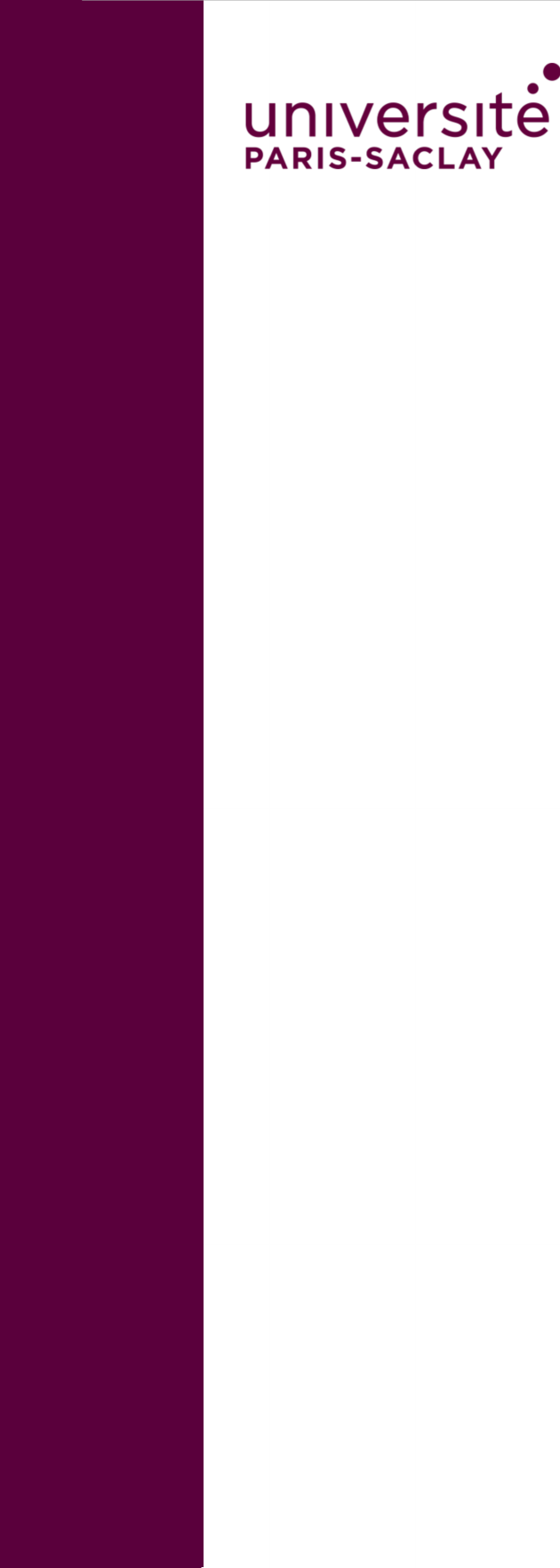}};

\fontfamily{fvs}\fontseries{m}\selectfont

\color{white}

\vspace{-10mm}
\flushright
\vspace{5cm}
\color{Prune}
\fontfamily{cmss}\fontseries{m}\fontsize{22}{26}\selectfont%
On Quantum Programming Languages
 
\vspace{2cm}
\normalsize

{\fontseries{bx}\selectfont\textcolor{black}{Habilitation à diriger des recherches \\
de l'Université Paris-Saclay}}

\vspace{2cm}
{\fontseries{bx}\selectfont\textcolor{black}{présentée et soutenue à Gif-sur-Yvette,\\le 24 Septembre 2024, par}}\\
\bigskip
\Large {\color{Prune} {\rm Benoît \textsc{Valiron}}}

\vspace{\fill}

\flushleft
\newcommand{\textbx}[1]{{\fontseries{bx}\selectfont#1}}
\scriptsize
\begin{tabular}{|p{7cm}l}
\arrayrulecolor{Prune}
{\footnotesize {\fontseries{bx}Composition du jury}}\\
& \\
\textbx{Pascale {\rm\bfseries\textsc{Le Gall}}} &   Présidente\\ 
Professeure des Universités, CentraleSupélec & \\
\textbx{Luís {\rm\bfseries\textsc{Soares Barbosa}}} &  Rapporteur \\ 
Full Professor, Universidade do Minho &   \\ 
\textbx{Chris {\rm\bfseries\textsc{Heunen}}} &  Rapporteur \\ 
Professor, University of Edinburgh &   \\ 
\textbx{Emmanuel {\rm\bfseries\textsc{Jeandel}}} &  Rapporteur \\ 
Professeur des Universités, Université de Lorraine   &   \\ 
\textbx{Cyril {\rm\bfseries\textsc{Branciard}}} &  Examinateur \\ 
Chargé de Recherche CNRS, Institut Neel de Grenoble    &   \\ 
\textbx{Delia {\rm\bfseries\textsc{Kesner}}} &  Examinatrice \\ 
Professeure des Universités, Université Paris Cité   &   \\ 
\textbx{Sophie {\rm\bfseries\textsc{Laplante}}} &  Examinatrice \\ 
Professeure des Universités, Université Paris Cité &  \\ 
\end{tabular} 

\medskip
\begin{tabular}{|p{7cm}l}
\arrayrulecolor{Prune}
{\footnotesize {Parrain}}\\
& \\
\textbx{Pablo {\rm\textsc{Arrighi}}} & \\ 
Directeur de Recherche, {\rm\bfseries\textsc{Inria}} & 
\end{tabular} 
\end{titlepage}

\clearpage{\thispagestyle{empty}\cleardoublepage}

\thispagestyle{empty}
\newgeometry{margin=0cm, top=1.5cm, bottom=1.25cm, left=2cm, right=2cm}
\fancyhfoffset[E,O]{0pt}

\noindent 
\includegraphics[width=4cm]{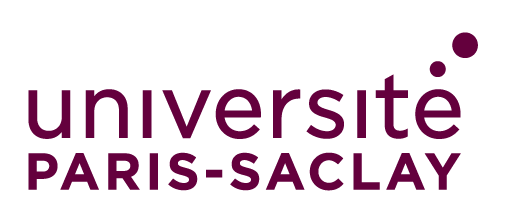}
\vspace{5mm}

\begingroup

\sffamily
\small

\begin{mdframed}[linecolor=Prune,linewidth=1]

  {\fontseries{bx}Titre\,:} De l'étude des langages de programmations quantiques
  \hfill

\vspace{-4mm}
\begin{multicols}{2}
  \begin{otherlanguage}{french}
    \noindent {\fontseries{bx}Résumé\,:}    %
    Cette thèse présente mes contributions à la recherche depuis la
    soutenance de mon doctorat en 2008. J'ai eu la chance de
    participer au développement des langages de programmation
    quantique depuis le début\,: le document a pour but de présenter
    mon point de vue sur l'évolution du sujet, mes contributions, et
    les tendances actuelles dans la communauté. Le document est
    destiné à un doctorant spécialisé dans les méthodes formelles.

    Depuis 2008, la programmation quantique a énormément évolué. Alors
    que le matériel est passé d'expériences en laboratoire à des
    coprocesseurs quantiques prêts à l'emploi, les langages quantique
    sont passés de principes mathématiques abstraits à des
    environnements de développement complet. Leur conception doit
    désormais tenir compte du matériel ainsi que des cas d'usage. En
    outre, de nouveaux paradigmes de calcul non standard émergent,
    basés sur la possibilité de considérer la superposition des
    exécutions en plus de la superposition des données. Tout cela
    soulève des défis passionnants pour l'avenir.
  
    La présentation commence par discuter l'état du domaine en 2008,
    avec une discussion sur les différentes approches telle que le
    lambda-calcul quantique---l'une de mes contributions avant
    2008. Le manuscrit se penche ensuite sur trois sujets pour
    lesquels j'ai été activement impliqué.

    La thèse se concentre d'abord sur l'avènement de langages de
    programmation quantique évolutifs. Dans ce contexte, j'ai
    participé au développement de Quipper, un langage de description
    de circuits inscrit dans Haskell, et de Qbricks, un outil de
    vérification déductive de programmes quantiques. Le deuxième thème
    principal de la thèse est la sémantique des lambda-calculs
    quantique\,: mes contributions concernent la description de
    sémantiques inspirées par des modèles de logique linéaire. La
    troisième partie de la thèse analyse la superposition d'exécutions
    de programmes\,: mes contributions sont le SWITCH quantique,
    montrant comment ce modèle ne peut pas être réduit à des circuits
    quantiques, et la conception de langages pour des programmes en
    superposition.

    La thèse se termine par quelques tendances actuelles de la
    recherche\,: l'essor des langages graphiques, l'unification des
    contrôles quantiques et classiques, le développement de chaînes
    d'outils de compilation quantique, et l'analyse statique pour les
    programmes quantiques.
  \end{otherlanguage}  
\end{multicols}
\end{mdframed}

\begin{mdframed}[linecolor=Prune,linewidth=1]

{\fontseries{bx}Title:} On Quantum Programming Languages

\vspace{-4mm}
\begin{multicols}{2}
  \begin{otherlanguage}{english}
    \noindent {\fontseries{bx}Abstract:}
    This thesis ---\textit{Habilitation à diriger des recherches}---
    presents some of my research contributions since my Ph.D defense
    in 2008. I have had the chance to participate in the development
    of quantum programming languages since their early developments:
    the presentation aims to present my point of view on the evolution
    of the subject, my contributions, and the current research trends
    in the community. The target audience is a graduate student
    interested in pointers to the field of quantum programming
    languages

    Since 2008, the state of quantum programming has evolved
    tremendously. As quantum hardware moved from physical artifacts in
    bench labs to ready-to-use quantum coprocessors, quantum
    programming languages evolved from abstract mathematical
    principles to scalable proposals. Language designs now need to
    consider constraints coming from both hardware and use
    cases. Furthermore, novel, non-standard computational paradigms
    are emerging based on the possibility of considering the
    superposition of executions on top of the superposition of
    data. All of this raises exciting challenges for the years to
    come.
  
    The presentation starts with a discussion of the state of the
    field of quantum programming language in 2008, with a discussion
    on the attempts at toy languages, and in particular, the quantum
    lambda calculus---one of my contributions before 2008. The
    manuscript then dives into three main subjects I have been
    actively involved in.

    The thesis first focuses on the advent of scalable quantum
    programming languages. In this context, I have participated in
    developing Quipper, a domain-specific, circuit-description
    language embedded in Haskell, and Qbricks, a tool for deductive
    verification of quantum programs. The second main topic of the
    thesis consists of the semantics of quantum lambda-calculi: My
    contributions concern the description of semantics inspired by
    models of linear logic. The third part of the thesis analyses the
    superposition of program executions: my contributions are the
    quantum SWITCH, showing how this model cannot be reduced to
    quantum circuits, and the design of languages for programs in
    superposition.

    The thesis concludes with a few current research trends: the rise
    of graphical languages, the reconciliation of quantum and
    classical control, the development of quantum compilation
    toolchains, and static analysis for quantum programs.
  \end{otherlanguage}  
\end{multicols}
\end{mdframed}

\endgroup
\restoregeometry
\clearpage{\thispagestyle{empty}\cleardoublepage}

\lhead{Remerciements}
\chapter*{Remerciements}

\begin{otherlanguage}{french}
Je tiens tout d'abord à remercier chaleureusement Luis Soares Barbosa,
Chris Heunen et Emmanuel Jeandel d'avoir accepté de rapporter ce
manuscrit, ainsi que Cyril Branciard, Delia Kesner, Sophie Laplante et
Pascale Le Gall pour leur présence sur le jury. Je remercie finalement
Pablo Arrighi pour son role de parrain qui fut décisif dans la
finalisation de ce manuscrit.

Cette habilitation est une histoire qui a commencée en 2018, quand
Frédéric Boulanger a accepté de donner sa casquette HDR pour
l'encadrement de
이동호
(Dongho Lee). Il était entendu que je
passerais mon HDR d'ici à la soutenance. Cela n'a pas vraiment été le
cas, et j'ai eu recours de nombreuses fois à Frédéric, mais aussi Marc
Baboulin, Gilles Dowek et Pablo Arrighi pour servir de co-encadrants.
Je tiens à les remercier ici pour leur aide.

Grâce à leur soutien, j'ai eu l'honneur d'accompagner jusqu'à leur
soutenance Timothée Goubault de Brugière,
이동호
(Dongho Lee), Kostia
Chardonnet, Agustin Borgna et Louis Lemonnier. À l'heure où j'écris
ces lignes, Jérome Ricciardi, Nicolas Heurtel et Julien Lamiroy sont
toujours en thèse. J'espère pouvoir assister à leur soutenance en
ayant finalement terminer cette habilitation qui est devenue au fil du
temps l'arlésienne de l'équipe. Je remercie tous mes étudiants, qu'ils
aient fini ou non: l'interaction avec eux forment le cœur de mon
attachement au travail de recherche.

Je remercie aussi les collègues de l'équipe QuaCS qui a vu le jour
pendant la (longue)
gestation de cette habilitation. L'arlésienne qu'est devenue cette
habilitation a généré un fond de blagues qui ont maintenu une pression
salutaire.

En dernier lieu, je remercie ma famille
pour leur compréhension et leur soutien, pendant tous
ces soirs et week-ends
durant lesquels j'ai été particulièrement peu disponible.


\end{otherlanguage}
\clearpage{\thispagestyle{empty}\cleardoublepage}

\lhead{}
\chapter*{}
\thispagestyle{empty}

\begin{otherlanguage}{french}
\hfill\itshape Pour Caroline, sans qui la vie serait moins belle.
\end{otherlanguage}
\clearpage{\thispagestyle{empty}\cleardoublepage}

\pagestyle{fancy}

\lhead{Table of Contents}

\tableofcontents

\clearpage{\thispagestyle{empty}\cleardoublepage}

\chapter{Introduction}

\lhead{Chapter \thechapter}
\rhead{Introduction}

The birthdate of quantum computing can be traced back to 1982 when
Feynman \cite{feynman1982simulating} envisioned using controllable
quantum systems to simulate physical behavior. Since then, quantum
computing has grown into a thriving research field, with foreseen
applications in a wide range of topics: quantum simulation for
pharmaceuticals and chemicals applications, quantum linear algebra for
AI and machine learning, quantum optimization and search, quantum
factorization, \etc. As of December 2022, more than 430 algorithms
were registered on the Quantum Algorithm Zoo \cite{quantum-zoo}, a
comprehensive catalog of quantum algorithms.

In parallel to this boom in algorithm design, the investment in
quantum computing is reaching an all-time high, both from public and
private actors. France set up the ``\emph{Plan Quantique}'' last year,
while Europe launched the ``Quantum Flagship'' a few years ago. In the
previous five years, both the number of startups and the investment in
quantum computation have skyrocketed to more than 200 startups
worldwide, and 25 billion dollars of public investment as of 2021
\cite{mckinsey2021}.

The idea behind quantum computing is to code information on the state
of objects governed by the laws of quantum physics. The mathematical
theory is well established \cite{nielsen02quantum} and makes it
possible to reason on what is doable---and what is not --- in this
computation paradigm without having to rely on concrete hardware. The
state of a quantum memory can be regarded as a complex, linear
combination of bit-strings. This mathematical formalism entails two of
the main features of quantum computation: superposition of data and
entanglement. On the downside, reading information from a quantum
memory is a destructive and probabilistic operation, modifying the
global state of the memory.

One of the first and maybe most notorious quantum algorithms, Shor's
factoring algorithm \cite{shor94algorithms}, placed quantum
computation on the radar for potentially disruptive
technologies. However, in part due to the lack of scalable hardware at
the time, later quantum algorithms were typically developed as tools
to explore the complexity speedup entailed by using a quantum
memory. If this purely theoretical approach to quantum computation
uncovered interesting foundational results, it remained somehow far
from concrete instantiations.

Quantum algorithms such as the one devised by Shor require a level of
abstraction higher than what can be needed by Feynman's vision of
quantum simulation. Knill \cite{knill96conventions} proposed in 1996 a
rationale for what a quantum coprocessor should permit to implement
such algorithms. This analysis is the basis of a wide range of works
on the \emph{computer science} aspect of quantum computation. On one
end of the spectrum, a series of research developments discuss
concrete quantum programming languages or libraries for interacting
with a quantum memory \cite{bettelli03architecture, omer2000quantum,
  omer2003structured}. At the other end of the spectrum, Knill's
description seeded lines of research on the \emph{semantics} of
quantum programming language based on models of linear logic, domain
theory and category theory \cite{selinger04quantum,
  selinger04semantics, tonder04lambda, valiron2004msc,
  valiron2008phd}, the latter cross-fertilizing with early
formalizations of quantum information and quantum protocols
\cite{coecke04informationflow, abramsky04categorical}.

The 2010s have seen a rapid upscaling of the proposals pertaining to
quantum programming. Stirred by the active development of hardware and
integrated quantum coprocessors, the field has moved from
blackboards and lab benches to
industrial use cases. Although the hardware is still in the so-called
\emphidx{Noisy Intermediate-Scale Quantum} (\emphidx{NISQ}) state with
quantum memory of a few hundred uncorrected qubits subject to
decoherence \cite{preskill2018nisq},
the research in quantum programming and compilation is thriving in
proposals for making use of these memories \cite{bharti2022noisy}.
Indeed, the memories are
large enough to make it impractible to manipulate by hand, yet
constrained enough to require dedicated languages and sophisticated
compilation and optimization techniques.
The field is furthermore
opening the path toward \emphidx{Large Scale Quantum} (\emphidx{LSQ})
computation, for when the hardware will be able to support error
correction.

The arrival of concrete machines with more than a handful of qubits
has created a strong pull effect for the development of quantum
programming languages \cite{list-tools}.
Indeed, if hand-writing circuits is feasable
for a few qubits, it quickly becomes cumbersome and error-prone even
for a few dozen of qubits. Turning a pen-and-paper algorithm to an
effective, runnable sequence of gates on a quantum coprocessor with a
hundred qubits
requires a programming language to describe and manipulate the structure of
the circuit. Whether this language is standalone or embedded, it has to
be formalized enough to support analysis techniques and tools 
to assess the validity of quantum programs.
Moreover, whether in the NISQ or the LSQ era, quantum coprocessors
are and will arguably be both expensive to run and limited in
resources. Optimizing the resource footprint of an algorithm is then
critical, opening the field to the development of synthesis and
optimization techniques for quantum programs.

\begin{mylife}
  I entered the game in 2002 when I started a Master's program at the
  University of Ottawa under the supervision of Peter Selinger. I had
  the opportunity to design a quantum lambda calculus and work on its
  semantics, first as a Master's student and then as a PhD student,
  still under Peter Selinger's supervision. I had, therefore, the
  chance to be on the frontline of the research on quantum programming
  languages.
  
  Since my Ph.D. defense in 2008, the field has evolved and
  significantly matured, and I was lucky to be part of the
  process. The rest of the manuscript summarizes this evolution, seen
  from my own view, and focuses on my contributions until now.
  Following the rules of the \emph{Habilitation à diriger des
    Recherche (HDR)}, I take the year 2008---time of my
  Ph.D. defense--- as a pivot and focus on what happened next. The
  notion of ``present time'' being a moving target, I chose 2020 (give
  or take) as the end of the past and the beginning of the
  future---this is the time where I started to write this thesis.
  
  I shall use blue text boxes like the one encapsulating this
  paragraph to reflect on my experience with the topics and subjects
  discussed.
\end{mylife}

\paragraph{Scope and plan of the Manuscript.}
This thesis is concerned with quantum programming languages from a
formal perspective. The scope is deliberately skewed towards the
research interests---and the work---of the author. The document
focuses on three emblematic research threads
(Chapters~\ref{ch:compil}, \ref{ch:sem} and~\ref{ch:qcont}) in which
the author participated between 2008 and 2020.  For each chapter, the
related publications are summarized at the end in a separated table:
Tables~\ref{tab:publis-sem}, \ref{tab:publis-compil}
and~\ref{tab:publis-qcont}. The later publications are in
Table~\ref{tab:publis-opening}.

Chapter~\ref{ch:backthen} briefly presents some background material
and reviews the state of quantum programming languages as it was in
2008. The discussion covers the preliminary design proposals for
quantum languages with a focus on the quantum lambda calculus, one of
our contributions at the time.

Chapter~\ref{ch:compil} discusses the design of quantum programming
languages within the coprocessor model and the shift from toy languages
to scalable programming environments. We have, in particular, been involved in the
development of \quipper, a domain-specific, circuit-description
language embedded in Haskell. The language comes with sound design
principles that are still state of the art nowadays. We then discuss
two related aspects: circuit synthesis and quantum program
certification. Our contributions to circuit synthesis are
twofold: a strategy for generating oracles by turning a classical
piece of code into a family of reversible circuits and
approaches using numerical analysis to automate circuit synthesis
based on a matrix-like description. Regarding program verification, we
have been involved in deductive verification, particularly with developing
the tool Qbricks.

Chapter~\ref{ch:sem} explores the semantics of quantum lambda calculus
and its extension with circuit-description capabilities, bridging with
Chapter~\ref{ch:compil}. A semantics is a formal description of some
of the properties of a programming language: its structures, its
behavior, its action, etc. A semantics usually shares a strong
connection with a model of some logic through a Curry-Howard
correspondence. In the case of quantum computation, one of our
contributions before 2008 has been to connect quantum lambda calculus
with linear logic. This resource-sensitive logic forms a natural
framework for reasoning on the non-duplicability of quantum
registers. The chapter presents such Curry-Howard correspondence,
linking linear logic and the quantum lambda calculus. Three aspects of
the problem are then discussed: first, the quest for a denotational
semantics for the corresponding typed quantum lambda calculus; then
the link between the typed quantum lambda calculus and an
interaction-based model of linear logic: Geometry of Interaction;
finally, the description of denotational semantics for quantum
lambda calculi with circuit-description features.

Chapter~\ref{ch:qcont} examines an effect specific to quantum
computation: quantum control. This non-standard model of
computation generalizes the notion of superposition: On top of
the superposition of data, we consider the possibility of
superposition of executions. All of this opens several challenges, such as the
expressivity of this new computational paradigm and the design of
syntactic languages to describe superpositions of programs. Concerning
the expressivity of the model, our seminal contribution is the quantum
{\qswitch}: a minimal algorithm featuring a superposition of execution that
quantum circuits cannot represent. On the syntactic aspect, we
explored several approaches for functional languages with
superposition of programs, using not only lambda calculus but also
specific syntax based on pattern-matching.

Finally, Chapter~\ref{ch:opening} describes a few research trends in
the community corresponding to the interest of the author: the rise of
graphical languages, the problem of the unification of quantum and
classical control, the definition of quantum compilation toolchains,
and the challenge of quantum program certification.

\paragraph{Blind Spot.}
Many important aspects pertaining to the field of quantum languages
have been left undiscussed in this document. In particular, we shall
not consider the problem of error correction, the development of
quantum algorithms, or (most) practical aspects of quantum
compilation. Graphical languages and ZX in particular
will not be discussed to the extent
they deserve---again, the scope of the document is only the author's
existing research production. It should also be understood that we
shall not discuss industrial-scale quantum programming languages; the
presentation stays at a formal, theoretical level.

\paragraph{Audience.}
The target audience for this document is a graduate student or a
researcher in formal methods interested in acquiring pointers to the
field of quantum programming languages. More precisely, the main
techniques used in the various works described in the document are
rewriting, logical and type systems, and category theory. The reader
should not expect complete proofs of results; instead, the
presentation tries to stay high-level, and the main results and
constructions are, in general, given through the use of examples.


\clearpage{\thispagestyle{empty}\cleardoublepage}

\chapter[Quantum Programming back in 2008]{%
  Quantum Programming\\*back in 2008%
}
\label{ch:backthen}
\rhead{Back in 2008}

Since 2008, the state of the art in quantum programming languages has
evolved a lot, particularly with new actors shaping
the field in inconceivable ways back then. This chapter is devoted to
summarizing the state of the field back then.
\begin{itemize}
\item Section~\ref{sec:primer-qc} offers a quick introduction to
  quantum computation in general. On top of the mathematical
  foundations, the section describes the standard computational
  paradigm for quantum computation: the quantum coprocessor
  model. The section also discusses an emerging language at the time
  with a graphical notation: the ZX calculus. This section sets the
  playground for the rest of the thesis.
\item Section~\ref{sec:q-prog-models} presents the state of quantum
  programming languages around 2008. The section discusses a few
  emblematic approaches and design strategies. Indeed, some of them
  are at the root of the subsequent developments. For instance, in
  hindsight, the notion of circuit-description language was already
  there in the quantum IO monad of Altenkirch\&Green's
  \cite{green2009qio}, most of the challenges about quantum control
  were established in for QML \cite{grattage05functional}, and Bettelli
  and Ömer had already started pondering the interaction with the
  quantum coprocessor
  \cite{bettelli03architecture,omer2003structured}.
\item Section~\ref{sec:qlc} introduces one of our main contributions at
  the time: the quantum lambda calculus
  \cite{valiron2004msc,valiron2008phd,selinger2006lambda}.
  This extension of the lambda
  calculus provides a theoretical framework for reasoning on
  functional programming languages accommodating quantum
  computation. The section discusses the necessity for a linear type
  system due to the non-duplicability of quantum information and
  sketches the existing denotational semantics at the time. As the
  section shows, existing semantics for the quantum lambda calculus
  were still very modest. In particular, they could not capture both
  duplicable and non-duplicable data at the same time
  \cite{selinger06fully,valiron2008categorical}.
\end{itemize}

\section{Primer on Quantum Computation}
\label{sec:primer-qc}

In this section, we lay out the background on quantum computation
needed for the rest of the thesis. We invite the reader to consult
e.g.~\cite{nielsen02quantum} for more details.

\subsection{Quantum Memory}

In the standard model of quantum computation, one has access to a
quantum coprocessor\index{coprocessor} holding a special kind of
\emph{quantum} memory\index{quantum memory}. It consists of data
encoded on the state of quantum particles: photons, ions,
\emph{etc}. The behavior of the quantum memory is derived from the
rules of quantum mechanics. From a mathematical standpoint, the
\emph{state}\index{state}\index{quantum state} of the quantum memory
is a normalized vector in a Hilbert space~\cite{nielsen02quantum},
usually considered modulo a global phase\index{global phase}
---i.e. multiplication by a (non-zero) scalar.  The smallest piece of
quantum information is the \emph{quantum bit}\index{quantum bit}, or
\emph{qubit}\index{qubit}: its state is represented by a normalized vector in the
Hilbert space $\qH$ of dimension 2. One chooses a basis
$\{\ket{0},\ket{1}\}$ of two orthonormal vectors: a canonical
representation for the state $\ket\phi$ of a qubit
is therefore of the general form
\[
  \ket\phi = \rho_0\ket0 + \rho_1e^{i\theta}\ket1,
\]
where $\rho_0$ and $\rho_1$ are positive reals such that
$\rho_0^2+\rho_1^2=1$ and $\theta$ is an angle between $0$ and $2\pi$.
The value $\theta$ is a \emph{phase}\index{phase}, whereas $\rho_0$
and $\rho_1$ are called \emph{amplitudes}\index{amplitude}.
In general, we however simply consider a representative
$\alpha\ket0+\beta\ket1$ with $\alpha,\beta\in\Cx$ and
$|\alpha|^2+|\beta|^2=1$, keeping in mind that the \emph{global
  phase}\index{global phase} $\frac{\alpha}{|\alpha|^2}$ is not
relevant.

The notation $\ket\phi$ is called a \emph{ket}\index{ket}: if we
choose the lexicographic ordering on the basis $\ket0,\ket1$, then
$\ket\phi$ stands for the column vector
\[
  \ket\phi = \alpha\ket0+\beta\ket1 =
  \left(\begin{matrix}\alpha\\\beta\end{matrix}\right).
\]
There is a dual notation for row-vectors ---or functionals---, the
\emph{bra}\index{bra}. The conjugate transpose of $\ket\phi$ is
therefore
\[
  \bra\phi = \left(\bar\alpha~\bar\beta\right).
\]
The notation is consistent with the \emph{scalar product}, as follows:
\[
  \braket{\phi\mid\phi} = (\bra\phi)(\ket\phi) =
  \left(\bar\alpha~\bar\beta\right)
  \left(\begin{matrix}\alpha\\\beta\end{matrix}\right)
  =
  |\alpha|^2 + |\beta|^2 = 1.
\]
The basis $\ket0,\ket1$ therefore forms an \emphidx{orthonormal
  basis}. Another standard orthonormal basis is $\ket+,\ket-$, where
$\ket+=\frac1{\sqrt2}(\ket0+\ket1)$ and
$\ket-=\frac1{\sqrt2}(\ket0-\ket1)$.

\paragraph{Kronecker product}
When considering several quantum registers simultaneously, the state
of the overall system lies within the \emphidx{Kronecker product}, or
\emphidx{tensor product} of the individual state spaces. If Kronecker
products can be defined through a universal
property~\cite[Ch.\,XVI]{lang2002algebra}, a
simple presentation can be done with spaces equipped with bases.
Consider the two Hilbert
spaces $\hilb{E}$ and $\hilb{F}$ of respective bases
$B_{\hilb{E}}=\{e_i\}_i$ and $B_{\hilb{F}}\{f_j\}_j$. The space
$\hilb{E}\tensor\hilb{F}$ is defined as the Hilbert space with
(formal) basis
$B_{\hilb{E}}\times B_{\hilb{F}} = \{(e_i,f_j)\}_{i,j}$.  We write the
pair $(e_i,f_j)$ as $e_i\tensor f_j$, and we bi-linearly extend the
notation $-\tensor-$ to linear combinations as follows:
\[
  \left(\sum_i\alpha_i\cdot e_i\right)
  \tensor
  \left(\sum_j\beta_j\cdot f_j\right)
  =
  \sum_{i,j}(\alpha_i\beta_j)\cdot(e_i\tensor f_j).
\]
The ket- and bra-notations make it possible to shorten the tensor
notation: we write $\ket{00}$ for $\ket{0}\tensor\ket{0}$. The
canonical basis for a 2-qubit system is then in lexicographic ordering
$\ket{00}$, $\ket{01}$, $\ket{10}$, $\ket{11}$. Unless stated otherwise, the
convention is to always use the lexicographic ordering for basis.

Consider two registers $A$ and $B$ of respective states
$\ket{\phi}_A\in\hilb{H}_A$ and $\ket{\psi}_B\in\hilb{H}_B$. The state
of the joint system $AB$ is then
$\ket{\phi}_A\tensor\ket{\psi}_B\in\hilb{H}_{AB} =
\hilb{H}_A\tensor\hilb{H}_B$. Such a state is called
\emphidx{separable}. Every state is however not necessarily separable:
suppose for instance that $A$ and $B$ are both single qubits. A
valid state for the 2-qubit system $AB$ is
\[
\frac1{\sqrt2}(\ket{00}+\ket{11})
\]
which cannot be factorized as $\ket\phi\tensor\ket\psi$: such a state
is called \emphidx{entangled}.

\subsection{Quantum Operations}
\label{sec:q-op}

The operations one can perform on a quantum memory are of two kinds:
\emph{unitary operations} and
\emph{measurements}\index{measurement}. The former correspond to
actions internal to the quantum memory, without feedback, while the
latter models classical information retrieval from the quantum memory.

\paragraph{Unitary Operations.}
A unitary operation corresponds to a unitary endomorphism on the space
of states of the quantum memory ---as in linear algebra---. In
particular, such an operation is linear, invertible, and sends
orthonormal bases to orthonormal bases.

In general, a quantum coprocessor only supports a small set of unitary
operations, called \emph{unitary gates}, or \emph{quantum
  gates}\index{gates}. They usually only act on one or two qubits at a
time. A standard list of such gates acting on one qubit can be found in
Table~\ref{tab:1q-gates}. Standard gates acting on two qubits are the
control-NOT and the SWAP gate 
\[
 \CNOT = 
 \left(\begin{matrix}1&0&0&0\\0&1&0&0\\0&0&0&1\\0&0&1&0\end{matrix}\right),
 \qquad
 \SWAP = 
 \left(\begin{matrix}1&0&0&0\\0&0&1&0\\0&1&0&0\\0&0&0&1\end{matrix}\right),
\]
$\CNOT$ sends $\ket{0x}$ to $\ket{0x}$ and $\ket1\otimes\ket{x}$ to
$\ket1\tensor\ket{\neg x}$: it leaves invariant the subspace
$\ket0\tensor\qH$ and acts as $X$ on the second qubit in the subspace
$\ket1\tensor\qH$. The SWAP-gate is sending $\ket{xy}$ to $\ket{yx}$:
Note how we use the lexicographic ordering on bases to represent the
matrices.

\begin{table}
  \[
    \begin{array}{r@{~\triangleq~}l@{\qquad\qquad}r@{~\triangleq~}l}
      H
      & \frac1{\sqrt2}\left(\begin{matrix}1&1\\1&-1\end{matrix}\right)
      & R_Y(\theta)
      & \left(\begin{matrix}\cos(\frac\theta2)&-\sin(\frac\theta2)\\\sin(\frac\theta2)&\cos(\frac\theta2)\end{matrix}\right)
      \\
      X, \textit{NOT}
      &  \left(\begin{matrix}0&1\\1&0\end{matrix}\right)
      & R_X(\theta)
      & \left(\begin{matrix}\cos(\frac\theta2)&-i\sin(\frac\theta2)\\-i\sin(\frac\theta2)&\cos(\frac\theta2)\end{matrix}\right)
      \\
      Z
      & \left(\begin{matrix}1&0\\0&-1\end{matrix}\right)
      & R_Z(\theta)
      & \left(\begin{matrix}e^{-i\frac\theta2}&0\\0&e^{i\frac\theta2}\end{matrix}\right)
      \\
      S
      & \left(\begin{matrix}1&0\\0&i\end{matrix}\right)
      &
      T
      & \left(\begin{matrix}1&0\\0&e^{\frac{i\pi}4}\end{matrix}\right)
    \end{array}
  \]
  \caption{Examples of 1-qubit Quantum Gates}
  \label{tab:1q-gates}
\end{table}

In general, given a unitary $U$ acting on $n$ qubits, the
\emphidx{control} of $U$ is the gate $C-U$ acting on
$\qH\otimes\qH^{\otimes n}$ and defined as
$\ket0\otimes\ket\phi\mapsto \ket0\otimes\ket\phi$ and
$\ket1\otimes\ket\phi\mapsto\ket1\otimes(U\ket\phi)$. Using the
lexicographic representation of basis states, the gate $C-U$ can be
represented as the block-matrix
\[
 C-U = 
 \left(\begin{matrix}1&0\\0&U\end{matrix}\right).
\]
As an example, the gate $\CNOT$ is $C-X$. We can also define the
\emphidx{Toffoli gate} that acts on 3 qubits and which is
defined as $C-CNOT$. It sends $\ket{xy}\tensor\ket{z}$ to
$\ket{xy}\tensor\ket{z\oplus(x\wedge y)}$.

A quantum gate-set
is suitable for general quantum computation if
it is \emphidx{universal}, i.e. if any unitary map acting on $n$
qubits can be realized with composition and tensors of elementary
gates. Depending on the gate-set, this realization can be exact,
or approximate up to an arbitrary small error.

\paragraph{Measurement.}
A measurement corresponds to the (classical) observation of a quantum
system whose state lives in the Hilbert space $\hilb{H}$ to retrieve a
classical piece of information. Operationally, it consists in choosing
two orthogonal subspaces $\hilb{H}_0$ and $\hilb{H}_1$ spanning
$\hilb{H}$ and determining whether the state of the system belongs to
$\hilb{H}_0$ or $\hilb{H}_1$. In this case, any vector
$\ket{\phi}\in\hilb{H}$ can be decomposed in
$\ket{\phi} = \alpha\ket{\phi_0}+\beta\ket{\phi_1}$, with
$\ket{\phi_0}\in\hilb{H}_0$ and $\ket{\phi_1}\in\hilb{H}_1$, and
$|\alpha|^2+|\beta|^2 = 1$.  A measurement against the decomposition
$\hilb{H}=\hilb{H}_0\oplus\hilb{H}_1$ will project $\ket\phi$ onto one
of the two subspaces with some probability: $\ket\phi$ is changed to
$\ket\phi_0$ or $\ket\phi_1$ (modulo renormalization) with probability respectively
$|\alpha|^2$ or $|\beta|^2$. The \emph{classical result} of the
measurement is the subspace to which the state now belongs.

For instance, measuring a qubit $\alpha\ket0+\beta\ket1$ along the
basis $\{\ket0,\ket1\}$ ---that is, the decomposition
$\qH=(\Cx\ket0)\oplus(\Cx\ket1)$--- turns the qubit into $\ket0$ with
probability $|\alpha|^2$, in which case we get the classical result
``$0$'', or into $\ket1$ with probability $|\beta|^2$, in which case we
get the classical result~``$1$''. By convention, the result ``$1$''
stands for ``True'' and ``$0$'' for ``False''.

\begin{figure}[tb]
  \centering
  \includegraphics[page=1]{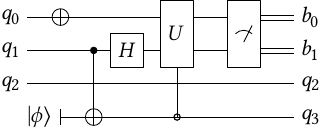}
  \caption{Example of quantum circuit}
  \label{fig:ex-q-circ}
\end{figure}

\paragraph{Quantum Circuits.}
Unitary gates are used to realize a global unitary operation on the
memory state-space. Such operations are historically
represented using the ad-hoc, graphical notation of \emphidx{quantum circuit}
\cite{yao1993quantum}. Quantum circuits form the quantum counterpart of classical, Boolean
circuits. Due to the peculiar nature of quantum data, they
are however much simpler than Boolean circuits: there is no branching nor
possible loop-back. Simple horizontal lines read from left to right
represents the life-span of a qubit or a quantum register, and boxes
on wires represents operations on them. An example is shown in
Figure~\ref{fig:ex-q-circ}. Wires can be labeled. In the example circuit,
$q_0, q_1$ and $q_2$ are input wires of the circuit, while $q_3$ is
initialized with $\ket{\phi}$. The NOT-gate is represented with a
$\oplus$-symbol, and generic boxes are rectangles: $H$ acts on $q_1$,
while $U$ acts on $q_0$ and $q_1$. $U$ is negatively controlled by
$q_3$, while the NOT-gate on $q_3$ is positively controlled by
$q_1$. Circuits can also contains measurements, shown in
Figure~\ref{fig:ex-q-circ} as the last box on the right. Boolean
results are represented with double-wires and can be labeled for easy
referring.

Quantum circuits can be extended with more constructs: measurements,
wire initialization (such as shown in Figure~\ref{fig:ex-q-circ}),
wire termination, \etc. In a quantum circuit wires that are
initialized and then terminated inside the circuit correspond to
temporary registers. They are called
\emph{auxiliary}\index{auxiliary} wires,
or \emph{ancillas}\index{ancilla}. Ancillas are more subtle to use than
conventional, local variables: terminating an ancillas amounts to
measure the corresponding qubit. If needed, special care must
therefore be taken to retain unitarity.

\paragraph{Hardware}
The mathematical model is an ideal representation of the memory setup
at the hardware level. Indeed, physical qubits are noisy, as they are
subject to decoherence \cite{schlosshauer2008decoherence}.
The hardware also entails
topological constraints ---it might not be possible to act on two
physically distant qubits--- or limitations on the gate set. From a
programming perspective, these problems are to be addressed in the
context of a quantum compilation toolchain.

\subsection{Mixed States}
\label{sec:mixed-states}

An arbitrary sequence of operations sent to a quantum memory
interleaves unitaries and measurement. In general, the resulting state
of the quantum memory at the end of the computation is therefore not a
single quantum state $\ket{\phi}$ ---a \emphidx{pure state}--- but a
\emphidx{mixed state} ---whose naive representation would be a
probability distribution of pure states. Does this mean that one can
use the set of such probability distributions as a valid model for
mixed states? The answer is not so clear: it depends on what
\emphidx{observations} are allowed.

\paragraph{Superoperators.}
If a quantum computation is understood as a quantum experiment, the
only available operations are unitaries and measurements, and the only
possible classical outcome of an observation is the (classical) result
of a measurement (did we measure 0 or 1?). In this configuration, as
physicists already noticed \cite{nielsen02quantum} probability
distributions do not make a sound model for mixed states. Instead of
considering $\sum_i p_i\{\ket{\phi_i}\}$, a semantics matching the
observational equivalence given by unitaries and measurements is the
\emphidx{positive matrix}
\[
  \rho = \sum_i p_i\ket{\phi_i}\bra{\phi_i}
\]
A positive matrix of trace 1 (such as this one) is also called a
\emphidx{density matrix}.
Positive matrices form a semantics for mixed states supporting both
unitary operations and measurements. In this framework, a general
quantum computation inputting $n$ qubits and outputting $m$ quantum bits
is represented by a trace-preserving, \emphidx{completely positive
  map} ---also known as \emphidx{superoperator}---
\begin{equation}\label{eq:cpm}
  F : \Cx^{2^n\times2^n}\to\Cx^{2^m\times2^m}.
\end{equation}
A completely positive map (CPM)\index{CPM} is a linear map such that for all
$k\in\mathbb{N}$, $F\otimes\text{id}_{\Cx^{k\times k}}$ sends positive
matrices to positive matrices.


\paragraph{Löwner order.}
Positive matrices (and by extension completely positive maps) feature
an ordering relation, the \emphidx{Löwner order}
\cite{lowner1934order,loewner1950some,beckenbach1983inequalities}:
$A\sqsubseteq B$
whenever $B-A$ is positive. This order is stable under sum and
(non-negative) scalar multiple. It makes the cone of positive $n\times
n$ matrices a \emphidx{bounded dcpo}: any bounded, directed set of
positive matrices under the Löwner order\index{Löwner order} admits a least upper
bound. This relation is consistent with the trace: if $A\sqsubseteq B$
then $\text{tr}(A)\leq\text{tr}(B)$.

The Löwner order makes it possible to interpret possibly
non-terminating quantum programs using positive matrices of trace less
or equal to $1$ and trace \emph{non-increasing} completely positive maps
\cite{selinger04quantum}. Following the standard domain interpretation
\cite{plotkin83domains}, the 0-valued element ---bottom of the cone---
corresponds to the diverging program.

\subsection{Quantum Coprocessor Model}
\label{sec:q-models-comp-2008}

In order to move from a mathematical model based on Hilbert
spaces---or from physics ex\-periments---to a programmable model of
computation, Knill proposes in 1996 the \emphidx{QRAM
  model} \cite{knill96conventions}, with a few basic pseudo-code
constructs to express quantum algorithms.

Although other computational paradigms exist such as
Measurement-based computation
\cite{mraussendorf2003mbqc,raussendorf2001one-way} or adiabatic
computation \cite{farhi2000quantum,farhi2001quantum}, Knill's QRAM
model has so far remained the standard model used in the design of
quantum algorithms \cite{quantum-zoo}.
Indeed, from the perspective of the quantum coprocessor, the run of a
quantum algorithm can be summarized with three classes of operations:
quantum register initializations; application of elementary gates on
arbitrary qubits; measurements of arbitrary qubits.

Knill's QRAM model requires these low-level operations since they form
the core of the interactions between the classical machine and
the quantum coprocessor. 
In addition to these elementary building blocks, Knill proposes a few
high-level constructs such as subroutine inversion or controlling.

\begin{figure}[tb]
  \centering
  \includegraphics[page=2]{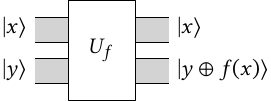}
  \caption{Structure of the typical oracle}
  \label{fig:qoracle}
\end{figure}

He also proposes to support the invocation of classical operations as
oracles.
An \emphidx{oracle} is the translation of the classical
structure of the program
into a quantum unitary. It can for instance be the structure of a
graph, an arithmetic operation, \etc. In general, provided that the
description of the problem is written as a (classical, irreversible)
map $f : \bool^n\to\bool^m$, one can always build the quantum unitary
$U_f:\qH^{\otimes n}\otimes\qH^{\otimes m}\to\qH^{\otimes
  n}\otimes\qH^{\otimes m}$ shown in Figure~\ref{fig:qoracle} and
defined as
\[
  U_f : \ket{x}\otimes\ket{y}\mapsto\ket{x}\otimes\ket{y\oplus f(y)}.
\]
Oracle generation is the topic of Section~\ref{sec:oracle-gen}.

\subsection{ZX calculus}
\label{sec:graph-calc}

Nowadays, graphical calculi for quantum computation are commonplace
\cite{coecke2017picturing}. However, if physicists were already making
good use of graphical representations with e.g. Feynman diagrams
\cite{wuthrich2011genesis}, in 2008 there were seldom graphical
languages for quantum computation.

Apart from the ad-hoc representation of quantum circuits, in the late
2000's a steady trend was however taking off. Building on category
theory and led by Abramski and Coecke
\cite{coecke04informationflow,abramsky04categorical}, the computer
science research group at Oxford became a thriving center for a
novel graphical representation based on the interaction of pairs of
mutually unbiased observables: the ZX calculus
\cite{coecke2007graphical, coecke2008interacting, zx-publications,
  coecke2017picturing}.

Falling within the large class of tensor network representation
\cite{biamonte2017tensor}, the ZX calculus can be regarded as a
graphical language for a special kind of dagger, compact closed
category with two commutative $\dagger$-Frobenius monoids
\cite{coecke2013description}:
\begin{equation}\label{eq:zx}
  \includegraphics[page=1]{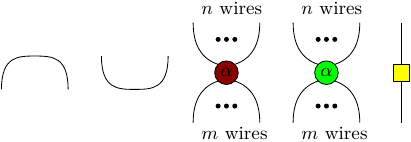}
\end{equation}
ZX-diagrams are composed from these constructs and read from top to
bottom. The \emphidx{green spider} corresponds to the basis
$\left\{\ket0,\ket1\right\}$ and the \emphidx{red spider}
to the basis $\left\{\ket+,\ket-\right\}$. The $\alpha$ is a phase.
These two algebras form a bialgebra satisfying the Hopf law
\cite{coecke2008interacting,coecke2011interacting}, so for instance
\begin{center}
  \includegraphics[page=2]{fig/zx.pdf}
\end{center}
where a node with no label corresponds to the phase $0$.

The ZX calculus aims at abstracting away the structure of finite
Hilbert spaces. A ZX term indeed admits a standard representation
as general, linear function acting on Hilbert spaces. A diagram with $n$
input and $m$ output wires corresponds to a linear function
$\qH^{\otimes n}\to\qH^{\otimes m}$. For instance, the ``cap'' in
Eq.~\eqref{eq:zx} is the map $\Cx\to\qH\otimes\qH$ sending $1$ to
$\ket{00} + \ket{11}$, while the green node is sending
$\ket{1\cdots1}$ to
$e^{i\alpha}\ket{1\cdots1}$ and any other basis state to itself.

The ZX calculus is mentioned later in Section~\ref{sec:zx-ir}.

\section{Quantum Programming Paradigms}
\label{sec:q-prog-models}

Before 2010, with the lack of concrete quantum coprocessors and
use-cases for quantum algorithms, quantum programming was mostly a
theoretical playground \cite{gay2006quantum}.
Nonetheless, with hindsight, much of the
current state-of-the-art techniques in quantum programming were
already latent. We present them in this section, building on five works
spanning the approaches at the time: Bettelli's \Cpp
library~\cite{bettelli03architecture}, Ömer's QCL
language~\cite{omer2003structured}, Altenkirch\&Grattage's QML
for quantum control \cite{grattage05functional},
Altenkirch\&Green's quantum IO monad~\cite{green2009qio}, Vizzotto's
quantum arrows~\cite{vizzotto06structuring}. An important paradigm for
this thesis is the quantum lambda calculus~\cite{selinger2006lambda}:
we discuss it in Section~\ref{sec:qlc}.

\paragraph{Quantum Programming within Classical Environment.}
A quantum algorithm aims at solving a classical problem instance, and
it is meant to run on a classical computer, piloting a quantum
coprocessor. As such, the \emphidx{control flow} of the program is
purely classical. It therefore makes sense to package the interaction
with the quantum coprocessor into a dedicated library of an existing
programming language. Bettelli \textit{et
  al}~\cite{bettelli03architecture} proposes such a library within
\Cpp, capitalizing on \Cpp object model to build circuit
abstraction. Although this particular implementation has not spurred
any spin offs, the concept of using an existing (classical) programming
language to host a quantum language has been very successful and is
still in use in most current programming environment for quantum
computation such as \qiskit. The limit of this approach is however the
ability to reason about quantum programs and to offer tools for certified
quantum programming.

\paragraph{Circuits as Side-Effects.}
Instead of using \Cpp as a host language,
Altenkirch\&Green \cite{green2009qio} proposes a
Haskell \emphidx{domain-specific language} (\emphidx{DSL}),
building on Haskell's monadic paradigm to abstract away
the interaction with the quantum coprocessor.
Altenkirch\&Green~\cite{green2009qio} in fact presents the \emph{first}
\emph{formal} baseline for a sound understanding of the interaction with the
quantum coprocessor: it can be understood as an input/output
side-effect. A quantum program \emph{outputs} gates to the coprocessor,
while it \emph{inputs} results of measurements. Haskell makes it
possible to give a clear interface to a side effect. With the
\emphidx{quantum IO} monad \QIO, one can therefore type qubit
initialization and measurements as
\begin{haskell}
qinit :: Bool -> QIO Qbit
meas  :: Qbit -> QIO Bool
\end{haskell}
\noindent
that is, \texttt{qinit} inputs a Boolean value and returns a qubit object
within the \QIO interface: such an operation only makes sense within
the context described by the interface. An implementation can be a
real quantum coprocessor, or a simulator, or some more exotic
implementation for instance recording all possible execution
traces. The quantum IO monad framework forms the
baseline for the development of \quipper discussed in
Section~\ref{sec:quipper}.

\paragraph{Circuits as Functions.}
Unlike Bettelli's approach \cite{bettelli03architecture} a quantum
circuit in the \QIO framework is a regular function in the host
language. A circuit on one wire is typed with
\begin{center}
  \texttt{Qbit -> Circ Qbit}.
\end{center}
The input wire of the circuit is the input of the function, and the
output wire of the circuit is the output of the function. As a
side-effect, the function generates a piece of circuit. The
\texttt{Circ} type constructor encapsulates all the interaction
between the program and the coprocessor.

Using regular functions to model circuits might limit the amount of
manipulation allowed on circuits. Some operations such inversion
cannot be easily formulated in such a general context. An alternative
proposal has been formulated by Vizzotto \textit{et
  al}~\cite{vizzotto06structuring}. The proposal builds on Haskell's
implementation of \emph{arrows}: a contrived notion of function,
distinct from Haskell's usual function-type. This \emphidx{quantum
  arrow} can therefore encapsulate all of the interaction with the
quantum coprocessor, and it offers an alternative approach to the
\QIO monad---albeit arguably less intuitive to
program. Nonetheless, the two layers of arrows (the special
quantum arrow and the regular, Haskell arrow-type) are very versatile,
and can be seen as a foundation for Theseus~\cite{james2014theseus}
and the contributions presented in Section~\ref{sec:rev-pat}.

\paragraph{Programming Constructs of the Quantum Coprocessor.}
Although in the 2000's quantum coprocessors were still a very
theoretical notion, there were already attempts at exploring their
programming capabilities. Instead of simply stacking gates into a
circuit, Ömer proposes with the imperative language \simpleidx{QCL}
\cite{omer2003structured} quantum-specific subroutines, making it
possible to distinguish features only available classically or only
available quantumly. QCL can in a sense be regarded as a preliminary
exploration of the current trend of hybrid quantum programming.

The language QCL is very imperative and the approach fails to catch
the control flow hidden
\emph{inside} a quantum circuit. A naïve \emphidx{quantum control}
consists in the usual control of unitary, seen as a \emphidx{quantum
  test}: an operator $U$ acting on wire $q_0$ and controlled by wire
$q_1$ can be regarded as a test on $q_1$.
The language QML~\cite{grattage05functional}
is arguably the first one to offer such syntactic,
purely quantum  test. The authors derive a small first-order
language in which such a control can be written with an
\texttt{if-then-else} construct: the test is quantum in the sense that
the value tested upon is never measured, and both branches of the test
fire in superposition.
Quantum control is however an elusive notion, and besides simple
tests, allowing general superpositions of execution paths has been
shown highly non-trivial and is still an active research
area, discussed in Chapter~\ref{ch:qcont}.

\section{Quantum Lambda Calculus}
\label{sec:qlc}

One of the main topics of this thesis is the quantum
lambda calculus \cite{selinger05lambda:conf, selinger2006lambda,
  valiron2004msc, selinger2009quantum, valiron2008phd}. This language
formalizes the notion of quantum, higher-order functional programming
language with classical control, where a program is running on a
classical computer with access to a quantum coprocessor. The language
is equipped with a set of constructs and an operational semantics to
formalize the interaction with the coprocessor. This approach
has been shaping what is now the state of the art in term of quantum
programming and quantum program certification.

This section can be regarded as a quick introduction to
Chapter~\ref{ch:sem}.  We first briefly recall the lambda
calculus. We then discuss the strategy employed in
\cite{selinger2006lambda} to extend it to support quantum computation:
the resulting formal language is the quantum lambda calculus, our main
contribution before 2008. We then present the notion of type system we
developed using linear logic, discussing why it makes
a suitable framework for typing
quantum data. We finally quickly sketch the state of denotational
semantics of the quantum lambda calculus in 2008.

\begin{mylife}
  This has been the subject of my M.Sc.~\cite{valiron2004msc} and my
  Ph.D.  thesis~\cite{valiron2008phd}: my main contribution at the
  time has been the study of the quantum lambda calculus and of its
  semantics.
\end{mylife}

\subsection{Lambda Calculus}
\label{sec:lc}

The lambda calculus\index{lambda calculus}~\cite{barendregt84lambda}
is a versatile model of higher-order programming languages, where
functions are first-class terms that can be returned or passed along
as arguments. The basic constructs consist of variables:
$x,y,\ldots$, lambda-abstractions $\lambda x.M$, standing for
functions of argument $x$ and body $M$, and applications $MN$: the
term $N$ is
an argument fed to the function $M$. Terms of the language are called
\emphidx{lambda-terms}. A variable $x$ in a term $M$ may be
\emph{bound}\index{bound variable} by a lambda; otherwise it is called
\emph{free}\index{free variable}. Computation is typically defined
with a rewrite-system based on the so-called \emphidx{beta-reduction}:
\[
  (\lambda x.M)N \to M[x:=N].
\]
Various constraints can be set, yielding \emph{evaluation
  strategies}\index{evaluation strategy}: call-by-value, call-by-name,
call-by-need, \etc \cite{plotkin75callbyname,maraist1995call-by-name},
\etc.
The language can furthermore be extended with
constants and other constructs to natively support other programming
features and/or side-effects.

Lambda-terms can be typed~\cite{barendregt2013lambda}: the grammar of
types consists at least of one constant type $\alpha$ and an arrow
constructor $A\tto B$. A term $\lambda x.M$ being a
function, its type is of the form $A\tto B$, when $x$ is meant to be
of type $A$ and $M$ of type $B$. A set of \emphidx{typing rules}
then specify what is a valid type for a given term. For instance, if
$M$ is of type $A\tto B$ and $N$ is of type $A$, then $MN$ can be
specified of type $B$:
\begin{equation}
  \infer{MN:B}{M:A\tto B & N:A}.\label{eq:typ-app}
\end{equation}

Typed lambda calculi form the canonical medium for the
\emphidx{Curry-Howard isomorphism}: a correspondence identifying types
with logic formulas, and terms with proofs in the
logic \cite{girard89proofs}. For instance, the rule \simpleidx{Modus-Ponens}
\[
\infer{B}{A\tto B & A}
\]
corresponds to the typing rule of the application shown in
Eq.~\eqref{eq:typ-app}. In a similar way as lambda calculus can be
extended with constants and constructs, type systems can be very
sophisticated to capture many properties of the underlying
language~\cite{pierce02types}, drawing deep connections with
expressive logics. In the context of a quantum extension to the
lambda calculus, such a relevant logic turns out to be linear
logic~\cite{girard87linear}: this is discussed in
Section~\ref{sec:lin-typ-sys}.

\subsection{Quantum Extension to the Lambda Calculus}
\label{sec:qext-lc}

The idea behind the quantum lambda calculus is to offer an interface
to the quantum coprocessor. To this end, it is natural to endow the
language with two constant types $\bit$ and $\qbit$, respectively
standing for classical Boolean values and quantum bits. Three term constants
can then be added: $\meas$ for measuring a qubit, $\qinit$ for
initializing new qubits, and a family of constants $U$, ranging over
a set of unitary gates. 

The question is how to incorporate qubit objects in the language. A
naïve approach consists in adding one constant for each quantum state:
one could then write for instance
\begin{equation}
  \lambda f.\lambda g.(f\ket0)(g\ket1)
  \label{eq:term-bog}
\end{equation}
The problem with such an approach is entanglement: What if the
two-qubit system in state $\ket{01}$ used in Eq~\eqref{eq:term-bog}
where instead in state $\frac1{\sqrt2}(\ket{00}+\ket{11})$? As
proposed by van Tonder~\cite{tonder04lambda}, one could imagine a
quantum superposition of terms. But this turns out to be in fact
equivalent to simply storing the quantum state on the side, and using
pointers to qubits in the term, as follows:
\begin{equation}
  \left[ ~~
    \frac1{\sqrt2}(\ket{00}+\ket{11}),\quad
    \ket{xy},\quad
    \lambda f.\lambda g.(f\,x)(g\,y)~~
  \right]
  \label{eq:ex-q-state}
\end{equation}
with $\ket{xy}$ being a compact representation for a function sending
$x$ to qubit position $0$ and $y$ to qubit position $1$.

In a series of papers \cite{selinger05lambda:conf, selinger2006lambda,
  valiron2004msc, selinger2009quantum, valiron2008phd}, we follow this
now standard procedure to define a quantum lambda calculus and its
operational semantics. A program is then a triple $[Q,L,M]$ mimicking
a simple quantum coprocessor where gates are sent one-at-a-time. In
the triple, the element $Q$ is the state of the quantum memory, $M$ is
a lambda-term with free variables standing for pointers to qubits in
the memory, and $L$ is a \emphidx{linking function} addressing each
pointer to their qubit position in the memory. Due to the nature of
the measurement, the evaluation ends up being probabilistic: there is
the need for a choice of \emphidx{reduction strategy}, since tossing a
coin and duplicating the result is not the same thing as duplicating
the coin and tossing (once) each copy. In the case of the quantum
lambda calculus, following effectful higher-order languages such as
Ocaml, the original choice has been a call-by-value reduction
strategy: an argument is reduced to a value before being passed along
to the function.

This standard abstract machine and reduction procedure is described in
more details in Section~\ref{sec:qlc-ll} together with our later
contributions.

\subsection{Linear Type System}
\label{sec:lin-typ-sys}

In the quantum lambda calculus, qubits have a special property: they
are \emph{non-duplicable}. Indeed, if the function
\[
  \lambda x.(\CNOT\,x)\,x
\]
inputting a qubit and passing it to the control-NOT both as control
and as active qubit is not valid. Similarly, the behavior of the term
\[
  \lambda x.(M(\meas\,x))(U\,x))
\]
heavily depends on the evaluation ordering of arguments. In order to
feature the usual safety properties, a type system for the quantum
lambda calculus has to enforce non-duplicability, i.e. linearity of
qubits. 

Because of the higher-order nature of the language, non-duplicability
is not restrained to qubits. For instance, the term
\[
  (\lambda x.(\lambda f.f\,x))(\qinit\,\ttrue)
\]
(where $\ttrue$ stands for the Boolean True) initializes a new qubit
and makes a function using this qubit. The function is thus
non-duplicable as it contains a qubit inside its body. Non-trivial
examples can be constructed based on the teleportation
algorithm~\cite{selinger2006lambda} or the Bell
experiment~\cite{valiron2008quantum}.

A suitable resource-sensitive logic is \emphidx{linear
  logic}~\cite{girard87linear}: objects are strictly linear by
default, and a special type constructor ``$!$'' is added to the logic
to tag duplicable and erasable elements. The original type systems for
the quantum lambda calculus builds upon linear logic:
In Section~\ref{sec:qlc-ll}, we present an instantiation following
intuitionistic affine linear logic.

\subsection{Towards a Denotational Semantics}
\label{sec:toward-den-sem}

A \emphidx{denotational semantics} is a mathematical---or
categorical---models characterizing the behavior of programs
\cite{stoy1977denotational,schmidt1986denotational,lambek89introduction}. A
denotational semantics attaches to each type a mathematical
space---or an object of a category---and to each well-typed term a
suitable function ---or morphism.

\paragraph{The strictly linear fragment.}
In the context of quantum computation, the natural mathematical
framework consists in density matrices and superoperators ---or more
generally, positive matrices and completely positive
maps---.
Capitalizing on the Choi theorem~\cite{choi1975completely}, 
Selinger~\cite{selinger04quantum,selinger04semantics}
describes a (concrete) compact closed category based on cones of positive
matrices and completely positive maps. This category has been shown to
provide a fully-abstract model for strictly linear quantum computation
in~\cite{selinger06fully}.
However, as it is based on finite-dimensional vector
spaces it cannot handle inductive types such as natural numbers or
lists. Similarly, it is not expressive enough to model the type
constructor ``$!$'': this will be the subject of
Section~\ref{sec:ll-cpm}.

\paragraph{Towards duplication.}
In \cite{valiron2008categorical}, we made a preliminary, abstract proposal
for the structure required for a model of a full quantum
lambda calculus. The proposed structure
is based on 2 categories and a strong monad:
\begin{itemize}
\item A symmetric monoidal category $\cat{C}$, standing for the computations
  available inside the quantum coprocessor,
\item A cartesian closed category $\cat{D}$ for classical,
  effect-free higher-order computation,
\item A strong monad on $\cat{D}$ abstracting the probabilistic
  side-effect
\end{itemize}
The two categories $\cat{C}$ and $\cat{D}$ form a linear-non-linear
model~\cite{bierman93intuitionistic,benton94mixed}, therefore giving
rise to a semantics for the ``$!$'' operator as a comonad.
Linear-non-linear models
form the root of all existing semantics for state-of-the-art circuit-description
languages \cite{paykin2017qwire, rios2017categorical,
  lindenhovius2018enriching, fu2020linear}.
Accommodating duplication and circuit construction is the topic of
Chapter~\ref{ch:sem}.


\clearpage{\thispagestyle{empty}\cleardoublepage}

\chapter[Quantum Languages and Compilation Toolchain]{%
  Quantum Languages and\\*Compilation Toolchain%
}
\label{ch:compil}
\rhead{Quantum Compilation}

At the turn of the 2010s, quantum coprocessors started to be
considered mature enough for quantum algorithms to be competitive
compared to purely classical ones \cite{iarpa-qcs}.

The problem was to connect two distinct lines of work. On one side,
the design of quantum algorithms, focusing on their asymptotic
behavior, and the other quantum programming languages, very
minimalist at the time. Furthermore, leaning toward the concrete
use of quantum algorithms requires to conceptualize the compilation of
the language onto concrete hardware. A quantum compilation toolchain
needs to take into account the constraints of the coprocessor: the
small memory size, the structure of the memory, and possibly the noise of
the backend. Because of the many ways quantum algorithms are
described, a compilation frameworks has to be equipped with robust
methods for synthesizing and optimizing circuits out of classical
specifications---whether provided as matrices or given as classical
code. Finally, the counter-intuitive behavior of quantum computation
added to the difficulty of testing programs hints toward the development
of a dedicated set of formal methods and analysis techniques for
quantum program.

This chapter is devoted to a presentation of the author's work on
these aspects: design of a scalable quantum programming language,
circuit synthesis techniques, and analysis tools from a practical
point of view. Each of them covers a section.
\begin{itemize}
\item Section~\ref{sec:circ-desc-lang} presents our main contribution
  on the topic of scalable quantum programming language: the design of
  the language \quipper \cite{green2013quipper}.
  We first discuss the concept of
  circuit-description language and how it offers a sound, formal
  paradigm for interacting with the coprocessor. We then introduce
  \quipper, a domain-specific language embedded in Haskell and
  following this principle. We finally present a use-case enlightening
  the effectiveness of the approach: The logical resource estimation
  of an instance of the Quantum Linear System Algorithm
  \cite{scherer2017concrete}.
\item Section~\ref{sec:circ-syntesis} discusses three of our lines of
  works concerned with circuit synthesis and optimization. We first
  present a technique, novel at the time, to automatically construct
  an oracle (the circuit $U_f$ of Figure~\ref{fig:qoracle}) from the
  code of a classical function (the function $f$ of
  Figure~\ref{fig:qoracle}) \cite{valiron2016generating}.
  We then discuss circuit synthesis out of
  the description of a unitary matrix---an array of complex
  numbers \cite{brugiere2020householder}.
  We finally turn to the question of the use of the ZX
  calculus as a tool for describing and optimizing quantum circuits
  \cite{borgna2021hybrid}.
\item Section~\ref{sec:spec-verif} considers the problem of quantum
  program certification. Testing being hard ---if not impossible---
  when manipulating quantum information, certifying that a quantum
  program behaves as expected requires formal methods and proof
  techniques. In this section, we discuss the problems this raises and
  argue that deductive verification is a suitable technique for the
  problem.  We then present our contributions: a Floyd-Hoare logic for
  recursive quantum programs \cite{xu2021reasoning},
  and \qbrick, our proposal for proving
  properties of quantum programs in a scalable manner
  \cite{chareton2021automated}.
\end{itemize}

\begin{mylife}
  My contributions to the field are summarized by the sectioning of
  the chapter. Section~\ref{sec:circ-desc-lang} covers the series of
  works on the design of quantum programming languages derived from my
  post-doc in the US in 2011-2013. Section~\ref{sec:circ-syntesis}
  highlights some of the results I participated in developing, in
  particular with two of my former Ph.D students: Timothée Goubault de
  Brugière, Ph.D student in CIFRE co-supervised with Marc Baboulin
  (LMF) and Cyril Allouche (Atos), and Agustin Borgna, Ph.D student
  co-supervised with Simon Perdrix (LORIA). Finally,
  Section~\ref{sec:spec-verif} skims through the problem of
  specification and verification, and my contributions to the field,
  some of it coming from a collaboration with the quantum group at
  CEA-LIST/LSL. The collaboration is still ongoing with a Ph.D student:
  Jérome Ricciardi.
\end{mylife}


\section{\texorpdfstring{\quipper}{Quipper}: a Circuit-Description Language}
\label{sec:circ-desc-lang}

This section is devoted to one of our main contributions: the design of
\quipper, the first scalable quantum programming language. Before
\quipper, the state of the field, described in
Section~\ref{sec:q-prog-models}, showed little connection between
algorithm use-cases and quantum programming languages. This was a
serious roadblock for investigating the concrete applicability of
quantum algorithms.

The main formal realization we made while working on \quipper is the
fact that realistic quantum algorithms require a
\emphidx{circuit-description} language with both low-level and
high-level circuit constructors. With \quipper, we propose a formal,
sound setting for representing quantum programming, opening the door to
program verification and certification. This section is devoted to the
presentation of \quipper. Section~\ref{sec:qpl-design} discusses the
main design principles we developed. Section~\ref{sec:quipper}
presents \quipper, and Section~\ref{sec:lre} sketches one of our
contribution using \quipper: the logical resource estimation of an
instance of a quantum algorithm for solving linear systems of equations.

\begin{mylife}
  From 2011 to 2013 I was postdoc at the University of
  Philadelphia, in the US, employed by the large pan-American
  QCS project \cite{iarpa-qcs} funded by IARPA. The project spanned
  physics and computer science; I was hired to work on the
  language aspect. 

  One of the goals of the QCS project was to provide a \emph{concrete}
  logical resource estimation for quantum algorithms. Seven algorithms
  were chosen by IARPA:
  \begin{compactenum}
  \item \cite{childs2003exponential} to find a labeled node in a graph;
  \item \cite{ambainis2010andor} to evaluate a NAND formula;
  \item \cite{hallgren2007polynomial-time} to approximate the class group of a real
    quadratic number field;
  \item \cite{whitfield2011simulation} to compute the ground state
    energy level of a particular molecule;
  \item \cite{harrow2009quantum, ambainis2012variable,
      clader2013preconditioned} to solve a linear system of
    equations;
  \item \cite{regev2004quantum} to choose the shortest vector
    among a given set;
  \item \cite{magniez2007quantum} to exhibit a triangle inside a dense
    graph.
  \end{compactenum}
  The objective was to span a reasonably representative set of the
  existing algorithms of the time\footnote{Note
  however how the later trend of variational algorithms is
  ---obviously--- not  represented.}. The chosen algorithms make use of a
  wide variety of quantum primitives such as amplitude amplification,
  quantum walks, quantum Fourier transform (QFT)\index{QFT}, quantum
  phase estimation (QPE)\index{QPE}, quantum simulation, \etc.
  Several
  of the algorithms also require the implementation of sophisticated
  classical oracles. The starting point for each of our algorithm
  implementations was a detailed description of the algorithm provided
  by IARPA.

  As part of the project, we developed \quipper as a tool to answer
  the particular problematics of \emph{coding} the aforementioned
  algorithms in the context of the IARPA project, and the research
  spurred a series of papers: \cite{green2012report, green2013quipper,
    green2013introduction, smith2014quipper, valiron2015programming,
    scherer2017concrete}.  Along the line we conceptualized the
  language design principles presented in
  Section~\ref{sec:qpl-design}, and we used \quipper for concrete
  logical resource estimation. My contribution on the latter part is
  presented in Section~\ref{sec:lre}.
\end{mylife}

\subsection{Discussion: Quantum Programming Language Design}
\label{sec:qpl-design}

In Section~\ref{sec:q-models-comp-2008}, circuits were merely seen as
sequences of elementary gates. However, in most quantum algorithms
circuits are more complex structures, built compositionally from
smaller sub-circuits and circuit combinators. If they are usually
static objects, buffered until complete before being flushed to the
quantum coprocessor, in some algorithms, circuits are even
\emph{dynamically} generated: the tail of the circuit depending on the
result of former measurements.

In this section, we discuss the high-level structure of quantum
algorithms, the requirements for a quantum programming language, and
review some of the existing proposals.\footnote{This section is heavily inspired from my own contribution in \cite{chareton2021automated}.}

\subsubsection{Structure of Quantum Algorithms}
\label{sec:struct-quantum-algo}

\begin{figure}[t]
  \begin{subfigure}[t]{2.4in}
    \centering
    \fbox{\includegraphics[scale=.8]{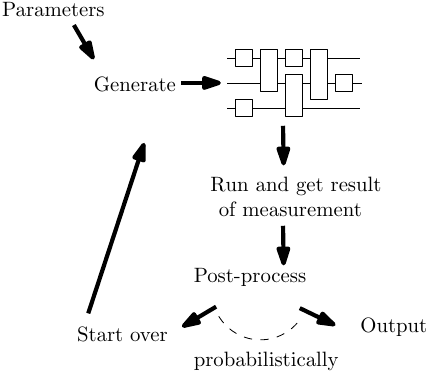}}
    \caption{Static scheme}
    \label{fig:wf-static}
  \end{subfigure}
  \hfill
  \begin{subfigure}[t]{2.2in}
    \fbox{\includegraphics[scale=.8]{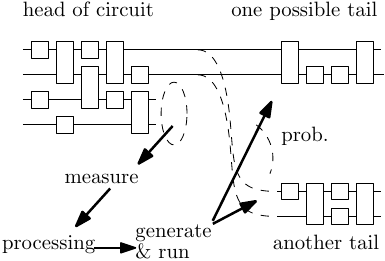}}
    \caption{Dynamic scheme}
    \label{fig:wf-dyn}
  \end{subfigure}
  \caption{Workflows for quantum algorithm}
  \label{fig:circ-workflow}
\end{figure}

The usual model for quantum computation was discussed in
Section~\ref{sec:q-models-comp-2008}:
a classical computer controls a quantum
coprocessor, whose role is to hold a quantum memory. A programmatic
interface for interacting with the coprocessor is provided to the
programmer sitting in front of the classical computer. The interface
gives methods to send instructions to the quantum memory to allocate
and initialize new quantum registers, apply unitary gates on qubits,
and eventually perform measurements. If the set of instructions is
commonly represented as a circuit, it is merely the result of a
\emph{trace of a classical execution} of a classical program on the
classical computer.

Figure~\ref{fig:circ-workflow} presents two standard workflows with a
quantum coprocessor. In Figure~\ref{fig:wf-static}, the classical
execution inputs some (classical) parameters, performs some
pre-processing, generates a circuit, sends the circuit to the
coprocessor, collects the result of the measurement, and finally
performs some post-processing to decide whether an output can be
produced or if one needs to start over. Shor's factoring
algorithm~\cite{shor97polynomial} and Grover's
algorithm~\cite{grover1996fast} fall into this scheme: the circuit is
used as a fancy probabilistic oracle. Most of the recent variational
algorithms~\cite{mcclean2016theory,cerezo2021variational} such as VQE
\cite{peruzzo2014variational} or QAOA
\cite{farhi2001quantum, farhi2014quantum}
also fall into this scheme, with
the subtlety that the circuit might be updated at each step.  The
other, less standard workflow is presented in
Figure~\ref{fig:wf-dyn}. In this scheme, the circuit is built ``on the
fly'', and measurements might be performed on a sub-part of the memory
along the course of execution of the circuit. The latter part of the
circuit might then depend on the result of classical processing in
the middle of the computation. One can for example cite the Unique
Shortest Vector algorithm \cite{regev2004quantum}, or the more
standard repeat-until-success procedures
\cite{lim2005repeat-until-success, paetznick2014repeat-until-success}.

Understanding quantum circuits as a by-product of the execution of
classical programs shines a fresh light on quantum algorithms. Unlike
a naive interpretation, a quantum algorithm cannot 
be identified with a quantum circuit. Instead, in general, at the very
least a quantum algorithm describes a \emph{family} of quantum
circuits. Indeed, consider the setting of
Figure~\ref{fig:wf-static}. The algorithm is fed with some parameters
and then builds a circuit: the circuit will depend on the shape of the
parameters. If for instance we were using Shor's factoring algorithm,
we would not build the same circuit for factoring 15 or
110,423,192,017. A quantum programming language
should therefore be able to describe \emphidx{parametric families} of circuits.

\begin{figure}
  \centering
  {\includegraphics[scale=.8]{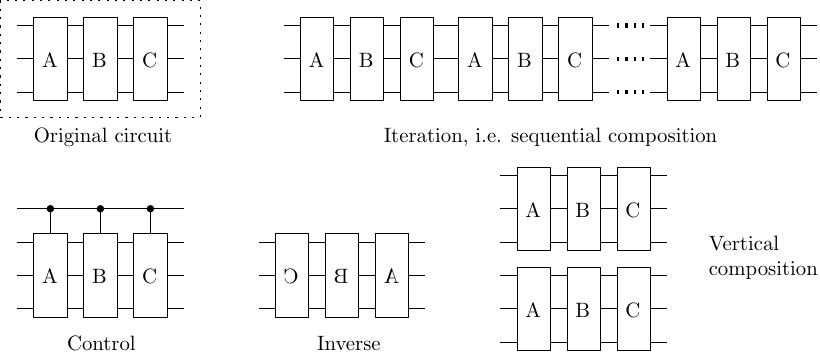}}
  \caption{Standard Circuits Combinators}
  \label{fig:hl-circ-comb}
\end{figure}

\begin{figure}
  \centering
  {\includegraphics[scale=.8]{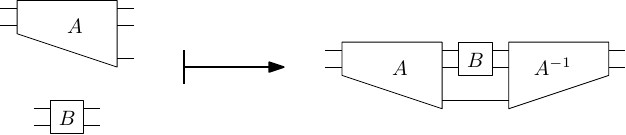}}
  \caption{Example of derived circuit combinator}
  \label{fig:hl-circ-comb-der}
\end{figure}

The circuits described by quantum algorithms are potentially very
large. We show for instance in
\cite{scherer2017concrete} how a concrete instance of the HHL
algorithm \cite{harrow2009quantum} for solving linear systems of
equations can require
as many as ${\sim}10^{40}$ elementary gates, if not optimized--- see
Section~\ref{sec:lre} for details.
To handle the scalability, quantum algorithms describe circuits
by composing sub-circuits---possibly described as
list of elementary gates but not only---using high-level circuit
combinators. These combinators build circuits
by (classically) processing possibly large sub-circuits. Some
standard combinators are shown in Figure~\ref{fig:hl-circ-comb}
(where we represent inverse with reflected letters).
Note that there is a distinction
to be made between a combinator, applied on a sub-circuit, and its
semantics, which is an action on each elementary gate. Combinators are
abstractions that can be composed to build larger combinators, such as
the one presented in Figure~\ref{fig:hl-circ-comb-der} built from
inversion, controlling and sequential composition.

\subsubsection{Requirements for Quantum Programming Languages}
\label{sec:qpl-req}

Any scalable quantum programming language should therefore allow the
following operations within a common framework.
\begin{itemize}
\item Manipulation of quantum registers and quantum circuits as
  first-class objects. The programmer should both be able to refer to
  ``wires'' in a natural manner and handle circuits as independent
  objects.
\item Description of \emph{parametric families} of quantum circuits,
  both in a procedural manner as sequence of operations---gates or
  subcircuits---and in an applicative manner, using circuit
  combinators;
\item Classical processing. In our
  experience~\cite{green2013quipper}, quantum algorithms mostly
  consists of classical processing: processing the parameters,
  building the circuits, processing the result of the measurement.
\end{itemize}
This broad description might call for refinements. For instance,
some of the classical processing might be performed on the quantum
coprocessor ---such as the simple classical controls
required for quantum error correction. The level of classical
processing performed on the classical computer ---therefore requiring
communication through the interface--- and performed on the quantum
coprocessor ---requiring a more or less sophisticated device--- is
dependent on the physical implementation. If some recent proposals
such as Quingo~\cite{quingo} discuss the design of quantum programming
languages aware of the two levels of classical processing ---in and out
of the coprocessor--- this is still work in progress.

\subsubsection{Review of the Existing Approaches}
\label{sec:rev_of_ex_app}

Most of the current existing quantum programming languages 
follow the requirements discussed in Section~\ref{sec:qpl-req}.
In this section, we review some typical approaches followed both in
academic and in industrial settings. This review is by no means meant to
be exhaustive: its only purpose is to discuss the possible strategies
for the design of quantum programming languages (QPLs)\index{QPL}.

When designing a realistic programming language from scratch, the main
problem is the access to existing libraries and tools. In the context
of quantum computation, one would need for instance to access the
filesystem, make use of specific libraries,
\emph{etc}. In order to quickly come
up with a scalable language, the easiest strategy consists in
\emph{embedding} the target language in a host language.\index{host
  language}\index{embedded language}
Indeed, a
quantum programming language can be seen a domain-specific language
(DSL)\index{DSL}\index{domain-specific language},
and it can be built over a regular language.
One can then rely on the possibly well-maintained and optimized
compiler or interpreter of the host language. 

If the advantages of working inside a host language are clear, there
are two main drawbacks, The first one is the potential rigidity of the
host language: there might be constructs natural to the DSL that are
hardly realizable inside the host language. The second drawback has to
do with the compilation toolchain: the shallow embedding of the DSL
makes it impossible to access its abstract syntax tree (AST),
therefore rendering its manipulation impossible.

\paragraph{Embedded QPLs.}
The first scalable embedded proposal\index{embedded language} is
\quipper~\cite{green2013quipper,green2013introduction}\index{Quipper}.
Embedded in Haskell,\index{Haskell}
it capitalizes on \emphidx{monads} to model the interaction with
the quantum coprocessor. \quipper's monadic semantics is meant to be
easily abstracted and reasoned over: it is the subject of
Section~\ref{sec:quipper}. Since \quipper, there has been a steady
stream of embedded quantum programming languages, often dedicated to a
specific quantum coprocessor or attached to a specific vendor, and
mostly in Python: \qiskit~\cite{qiskit} and
ProjectQ~\cite{projectq} for IBM, CirQ for Google,
Strawberry Fields~\cite{strawberry-fields} for Xanadu,
PyQuil and Forest for Rigetti~\cite{smith2017practical},
AQASM for Atos~\cite{aqasm-github},
\emph{etc}. From a language-design perspective,
most of these approaches make heavy use of Python objects to
represent circuits and operations. The focus is on usability and
versatility more than safety and well-foundness.

\paragraph{Standalone QPLs.}
On the other side of the spectrum, quantum programming languages have
been designed as standalone languages, with their own parser, and
therefore abstract syntactic tree. Maybe the first proposed scalable
language was Ömer's QCL~\cite{omer2003structured}. Ömer experimented
with several features such circuit-as-function, automatic inversion and
oracle generation. However, due to its non-modular approach the
language did not have successors.
Liqui$|\rangle$~\cite{wecker2014liquid} and its sequel Q\#
\cite{svore2018qsharp}, developed by Microsoft are good examples of an
attempt at building a standalone language while keeping a tight link
with an existing programming environment: Q\# is based on
the F\# framework, making it possible to easily ``use'' library
functions from within a Q\# piece of code. On the other hand, Q\# has
his own syntax and type system. This makes it possible
to capture run-time errors specific to quantum computation.
ScaffCC~\cite{javadiabhari2015scaffcc} is another example of a QPL
with its dedicated parser. If the language is rather low-level its
compiler has been heavily optimized and experimented over, and it
serves as support for a long stream of research on quantum compiler
optimizations \cite{chong2017programming,litteken2020updated}.
The last noteworthy language to cite in series is \silq~\cite{silq},
as it serves as a good interface with the next paragraph: aimed at
capturing most of the best practice in term of soundness and safety,
it is nonetheless targeted toward usability.

\subsection{Our Proposal: \quipper}
\label{sec:quipper}

This section is devoted to the presentation of the language \quipper:
a circuit-description language based on the design principles described in
Section~\ref{sec:qpl-design}. \quipper is a language embedded in the
host language Haskell and uses a monadic semantics to enforce the
desired operational semantics---that is, circuit construction.
Section~\ref{sec:monsem} quickly presents what is a monadic semantics and
Section~\ref{sec:quipper-desc} describes how \quipper makes use of it.

\subsubsection{Circuit Construction with a Monadic Approach}
\label{sec:monsem}

The solution devised by \quipper consists in relying on a special
language feature from Haskell called \emph{monad}. A monad is a type
operator encapsulating a side effect. Consider for instance a
probabilistic side effect. The monad is regarded as a type operator,
e.g. {\tt P}. There are two classes of terms:
terms without side-effect, with types e.g. {\tt Bool}, or {\tt Int},
and terms with side-effect, with types e.g. {\tt P(Bool)}, standing for
``term evaluating to a boolean, possibly with a probabilistic
effect'', or {\tt P(Int)}. The operator {\tt P(-)} captures the
probabilistic side effect.

A monad comes with two standard maps: {\tt return} regards a value as
a ``term with a (trivial) side-effect'', and {\tt eval}, for applying
a function to an effectful term. For {\tt P} we would have
\[\texttt{return} :: \texttt{a} \to \texttt{P(a)}\qquad
  \texttt{eval} :: (\texttt{a} \to \texttt{P(b)}) \to \texttt{P(a)} \to \texttt{P(b)}
\]
A few equations have to be satisfied by {\tt return} and {\tt eval}
for them to describe a monad. For instance, {\tt eval\,return} is the
identity on {\tt P(a)}. There can of course be more operations: for
instance, we can add to the signature of {\tt P} an operator {\tt
  coin} of type {\tt () -> P(Bool)}, whose semantics would be to
return $\ttrue$ or $\ffalse$ with equal probability.\footnote{In
  Haskell, the unit type is denoted with {\tt ()}.}.

A nice property of monads is that effectful operations can be written
with syntactic sugar in an imperative style, as follows.
\begin{haskell}
do
  x <- coin ()
  if x then return 0 else return 1
\end{haskell}

The program above is of type {\tt Int} and is equal to
\begin{center}
  \texttt{eval ($\lambda$\,x.if x then return 0 else return 1)~(coin ())}
\end{center}
once the syntactic sugar has been removed.

Following this approach, quantum computation can be understood as
a side-effect: it combines both (1) read/write effects, since gates are
sent to the coprocessor, and results of measurements are received; (2)
Probabilistic effects, since measurement is a probabilistic
operation.  The first attempt at formalizing this monad is Green's
quantum IO monad~\cite{green2009qio}: it has then been
further developed in \quipper~\cite{green2013quipper}, subject of
Section~\ref{sec:quipper-desc}.

\subsubsection{Design principles for \quipper}
\label{sec:quipper-desc}

\quipper is built as an embedded language in Haskell. It makes heavy
use of Haskell's type classes \cite{wadler1989type-classes} and type
families \cite{kiselyov2010type-families} to enhance
parametricity. Monads freely come in Haskell as a particular type
class.
As discussed in Section~\ref{sec:monsem}, \quipper's operational
semantics relies on a specific \emph{monad} encapsulating circuit
construction: the {\tt Circ} monad. The interaction with the
coprocessor can be modeled with an I/O interface: gates are emitted,
while branching occurs following a read operation. The {\tt Circ}
monad is then based on an inductive construction akin to the following.
\begin{haskell}
data CircIO a =
  Empty a
| Unit Gate
| Meas (Bool -> CircIO a)
\end{haskell}

Circuits in \quipper feature
wires of type {\tt Qubit} and {\tt Bit} ---i.e. bit-wires resulting
from a measure. The signature of the {\tt Circ} monad includes the
following operations.
\begin{haskell}
qinit :: Bool -> Circ(Qubit)
measure :: Qubit -> Circ(Bit)
dynamic_lift :: Bit -> Circ(Bool)
had :: Qubit -> Circ(Qubit)
\end{haskell}
\noindent
The coin-toss of Section~\ref{sec:monsem} can be written as
\begin{haskell}
coin () = do
  q <- qinit True
  q' <- had q
  r <- measure q'
  dynamic_lift r
\end{haskell}
\noindent
The function {\tt coin} is of type {\tt () -> Circ(Bool)}: running
{\tt coin} will merely generate a computation ---a circuit to be
executed--- waiting to be executed.

Thanks to the monadic encapsulation, circuits can be manipulated
within Haskell. For instance, inversion and control can be coded in
Haskell as circuit combinators with the following types.
\begin{haskell}
inverse :: (a -> Circ b) -> (b -> Circ a)
control :: (a -> Circ b) -> ((a,Qubit) -> Circ (b,Qubit))
\end{haskell}

One can also define operators to interact with circuits, such as
\begin{haskell}
count :: (Circ a) -> Int
simulate :: (Circ [Qubit]) -> Prob [Bool]
\end{haskell}
\noindent
where \texttt{count} returns the number of gates in the circuit, and
\texttt{simulate} classically emulates the input circuit followed with
a measurement and returns the probability distribution.

The strength of Haskell's monadic approach is the ability to capture
\emph{parametric families} of circuits within the framework. For
instance, a tower $H^{\otimes n}$ of Hadamard gates (parameterized by
$n$) can be defined as
\begin{haskell}
  mapM hadamard :: [Qubit] -> Circ [Qubit]
\end{haskell}
\noindent
where \texttt{[Qubit]} stands for the type of list of qubits. When fed
with one specific list of qubits, the program generates the
corresponding circuit. This program is then indeed the description of a
\emph{family} of circuits.

\subsection{Use-Case: Logical Resource Estimation of the QLS Algorithm}
\label{sec:lre}

In this section, we present one of the concrete application of the
language \quipper: the first complete
\emphidx{logical resource estimation} for one particular, concrete algorithm.

Indeed, before 2013, quantum algorithms were still theoretical apparatuses
meant to study the inherent asymptotic complexity of problems. While
moving towards concrete use-cases, one of the problem that arises is
the discrepancy between the theoretical efficiency of an algorithm and
its particular implementation on a concrete problem
instance. In particular, as a quantum algorithm builds a circuit, what is
the size of this circuit, provided a given problem instance?

In a journal publication \cite{scherer2017concrete}, we perform a
logical resource estimation for the \emphidx{quantum linear system}
problem (QLS)\index{QLS}, for solving linear systems of equations.
If the original algorithm
has been laid out by Harrow, Hassidim, and Lloyd
\cite{harrow2009quantum}---thus the common name for the algorithm: HHL---the
algorithm went through several
refinements: first by Ambainis \cite{ambainis2012variable} and then by
Clader \etal \cite{clader2013preconditioned}. The latter was the focus
of the work that was analyzed in the journal publication
\cite{scherer2017concrete}.

\paragraph{Statement of the Problem}
The QLS algorithm aims at solving a system of linear equations of the
form $A\vec{x}=\vec{b}$, where $A$ is a Hermitian $N\times N$ matrix
of complex numbers, $\vec{b}$ is a $\Cx$-vector of dimension $N$, and
$\vec{x}$ is the unknown vector. Solving the equation morally
corresponds to inverting $A$. 

The basic idea of the QLS algorithm is the following. Provided that
$\lambda_i$ and $u_i$ are respectively the eigenvalues and
eigenvectors of $A$, if $\vec{b}=\sum_i\beta_i u_i$, with suitable side conditions the solution of the
equation is simply
\[
  \vec{x} = \sum_{i=1}^N\frac{\beta_i}{\lambda_i} u_i.
\]
The algorithm then relies on several non-trivial pieces: the
\simpleidx{Quantum Phase Estimation} (\simpleidx{QPE}) to retrieve the
$\lambda_i$'s; oracles for $A$, $\vec{b}$ and the inversion; an
\simpleidx{Hamiltonian simulation} \cite{berry2007efficient} to build
a circuit for $e^{itA}$.

\paragraph{Summary of the Complexity Analysis.}
The original QLS algorithm and its subsequent refinements
\cite{harrow2009quantum, ambainis2012variable,
  clader2013preconditioned} leave aside the implementation details and
only focus on the general asymptotic complexity of the algorithm. It
uses several parameters of the problem instance: the size $N$ of the
matrix\,; the maximal error allowed $\varepsilon$\,; the
\emph{sparseness}\index{sparse} $d$ of the matrix $A$, that is, the maximum number
of non-zero entries per row and column\,; the \emphidx{condition number}
$\kappa$, defined as the ratio between the largest and smallest
eigenvalues of $A$. The smaller $\kappa$ is, the closer it is to be
invertible: $\kappa$ gives information on the stability of the
solution $\vec{x}$.

That being said, the best known classical algorithm for solving linear
systems of equations are based on the conjugate gradient method
\cite{shewchuk1994introduction,saad2003iterative}, and they have a
run-time complexity of $O(Nd\kappa\log(1/\varepsilon))$. By contrast,
the HHL algorithm \cite{harrow2009quantum} attains
\begin{equation}\label{eq:bigO-hhl}
  \widetilde{O}(\kappa^2d^2\log(N)/\varepsilon)
\end{equation}
(where $\widetilde{O}(-)$ suppresses more slowly growing terms
compared to $O(-)$). Provided that the matrix is well-conditioned and
sparse enough, we theoretically get an exponential improvement over
the classical algorithm. The improvement proposed by Clader \etal
\cite{clader2013preconditioned} provides a complexity
\begin{equation}\label{eq:bigO-clader}
  \widetilde{O}(\kappa d^7\log(N)/\varepsilon^2).
\end{equation}
For very sparse
matrices, this algorithm is likely to beat the original HHL algorithm.

However, all of these ``big-O'' complexities are blind to the
structure of the concrete description of the matrix $A$ and of the
vector $\vec{b}$: they do not help with logical resource estimates for
concrete problem instances.

\paragraph{Logical Resource Estimation for a Problem Instance.}
In the project QCS---and the resulting paper
\cite{scherer2017concrete}---we applied the QLS algorithm to
a linear system of equations coming from the discretization of
Maxwell's equations using the \simpleidx{finite-element method}
(\simpleidx{FEM}), to
determine the \simpleidx{electromagnetic scattering}
cross-section of a specified
target object \cite{chatterjee1993edge-based}.  FEM tends to generate
\emphidx{sparse} matrices, one of the conditions for the QLS
algorithm.

To decide on the size $N$ of the matrix, using the big-O estimates, we
came to $N \sim 4\cdot 10^7$ as the ``cross-over point'' at which the
quantum algorithm would beat the classical algorithm. We chose the
somewhat larger value $N=332,020,680$ to stay on the safe side: it is
reasonable to expect such a problem size to be hardly tractable
classically. The oracles for $A$ and $\vec{b}$ can be derived from the
problem instance. Their description is \emph{classical}: see
e.g. Fig.~\ref{fig:oracle-r-spec} for a piece of the specification of
the oracle $\vec{r}$, coded in Haskell. Note in particular how this
requires high-level libraries such as trigonometric functions.

\begin{figure}
  \newsavebox{\mylistingbox}
  \begin{lrbox}{\mylistingbox}
    \begin{minipage}{\textwidth}
\begin{haskell}
calcRweights y nx ny lx ly k theta phi =
     let (xc',yc') = edgetoxy y nx ny in
     let xc = (xc'-1.0)*lx - ((fromIntegral nx)-1.0)*lx/2.0 in
     let yc = (yc'-1.0)*ly - ((fromIntegral ny)-1.0)*ly/2.0 in
     let (xg,yg) = itoxy y nx ny in
     if (xg == nx) then
         let i = (mkPolar ly (k*xc*(cos phi)))*
                 (mkPolar 1.0 (k*yc*(sin phi)))*
                 ((sinc (k*ly*(sin phi)/2.0)) :+ 0.0) in
         let r = ( cos(phi) :+ k*lx )*((cos (theta - phi))/lx :+ 0.0) in i * r
     else if (xg==2*nx-1) then
         let i = (mkPolar ly (k*xc*cos(phi)))*
                 (mkPolar 1.0 (k*yc*sin(phi)))*
                 ((sinc (k*ly*sin(phi)/2.0)) :+ 0.0) in
         let r = ( cos(phi) :+ (- k*lx))*((cos (theta - phi))/lx :+ 0.0) in i * r
     else if ( (yg==1) && (xg<nx) ) then 
         let i = (mkPolar lx (k*yc*sin(phi)))*
                 (mkPolar 1.0 (k*xc*cos(phi)))*
                 ((sinc (k*lx*(cos phi)/2.0)) :+ 0.0) in
         let r = ( (- sin phi) :+ k*ly )*((cos(theta - phi))/ly :+ 0.0) in i * r
     else if ( (yg==ny) && (xg<nx) ) then 
         let i = (mkPolar lx (k*yc*sin(phi)))*
                 (mkPolar 1.0 (k*xc*cos(phi)))*
                 ((sinc (k*lx*(cos phi)/2.0)) :+ 0.0) in
         let r = ( (- sin phi) :+ (- k*ly) )*((cos(theta - phi)/ly) :+ 0.0) in i * r
     else 0.0 :+ 0.0
\end{haskell}
    \end{minipage}
  \end{lrbox}
  \centering
  \scalebox{0.9}{\usebox{\mylistingbox}}
  \caption{Part of the specification of the $\vec{r}$ oracle}
  \label{fig:oracle-r-spec}
\end{figure}

\paragraph{Discussion}
For the chosen problem instance, the other parameters governing the
complexity yield $d=7$, $\kappa = 10^4$ and $\varepsilon = 0.01$. The
logical resource estimation for $N=332,020,680$ is shown in
\cite[Table\,2, p.\,42]{scherer2017concrete}: the circuit generated by
the algorithm consists of $2.37\cdot10^{29}$ elementary gates amongst
$H, T, S, X, Z$ and $\CNOT$, and $3\cdot 10^8$ total qubits (most of
them being ancillas required for the oracles). Not counting the
oracles, the number of gates falls to $3.34\cdot10^{25}$ with only
$281$ qubits. The bottom line is that a big-O resource estimate is not
enough for deciding on the usability of a particular quantum
algorithm. Another conclusion is that optimization techniques are
going to be an essential tool in a quantum compilation toolchain.

The analysis performed in \cite{scherer2017concrete} was novel at the
time: it was the first concrete analysis of the resources needed to
run a quantum algorithm, without relying on ``big-O'' estimates.
Since the coding of the QLS algorithm in \quipper there has been
a steady stream of research on Hamiltonian simulation, see e.g.
\cite{berry2014exponential,berry2015hamiltonian,low2019hamiltonian}.

\section{Circuit Synthesis and Optimization}
\label{sec:circ-syntesis}

A quantum circuit serves two purposes. First of all, it serves as the
description of a linear operation on the memory state
space. Furthermore, it gives a procedure to implement this linear map,
with informations on the resources required to realize it.

Along the description of a quantum circuit, some subcircuits might
only be specified by the linear map they implement. The designers of
the quantum algorithm relies on an external authority to attest
that the corresponding subcircuit is indeed realizable within the
required framework, and leaves the generation of the quantum circuit
to a hypothetical compiler.

Circuit synthesis tools are therefore crucial tools for a quantum
compilation toolchain. This section explores three
cases. Section~\ref{sec:oracle-gen} considers the problem of
synthesizing oracles. A typical oracle is given as a classical
description such as the structure of a graph to explore or an
arithmetic operation to perform. The section presents the solution we
implemented for \quipper, automatically turning a classical code into
a reversible circuit. Section~\ref{sec:num-qcirc-synth} focuses on the
synthesis of circuits corresponding to linear maps given as matrices
of complex numbers. We present two solutions based on numerical
techniques to answer the problem, and discuss the sizes of the
generated circuits. Finally, Section~\ref{sec:zx-ir} sketches our
contribution for circuit generation out of a ZX description, in an
hybrid quantum and classical setting.

\begin{mylife}
  Section~\ref{sec:oracle-gen} is devoted to my contribution on
  automatic generation in \quipper for the QLS algorithm, to be able
  to handle the oracle of
  Figure~\ref{fig:oracle-r-spec}. Section~\ref{sec:num-qcirc-synth}
  presents works stemming out of the Ph.D supervision of Timothée
  Goubault de Brugière, CIFRE co-supervised by Marc Baboulin (LRI) and
  Cyril Allouche (Atos). Section~\ref{sec:zx-ir} discusses results
  from the Ph.D of Agustin Borgna, co-supervised with Simon Perdrix
  (LORIA, Nancy).
\end{mylife}

\subsection{Circuit Synthesis from Oracle Specification}
\label{sec:oracle-gen}

My first contribution to the field of circuit synthesis consists in
the development of an automated procedure to translate a functional
program working with Boolean values to a circuit realizing the same
computation. Written in Haskell using Template Haskell
\cite{template-haskell}, the tool inputs a Haskell, first-order
function and produces an object in the \texttt{Circ} monad,
encapsulating a circuit realizing the input function. The tool is one
of the libraries available with \quipper, and it has been extensively
used for the oracle of the QLS algorithm, as discussed in
Section~\ref{sec:lre}. In particular, it was what made it
possible to realize the trigonometric functions using fixed-point real
numbers (see e.g. Figure~\ref{fig:oracle-r-spec}).

The tool makes heavy use of Haskell's monad feature. I developed the
formalism in a publication \cite{valiron2016generating}: This section
is devoted to its presentation.

\paragraph{Irreversible to Reversible Computation}

The study of reversible computation and how it relates to irreversible
computation has been a subject of research since the 1960s. 
Landauer \cite{landauer1961irreversibility} follows a trend of
research discussing how
irreversible computation dissipates energy (and heat) and how
reversible computation could be a way to reduce computational energy
consumption. In the following years, several
models of reversible computation have been proposed: reversible Turing
machines \cite{bennett1973logical}, reversible cellular automata
\cite{moore1962machine,toffoli1977computation,durand-lose2002computing},
reversible boolean circuits \cite{toffoli1980reversible}, billiard ball
models \cite{fredkin1982conservative}, \etc. Various concrete,
physical reversible processors have also been proposed in the
literature, aiming at being more efficient than their irreversible
counterparts \cite{hall1992electroid,frank1999reversibility}.
The interest for the subject has
not declined \cite{bennett2000notes,adamatzky2002collision-based,frank2020special}, as
for instance shown by the recent ICT COST Action IC1405
\cite{cost-rev} and the series of conferences on reversible
computation \cite{rc2021}.

Although the two subjects stem from distinct origins, reversible
computation has seen an unexpected use in quantum computation. Indeed,
as discussed in Section~\ref{sec:qpl-design} one of the necessary building block
of quantum algorithms is \emph{oracles}: unitary maps realizing
classical, irreversible computations. As a unitary map is first and
foremost a reversible operation, all of the machinery developed for
reversible computation can be used for oracle synthesis. And indeed,
in the literature most of the complexity analysis of quantum algorithm
relies on the seminal papers of Fredkin, Toffoli and Bennett
\cite{toffoli1980reversible,fredkin1982conservative,bennett1973logical}
to assert the existence of efficient oracle synthesis.

Fredkin and Toffoli 
\cite{toffoli1980reversible,fredkin1982conservative} were amongst the
first ones to state the problem of reversible computation using a
circuit formalism. Within this framework, they integrate the
so-called \emphidx{Landauer's embedding} with \emphidx{Bennett's
  trick} to turn a classical, irreversible function
\[
  f:\bit^n\to\bit^m
\]
into a reversible circuit in the shape of the oracle shown in
Figure~\ref{fig:qoracle}, computing
\begin{equation}\label{eq:bennett}
  \begin{array}{llll}
    \widetilde{f}
    :{}& \bit^n\times\bit^r\times\bit^m&\to&\bit^n\times\bit^r\times\bit^m
    \\ &(\vec{x},~\vec{0},~\vec{y})&\mapsto&(\vec{x},~\vec{0},~\vec{y}\oplus
                                    f(\vec{x})).
  \end{array}
\end{equation}
Recall that $\oplus$ stands for bitwise XOR boolean gate.  Provided
that the function $f$ is described by a (boolean) formula, Toffoli
shows that the function $\widetilde{f}$ can be realized by a circuit
of linear size compared to the number of logical gates in the
description of $f$. The register of $r$ bits in the middle is used for
storing intermediate results, and the computation sets it back to
$\vec{0}$ when done.

\begin{figure}[tbh]
  \centering
  \includegraphics[page=3,scale=.8]{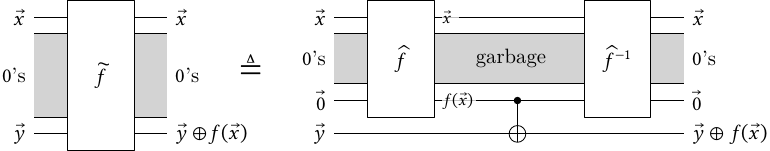}
  \caption{Bennett's trick}
  \label{fig:bennett-trick}
\end{figure}

The idea consists in using Bennett's trick \cite{bennett1973logical} to
first build a (reversible) circuit $\widehat{f}$ computing the
\emph{Landauer embedding} of $f$
\[
  \begin{array}{llll}
    \widehat{f}
    :{}& \bit^n\times\bit^r\times\bit^m&\to&\bit^n\times\bit^r\times\bit^m
    \\ &(\vec{x},~\vec{0},~\vec{0})&\mapsto&(\vec{x},~\text{garbage},~f(\vec{x}))
  \end{array}
\]
in a compositional manner. Note that compared to
Eq.~\eqref{eq:bennett}, the middle register is not cleaned after use,
rendering the computation irreversible if we were to throw away the
garbage. However, as discussed below the map $\widehat{f}$ is only
built from reversible components: one can recover the map
$\widetilde{f}$ of Eq~\eqref{eq:bennett} using the construction shown
in Figure~\ref{fig:bennett-trick}.

\begin{figure}[tbh]
  \centering
  \includegraphics[page=1,scale=.8]{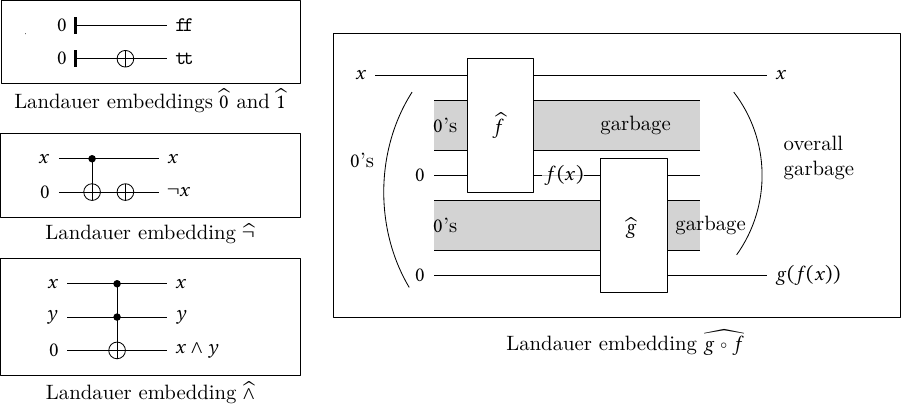}
  \caption{Landauer embeddings of elementary logical blocks}
  \label{fig:landauer}
\end{figure}

\begin{figure}[tbh]
  \centering
  \begin{subfigure}[t]{.68\textwidth}
    \centering
    \includegraphics[page=2,scale=.75]{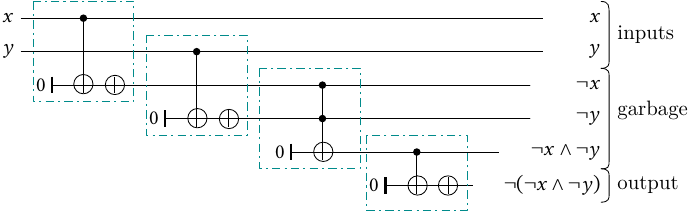}
    \caption{Landauer embeddings}
    \label{fig:landauer-or}
  \end{subfigure}
  \hfill
  \begin{subfigure}[t]{.3\textwidth}
    \centering
    \includegraphics[page=4,scale=.75]{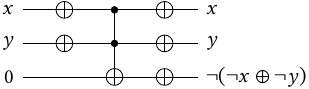}
    \caption{Direct implementation}
    \label{fig:better-or}
  \end{subfigure}
  \caption{Circuits for $(x,y)\mapsto \neg((\neg x)\wedge(\neg y))$}
\end{figure}

To understand how to compositionally build $\widehat{f}$,
assume that $m=1$, and that
$f$ is built from boolean constants, conjunction ($\wedge$), negation
($\neg$), and composition thereof. The corresponding Landauer
embeddings are shown in Fig.~\ref{fig:landauer}, and the embedding of
the function $(x,y)\mapsto \neg((\neg x)\wedge(\neg y))$ is presented
in Figure~\ref{fig:landauer-or}. Although it gives a verbose circuit
---see the equivalent, shorter circuit in
Figure~\ref{fig:better-or}--- it is efficient in the sense that the
size of the circuit is linear on the size of the formula: each dashed
sub-circuit corresponds to one logical operator.

\paragraph{Formalization of the Specification Language.}
The compositional procedure presented above
can be formalized and generalized to higher-order functions, with the
use of a monad to store the circuit under construction. For sake of
conciseness, in the following, we present a representative subset of
the language described in \cite{valiron2016generating} ---we invite
the reader to consult the original paper for details.

The idea consists in considering a simply-typed lambda calculus with
Boolean values, and in relating two possible operational semantics for
it.  One semantics is the usual one, where a term of Boolean type
rewrites to a Boolean value. The other one is instead a partial
evaluation strategy, where the operations to perform are stored in a
circuit: the circuit to be evaluated.

If the language is denoted with $\Toylang$, we can consider for
instance the definition of terms and types as
\[
  \begin{array}{lll}
    M,N &::=& x\bor \lambda x.M\bor MN\bor
              \ttrue \bor\ffalse\bor \texttt{not}\bor\texttt{and},
    \\
    A,B &::=& \bool\bor A\to B.
  \end{array}
\]
The language in \cite{valiron2016generating} also contains lists,
pairing, if-then-else and fixpoints, but these constructs do not need
a substantially different approach. In any case, typing judgment are
standard: they consist of a typing context , i.e. a set of typed
variables $\Delta=x_1:A_1,\ldots,x_n:A_n$, and a term $M$ of type $A$,
written $\Delta\vdash M:A$. The typing rules for typing derivations
describing valid typing judgments are as expected: the conjunction is
for instance typed as $\texttt{and}:\bool\to\bool\to\bool$.

The first operational semantics is a standard call-by-value reduction
strategy: we define values $V,W$ and application contexts $S[-]$ as
usual, and we for instance have the rule stating that $S[(\lambda x.M)V]
\to_\beta S[M\{x:=V\}]$ and that $S[(\texttt{and}\,\ffalse)\,\ttrue]
\to_\beta S[\ffalse]$.
The second operational semantics consists in a partial evaluation:
instead of evaluating \texttt{not} and $\texttt{and}$, the semantics
``stores'' the operations to be performed inside a (reversible)
circuit. The semantics is therefore based on an abstract machine of
the form
\begin{equation}\label{eq:am-rev-circ}
  \left(\cincludegraphics{page=1}{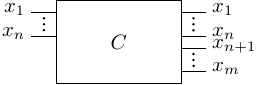}, ~M\right)
\end{equation}
where $C$ is a reversible circuit consisting in wire initializations,
NOT, CNOT and Toffoli gates, and where the free variables of $M$ are
within $\{x_1,\cdots,x_m\}$. For the detailed definition of circuits,
we refer the reader to \cite{valiron2016generating}.  The rule for
{\tt and} is for instance shown in Figure~\ref{fig:and-monadic-sem},
where $z$ is a fresh variable.

\begin{figure}[tbh]
  \centering
  \[
    \left(\cincludegraphics{page=2}{fig/circ-op-sem.pdf},~S[\texttt{and}\,x_i\,x_j]\right)
    \quad\to\quad
    \left(\cincludegraphics{page=3}{fig/circ-op-sem.pdf},~S[z]\right)
  \]
  \caption{Rule for {\tt and} in the monadic semantics}
  \label{fig:and-monadic-sem}
\end{figure}

The two semantics feature usual safety properties, and they
\emph{coincide} \cite[Th.\,17]{valiron2016generating}: the circuit
generated from the partial evaluation of a lambda-term $M$ realizes
the function described by $M$.  For instance, the (non-closed) term
\begin{equation*}
  x:\bool,y:\bool\vdash
  (\lambda f.f\,(\texttt{and}\,(f\,x)\,(f\,y)))\,\texttt{not}:\bool
\end{equation*}
reduces to the circuit shown in Figure~\ref{fig:landauer-or}.

\paragraph{Automated Oracle Synthesis as Monadic Lifting.}
The language $\Toylang$ and the two semantics can be extended with
pairs, coproducts, lists, fixpoints and tests ---\,see
\cite{valiron2016generating}. Together with these extensions, 
one can internalize the definition of circuit within the
language $\Toylang$ itself. The abstract-machine semantics can then be simulated
inside $\Toylang$ using a generic \emphidx{monadic lifting}, close to what was
proposed in \cite{swamy2011lightweight}.  It is the transposition of
Haskell’s monads to our language $\Toylang$ ---and of the strategy
used in \quipper for automatic oracle synthesis. The main
characteristic of the reversible abstract-machine is to change the
operational behavior of the type \bool: the terms {$\ttrue$}, {$\ffalse$} and
the inline Boolean combinators do not reduce as regular
lambda-terms. Instead, they trigger a side-effect, which can be
simulated within a suitable monad.

The main strength of this approach---and its instantiation with
Template Haskell in Quip\-per---is to allow the parametric
description of \emphidx{families of circuits}. Indeed, in \quipper, a program of
type {\tt [Bool] ->[Bool]} is lifted to
{\tt [Qubit] -> Circ\,[Qubit]}: the resulting code generates a
circuit whose shape depends on the size of the input list. The circuit
is provably \emph{equivalent to its specification}: the classical
program being lifted.
Such an approach was---and to this day still is---novel.

\paragraph{Discussion of Other Approaches.}
In the literature, the design of a reversible circuit from the
description of a conventional function has conventionally been
approached through its truth table or properties thereof. Several
methods have been designed to generate compact circuits, although only
for fixed-size circuits. It would be interesting to see how to merge
the two approaches.

One can for instance
consider local, peep-hole optimizations based on
templates~\cite{maslov2003fredkintoffoli, maslov2005toffoli,
  saeedi2013synthesis}, or rely on SAT solvers \cite{haaswijk2018sat}.
Standard classical synthesis techniques based
on BDD \cite{wille2010effect}, on LUT mapper \cite{meuli2019ros} or on
ESOP and Reed-Muller techniques
\cite{fazel2007esop-based,gupta2006algorithm,miller2009synthesizing}
have been used with some success. Approaches such as QMDDs
---quantum versions of binary decision diagrams--- have also been
considered and shown rather efficient \cite{zulehner2017improving}.
At a high-level approach, one could also make
use of efficient libraries of reversible circuits for arithmetic
operations \cite{vedral1996quantum, takahashi2005linear-size,
  draper2006logarithmic-depth, takahashi2008fast, wiebe2016quantum,
  ruiz-perez2017quantum, mogensen2019reversible} or real
analysis~\cite{nachtigal2011design, wiebe2013floating,
  nguyen2013space-efficient, soeken2017hierarchical, haner2018quantum,
  haner2018optimizing}.

However, if these techniques make it possible
to write reversible functions
with arbitrary truth tables \cite{wille2008revlib}, they do not
usually scale well with the size of input \cite{haaswijk2018sat}.

Synthesis of reversible circuits can be seen as a small branch of the
vast area of hardware synthesis. In general, hardware synthesis can be
structural (description of the structure of the circuit) or behavioral
(description of algorithm to encode).
In this context, Bennett's Pebble game \cite{bennett1989timespace,
  levin1990note} have been used with success to optimize the width and
depth of circuits \cite{amy2017verified, bhattacharjee2019reversible}.
Functional programming languages have been used for both structural
and behavioral descriptions. On the more structural side one finds
Lava \cite{claessen2001embedded}, BlueSpec \cite{nikhil2004bluespec},
functional netlists \cite{park2008functional}, \etc. On the behavioral
side we have the Geometry of Synthesis \cite{ghica2012geometry},
Esterel \cite{berry2000foundations}, ForSyDe \cite{sander2017forsyde},
\etc. Two proposals sitting in between structural and behavioral
approaches are worth mentioning. First, the imperative, reversible
synthesis language SyRec \cite{wille2010syrec}, specialized for
reversible circuits.  Then, Thomsen’s proposal
\cite{thomsen2012functional}, allowing to represent a circuit in a
functional manner, highlighting the behavior of the circuit out of its
structure.

On the logic side, Geometry of Interaction
\cite{girard1988goiI,girard1988goiII,girard1995goiIII,girard2003goiIV,
  mackie1994geometry,mackie1995geometry,ghica2007geometry}
is a framework that can be adapted to turn functional
programs into reversible computation
\cite{abramsky2005structural,danos1999reversible}, using the 
idea of turning a typing derivation into a reversible automaton.
There have also been attempts to design reversible abstract machines
and to compile regular programs into reversible computation, e.g. a
reversible version of the SEMCD machine \cite{kluge1999reversible}.
More recently, the compiler REVS \cite{parent2017revs} aims at compiling
conventional computation into reversible circuits.

Monadic semantics for representing circuits is something relatively
common, specially among the DSL community: apart from \quipper
discussed in Section~\ref{sec:quipper}, one can name Lava
\cite{claessen2001embedded}, Fe-Si \cite{braibant2013formal},
\etc. Other approaches use more sophisticated constructions, with type
systems based on arrows \cite{james2012information} in order to
capture reversibility: these approaches point towards full-fledged
reversible programming languages, discussed in Section~\ref{sec:back-rev-lang}.

\subsection{Circuit Synthesis from General Unitary Matrices}
\label{sec:num-qcirc-synth}

In the very general case, a unitary on $n$ quantum bits is
characterized by a matrix consisting of $(2^n)^2$ complex
numbers. Since the matrix is unitary, the number of parameters is
slightly smaller than $4^n$: it is however still very much
exponential on the number of qubits.

In the case of an intentional description, such as a formula or a
program, this description might reduce the number of degrees of
freedom of the problem and a quantum circuit of polynomial size on
the number of qubits might be obtained (as e.g. in the case presented
in Section~\ref{sec:oracle-gen}). However, for a given gate set on 1
and 2 wires, if the matrix is only given in term of its (complex)
coefficients, in general the size of a quantum circuit corresponding
to the matrix is bound to be exponential.

One question that can however nonetheless be posed is how to get a
circuit out of this array-based description, and how to obtain it in
an as efficient as possible way. In this section, we describe two
results we obtained, together with Marc Baboulin, our Ph.D student
Timothée Goubault de Brugière, and Cyril Allouche
\cite{brugiere2019synthesizing, brugiere2020householder,
  brugiere2020phd}.

\paragraph{Circuit Synthesis via Householder Transformations}

An operator acting on $n$ qubits is represented by a matrix of size
$2^n\times2^n$.  Generating a circuit from an arbitrary matrix is
a problem that scales exponentially in $n$ in general, and
the problem of finding the smallest possible circuit for a particular
operator remains challenging: Knill \cite{knill1995approximation}
asserts the necessity of an exponential number of gates.
If several decomposition techniques have been
developed 
\cite{barenco1995elementary, cybenko2001reducing,
  mottonen2006decompositions, reck1994experimental,
  shende2006synthesis}, in all of them 
the resulting number of gates however still
lies within a factor of 2 of the theoretical lower bound
\cite{bullock2004asymptotically}.

In \cite{brugiere2020householder}, the circuit synthesis problem is
analyzed with a focus on both the size of the generated circuit
\emph{and} the time needed to generate it
\cite{amy2013meet-in-the-middle, heyfron2018efficient,
  matteo2016parallelizing, nam2018automated}. We rely on a
\emphidx{Householder decomposition} of the matrix to construct the
circuit.

In general, the Householder decomposition of any
matrix $A$ is of the form $QR$, where $Q$ is a unitary and $R$ is
upper triangular. $Q$ is obtained iteratively by zero-ing out $A$
column by column, applying \emph{Householder transformations} of the
form
\[
  H_k = I_k - a_k\cdot\ket{u_k}\bra{u_k}
\]
for a well-chosen scalar $a_k$ and vector $\ket{u_k}$.
At the end of the procedure, $Q$ consists in the product of the
$H_k$'s. 

Thanks to the specific structure of unitary matrices, one can derive a
significant theoretical and practical speedup for this specific QR
algorithm compared to the unmodified QR routine and the usual
technique for quantum circuit synthesis based on the quantum
\emphidx{Shannon decomposition} (\simpleidx{QSD})
\cite{shende2006synthesis}.

From a Householder decomposition, we then propose a circuit synthesis
procedure based on CNOT gates and rotations. The asymptotic counts
\cite[Tab.2 and Tab.3]{brugiere2020householder} are summarized in
Table~\ref{tab:hous}. Overall, this technique turns out to be faster
than the QSD-based method, although it provides circuits twice as
large. One of the interest of this work is to highlight the tread-off
in circuit synthesis: reducing the circuit size renders circuit
generation more costly.

\begin{table}
\centering
\begin{tabular}{|l|lll|}
  \hline
  Method & CNOT count & Rotation count & Flops
  \\\hline\hline
  QSD & $23/48\times4^n$ & $9/8\times4^n$ & $19\times 8^n$
  \\\hline
  Householder & $2\times4^n$ &  $2\times4^n$ & $2/3\times 8^n$
  \\\hline
  Lower Bound & $1/4\times4^n$ & $4^n$ & (unavailable)
  \\\hline
\end{tabular}
\caption{Asymptotic counts for QSD and Householder decomposition}\label{tab:hous}
\end{table}

\begin{figure}
  \centering
  \includegraphics{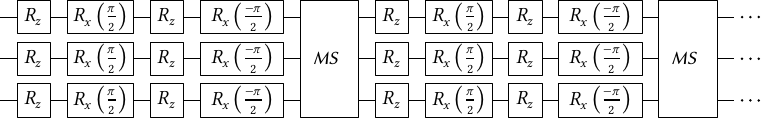}
  \caption{Structure of trapped-ions quantum circuits}
  \label{fig:trapped-ions}
\end{figure}

\paragraph{Circuit Synthesis with Gradient Descent.}
The focus of \cite{brugiere2019synthesizing} is the question of the
synthesis of \emphidx{trapped-ions} quantum circuits. The generic
structure of such circuits is a sequence of layers of the form shown in
Figure~\ref{fig:trapped-ions}.
The gates $R_z$ are each parameterized by a different angle, while the
gates $\textit{MS}$ are the entangling \simpleidx{Mølmer–Sørensen} gate
\cite{molmer1999multiparticle} defined by
\[
  \textit{MS}(\theta) \triangleq e^{i\theta(\sum_{i=1}^n\sigma^i_X)^2/4},
\]
with $\sigma^i_X$ the operator $X$ applied to the $i$-th qubit.

The question is then: for a given unitary, how many layers are needed,
and, for each layer, what parameters to choose for the $\textit{MS}$ gate
and the $R_z$ gates?
The paper \cite{brugiere2019synthesizing} offers two answers: a
theoretical lower bound and an experimental analysis of its
optimality.

For the theoretical lower bounds, we build on a previous approach
\cite{shende2004minimal}, proposing such a lower bound in the case of
circuits built from $\{\textit{SU}(2), \textit{CNOT}\}$.
The idea is to count the number of degrees of freedom in a quantum
circuit with a fixed structure and to show that this number has to
exceed a certain threshold to be sure that an exact synthesis is
possible for any operator.

\begin{figure}
  \centering
  \includegraphics{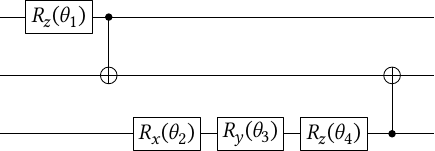}
  \caption{Example of parameterized circuit}
  \label{fig:param-circ-top}
\end{figure}

Consider for instance 
the circuit shape given in Figure~\ref{fig:param-circ-top}. It can be 
understood as a family of circuit with at most 4 degrees of freedom (one for
each rotation). In general a circuit structure
with $k$ degrees of freedom can be seen as a smooth function
\(
f:\mathbb{R}^k\to\textit{U}(2^n)
\)
mapping the values of angles to the space of unitary matrices of
size $2^n$.
We are interested in the image of the function $f$. If for any
operator $U$ on $n$ qubits there exists a vector of angles
$\vec{\theta}$ such that $f(\vec{\theta}) = U$, then we say that the
circuit shape is universal.

The contribution of \cite{brugiere2019synthesizing} consists in
deriving a lower bound on the number of layers required for
trapped-ions circuits of the shape shown in
Figure~\ref{fig:trapped-ions}, using a similar reasoning. The result
is that, to be universal, a topology \emph{must} in fact have \emph{at
  least}
\[
  \left\lceil
    \frac{
      2^{n+1}-2n-2
    }{
      2n+1
    }
  \right\rceil
\]
layers of $\textit{MS}$ gates.

In order to address the problem of the tightness of the bound,
we rely on a numerical method: The
\simpleidx{BFGS algorithm} \cite{nocedal2006numerical} (named after Broyden
\cite{broyden1970convergence}, Fletcher \cite{fletcher1970approach},
Goldfarb \cite{goldfarb1970family} and Shanno
\cite{shanno1970conditioning}).  If using heuristics or classical
optimization methods to synthesize circuits had already been tried before
\cite{martinez2016compiling, arrazola2019machine}, numerical
methods had however never been used to estimate the \emph{minimum
  quantum resources} required to synthesize a quantum circuit.

\begin{figure}
  \centering
  \scalebox{.8}{\begin{tikzpicture}[x=2mm,y=.2mm]
    \def\setplot#1#2#3{\node [#3] at #2 {#1}; \def\lab{#1}\def\r{#2}\def\pc{#3}}
    \def\plotdata#1{\node [\pc] at #1 {\lab}; \draw[thick, dotted, color=\pc] \r -- #1; \def\r{#1}}
    \draw[xstep=5,ystep=44,gray!50,very thin] (-2,-10) grid (40,360);
    \draw[thick,-] (-2,-10) -- node [anchor=north, inner sep=7mm] {Number of MS gates} (40,-10);
    \draw[thick,-] (-2,-10) -- node[anchor=east, inner sep=9mm] {\rotatebox{90}{Synthesis error}} (-2,360);
    \draw[thick,-] (40,-10) -- node[anchor=west, inner sep=9mm] {\rotatebox{90}{\# Iterations}} (40,360);
    \draw[color=red, very thick,-] (27,-10) -- (27,360);
    \foreach \x in {0,5,10,15,20,25,30,35}
    \draw[very thick] (\x,-12) -- (\x,-8) node[anchor=north] {\scalebox{.8}{$\x$}};
    \foreach \x in {0,1,2,3,4,5,6}
    \pgfmathtruncatemacro\result{200*\x}
    \pgfmathtruncatemacro\pos{52*\x-10}
    \draw[very thick] (39.8,\pos) -- (40.2,\pos) node[anchor=west] {\scalebox{.8}{$\result$}};
    \foreach \x in {0,1,2,3,4,5,6,7,8}
    \pgfmathtruncatemacro\y{44*\x}    
    \pgfmathparse{0.2*\x} \let \z \pgfmathresult
    \draw[very thick] (-1.8,\y) -- (-2.2,\y) node[anchor=east] {\scalebox{.8}{$\pgfmathprintnumber[fixed,precision=1]{\z}$}} ;
    \setplot{+}{(0,0)}{blue}
    \plotdata{(1,9)}\plotdata{(3,23)}\plotdata{(6,42)}
    \plotdata{(9,62)}\plotdata{(12,82)}\plotdata{(15,112)}
    \plotdata{(18,146)}\plotdata{(21,196)}\plotdata{(24,265)}\plotdata{(26,336)}
    \plotdata{(27,329)}\plotdata{(28,235)}\plotdata{(29,179)}\plotdata{(30,144)}
    \plotdata{(31,116)}\plotdata{(32,94)}\plotdata{(33,76)}\plotdata{(34,64)}
    \plotdata{(35,54)}\plotdata{(36,46)}\plotdata{(37,42)}\plotdata{(38,38)}
    \setplot{$\star$}{(0,336)}{black}\plotdata{(1,292)}\plotdata{(3,227)}
    \plotdata{(6,146)}\plotdata{(9,91)}\plotdata{(12,50)}\plotdata{(15,25)}
    \plotdata{(18,10)}\plotdata{(21,3)}\plotdata{(24,1)}\plotdata{(26,0)}
    \plotdata{(27,0)}\plotdata{(28,0)}\plotdata{(29,0)}\plotdata{(30,0)}
    \plotdata{(31,0)}\plotdata{(32,0)}\plotdata{(33,0)}\plotdata{(34,0)}
    \plotdata{(35,0)}\plotdata{(36,0)}\plotdata{(37,0)}\plotdata{(38,0)}
    \draw[fill=white] (6,355) rectangle (21,297) ;
    \node at (6,358) [anchor=north west]
    {\scalebox{.7}{\begin{minipage}{4cm}\begin{tabular}{@{}ll@{}}
            \textcolor{blue}{${\cdot}{\cdot}{\cdot}$+${\cdot}{\cdot}{\cdot}$} & \# Iterations
            \\[-.2ex]
            \textcolor{black}{${\cdot}{\cdot}{\cdot}{\star}{\cdot}{\cdot}{\cdot}$} & Error
            \\[-.2ex]
            \textcolor{red}{\raisebox{.5ex}{\rule{5ex}{1.5pt}}} & Lower bound
          \end{tabular}
        \end{minipage}}} ;
  \end{tikzpicture}}
\caption{4-qubits quantum circuit synthesis problem \cite{brugiere2019synthesizing}.
  \label{fig:bfgs}}
\end{figure}
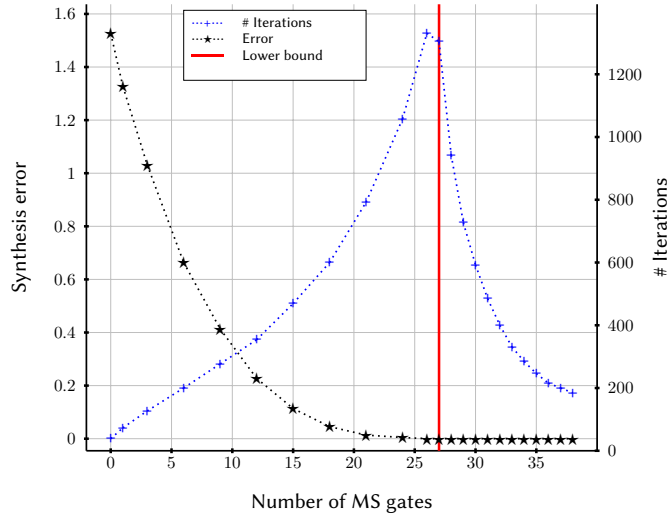

If the paper \cite{brugiere2019synthesizing} discusses the results in
details, here we only want to discuss the benchmark reproduced in
Figure~\ref{fig:bfgs}. The plot is realized as follows. One picks
topologies with increasing numbers of layers. For each topology, one
picks 50 random unitary matrices on $4$ qubits, and the optimization
process is run. The resulting plot is the obtained number of
iterations and the synthesis error.

First, one can note that the error decreases exponentially with the
number of MS gates: this is explained by the fact that the more
layers, the larger is the number of matrices that can be reached. The
interesting curve is the number of iterations required for
convergence: as the number of layers increases, so does the number of
required iterations for convergence,
until the vertical red line. This line represents the
theoretical lower bound: After this point, the number of iterations
sharply decreases.

The behavior of the curve corresponding to the number of iterations is
a good indication that the theoretical lower bound is indeed tight:
after this point, converging becomes easier and easier as we have more
degrees of freedom than necessary.

\subsection{Circuit Synthesis from ZX Specification}
\label{sec:zx-ir}

Quantum ---and reversible--- circuits are not the only graphical
language for representing quantum computation.
In the late 2000's, the \simpleidx{ZX} calculus of Section~\ref{sec:graph-calc} has turned into
popular alternative representation of quantum circuit
\cite{coecke2008interacting,coecke2011interacting,coecke2017picturing}.
This formal diagrammatic language with a more granular
representation than quantum circuits has been successfully used in applications
such as MBQC \cite{duncan2010rewriting,duncan2013graphical},
topological quantum computation \cite{horsman2011quantum},
Lattice-code surgery \cite{beaudrap2017zx} and
Pauli fusion \cite{beaudrap2019pauli},
as well as for
circuit simplification \cite{duncan2020graphtheoretic} and
verification of QEC \cite{chancellor2018graphical} such as 
Steane \cite{duncan2013verifying} and Color code \cite{garvie2017verifying}.


Another strength of the ZX calculus is its versatility. One can for
example find variants of the ZX calculus \cite{carette2020recipe}
such as the ZW calculus
\cite{hadzihasanovic2015diagrammatic,hadzihasanovic2017algebra} or the
ZH calculus aiming at the fragment Toffoli+H
\cite{backens2018zh}. But
the ZX calculus can also easily be equipped with extensions. One can
for instance quote the SZX calculus, for reasoning on \emph{arrays} of
qubits \cite{carette2019szx}, graphical calculi for qudits
\cite{ranchin2014depicting} and qutrits
\cite{wang2014qutrit,bian2015graphical,wang2017qutrit,townsend2022simplification,van2022building},
and the {\ZXg} calculus\index{ZX${}_{\subground}$}, for manipulating \simpleidx{mixed states}
\cite{carette2019completeness}.

\paragraph{Hybrid Quantum-Classical Synthesis from {\ZXg} Terms.}
In a paper co-authored with Simon Perdrix and our Ph.D student
Agustin Borgna \cite{borgna2021hybrid}, we propose a circuit
simplification technique in the context of quantum and classical
operations, using the {\ZXg}-calculus. This extension of the
ZX-calculus adds a discarding generator---a ``ground'', thus the
symbol ``\ground''---to the diagrams. making
it possible to represent operations interacting with the classical
environment \cite{carette2019completeness}.

The novelty of our approach comes from the fact that common
optimization strategies focus solely on the purely quantum aspect of
quantum computation
\cite{amy2014polynomial-time,heyfron2018efficient}.  Introduced by
Duncan {\etal} \cite{duncan2020graphtheoretic}, one of these
optimization techniques uses the ZX-calculus to apply granular
rewriting rules that ignore the boundaries of each quantum gate.
Their rewriting steps preserve a diagram property called
\emphidx{gFlow} that is required for the final extraction
of the ZX diagrams into circuits. Duncan's ZX optimization method was
latter used by Kissinger and van de Wetering
\cite{kissinger2020reducing} to reduce the number of
T-gates in quantum circuits.

In \cite{borgna2021hybrid}, we define the natural extension of the
pure Clifford optimization algorithm by Duncan \etal to hybrid
quantum-classical circuits using the {\ZXg} calculus. Our circuit
optimization procedure forgets the difference between quantum and
classical wires during the simplification process, representing
connections as a single type of edge. This allows it to optimize the
complete hybrid system as an homogeneous diagram, and results in
similar representations for operations that can be done either
quantumly or classically. We then propose a strategy to automatically
recover the part of the optimized circuit that can be treated in a
classical manner. Generally, in a physical quantum computer, the
classical operations are simpler to implement than their quantum
counterparts, and quantum simulators can exploit the knowledge of
which wires carry classical data to simplify their operation. As such,
it is beneficial to extract classical gates in the resulting circuit
where possible.

\section{Specification and Verification of Quantum Programs}
\label{sec:spec-verif}

In classical programming, a common verification technique consists
in testing and debugging \cite{brooks1995mythical}. In the case of
quantum programs, this standard approach is hard to implement and
bound to be insufficient \cite{huang2018qdb,huang2019statistical}. A
first problem is the probabilistic nature of quantum algorithms:
although feasible \cite{li2020projection-based}, assertion testing is
very intrusive, expensive resource-wise, and limited in its
expressiveness. A second, more fundamental problem comes from the cost of
running a quantum program. The cost can be monetary when running the code on
a physical machine, or resource-wise when emulating it, as
emulation requires an exponential quantity of classical resources.
In short, we may simply not be able to afford to perform hundreds of
runs of a piece of code just for testing.

If testing and debugging may not be a viable solution,
a wide range of \emphidx{formal verification} techniques
\cite{clarke1996formal} have been shown to be versatile tools for quantum
computation, amenable to many situations.
Several recent experiments have
successfully adapted known formal methodologies to the quantum
setting: Floyd--Hoare logics \cite{unruh2019qhl-ghost, unruh2019rqhl,
  ying2011floyd-hoare, ying2019automatic}, use of proof-assistants
\cite{boender2015formalization, paykin2017qwire, rand2018reqwire,
  rand2018phd, hietala2021proving, hietala2021verified}, abstract
interpretation \cite{perdrix2008quantum}, model checking
\cite{gay2008qmc,ying2014model-checking,feng2015qpmc}, equational
theories \cite{jeandel2017complete, kissinger2015quantomatic,
  fagan2018optimising, amy2018large-scale, amy2013algorithms,
  amy2019formal},
and deductive verification \cite{chareton2021automated}.

Section~\ref{sec:challenges-qverif} discusses the difficulties
regarding formal verification of quantum programs.
Section~\ref{sec:hoare-logic} then presents Floyd-Hoare logic and the
corresponding deductive verification techniques.  Finally,
Sections~\ref{sec:qhoare} and~\ref{sec:qbrick} presents our
contribution on deductive verification of quantum
programs. Section~\ref{sec:qhoare} discusses a quantum Floyd-Hoare
logic for first-order quantum programs supporting recursive calls and
measurements, while Section~\ref{sec:qbrick} presents the deductive
verification framework \qbrick, based on a recent, compact semantics for
quantum computation: \simpleidx{sum-over-paths}.

\subsection{Challenges for Quantum Formal Verification}
\label{sec:challenges-qverif}

Compared to classical computation, quantum computation raises a series
of problems for verification in general~\cite{chareton2021formal}.

\paragraph{Hybrid, probabilistic model}
Quantum algorithms are not monolithic, linear processes: as discussed
in Section~\ref{sec:qpl-design}, a quantum algorithm is a subtle interaction
between a classical computer and a quantum coprocessor, each having
their own properties and control flow. Validating the concrete
implementation of a quantum algorithm requires a suitable semantics
for this hybrid model.

Furthermore, gathering classical data from the quantum memory is an
inherently probabilistic procedure. In a sense, quantum computation
supersedes probabilistic computation: all of the issues coming from
the probabilistic settings also occur within quantum computation.

\paragraph{Limited resources}
A quantum algorithm describes \emph{logical} quantum circuits without
much care for the available resources. However, in the current NISQ
era \cite{preskill2018nisq} memory is expensive, with hardware
constraints such as limited connectivity. The number
of coprocessor cycles might also
be limited in case of absent or limited error-correction:
the evaluated circuit therefore has to be kept under a certain depth
\cite{cross2019validating}. In the current state of the technique,
adapting algorithms to the noise constraints can be challenging
\cite{gidney2021howtofactor}.
This makes quantum coprocessors akin to embedded systems: there is a
need for a fine-grained resource management.
If programming languages
such as \quipper \cite{green2013quipper} can help with resource
estimation, dedicated compilation tools have been developed to
automate circuit optimization and physical qubit layout
\cite{amy2014polynomial-time,parent2017revs,
  meuli2020enabling,hietala2021verified,sivarajah2020tket}.

\paragraph{Functional specifications}
A quantum algorithm comes with a \emph{functional specification}
describing its behavior. There are two kinds of functional
specifications. On one hand, an \emph{intentional} specification considers the
algorithm as an opaque instantiation of a mathematical function, and
only discusses the relationship between the input and the output of
the algorithm. Ying's quantum Floyd--Hoare logic
\cite{liu2019formal,ying2011floyd-hoare} is a typical approach leaning
towards intentional presentations.
On the other hand, an \emph{extentional} specification ``opens the box'' and also
describes \emph{how} the computation gets to the result. An
extentional specification might then for instance give requirements
on the size and shape of a circuit produced by the
algorithm. Approaches for extentional presentations typically use
dependent type systems \cite{paolini2019qpcf} or embeds into program
verification tools such as Coq \cite{hietala2021verified,
  paykin2017qwire} or Why3 \cite{filliatre2013why3,
  chareton2021automated}.

\paragraph{Compilation Toolchain}
As discussed in Section~\ref{sec:qpl-design}, a quantum program is not only
the description of \emph{one} quantum circuit: at a minimum it describes a
\emph{family} of quantum circuits, parameterized by the problem
instance. A specification concerns this family ---a versatile
verification tool should be able to handle parametricity.

Furthermore, along the compilation process the generated (families of)
circuits are transformed and optimized according to various
constraints coming from the error model, the hardware connectivity,
the cost of each gates, etc. In general, these transformations also
need to be validated: they should not modify the \emph{semantics} of
the circuit. This semantics in general involves linear algebra: the
validation tools should therefore handle it
\cite{hietala2021verified,chareton2021automated}, or restrict to
subsets such as reversible circuits~\cite{parent2017revs}.

\subsection{Floyd--Hoare Logic and Deductive Verification}
\label{sec:hoare-logic}

Deductive program verification is probably the oldest formal method
technique, dating back to Floyd and Hoare in the 1960's
\cite{floyd1967assigning,hoare1969axiomatic}. In this approach ---the
so-called \emphidx{Floyd--Hoare logic}--- a piece of code $C$ is
annotated with a logical contract \cite{meyer1992applying}
consisting of a pre-condition $P$
and a post-condition $Q$. The tuple $\{P\}C\{Q\}$ is valid if
whenever $P$ is true, executing the code $C$ makes the post-condition
$Q$ valid.

If one of the parameters of a Floyd--Hoare logic is the programming
language, the other parameter is the chosen logic. The objective is to
allow for a logic as expressive as possible, while being able to give
complete set of syntactic deduction rules together with an algorithm
as efficient as possible for inferring a valid proof of a
contract. The seminal works underlying the whole development of the
field are Dijkstra's algorithm for \emphidx{weakest precondition}
inference \cite{dijkstra1976discipline} and Burstall's proposal for
the addition of intermittent assertions \cite{burstall1974program}.

The weakest precondition inference algorithm is at the core of
the automation permitted by deductive verification based on
Floyd--Hoare logic. Consider the rule for sequential composition:
\begin{equation}\label{eq:hoare-seq}
  \infer{\{P\}C_1;C_2\{R\}.}{
    \{P\}C_1\{Q\}
    &
    \{Q\}C_2\{R\}
  }
\end{equation}
It states that for $R$ to be a valid-postcondition for the program
$C_1;C_2$ under the pre-condition $P$, one simply has to find $Q$ that
is both pre-condition for $C_2$ and post-condition for
$C_1$. Dijkstra's algorithm automates the discovery of a most-general
pre-condition $\textsf{wp}(F,C)$ for a code $C$ with post-condition
$F$. Rule \eqref{eq:hoare-seq} dictates that one can pick $Q$ to be
$\textsf{wp}(R,C_2)$. The weakest pre-condition
for $C_1;C_2$ then becomes $\textsf{wp}(\textsf{wp}(Q,C_2),C_1)$. To
recover $\{P\}C_1;C_2\{R\}$, a \emphidx{proof obligation} is
generated:
\[
  \textsf{wp}(\textsf{wp}(Q,C_2),C_1)\Rightarrow P.
\]
This formula can either be proven in a proof assistant or discharged
with an SMT-solver.

In the context of classical programming, these techniques have been
applied in academic or industrial contexts for many languages
\cite{hahnle2019deductive}. One can cite frameworks for the Pascal
language
\cite{deussen1995verification}, Ada \cite{luckham1985overview,spark},
Modula-3 \cite{leino1998extended},
Java \cite{huisman2016formal,filliatre2007whykrakatoacaduceus,esc-java-manual},
C
\cite{norrish1998c,filliatre2007whykrakatoacaduceus,kirchner2015frama-c},
the Method B capitalizing on Dijkstra's weakest precondition algorithm
\cite{robinson1997b,leuschel2011automated}, and 
the versatile Why3 environment for WhyML \cite{filliatre2013why3}.

\subsection{Quantum Floyd--Hoare Logic Handling Measurements}
\label{sec:qhoare}

Two main Floyd--Hoare logics specific to quantum computation have
emerged in recent years. The first line of work \cite{unruh2019rqhl,
  unruh2019qhl-ghost, barthe2020relational} proposes a Floyd--Hoare
logic for reasoning about programs implementing quantum protocols. The
framework is based on regular, classical logical constructors. The
logic is extended with the capability to reason about variables holding
quantum states, such as ``$x$ holds the qubit state
$\frac1{\sqrt2}(\ket{0}+\ket{1})$''.

The second approach to quantum Floyd--Hoare logic is now more than 10
years old and stems from Ying's research group
\cite{ying2011floyd-hoare, ying2013verification, ying2017invariants,
  liu2018qsi, li2018algorithmic, ying2019automatic, zhou2019applied,
  liu2019formal, hung2019quantitative, liu2019quantum,
  barthe2020relational, liu2022quantum, feng2022verification,
  ying2022proof}.
This prolific research avenue can be traced back to d'Hondt and
Panangaden's work on \emphidx{quantum weakest precondition}
\cite{dhondt2006quantum,hondt04weakest}, quantum equivalent to
Dijkstra's notion.
D'Hondt and Panangaden's idea consists in regarding
positive operators as (probabilistic) formulas on states. Remember
that the probability of measuring the density matrix $\rho$ in state
$\ket\phi$ is $\bra\phi\rho\ket\phi$. This can be rewritten as
$\trace(M\rho)$, with $M=\ket\phi\bra\phi$. In general, $M$ can be any
\emphidx{observable} operator, which for our purpose we can consider
as a density matrix. For instance, the observable
$\frac13\ket0\bra0+\frac23\ket1\bra1$ assesses the probability of
$\rho$ to be in the mixed state $\frac13\{\ket0\}+\frac23\{\ket1\}$.

Quantum programs in Ying's approach are the quantum equivalent of
a textbook while-language:
a fixed set of possible variables, all quantum, and each
spanning a given Hilbert space, and a few imperative
constructs with sequential composition for
acting on the state of the variables: assertion, tests, while-loop.
Since the only available types are quantum, branching is
probabilistic and based on the result of a measurement.

One can for instance write the program
\begin{equation}\label{eq:hoare-prog}
  x:= H\,x; \texttt{while}\,({\ket0}{\bra0}x = \ket0)\,\{ x:= H\,x \}
\end{equation}
which repeatedly measure $x$ against $\ket+$ until $\ket{1}$ is
obtained. An observable serving as formula is in this case acting on
the state of $x$ (i.e. it is acting on a qubit). 

Given a post-condition $Q$ and a program $C$, a \emphidx{quantum
  precondition} is an operator $P$ such that for any density matrix
$\rho$ representing a state of the memory of $C$,
$\trace(P\,\rho)\leq\trace(Q\,(C\,\rho))$, with $C\,\rho$ the state of
the system after the action of $C$. The operator $P$ is
\emphidx{weakest precondition} for $Q$ and $P$ if for all other
precondition $P'$, we have $P'\sqsubseteq P$ (using the Löwner order).
For instance, a precondition for the program shown in
Eq~\eqref{eq:hoare-prog} and the postcondition $\ket1\bra1$ is
$\ket0\bra0$. It is of course not unique ---the operator $0$ is also a
pre-condition--- but it is the weakest one.

Minsheng Ying's group has extensively worked on this approach, with
special attention to the structure of invariants required for the
while-loop. The group studied various extensions and problems such as
non deterministism \cite{li2014termination},
testing and debugging \cite{li2020projection-based},
linear-time properties \cite{ying2014model-checking},
termination and expected run-time \cite{li2018algorithmic,liu2022quantum},
parallel and distributed quantum programs \cite{feng2022verification,ying2022proof}.

At a high-level, if Ying's approach makes it easy to discuss
probabilistic behavior (since it is \emph{part of} the structure of
the logic), the shallow embedding inside operators ---similar to
Birkhoff and von Neumann's quantum logics
\cite{birkhoff1936logic,mittelstaedt1978quantum}--- limits its
expressiveness. For instance, It is hardly extensible to features such
as first-order or native manipulation of conventional types such as
natural numbers, lists, \etc.

\begin{mylife}
  In the context of the ANR project SoftQPro, I had the opportunity to
  dig into the subject with a former collaborator of Minsheng Ying,
  Zhaowei Xu, hired as a postdoc in our group. We worked on an
  extension of Ying's approach to quantum Hoare logic, to allow local
  variables and recursive subroutines. This collaboration yielded a
  paper to appear in TOCL \cite{xu2021reasoning}.
\end{mylife}

\subsection{\texorpdfstring{\qbrick}{Qbricks}: Deductive Verification with Parametrized Path Sums}
\label{sec:qbrick}

\begin{mylife}
  In 2017, I was invited at CEA-LIST/LSL by François Bobot and
  Sébastien Bardin to give a seminar to present \quipper. Along the
  discussion afterwards, we came to the conclusion that Why3
  \cite{filliatre2013why3} could very well serve as a host language
  for a quantum programming language, and that it could freely provide
  a means to certify and verify embedded quantum programs. Unlike
  Ying's quantum Hoare logic, the Why3 logic
  seemed expressive enough to state both intentional and extentional
  properties of programs.
  
  The project effectively started when they hired Christophe Chareton
  as postdoc to build on the idea. Moving from an hypothetical concept
  to a concrete tool able to prove Shor's algorithm properties took
  about 3 years. In \cite{chareton2021automated}, we present the
  outcome: Qbrick, a DSL embedded in Why3 coming with dedicated
  libraries of definitions and lemmas based on sum-over-paths
  \cite{amy2019formal}, dedicated to the formalization of quantum
  programs in a deductive verification framework. If I was involved in
  the theoretical development behind \qbrick, Christophe has been the
  kingpin of the development of the toolbox.
\end{mylife}

\noindent
In the situation described in Chapter~\ref{ch:compil}, a quantum
program might manipulate quantum registers of large dimension. For
specification and verification purposes, this renders the technique
presented in Section~\ref{sec:qhoare} hard to use: not only proofs
become sprawling but also positive operators in logical formulae
becomes daunting. This renders pen-and-paper proofs impossible. 

One solution consists in relying on a proof-assistant and to code
intentional properties inside the corresponding logic. This has been
done in Isabelle/HOL for Ying's Hoare logic \cite{liu2019formal}, and
in Coq for the QWIRE language \cite{paykin2017qwire,rand2018phd},
followed by the VOQC framework
\cite{hietala2021proving}.  However,
these approaches relies on (concrete) matrices, whether unitary
\cite{paykin2017qwire} or positive \cite{liu2019formal}. As it
turns out, matrices are not well-suited for automation, and long,
manual proofs are necessary for validating formal specification of
quantum programs in this formalism.

In the paper \cite{chareton2021automated}, we propose an alternative solution.
On one hand, instead of using a generic
proof-assistant such as Coq \cite{paykin2017qwire,rand2018phd} or
Isabelle/HOL \cite{liu2019formal,mahmoud2014formal}, we rely on Why3
\cite{filliatre2013why3}, a platform for deductive verification
dedicated to proof automation \cite{bobot2011why3}. On the other hand,
instead of using the hard-to-automate matrix formalism, we rely on a
compositional, functional semantics: \emphidx{sum-over-paths}
(or path-sums) \cite{amy2018large-scale,amy2019formal}.

\paragraph{Sum-over-Paths.}
Amy's \emphidx{path-sum} semantics offer an algebraic, intentional
presentation of quantum circuits, alternative to the matrix
presentation. The name comes from the correspondence with Feynman's
path integral \cite{feynman1965quantum}.
This very versatile framework is amenable to
other formalisms of quantum computation such as ZX calculus
\cite{coecke2017picturing,vilmart2021structure}.

The idea consists in
formalizing the standard function-style presentation of an operator $A$
\[
  \ket{x} \quad\longmapsto\quad \sum_{k=0}^{2^n}\alpha_{k,x}\ket{k}.
\]
Instead of listing all of the $\alpha_{k,x}$'s exhaustively, the
operator is written as a triple $(m,P,\phi)$, where $m$ is an integer
and $P$ and $\phi$ are integer polynomials such that $A$ is
\[
  \ket{x}\quad\longmapsto\quad
  \frac1{\sqrt{2^n}}\sum_{k=0}^{2^n-1}e^{\frac{2i\pi\cdot P_k(x)}{2^m}}\ket{\phi_k(x)}.
\]
For many ``interesting'' operators, the polynomials $P$ and $\phi$
form a more compact representation than the array of the
$\alpha_{k,x}$'s. Furthermore, this representation is closed
under functional composition and
Kronecker product, making it ideal for reasoning on quantum circuits.


One limitation of Amy's path sum is however that one cannot check
\emph{parameterized family} of circuits: akin to a model-checker, the
path-sum mechanism can only handle one \emph{fixed} circuit.

\paragraph{The Domain-Specific Language Qbricks}

In \cite{chareton2021automated} we propose Qbricks, a specification
and verification framework for quantum programs based on path-sums.
Before Qbricks, frameworks for proving properties of quantum programs
where either handling parametricity at the expense of automation
\cite{mahmoud2014formal, liu2019formal, paykin2017qwire,
  rand2018reqwire, hietala2021proving} or automated at the expense of
parametricity \cite{amy2018large-scale, amy2019formal,
  kissinger2015quantomatic}.
Our contribution to the field consists in reconciliating parametricity
and automation, with the development of a deductive verification
framework based on \emphidx{parameterized path sums}. 

Qbricks is embedded in Why3, inheriting its specification and
deductive verification features. The formalization comes with a
domain-specific language for circuit manipulation and a logic library
for manipulating path-sums. This gives a handle for
reasoning in terms of the WhyML language: our path-sums are naturally
parameterized.

Qbricks' domain specific language is following the qPCF's strategy for
circuit construction \cite{paolini2017qpcf} ---although qPCF is mainly
a theoretical exploration of dependent type systems in this
context. Unlike \quipper where wires are qubits that can be
instantiated and manipulated as variable and where circuits are
functions on qubits, circuits in Qbricks are opaque objects
manipulated with a few combinators: elementary gates, sequential and
parallel composition. 

The framework has been used to prove the first verified, parametric
implementation of the quantum part of Shor’s factoring algorithm
\cite{shor94algorithms,beauregard2003circuit}, including both the
polynomial complexity of the circuits and the probability
requirements. We also experimented with Grover \cite{grover1996fast}
and the quantum phase estimation subroutine (QPE)
\cite{kitaev1995quantum}. Our method
\cite[Sec\,8]{chareton2021automated} achieves a high level of proof
automation (96\% on Shor) and significantly reduces proof effort
(factor 13.6x compared with \cite{liu2019formal} on Grover, factors
7.7x and 6.4x compared with \cite{hietala2021proving} on respectively
QPE and Grover).

\paragraph{Example of Parametric Path-Sums}

Let us present an example to illustrate the interplay between the
language and the parametric path-sums. Consider the family of circuits
defined as an even number of Hadamard gates
\[
  \underbrace{\xymatrix@=2ex{
    \ar@{-}[r]&*++[F]{H}\ar@{-}[r]&*++[F]{H}\ar@{-}[r]&\cdots\ar@{-}[r]&*++[F]{H}\ar@{-}[r]&
  }}_{n\text{ gates ($n$ even)}}
\]
We can give a specification for a program generating such a circuit
family by
\begin{description}
\item[Precondition] $n\geq 0$ is even
\item[Postcondition] $C_n$ sends $\ket{x}$ to $\ket{x}$ and $C_n$
  consists of $n$ gates.
\end{description}
This contrived example is typical for the specification of a quantum
algorithm:
\begin{compactitem}
\item the description of the circuit family is parameterized by a
  classical parameter (here, the non-negative integer $n$);
\item The precondition imposes both constraints (here, the evenness of
  $n$) and soundness conditions (here, the non-negativeness of $n$) on
  the parameters;
\item The postcondition can both refer to the semantics of the circuit
  result and to its form and shape (here, its size).
\end{compactitem}
Regular path-sums are not adequate for representing the semantics of
the circuit family since the behavior of each circuit in the family
depends on its size: the path sum is
\[
  \ket{x}\mapsto
  \left\{
    \begin{array}{ll}
      \frac1{\sqrt{2^0}}\sum_{k=0}^{2^0-1}e^{2i\pi\cdot
      0}\ket{x}&\text{when $n$ is even}
      \\
      \frac1{\sqrt{2^1}}\sum_{k=0}^{2^1-1}e^{2i\pi\cdot
      \frac{kx}2}\ket{k}
      &\text{when $n$ is odd.}
    \end{array}
  \right.
\]
Compared to Amy's proposal, the phase and boolean
polynomials of path-sums are generalized to
generic, parameterized terms. In the case
of our example, the path-sum becomes
\[
  \ket{x}\mapsto
  \frac1{\sqrt{2^{n\%2}}}\sum_{k=0}^{2^{n\%2}-1}e^{2i\pi\cdot\frac{(n\%2)kx}2}\ket{\iftermx{\texttt{even}(x)}{x}{k}}.
\]
With Qbricks' framework, such a path-sum can be defined in the
language and reasoned upon in the logic.

\begin{table}[p]
  \centering
  \scalebox{.8}{\begin{minipage}{1.25\textwidth}
        \begin{mylife*}
          \begin{compactdesc}
            \itempaper{green2013quipper} \mycitegreenquipper
            \itempaper{green2013introduction} \mycitegreenintroduction
            \itempaper{smith2014quipper} \mycitesmithquipper
            \itempaper{valiron2015programming} \mycitevalironprogramming
            \itempaper{valiron2016generating} \mycitevalirongenerating
            \itempaper{valiron2017programmer} \mycitevalironprogrammer
            \itempaper{scherer2017concrete} \myciteschererconcrete
            \itempaper{allouche2018reuse} \myciteallouchereuse
            \itempaper{valiron2018formal} \mycitevalironformal
            \itempaper{brugiere2019synthesizing} \mycitebrugieresynthesizing
            \itempaper{brugiere2020quantum} \mycitebrugierequantum
            \itempaper{brugiere2020householder} \mycitebrugierehouseholder
            \itempaper{chareton2021automated} \mycitecharetonautomated
            \itempaper{borgna2021hybrid} \myciteborgnahybrid
            \itempaper{chareton2021formal} \mycitecharetonformal
          \end{compactdesc}
          \caption{Personal publications related to Chapter~\ref{ch:compil}.}
          \label{tab:publis-compil}
       \end{mylife*}
     \end{minipage}}
\end{table}


\clearpage{\thispagestyle{empty}\cleardoublepage}

\chapter[Semantics of Quantum Lambda-Calculi]{%
  Semantics of Quantum\\*Lambda-Calculi%
}
\label{ch:sem}
\rhead{Semantics}

Semantics can be considered the origin of all the formal tools
developed to analyze, certify, and verify programming languages
\cite{leeuwen1990formal}. It
consists in a formal description of the essence of programs aiming
at unearthing the structures underlying the capabilities of
programming language. Semantics draws links between the behavior of
programs---how they evolve and interact with their environment---,
their logical properties---how they are structured---, and the result
of their action---what they compute.

For classical, regular programming languages, semantics---and formal
methods---have been around for more than half a century. Based on
powerful frameworks such as category theory or the Curry-Howard
isomorphism \cite{curry1958combinatory,howard1980formulae},
semantics for classical programming languages gave birth
to a range of fine-grained analysis techniques of programs.

For classical programming languages, the underlying mathematical
structures are typically set-based, discrete structures
\cite{stoy1977denotational}.
Although the analysis of quantum programming languages can rely on and
adapt some of the work done in the classical setting,
several aspects fundamentally differ and require novel techniques. In
particular, in quantum computing, one deals with two kinds of objects:
regular, duplicable objects and quantum, non-duplicable
objects. Moreover, the canonical mathematical representation of
quantum states is based on vector spaces and operator algebras.

Developing a semantics for a quantum programming language then
requires a novel approach. In this chapter, we present our
contribution to the field, focusing on the quantum lambda calculus
and its extension as a circuit-description language.
\begin{itemize}
\item Section~\ref{sec:ll-tqlc} summarizes the base of our main
  approach: the fact that linear logic forms a suitable framework for
  a quantum type system, following Section~\ref{sec:lin-typ-sys}.
\item Section~\ref{sec:ll-cpm} discusses the procedure we followed for
  building a denotational semantics accounting for both quantum
  \emph{and} duplicable data. This kind of semantics interprets
  programs as functions. The semantics we propose is strongly inspired
  by quantitative semantics of linear logics \cite{pagani2014applying}.
\item Section~\ref{sec:goi} focuses on a complementary approach: the
  Geometry of Interaction. This technique provides executable
  semantics based on token-based automata. We show how quantum lambda-terms
  can be regarded as folded quantum circuits; the semantics gives a
  procedure for ``running'' them
  \cite{lago2015parallelism,lago2017geometry}.
\item Finally, Section~\ref{sec:red-ql} briefly discusses one of our
  recent results: a categorical semantics for \protoquipper, a
  circuit-description extension of the quantum lambda calculus
  supporting dynamic lifting: the ability to govern circuit generation
  based on the result of previous measurements \cite{lee2022concrete}.
\end{itemize}

\section{Linear Logic and Typed Quantum Lambda Calculus}
\label{sec:ll-tqlc}

As discussed in Section~\ref{sec:lin-typ-sys}, a natural logical framework for a
type system for quantum computation is linear logic. In this section,
we briefly introduce the logic and how it lays out a natural type
system for the quantum lambda calculus.

\subsection{Linear Logic}
\label{sec:ll}

The logic formula from \emphidx{linear logic} (LL)\index{LL} that we
shall be considering are
\[
  A,B~~{:}{:}{=}~~\alpha\bor A^\bot\bor\tunit\bor\bot\bor\punit\bor\top\bor A\tensor B\bor A\parr B\bor
  A\oplus B\bor A\& B\bor {!A}\bor ?A,
\]
where $\alpha$ ranges over a set of atomic formulas.
In linear logic there are two pairs of conjunctions/disjunctions:
a multiplicative version: $\tensor$ (with unit $\tunit$)
and $\parr$ (with unit $\bot$), and an additive version: $\&$ (with unit
$\top$) and $\oplus$ (with unit $\punit$). The connective
$(-)^\bot$ stands for the linear negation. It is extended to an
involution on formulas where $(A^\bot)^\bot = A$ as follows:
\begin{align*}
  \tunit^\bot &=\bot
  &({A\tensor B})^\bot &= A^\bot\parr B^\bot,
  & (!A)^\bot &= ?(A^\bot),\\
  \punit^\bot&=\top & (A\oplus B)^\bot &= A^\bot\&B^\bot,
\end{align*}
emphasizing the fact that ${\tensor}/{\parr}$, ${\oplus}/{\&}$, $!(-)/?(-)$,
$\tunit/\bot$ and $\punit/\top$ 
are \emphidx{dual} connectors. 
Intuitively, a negated formula stands for an \emph{hypothesis}
(i.e. an ``input'') while a non-negated formula for a
\emph{conclusion} (i.e. an ``output''). Following the intuition that
$\parr$ is a disjunction, we define a macro
$A\loli B = A^\bot\parr B$, then representing a (multiplicative)
linear implication.
In light of the duality of connectives, we can give a meaning to the
two last connectives, the modalities $!(-)$ (the
\emphidx{exponential}) and $?(-)$.  Indeed, in linear logic, formulas
are \emphidx{linear} by default: they correspond to resources that
have to be used exactly once. The connectives $!(-)$ and $?(-)$ make
it possible to relax this constraint: $!A$ stands for a duplicable and
erasable \emph{output} of type $A$, while $?A$ stands for a duplicable
and erasable \emph{input} of type $A$.

\begin{example}\label{ex:loli-dup-A}
  According to the intuitive meaning we gave to the linear logic
  connectives, without additional axioms 
  the formula $\alpha\loli \alpha$ should then be
  correct, while $\alpha\loli (\alpha\tensor \alpha)$ should not. On
  the other hand, ${!\alpha}\loli (\alpha\tensor \alpha)$ should be
  valid, since $!A$ is a ``duplicable'' resource.
\end{example}

As in classical logic, linear logic features a notion of
\emphidx{sequent}, that is, a sequence of formulas, denoted with
$\vdash A_1,\ldots, A_n$. We call ``$\vdash$'' a
\emphidx{turnstyle}. Generic sequences of formulas are denoted with
$\Delta,\Gamma,\ldots$. If $\Delta=x_1:A_1,\ldots,x_n:A_n$, and if
$\square$ is a unary connective, we write $\square\Delta$ for the
sequence $x_1:{\square}{A_1},\ldots,x_n:{\square}A_n$.

We say that a sequent is \emph{valid} if
it can be derived from the following rules.
\[
  \infer[\text{($\tunit$)}]{\vdash \tunit}{}
  \quad
  \infer[\text{($\bot$)}]{\vdash \Gamma,\bot}{\vdash\Gamma}
  \quad
  \infer[\text{($\top$)}]{\vdash \Gamma,\top}{}  
  \quad
  \infer[\text{(ax)}]{\vdash A,A^\bot}{}
  \quad
  \infer[\text{(ex${}_\sigma$)}]{\vdash A_{\sigma(1)},\ldots,A_{\sigma(n)}}{
    \vdash A_1,\ldots A_n
  }
\]
\[
  \infer[\text{(cut)}]{\vdash \Gamma,\Delta}{
    \vdash \Gamma, A
    &
    \vdash \Delta, A^\bot
  }
  \qquad
  \infer[\text{($\tensor$)}]{\vdash \Gamma,\Delta, A\tensor B}{
    \vdash \Gamma, A
    &
    \vdash \Delta, B
  }
  \qquad
  \infer[\text{($\parr$)}]{\vdash \Gamma, A\parr B}{
    \vdash \Gamma, A, B
  }
\]
\[
  \infer[\text{($\oplus_1$)}]{\vdash \Gamma,A\oplus B}{
    \vdash \Gamma, A
  }
  \qquad
  \infer[\text{($\oplus_2$)}]{\vdash \Gamma, A\oplus B}{
    \vdash \Gamma, B
  }
  \qquad
  \infer[\text{($\&$)}]{\vdash \Gamma,A\& B}{
    \vdash \Gamma, A
    &
    \vdash \Gamma, B
  }  
\]
\[
  \infer[\text{(p)}]{\vdash ?\Gamma,!A}{\vdash ?\Gamma, A}
  \qquad
  \infer[\text{(d)}]{\vdash \Gamma,?A}{\vdash \Gamma, A}
  \qquad
  \infer[\text{(w)}]{\vdash \Gamma,?A}{\vdash \Gamma}
  \qquad
  \infer[\text{(c)}]{\vdash \Gamma,?A}{\vdash \Gamma, ?A,?A}
\]
where $\sigma$ is a permutation over $\{1,\ldots n\}$.
Note that there is no rule for the unit $\punit$.
By abuse
of notation the rule (ex${}_\sigma$) is left implicit in the
description of proofs. From the rules one can check that $\tensor$
indeed behaves like a conjunction while $\parr$ behaves like a
disjunction. One can also see how $\otimes/\&$ has a multiplicative
flavor ---contexts are disjoints--- while $\oplus/\&$ has an
additive flavor ---contexts are shared---.

\begin{remark}\label{rem:necessary-exc}
  Note that the position of the formulas in a sequent is essential,
  as otherwise the following proof is ambiguous:
  \begin{equation}
    \infer[\text{(cut).}]{\vdash A,A^\bot}{
      \infer[\text{(ax)}]{\vdash A,A^\bot}{}
      &
      \infer[\text{(ax)}]{\vdash A,A^\bot}{}
    }
    \label{eq:ambiguous-pf}
  \end{equation}
  Which pair $A,A^\bot$ was canceled out by the (cut)-rule?
\end{remark}

\begin{remark}\label{rem:mix}
  Two derivable rules are often added; they are specially useful when
  considering proof-nets.
  \[
    \infer[\text{(empty)}]{\vdash}{} \qquad
    \infer[\text{(mix)}]{\vdash \Gamma,\Delta}{\vdash \Gamma&\vdash
      \Delta}
  \]
  in which case we refer to the logic as LL+mix.
\end{remark}

\begin{example}\label{ex:lin-mp}
  A linear \simpleidx{Modus-Ponens} can be derived as follows, where we add a
  dummy rule for highlighting the unfolding of $\multimap$:
  \begin{equation}
    \infer[\text{(cut)}]{\vdash B}{
      \infer*[\pi_1]{\vdash A}{}
      &
      \infer[\text{(cut)}]{\vdash B,A^\bot}{
        \infer*[\pi_2]{\vdash A\multimap B}{}
        &
        \infer[\text{(unfold)}]{\vdash (A\multimap B)^\bot,B,A^\bot}{
          \infer[\text{($\tensor$)}]{\vdash A\tensor B^\bot,B,A^\bot}{
            \infer[\text{(ax)}]{\vdash A,A^\bot}{}
            &
            \infer[\text{(ax)}]{\vdash B,B^\bot}{}
          }
        }
      }
    }
    \label{eq:pf-lin-mp}
  \end{equation}
  In the proof of Eq.~\eqref{eq:pf-lin-mp}, we omitted a call to the
  rule (ex${}_\sigma$) at the (unfold) position: a full proof with
  sequents is potentially verbose with many bureaucratic permutations
  of formulas.
\end{example}

\begin{remark}\label{rem:ill}
  The turnstyle notation for sequent can be extended by identifying
  $\Delta\vdash\Gamma$ and $\vdash \Delta^\bot,\Gamma$. The notation
  adds the meta-information that $\Delta$ is to be regarded as an
  input and $\Gamma$ as an output. This triggers one interesting
  variant of linear logic for this chapter: \emphidx{intuitionistic
    linear logic} (ILL)\index{ILL} \cite{troelstra1992lectures}. In
  ILL, we consider special sequents with exactly one formula as
  conclusion: sequents are of the form $\Delta\vdash A$. Additionally,
  the negation $(-)^\bot$ is not anymore an involution.
\end{remark}

Intuitionistic logic can be faithfully encoded inside linear
logic~\cite[Sec 5.1]{girard87linear}. Regular, classical simply-typed
programs can therefore be mapped to proofs of linear logics;
cut-elimination then corresponds to program evaluation.  Passing
through a linear-logic encoding gives a fine-grained handle on the
choice of evaluation strategy through the placement of the exponential
modality~\cite{simpson2005reduction}. The intuition is to consider a
term typed with $!A$ as a \emphidx{thunk}: a frozen computation. It
can be duplicated (with contraction), erased (with weakening), and run
with dereliction. Historically, there are two canonical encodings
building on this intuition: one implementing call-by-value, where
$-\to -$ is mapped to $!(-\loli -)$ the other one call-by-name, where
$-\to-$ is mapped to $(!-)\loli B$.

We conclude this section by mentioning interesting \emph{fragments}
of linear logic, each one with a intuitionistic and a classical
variant. The first one can be inferred from Example~\ref{ex:lin-mp}:
\emphidx{Multiplicative Linear Logic} (MLL)\index{MLL}, where formulas
are restricted to $\tensor$ and $\parr$ (and $\loli$). This logical
fragment is \emph{purely linear}. There is then
\emphidx{Multiplicative Exponential Linear Logic} (MELL)\index{MELL},
where formulas consists of $\tensor$, $\parr$ together with the
\emphidx{modalities} ``$!$'' and ``$?$''. These are the two fragments that we
shall be considering in this paper. We can nonetheless mention the
(strictly linear) fragment MALL\index{MALL} of
\emphidx{Multiplicative, Additive Linear Logic} with $\otimes/\parr$
and $\oplus/\&$.

\subsection{Quantum Lambda Calculus and Linear Logic}
\label{sec:qlc-ll}

We claimed in \ref{sec:lin-typ-sys}
that linear logic forms a natural framework
for a type system of quantum lambda calculi. In this section, we
present the instantiation described in \cite{pagani2014applying}: it
will serve as a support for the rest of the discussion in this
chapter.

The language is defined as follows.
\begin{alignat}{10}
  M,N,P
  ~~{:}{:}{=}~~
  & x\bor \lambda x.M\bor MN\bor \label{eq:qlc-lc}
  \\
  & \tuple{M,N} \bor \lettermx{\tuple{x,y}}{M}{N} \bor
     \unitterm \bor \lettermx{\unitterm}{M}{N}\bor
     \label{eq:qlc-pair}
  \\
  & \ttrue \bor \ffalse \bor \iftermx{M}{N}{P}\bor
     \label{eq:qlc-bool}
  \\ 
  & U \bor \qinit \bor \meas \bor
     \label{eq:qlc-q}
\intertext{ It consists of a regular
  lambda calculus~\eqref{eq:qlc-lc}, extended with: pairing
  constructs~\eqref{eq:qlc-pair}, where $\tuple{M,N}$ stands for
  the pair of $M$ and $N$ and $\unitterm$ is the unit-term;
  Boolean values and tests~\eqref{eq:qlc-bool}; constants for
  manipulating qubits~\eqref{eq:qlc-q}, where $U$ ranges over a
  fixed set of unitary maps. The language can also be extended with
  recursion, using the following construct:}
  & \letrectermx{f~x}{M}{N}.
     \label{eq:qlc-rec}
\end{alignat}

The type system for the language is as follows.
\begin{alignat}{10}
  A,B
  ~~{:}{:}{=}~~
  & \qbit\bor\bit\bor A\loli B\bor A\tensor B\bor \unittype\bor{!A}. 
  \label{eq:qlc-typ}
\end{alignat}
It consists of two constant types \qbit, for representing qubits, and
\bit, for the Boolean values {\ttrue} and {\ffalse}, and type constructors:
for pairing ($A\tensor B$), functions ($A\loli B$), unit-type
$\unittype$ for representing $\unitterm$, and duplicable elements
($!A$). We use the same notations as MELL to highlight the
relationship with the logic. The tensor is associative to the right:
$A\tensor B\tensor C = A\tensor(B\tensor C)$. We write $A^{\tensor n}$
for the $n$-th tensor of $A$.
Finally, we define a notion of \emphidx{value}: 
\[
  V,W\quad{:}{:}{=}\quad
  x\bor\ttrue\bor\ffalse\bor\lambda x.M\bor\tuple{}\bor\tuple{V,W}.
\]

We consider terms to be implicitly typed: every subterm comes with a
type. A \emph{typing judgment} is a triple written
$\Delta\vdash M:A$, where $M$ is a (typed) term, $A$ is a type and
$\Delta$ is an unordered list of typed variables:
$\Delta = x_1:A_1,\ldots,x_n:A_n$. A typing judgment is \emph{valid}
if it can be derived from the typing rules presented below. We also
require that whenever $\Delta\vdash M:A$ is valid, then $A$ is the
implicit type of $M$.

The typing rules follow the proof rules of \emph{intuitionistic}
linear logic (as discussed in Remark~\ref{rem:ill}).
The typing rules for the core
lambda calculus of~\eqref{eq:qlc-lc} are as follows.
\[
\infer[\text{(ax)}]{!\Delta, x:A\vdash x:A}{}
\qquad
\infer[\text{($\loli_I$)}]{\Delta\vdash\lambda x.M:A\loli B}{
  \Delta,x:A\vdash M:B
}
\]
\[
\infer[\text{($\loli_E$)}]{!\Delta,\Gamma_1,\Gamma_2\vdash MN:B}{
  !\Delta,\Gamma_1\vdash M:A\loli B
  &
  !\Delta,\Gamma_2\vdash N:A
}
\]
Note how contraction is included inside the ($\loli_E$)-rule: this
will be the case for every branching rule. Also note how a cut-rule
is implicitly used, as it will be the case for every elimination
rules.

The pairing constructs correspond to the proof rules of
$\tensor$ and $\tunit$.
\[
\infer[\text{($\tensor_I$)}]{!\Delta,\Gamma_1,\Gamma_2\vdash
  \tuple{M,N}:A\tensor B}{
  !\Delta,\Gamma_1\vdash M:A
  &
  !\Delta,\Gamma_2\vdash N:B
}
\qquad
\infer[\text{($\unittype_I$)}]{!\Delta\vdash\unitterm:\unittype}{}
\]
\[
\infer[\text{($\tensor_E$)}]{!\Delta,\Gamma_1,\Gamma_2\vdash
  \lettermx{\tuple{x,y}}{M}{N}:C}{
  !\Delta,\Gamma_1\vdash M:A\tensor B
  &
  !\Delta,\Gamma_2,x:A,y:B\vdash N:C
}
\]
\[
\infer[\text{($\unittype_E$)}]{!\Delta,\Gamma_1,\Gamma_2\vdash
  \lettermx{\tuple{}}{M}{N}:C}{
  !\Delta,\Gamma_1\vdash M:\unittype
  &
  !\Delta,\Gamma_2\vdash N:C
}
\]
If contraction is included inside branching rules, weakening is
handled at axiom rules. The two remaining rules for manipulating
modalities ---\emphidx{dereliction} and \emphidx{promotion}---
each feature an explicit
rule but with a caveat. We force dereliction to only happen at a leaf
of the typing derivation, in order to ensure uniqueness of typing
derivation. For technical convenience, we also restrict duplication to
function-types. Finally, promotion is constrained to \emph{values}
this is in line with the call-by-value operational semantics of the
languages, presented below.
\[
\infer[\text{(axd)}]{!\Delta, x:{!(A\loli B)}\vdash x:A\loli B}{}
\quad
\infer[\text{(p)}]{!\Delta\vdash V:{!(A\loli B)}}{
  !\Delta\vdash V:A\loli B & V \text{a value}
}
\]
If the language can be expanded with additive types ---and even
inductive types--- \cite{pagani2014applying}, in this chapter we only
consider Boolean values, which, albeit weaker, already capture much of the
intricacy of the additives.
\[
\infer[\text{($\ttrue,\ffalse$)}]{!\Delta\vdash \ttrue,\ffalse:\bit}{}
\quad
\infer[\text{(if)}]{!\Delta,\Gamma_1,\Gamma_2\vdash
  \iftermx{P}{M}{N}:C}{
  !\Delta,\Gamma_1\vdash P:\bit
  &
  !\Delta,\Gamma_2\vdash M,N:C
}
\]
Finally, the constants for manipulating qubits are typed as follows.
\[
  \infer{!\Delta\vdash\qinit:\qbit\loli \bit}{}
  \quad
  \infer{!\Delta\vdash\meas:\bit\loli \qbit}{}
  \quad
  \infer{!\Delta\vdash U:\qbit^{\otimes n}\loli \qbit^{\otimes n}}{}
\]
In the rule for $U$, the number $n$ stands for the arity of $U$.

\begin{remark}
  Note how there is no need for an exchange rule similar to
  (ex${}_\sigma$), since types are indexed with variables. This is one
  of the solutions to the problem of bureaucracy; the other one is to
  use proof-nets, discussed in Section~\ref{sec:goi}.
\end{remark}

The language is equipped with an operational semantics in the form of
the abstract machine described in Section~\ref{sec:qext-lc}: a program
is a triple $[Q,L,M]$, where $Q\in\qH^{\tensor n}$ is a normalized
vector of dimension $2^n$, $M$ is a term with $n$ free variables
$x_1,\ldots,x_n$ and $L$ is a bijection
$\{x_1,\ldots,x_n\}\to\{1,\ldots,n\}$. The variables $x_i$ represent
qubits inside the term $M$. A program is well-typed of type $A$,
written $[Q,L,M]:A$, whenever
\[
  x_1:\qbit,\ldots,x_n:\qbit\vdash M:A
\]
is valid.

Programs are equipped with a \emph{probabilistic} rewrite system
$(\to_p)$ ($p\in[0,1]$), extending the call-by-value evaluation of (regular)
lambda calculus.

An \emphidx{applicative context} is a ``term with a hole'' pointing
where evaluation can happen. For instance, in our setting we do not
allow rewriting under lambdas. Applicative contexts are defined
according to the following grammar.
\[
  \begin{array}{ll}
  C[-]
  ~~{:}{:}{=}~~
  & [-]\bor MC[-]\bor C[-]V\bor\tuple{C[-],N}\bor\tuple{V,C[-]} \bor
  \\
  & \lettermx{\tuple{x,y}}{C[-]}{N} \bor
     \lettermx{\unitterm}{C[-]}{N}\bor
  \\
    & \iftermx{C[-]}{N}{P}.
  \end{array}
\]
The rewrite system consists of two parts: the ``classical'' part, not
interacting with $Q$ and $L$, and the ``quantum'' part, whose goal is
to emulate the interaction with the quantum coprocessor. We define a
first rewrite system $\to_c$ on terms characterizing the classical
part of the evaluation, as follows.
\[
  \begin{array}{lll}
    C[(\lambda x.M)V]&\to_c& C[M[x:=V]]
    \\
    C[\lettermx{\tuple{}}{\tuple{}}{M}]&\to_c&C[M]
    \\
    C[\lettermx{\tuple{x,y}}{\tuple{V,W}}{M}]&\to_c&C[M[x:=V,y:=W]]
    \\
    C[\iftermx{\ttrue}{M}{N}]&\to_c&C[M]
    \\
    C[\iftermx{\ffalse}{M}{N}]&\to_c&C[N]
    \\
    C[\letrectermx{f~x}{M}{N}]&\to_c&C[N[f:=\lambda
                                    x.\letrectermx{f~x}{M}{M}]]
  \end{array}
\]
We can then define the rewrite system on programs as first 
\[
  M\to_c N \qquad\text{implies}\qquad [Q,L,M] \to_1 [Q,L,M']
\]
for the classical part, and for the quantum part, assuming
$Q\in\qH^{\tensor n}$ and $z$ is fresh:
\begin{align*}
  [Q,L,C[\qinit\,\ttrue]]
  &{}\to_1[Q\tensor\ket1,L\cup\{z\mapsto n+1\}, C[z]],
  \\
  [Q,L,C[\qinit\,\ffalse]]
  &{}\to_1[Q\tensor\ket0,L\cup\{z\mapsto n+1\}, C[z]],
  \\
  [Q,L,C[U\,x]]
  &{}\to_1[(U\tensor I)Q,L, C[x]]
  && \hspace{-4ex}\text{if $U$ is unary and $L(x)=1$,}
  \\
  [Q,L,C[U\,\tuple{x,y}]]
  &{}\to_1[(U\tensor I)Q,L, C[\tuple{x,y}]]
  && \hspace{-4ex}\text{if $U$ is binary, $L(x)=1$, $L(y)=2$,}
  \\
  [Q,L,C[\meas\,x]]
  &{}\to_{|\alpha_b|^n}[Q_b,L, C[b]],
\end{align*}
when $Q=\alpha_{\ffalse}\ket0\tensor Q_\ffalse +
\alpha_{\ttrue}\ket1\tensor Q_\ttrue$ with $Q_\ffalse$ and
$Q_\ttrue$ normalized.

The language satisfies the usual safety properties: subject reduction
and progress.

\begin{example}\label{ex:dup-qubit}
  The type system is designed to not allow the duplication of quantum
  bits. The type $!\qbit$ is therefore empty: there is no closed term
  $M$ such that $\vdash M:{!\qbit}$. This property heavily relies on
  the constraint we placed on the promotion rule (p): one can only
  duplicate \emph{values}. It is however
  possible to build a duplicable term of type
  $!(\unittype\loli\qbit)$, as for instance
  \[
    \vdash \lambda x.(\lettermx{\tuple{}}{x}{H(\qinit\,\ffalse)}):{!(\unittype\loli\qbit)}
  \]
  is derivable. We come back to this example in Example~\ref{ex:dup-qubit2}
\end{example}

\subsection{Cut-elimination and Curry-Howard Isomorphism}
\label{sec:cut-elim}

Besides the cut-rule (and (ex${}_\sigma$), which does not count), the
proof rules of linear logic
are structural: they construct a sequence of formulas out of
more primitive ones. One important question in logic is, given a proof
$\pi$, whether one can rewrite it to obtain a \emphidx{cut-free}
proof, using only structural rules (and exchange rules). This problem
is known as \emphidx{cut-elimination}.
In the case of LL (and LL+mix), one can equip the set of proofs
with a strongly normalizing and confluent rewriting system whose
normal forms are precisely cut-free proofs~\cite{girard89proofs}. For the sake of
the presentation, we only discuss two of them: the interaction between
(cut) and (ax)
\[
  \infer[\text{(cut)}]{\vdash A}{
    \infer[\text{(ax)}]{\vdash A,A^\bot}{}
    &
    \infer*{\vdash A}{\pi}
  }
  \qquad
  \to
  \qquad
  \infer*{\vdash A}{\pi}  
\]
and the rewriting of a cut between ($\tensor$) and ($\parr$)
\begin{equation}\label{eq:rw-tens-par}
  \infer[\text{(cut)}]{\vdash \Delta_1,\Delta_2,\Gamma}{
    \infer[\text{($\otimes$)}]{\vdash \Delta_1,\Delta_2,A\otimes B}{
      \infer*{\vdash \Delta_1,A}{\pi_1}
      &
      \infer*{\vdash \Delta_2,B}{\pi_2}
    }
    &
    \infer[\text{($\parr$)}]{\vdash \Gamma,A^\bot\parr B^\bot}{
      \infer*{\vdash \Gamma,A^\bot,B^\bot}{\pi_3}
    }
  }
  \quad
  \to
  \quad
  \infer[\text{(cut)}]{\vdash \Delta_1,\Delta_2,\Gamma}{
    \infer*{\vdash\Delta_1,A}{\pi_1}
    &
    \infer[\text{(cut)}]{\vdash \Gamma,\Delta_2, A^\bot}{
      \infer*{\vdash\Delta_2,B}{\pi_2}
      &
      \infer*{\vdash\Gamma,A^\bot,B^\bot}{\pi_3}
    }
  }
\end{equation}

Remembering that the linear implication $A\loli B$ is built as
$A^\bot\parr B$, note how the rewrite rule shown in
Eq.~\eqref{eq:rw-tens-par} corresponds to a form of
$\loli$-elimination. This in fact precisely corresponds to the
beta-rule of the lambda calculus, such as in the presentation of
Section~\ref{sec:lc} (Although the relation is slightly non-trivial
\cite{regnier1992phd} and linked to
explicit substitution \cite{cosmo1997strong,accattoli2015proof})

This \emphidx{Curry-Howard correspondence} for linear logic has been
analyzed by many authors \cite{abramsky93computational,
  benton92linear,benton93term,bierman93intuitionistic,wadler93syntax}.
Formalizing the intuition drawn in Example~\ref{ex:loli-dup-A}, type systems based
on linear logic make it possible to specify whether resources are
used only once: a function of type $A\loli B$ is guaranteed to use its
argument exactly once, while the type $\alpha\loli (\alpha\tensor \alpha)$ is
empty. Many refinements or extensions are possible. For instance, one
can relax the linearity constraint and allow weakening
---i.e. erasure--- of linear resources to get \emphidx{affine linear logic}~\cite{troelstra1992lectures}.
One can use the exponential
modality to characterize implicit computational
complexity~\cite{asperti2002intuitionistic,
  baillot04light,girard1998light,lafont04soft,lago2010quantum}, or add
annotations to exponentials to keep track of the number of uses of a
particular resource with bounded linear
logic~\cite{girard1992bounded,lago2010bounded} or discuss differential privacy \cite{reed2010distance,gaboardi2013linear}. 
And, as exemplified in~\cite{pagani2014applying} and the
sketch of Section~\ref{sec:qlc}, one
can distinguish between duplicable and non-duplicable data and apply
it to quantum computation and the manipulation of qubits.

\section{A Denotational Semantics}
\label{sec:ll-cpm}

This section is devoted to the study of a \emph{denotational}
semantics\index{denotational semantics}
for the quantum lambda calculus. A denotational semantics is
an interpretation of programs as mathematical functions, composition
of programs corresponding to function composition. Denotational
semantics are expressive tools to bridge programming languages with
logical theories through the Curry-Howard correspondence. By
exhibiting the compositional structures underlying a language, a
denotational semantics validates the soundness of its design.

One of the challenge in semantics is the \emph{compatibility} of quantum
and classical features when intertwined, as exemplified in the quantum
lambda calculus. On the one hand, the typical semantics for quantum
computation relies on linear maps and positive operators in finite
dimension. On the other hand, classical information should be
duplicable, therefore requiring some notion of
non-linearity. Finally, the mix of quantum information within
classical datatypes such as lists entails non-standard objects such as
infinite datatypes of list of qubits, hinting at the need for infinite
dimensional vector spaces.

In this section, we present our solution for such a denotational
semantics. We follow an iterative approach, starting with a quick
review of the previous existing approaches
(Section~\ref{ssec:cpm-before}). We then present our work: a simple
semantics based on completely positive maps
(Section~\ref{ssec:cpm-basic}) to which we progressively add constructs:
additives (Section~\ref{ssec:cpm-additive}), recursive datatypes
(Section~\ref{ssec:cpm-rec}), and duplication
(Section~\ref{ssec:cpm-dup}). We conclude with a discussion on
other possible approaches (Section~\ref{ssec:cpm-discussion}).

\begin{mylife}
  This section is the result of a long gestation. It started at the
  end of my Ph.D thesis with the design of a CPM-based semantics for a
  purely linear quantum lambda calculus \cite{selinger06fully}. The
  adjunction of recursive datatypes and duplication was then a
  roadblock for a long time: how to include them in a sound way within
  a finite-dimensional setting? The knot was untangled in 2014 with
  the development of general techniques for quantitative semantics of
  linear logic and semantics of probabilistic PCF
  \cite{ehrhard2014probabilistic}. With Michele Pagani, we were able
  to port these technique to the quantum case and answer the problem
  \cite{pagani2014applying}.
\end{mylife}

\subsection{Background on Denotational Semantics}
\label{ssec:cpm-before}

Modeling higher-order languages have historically been done using
\emph{domains} and continuous lattices. Algebraic effects, such as
probabilities, can be handled with the use of a suitable
monad~\cite{jones1990phd,gierz2003continuous} (although this
requires some care \cite{graham1987closure,jung1998troublesome}).
On the other hand, linear algebra and functional analysis have been
from the very beginning an extensional target model for linear logic.
Originally designed for System F \cite{girard1986system}, coherent
spaces ---at the root of the design of linear logic
\cite{girard87linear}--- have soon been generalized to support algebraic
effects \cite{girard1999coherent, ehrhard2002kothe, girard2004between,
  danos2011probabilistic, ehrhard2011computational}. The other
original semantics for linear logic, quantitative
domains~\cite{girard1988normal}, has also spurred many rich algebraic
models: Fock spaces \cite{blute94fock}, Hopf algebras
\cite{blute96hopf}, Köthe spaces~\cite{ehrhard2002kothe}, finiteness
spaces~\cite{ehrhard2005finiteness}, \textit{etc}. It makes it
possible to prove fine-grained quantitative properties of
programs~\cite{laird2013weighted}.

From a categorical perspective, building a model of linear logic
requires one to accommodate four components: the multiplicative fragment,
the additive fragment, the modalities and the involution ---the last
one being optional if the target is \emph{intuitionistic} linear
logic. Depending on the computational objective, one can also ask for
traces, and/or fixpoints, and/or allow affine behavior,
\textit{etc}. 

As the type system of quantum lambda calculus is based on linear
logic, it is reasonable to look for a suitable algebraic model of
linear logic capturing quantum effects. One of Girard's goals is to
bridge physics and logic: Instead of trying to organically extract
logical structure out of positive operators and quantum observables
\cite{birkhoff1936logic,mittelstaedt1978quantum,dalla2002quantum}
---a problematic approach from a computational perspective
\cite{abramsky2007temperleylieb}--- Girard started from the desired logical
structures and built a semantics inspired from quantum structures:
\emph{quantum coherent spaces} \cite{girard2004between}. Although such
a quantum-based semantics is expressive enough to model (restricted
forms of) modalities \cite{baratella2010quantum}, Selinger
\cite{selinger04semantics} showed that it is not adequate for modeling
quantum computation, as it is missing some entangled
states ---for instance $\frac1{\sqrt2}(\ket{00}+\ket{11})$.

\subsection{CPM as Compact Closed Category}
\label{ssec:cpm-basic}

As discussed in Section~\ref{sec:mixed-states}, linear distributions of pure states
are adequately represented with trace-1 positive matrices.  Selinger
discusses how possibly non-terminating quantum programs can then be
modeled with trace-non-increasing \emph{completely positive maps}:
a \emphidx{superoperator}. A
\simpleidx{completely positive map} (\simpleidx{CPM})
$f:\Cx^{n\times n}\to\Cx^{m\times m}$ is a linear map such that for
all $k$, the map $\text{id}_{\Cx^{k\times k}}\tensor f$ sends positive
matrices to positive matrices.
Let us discuss a few aspects of this definition.
\begin{enumerate}
\item A superoperator might therefore output a positive matrix
  of trace strictly less than one: this trace corresponds to the
  overall probability of termination of the corresponding algorithm.
\item Consider a valid quantum algorithm $P$ of input $A$ and of
  output $B$. One can construct another valid quantum algorithm with a
  dummy variable $C$: the resulting algorithm $P'$ inputs in
  $C\tensor A$ and outputs in $C\tensor B$. The denotation
  $\denot{P'}$ is equal to $\text{id}_{\denot{C}} \tensor
  \denot{P}$. This is the reason for the second constraint on
  superoperator.
\end{enumerate}
The trace-non-increasing constraint gives a fully complete model for
first-order quantum computation, as discussed
in~\cite{selinger04quantum}. In the case of higher-order quantum
computation, we drop this constraint and work instead
with general, completely positive maps. Indeed, in order to model
functions we can then rely on the Choi theorem \cite{choi1975completely},
stating that

\begin{theorem}[{\cite[Th. 2]{choi1975completely}}]\label{th:choi}
  Let $f$ be a linear map from $\Cx^{n\times n}$ to $\Cx^{m\times
    m}$. Then $f$ is completely positive if and only if
  $\chi_f\in\Cx^{mn\times mn}$ blockwise defined as
  \[
    \chi_f = \left(
      \begin{array}{ccc}
        f E_{1,1}&\cdots&f E_{1,n}
        \\
        \vdots & \ddots & \vdots
        \\ 
        f E_{n,1}&\cdots&f E_{n,n}
      \end{array}
    \right)
  \]
  is positive, where $E_{i,j}\in\Cx^{n\times n}$ is the matrix with
  $0$s everywhere apart for one $1$ on the $i$-th line and $j$-th
  column.\qed
\end{theorem}

Using Theorem~\ref{th:choi}, one can design a model of MLL using
positive matrices and completely positive maps, as follow.
We define the category \CPM with the following data:
\begin{itemize}
\item Objects: natural numbers;
\item Morphisms: $f:n\to m$ is a completely positive map
  $\Cx^{n\times n}\to\Cx^{m\times m}$.
\end{itemize}
\CPM can be equipped with a monoidal structure, behaving as the (usual)
multiplication on integers and as Kronecker product on morphisms.
Thanks to Theorem~\ref{th:choi}, the
functor $A\tensor(-)$ admits a right adjoint, according to the natural
isomorphism
\[
  \CPM(A\tensor B,C) \simeq \CPM(A,B\tensor C).
\]
This makes \CPM \emphidx{compact closed} \cite{selinger2007dagger},
model of MLL. Note however that the model is degenerated as $\parr$
and $\tensor$ coincide.

\subsection{Accommodating the Additives}
\label{ssec:cpm-additive}
In order to be able to at least manipulate Boolean values, we need to extend
\CPM to accommodate the additives. Since we are in the context of
finite dimensional vector spaces, the easiest is to consider the
\emphidx{biproduct completion} \CPMfinbip of \CPM:
\begin{itemize}
\item Objects: lists of natural numbers $\sigma=n_1,\ldots,n_k$
\item Morphisms: If $\sigma=n_1,\ldots,n_k$ and $\tau=m_1,\ldots,m_{k'}$, then $f:\sigma\to \tau$ is a family $f = \{f_{i,j}\}_{i,j}$ where $f_{i,j}:n_i\to m_j$ is a \CPM-morphism.
\item Composition is obtained with matrix multiplication: 
\begin{equation}\label{eq:comp-bip}
  \{f_{i,j}\}_{i,j}\circ\{g_{j,k}\}_{j,k}
  =
  \left\{
  \sum_j f_{i,j}\circ g_{j,k}
  \right\}_{i,k}
\end{equation}
whereas the identity is a diagonal matrix of identities.
\end{itemize}
The compact-closed structure of \CPM carries over to \CPMfinbip in a straightforward manner:
\[
  (n_1,\ldots,n_k)\tensor(m_1,\ldots,m_l)
  =
  n_1m_1,\ldots,n_1m_l,n_2m_1,\ldots,n_2m_l,\ldots\ldots,n_km_l
\]
The category \CPMfinbip makes a model of MALL, albeit
degenerate since both the multiplicatives and the additives
collapse. However, it makes a fully-abstract model of a \emph{strictly
  linear} lambda calculus, as shown in \cite{selinger06fully}.

\subsection{Accommodating Recursive Datatypes}
\label{ssec:cpm-rec}

If the category \CPMfinbip can accommodate additives, the system is
restricted to \emph{finite} biproducts
\(
  \bigoplus_{i=1}^nA_i
\). 
This limits the expressiveness of the system: recursive datatypes such as lists:
\begin{equation}\label{eq:list}
  [A] = \bigoplus_{n=0}^\infty A^{\otimes n}
\end{equation}
cannot be represented as they require \emph{infinite} biproducts.
Following the same intuition as for the construction of \CPMfinbip, an
infinite biproduct would correspond to having \emph{infinite} lists of
natural numbers for objects, and infinite-dimensional matrices. The
difficulty then comes with the composition, as we now end up with an
infinite sum in Eq.~\eqref{eq:comp-bip}. Indeed, in general, the sum
might not converge: consider for instance
$f : 1,1,1\ldots\to 1,1,1\ldots$ defined as \( f_{i,j}:1\to 1 \) being
the identity for all $i,j$. The composition of $f$ with itself does
not converge.

The solution we propose in \cite{pagani2014applying} consists in first
completing \CPM with ``all possible infinite'' elements.  Recall that
positive matrices admit a natural ordering: Löwner
order. This can be pointwise ported to completely positive
maps: each homset $\CPM(A,B)$ is a Löwner positive cone.
One interesting property of the Löwner order is that such a positive
cone is \emph{bounded} directed complete: (1) there is a minimum
element (the 0 function), and (2) any bounded directed subset
$D\in\CPM(A,B)$ admits a least upper bound.

Formally, the completion we consider is the \emphidx{D-completion}
\cite{zhao2010dcpo-completion}. For the purpose of the discussion, we
are interested in two properties: First, the D-completion is
functorial, and then it preserves existing least upper bounds. On
other words, the only additional elements are ``at infinity''
---precisely what we need.

We can therefore define the category \CPMd as follows: the objects are
those of \CPM, and the morphisms from $n$ and $m$ are exactly the
elements of the D-completion of $\CPM(n,m)$. The homsets $\CPMd(n,m)$
are now dcpos: generalized sums are always defined. This then makes it
possible to define the \emph{infinite biproduct completion} \CPMbip of
\CPMd exactly as desired: objects are infinite lists of objects of
\CPMd, and morphisms are infinite-dimensional matrices. The
composition is defined as in Eq.~\eqref{eq:comp-bip}, and the possibly
infinite sum resulting from the definition is well-defined.

\subsection{Accommodating Duplication}
\label{ssec:cpm-dup}

With infinite coproducts we can encode the behavior type $!A$ inside
the type of lists shown in Eq.~\eqref{eq:list}. Indeed, a duplicable
element of type $!A$ can be regarded as the biproduct of zero copies of the element, one copy of the element, two copies of the element, \etc.

The typical example is the program that inputs a coin, tosses it twice
and computes the conjunction of the results. In \CPM the type of a
coin is $1,1$: a pair of two probabilities $(a,b)$, where $a$ is the
probability of getting $\ttrue$ and $b$ the probability of getting
$\ffalse$. The aforementioned program therefore corresponds to the
(non-linear) map
\[
  (a,b)\longmapsto (a^2, 2ab + b^2).
\]
The reason for the non-linearity is the identification of $\bit$ and
$!\bit$. Instead, we can consider a more expressive representation for
the input, as
\[
(a_{*}, a_{\ttrue}, a_{\ffalse}, a_{\ttrue,\ttrue}, a_{\ttrue,\ffalse}, 
a_{\ffalse,\ttrue}, a_{\ffalse,\ffalse}, a_{\ttrue,\ttrue,\ttrue},
a_{\ttrue,\ttrue,\ffalse}, a_{\ttrue,\ffalse,\ttrue}\ldots)
\in
\tunit\oplus\bit\oplus(\bit\tensor\bit)\oplus\cdots.
\]
A duplicable coin producing $\ttrue$ with probability $a$ and $\ffalse$ with probability $b$ is now represented as the sequence
\[
(a+b,a,b,a^2,ab,ab,b^2,\ldots)
\]
and the aforementioned program has now for semantics
\[
  (a_{*}, a_{\ttrue}, a_{\ffalse}, a_{\ttrue,\ttrue},
  a_{\ttrue,\ffalse}, a_{\ffalse,\ttrue}, a_{\ffalse,\ffalse}, \ldots)
  \mapsto (a_{\ttrue,\ttrue},
  a_{\ttrue,\ffalse}+a_{\ffalse,\ttrue} + a_{\ffalse,\ffalse}),
\]
now a linear, completely positive map.

Such a construction is however failing in providing the required
categorical structure of comonoid. Indeed, if $!A$ is modeled with
$[A]$, two copies of $A$ can very well be distinct.

Instead of a plain tensor, what is needed for $!A$ is a
\emphidx{symmetric tensor} \cite{mellies2009explicit}, connected to Fock spaces and used e.g. for
modeling probabilistic programs in probabilistic coherent spaces
\cite{ehrhard2014probabilistic}. Considering the case of $!\bit$, it
corresponds to define
\[
!\bit \triangleq
\tunit\oplus\bit\oplus\bit^{\odot 2}\oplus\bit^{\odot 3}\oplus\cdots,
\]
where 
$\bit^{\odot n}$ is the equalizer of
\[
\xymatrix{
  \bit^{\odot n}\ar[r]&
  \bit^{\otimes n}
  \ar@/^2ex/[rr]^{\text{symmetry}}
  \ar@/_2ex/[rr]_{\text{symmetry}}
  \ar@{}[rr]|{\cdots}
  &&
  \bit^{\otimes n}
}
\]
For instance, $\bit^{\odot 2}$ corresponds to the subcone of
$1,1,1,1$ invariant under swap: this corresponds to
\[
\{\,(a,b,b,c)\bor a,b,c\in[0,1]\text{ and }a+2b+c\leq1\,\}
\]
or, equivalently,
\[
\{\,(a,b)\tensor(a,b)\bor a,b\in[0,1]\text{ and }a+b\leq1\,\}.
\]

In our case, this requires to modify the original category \CPM to
account for such equivalence classes: we invite the reader to consult
the paper for more information \cite{pagani2014applying}. In any case,
the resulting model is shown to be adequate for a quantum
lambda calculus of the form presented in Section~\ref{sec:qlc-ll}
together with coproducts and recursive types. The model is in fact
richer than what we originally showed: it is fully-abstract
\cite{clairambault2020full}.

\begin{example}\label{ex:dup-qubit2}
  In Example~\ref{ex:dup-qubit} we discuss the type $!(1\loli
  A)$. On the semantic side, in {\CPMd}, this type is literally equal
  $!A$. We can reconciliate this fact with the non-duplicability of
  qubits by realizing that the semantics {\CPMd} is richer than what
  can be expressed in the quantum lambda-calculus. In particular, the
  semantics can support not only call-by-value but also
  call-by-name. The semantics of $!\qbit$ then represents
  \emphidx{thunks} of terms computing a qubit. On the language side,
  in a call-by-value perspective such a term has to be encapsulated
  inside a lambda-abstraction: we fall back on $!(1\loli\qbit)$.
\end{example}

\subsection{Discussion}
\label{ssec:cpm-discussion}

To overcome the finite-dimensional limitation of \CPM, we used in
\cite{pagani2014applying} abstract constructions based on category and
domain theory. If one can argue that we only added ``infinite''
elements that are anyway not representable by programs, it can be
regarded as a limitation of our approach. Clairambault and de Visme
\cite{clairambault2020full} offer an alternative approach based on
event structures \cite{winskel1980events,winskel1987event} and game
semantics that does not require infinite dimensional spaces. It is
worth noting that their approach solves a long-standing issue in
quantum game semantics: capturing entangled elements within the tensor
\cite{delbecque2008game-store, delbecque2008game-data,
  delbecque2008quantum}.

Other approaches rely on generalizations of \CPM: C${}^*$ and von
Neumann algebras. Westerbaan \etal \cite{westerbaan2016quantum}
discusses how to recover the required structures, while
\cite[Sec. 4.3]{westerbaan2019category} describe a model for the
quantum lambda calculus in this framework. Finally, lately Pechoux
\etal \cite{pechoux2020quantum} discusses how to build recursive types
with von Neumann algebras. One can also mention the presheaf model of
Malherbe \cite{malherbe2010categorical,malherbe2013presheaf} and the
categorical construction of Hasuo and Hoshino based on Geometry of
Interaction \cite{hasuo2011semantics,hasuo2017semantics}.

\section{An Executable Semantics}
\label{sec:goi}

This section is devoted to the description of a low-level, operational
semantics for a quantum lambda calculus. As discussed in Section~\ref{sec:qlc-ll},
the standard computational
interpretation of the quantum lambda calculus
is a rewrite system based on variable substitution. We
discuss here how to retrieve a circuit-based interpretation of a
quantum lambda-term, using a technique stemming from the study of
linear logic: the \emphidx{Geometry of Interaction}
(GoI)\index{GoI}. Originating from Girard~\cite{girard1988goiI}, GoI
has shown useful in understanding the relationship between high-level
constructs and low-level, assembly-like
presentations~\cite{mackie1995geometry,ghica2011gos-IV}.

Section~\ref{sec:pf-nets} first presents the graphical notation of
proof-nets for representing proofs of linear logic.
Section~\ref{sec:net-lc} discusses how typed lambda-terms can be
interpreted as proof-nets.  Section~\ref{sec:tok-goi} describes the
token-machine presentation of GoI~\cite{danos1999reversible}, giving a
graph-based, executable semantics for programs when translated to
proof-nets.  Section~\ref{sec:goi-limits} exposes the limit of the
standard approach, as it does not support token synchronization:
this is required for quantum computation.  Section~\ref{sec:smll}
offers a generic solution, and Section~\ref{sec:cbv-goi} discusses the
solution we build specifically for quantum computation.

The strength of our proposal is to unfold the circuit-like structure
hidden inside quantum lambda-terms: tokens follow the tangled wires
of the circuit.

\subsection{Proof-Nets for MELL}
\label{sec:pf-nets}

One of the nagging issues with proofs of linear logic is the exchange
rules: as discussed in Remark~\ref{rem:necessary-exc} 
it is essential, yet it looks
like an unnecessary, bureaucratic construct.
An alternative, graphical presentation of proofs of linear logic
consists in
\emphidx{proof nets}. For more information on proof-nets, consult
  e.g. Laurent's notes~\cite{laurent2013introduction}, from which this
  section takes inspiration.

\begin{figure}[tbh]
  \begin{center}
    \includegraphics[scale=.8,page=1]{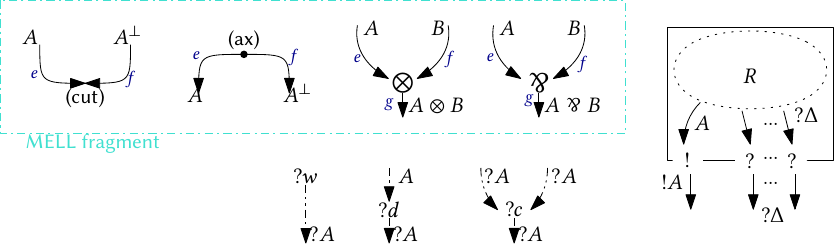}
  \end{center}  
  \caption{Constructors for MELL proof structures.}
  \label{fig:MELL-pfnets}
\end{figure}

A proof net is a \emphidx{proof structure} with a
\emphidx{validity criterion}. A proof structure is a directed graph,
possibly piecewise connected, with labeled edges and nodes. Thanks to
the graph structure, there is no need for permuting anything, and the
possible ambiguity of Eq~\eqref{eq:ambiguous-pf} disappears. 
In the context of this thesis, we will concentrate on the
multiplicative exponential fragment of linear logic (MELL), without
units. A proof
structure for MELL is built out of the nodes of
Figure~\ref{fig:MELL-pfnets}.

When clear, we shall omit the arrows symbol.  Input edges are called
\emphidx{premises} and output edges \emphidx{conclusions}; the edges
of a node are ordered. The nodes (cut) and (ax) corresponds to the
similarly named proof rules, while the nodes $\tensor$ and $\parr$
decompose multiplicative formulae. The edges for the MLL fragment (in
the Turquoise dashed box) are
named $e$, $f$ and $g$: they shall be used in Section~\ref{sec:tok-goi}.
The four other nodes are for
managing modalities.  The right-most node, the \emph{box-node}, stands
for the promotion proof rule (p). The box encapsulates a
proof-structure $R$. The wires going in the box goes through
\emph{doors}: left-most one is the \emph{principal} door while the
other are \emph{auxiliary} doors. The remaining nodes $?w$, $?d$ and $?c$ respectively stand for weakening, dereliction and promotion.
We extend
the notion of premises and conclusions to proof-structures: if the
premises of the structure $S$ are $A_1,\ldots A_n$ and the conclusions
are $B_1,\ldots, B_m$, we say that $S$ corresponds to the sequent
$\vdash A_1^\bot,\ldots A_n^\bot,B_1,\ldots,B_m$.

\begin{figure}
  \centering
  {\begin{subfigure}[t]{.47\textwidth}
    {\includegraphics[scale=.65,page=2]{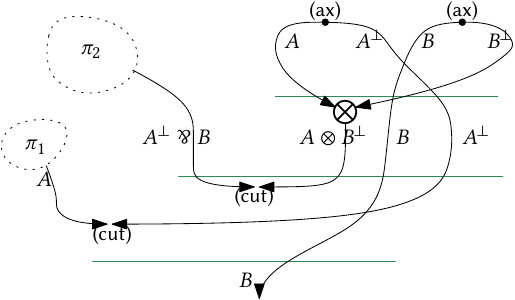}}
    \caption{Following the proof}
    \label{fig:pf-lin-mp-net-untangled}
  \end{subfigure}}
  \hfill
  {\begin{subfigure}[t]{.41\textwidth}
   {\includegraphics[scale=.65,page=3]{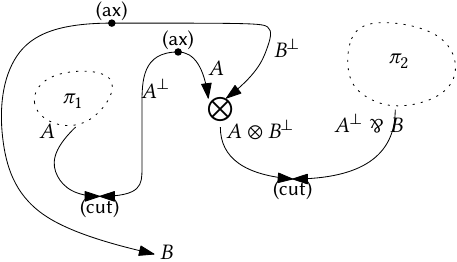}}
    \caption{Untangled}
    \label{fig:pf-lin-mp-net-tangled}
  \end{subfigure}}
  \caption{Proof-net corresponding to Eq.~\eqref{eq:pf-lin-mp}}
  \label{fig:pf-lin-mp-net}
\end{figure}

Note how each node corresponds to a proof rule (apart for the exchange
rule). A proof can then be directly transposed into a proof
structure. For instance, Fig.~\ref{fig:pf-lin-mp-net} the proof of
Eq.~\eqref{eq:pf-lin-mp} becomes the proof net shown in
Figure~\ref{fig:pf-lin-mp-net-tangled}. As it is a graph, it is the
same as the net in Figure~\ref{fig:pf-lin-mp-net-untangled}.

\begin{remark}\label{rem:constant-nodes}
  In the case where we have constant formulas $\alpha$ in the grammar
  of the logic, we can add specific nodes to reflect the corresponding
  proofs. 
\end{remark}

The fact that a sequent $\vdash\Delta$ admits a proof structure
does not however always imply that there exists a
proof for it: consider for instance the proof structure
\begin{equation}\label{eq:bug-net}
  \raisebox{-0.5\height}{\includegraphics[scale=.7,page=4]{fig/mll-pf-nets.pdf}}
\end{equation}
associated to the invalid sequent $\vdash A\tensor A^\bot$.

Characterizing \emphidx{proof nets}, i.e. proof structures
representing a valid proof for a given sequent, requires a
\emphidx{validity criterion}: many proposals
\cite{naurois2007correctness} have been proposed since the original
Girard's \emph{longtrip condition}~\cite{girard87linear}. Originally
developed for MLL~\cite{danos1989structure} but generalizable to
MELL~\cite{danos1990phd,guerrini2001parsing}, a versatile criterion is
Danos\&Regnier's \emph{switching condition}. It uses the notion of
\emphidx{path}: a path in a proof structure $\pi$ is a sequence of
nodes, pairwise connected with edges. In the case of MLL, it is called
\emphidx{switching} if its does not go through both premises of a
$\parr$-node. A proof-structure $\pi$ is called \emphidx{switching
  acyclic} when it does not contain switching cyclic paths. In this
case, we call it a \emphidx{proof net}.

\begin{example}
  The proof structure presented in Eq.~\eqref{eq:bug-net} is not
  switching acyclic: it does not
  correspond to a proof of $\vdash A\tensor A^\bot$.
\end{example}

In Section~\ref{sec:cut-elim}, we discussed cut-elimination for the
proofs of sequents: a similar procedure can be designed for
proof-nets. For instance, the rewrite rule shown in Eq.~\eqref{eq:rw-tens-par}
becomes for proof-nets
\begin{equation}\label{eq:tw-twn-par-pfnet}
  \raisebox{-0.5\height}{\includegraphics[scale=.7,page=5]{fig/mll-pf-nets.pdf}}
  \quad
  \longrightarrow
  \quad
  \raisebox{-0.5\height}{\includegraphics[scale=.7,page=6]{fig/mll-pf-nets.pdf}}
\end{equation}
The validity criterion are preserved through the reduction: a valid
proof-structure remains valid through rewriting.

There have been plenty of works and extensions of proof-nets:
interaction nets \cite{lafont1990interaction, lafont1995proofnets,
  mazza2006phd}, differential nets \cite{ehrhard2006differential,
  tranquilli2011intuitionistic}, \etc. It is a flexible structure
able to capture many logical aspects while leaving out much of the
bureaucracy of proofs.

\subsection{Encoding Higher-Order Languages}
\label{sec:net-lc}

Being a graphical representation, proof-nets make an easily
extensible, versatile representation for programs. Typically, as
mentioned in Remark~\ref{rem:constant-nodes} one can add to the
graphical language nodes representing constants and opaque operators,
and update the rewriting system accordingly. Similarly, one can extend
the boxing constructs to other situations, such as dealing with tests
\cite{dallago2014geometry} and recursion \cite{lago2015parallelism}.

\begin{figure}
  {\begin{subfigure}[t]{1.78in}
    {\includegraphics[scale=.765,page=1]{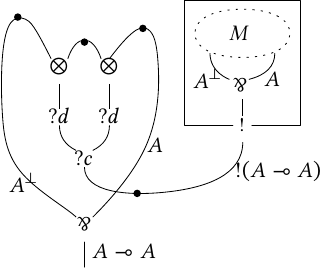}}
    \caption{Original}\label{fig:ex-prog-pf-net-1}
  \end{subfigure}}
  \hfill
  {\begin{subfigure}[t]{1.76in}
    {\includegraphics[scale=.65,page=2]{fig/ex-prog-pf-net.pdf}}
    \caption{Opening boxes}\label{fig:ex-prog-pf-net-2}
  \end{subfigure}}
  \hfill
  {\begin{subfigure}[t]{1.22in}
    {\includegraphics[scale=.65,page=3]{fig/ex-prog-pf-net.pdf}}
    \caption{Unfolding $\lambda$'s}\label{fig:ex-prog-pf-net-3}
  \end{subfigure}}
  \caption{Translation of $  \vdash (\lambda f.\lambda x.f(fx))\lambda w.M : A\loli A$}\label{fig:ex-prog-pf-net}
\end{figure}

A proof-net directly comes from the typing derivation: in the case
of the quantum lambda calculus, as the type system is based on linear
logic the translation is immediate. For instance, the term
\[
  \vdash (\lambda f.\lambda x.f(fx))\lambda w.M : A\loli A
\]
corresponds to the proof-net shown in
Figure~\ref{fig:ex-prog-pf-net}. Figure~\ref{fig:ex-prog-pf-net-1}
corresponds to the original term (modulo some yanking for
legibility). The duplicated subterm $\lambda w.M$ is the box on the
right of the cut, while the contraction on the left corresponds to the
duplication of the variable $f$. The derelictions ``open'' the two
copies of the box. The result of the copy and the opening of the boxes
is shown in Figure~\ref{fig:ex-prog-pf-net-2}: it corresponds to
$\lambda x.(\lambda w.M)((\lambda w.M)x)$. Finally,
Figure~\ref{fig:ex-prog-pf-net-3} shows the beginning of the unfolding
of the two ``$\lambda w.$''.

\subsection{Token-based Geometry of Interaction}
\label{sec:tok-goi}

Geometry of Interaction (GoI)\index{GoI}
\index{Geometry of Interaction} stands as one of Girard's flagship research
projects. Its main goal consists in extracting the
\emph{computational content} of a proof, stable under cut-elimination.
If Girard attacked this problem under many different
angles~\cite{girard1988goiI,
  girard1988goiII, girard1995goiIII, girard2003goiIV, girard2011goiV},
the one we consider in this chapter is the
\emphidx{token-based GoI}~\cite{girard1988goiI}. In this approach,
GoI can be seen as a procedure to construct a sequential data-flow
machine, with a token running in a proof net. In particular, it draws a
direct link between high-level programming constructs and low-level,
assembly languages~\cite{danos1999reversible, mackie1995geometry,
  mackie1994geometry} ---it has even been used as a backbone for
designing compilers, with support for higher-order
functions~\cite{ghica2011gos-III}, local, assignable
states~\cite{ghica2007geometry}, concurrency~\cite{ghica2010gos-II},
recursion~\cite{ghica2011gos-IV}, \textit{etc}.

In order to illustrate the difference between the standard approach
and the contribution presented in Section~\ref{sec:smll}, we propose a brief
introduction of the \emphidx{IAM} \cite{danos1999reversible} ---the \emphidx{Interaction
Abstract Machine}---. We focus for this presentation on the
multiplicative fragment MLL of MELL ---that is, without
modalities. In this section, to relate with
Section~\ref{sec:qlc-ll}, we follow the notation presented
in~\cite{lago2015parallelism}.

In the MLL fragment we consider, a formula is given by the grammar
\[
  A,B\quad{:}{:}{=}\quad
  \alpha\bor \alpha^\bot\bor A\tensor B\bor A\parr B.
\]
For the purpose of the discussion, and in line with
Section~\ref{sec:qlc-ll} we
replace the units with
constants $\alpha$, ranging over fixed set of identifiers.  A
proof-net consists of the nodes (cut), (ax), ($\tensor$),
($\parr$), together with a dummy node ($\alpha$) with one conclusion
of type $\alpha$ and no premises.

The state of an IAM on a proof-net $\pi$ consists of a triple $(e,s,d)$
where $e$ is an edge of $\pi$, $s$ an address and $d$ is a direction
$\uparrow$ or $\downarrow$. An address is a stack: a list of the
literals $l$ and $r$. Cons is denoted with ``$:$'', while the empty
stack is $\varepsilon$.
An address represents the position of a subformula inside a formula. For
instance, the address
$r:l:\varepsilon$ points to $\alpha_3$ in the formula
$(\alpha_1\otimes\alpha_2)\parr(\alpha_3\otimes \alpha_4)$.

\begin{table}
  \begin{align*}
    &\text{(cut)}
    &(e,s,\downarrow) &\to (f,s,\uparrow)
    &
      (f,s,\downarrow) &\to (e,s,\uparrow)
    \\
    &\text{(ax)}
    &(e,s,\uparrow) &\to (f,s,\downarrow)
    &
      (f,s,\uparrow) &\to (e,s,\downarrow)
    \\
    &\text{up $(\otimes)$}
    &(g,l{:}s,\uparrow) &\to (e,s,\uparrow)
    &(g,r{:}s,\uparrow) &\to (f,s,\uparrow)
    \\
    &\text{up $(\parr)$}
    &(g,l{:}s,\uparrow) &\to (e,s,\uparrow)
    &(g,r{:}s,\uparrow) &\to (f,s,\uparrow)
    \\
    &\text{down $(\otimes)$}
    &(e,s,\downarrow) &\to (g,l{:}s,\downarrow)
    &(f,s,\downarrow) &\to (g,r{:}s,\uparrow)
    \\
    &\text{down $(\parr)$}
    &(e,s,\downarrow) &\to (g,l{:}s,\downarrow)
    &(f,s,\downarrow) &\to (g,r{:}s,\uparrow)
  \end{align*}
  \caption{Rules for the IAM token Machine, MLL fragment.}
  \label{tab:rules-iam}
\end{table}

A (reversible) rewrite system for IAM states is then derived from the
structure of a net. Following the naming convention for edges shown in
Figure~\ref{fig:MELL-pfnets}, the rules for the token movements in
MLL proof nets are shown in Table~\ref{tab:rules-iam}.
An initial (resp. final) state of the IAM on an MLL-net $\pi$ consists
in a state of the form $(e,s,\uparrow)$ (resp. $(e,s,\downarrow)$),
where $e$ is a conclusion edge of $\pi$ and $s$ points to a constant
subformula of the formula attached to $e$. We write $\mathcal{I}$ the
set of initial states and $\mathcal{O}$ the state of final states.

\begin{proposition}
  If $\pi$ is an MLL-proof-net, the IAM machine deterministically sends
  initial states to final states: it induces a bijection
  $\Sigma_\pi:\mathcal{I}\to\mathcal{O}$. Furthermore, this bijection
  is invariant under cut-elimination.\qed
\end{proposition}

\begin{figure}[tbh]
  \centering
  \includegraphics[scale=.9,page=7]{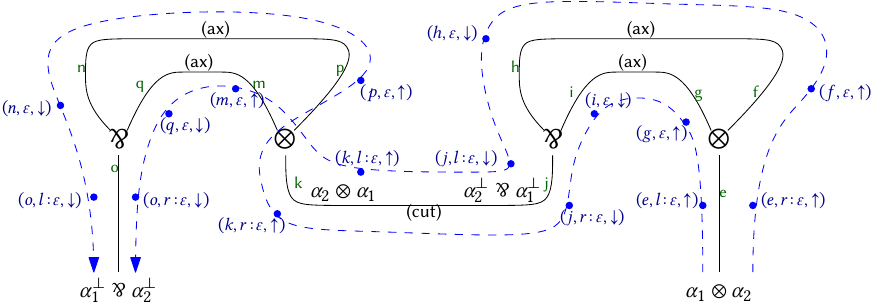}
  \caption{A run of the IAM token machine.}
  \label{fig:iam-swap}
\end{figure}

Figure~\ref{fig:iam-swap} shows the behavior of the IAM machine on the
proof-net corresponding to the cut of the proof of
$\alpha_1\otimes\alpha_2\vdash\alpha_2\otimes\alpha_1$ and the proof
of $\alpha_2\otimes\alpha_1\vdash\alpha_1\otimes\alpha_2$. This gives
the identity on $\alpha_1\otimes\alpha_2$. The token machine
``realizes'' the computation. Note for instance how the initial state
$(e,r{:}\varepsilon,\uparrow)$ sitting on $\alpha_2$ goes to the
terminal state $(o,r{:}\varepsilon,\downarrow)$, corresponding to a
token also sitting on $\alpha_2$. Note also how this is invariant
under the rewrite rule shown in Eq.~\eqref{eq:tw-twn-par-pfnet}.

This example generalizes: the IAM rewrite system on a proof-net $\pi$
\emph{realizes} the
computation described by the corresponding proof. The abstract machines
stemming from such an approach follow a \emphidx{call-by-name}
strategy~\cite{danos1999reversible, mackie1995geometry}: arguments are
passed to their calling functions without being evaluated. Formally,
\cite{danos1999reversible} makes a connection between the IAM and the
\emphidx{Krivine abstract machine} (KAM)\index{KAM}
\cite{krivine2007call-by-name}.

\subsection{Limits of the Conventional Approach}
\label{sec:goi-limits}

For our purpose, the two main limits to this conventional, stateless
token-based presentation of GoI are the strict sequentiality of the
machinery and the fact that it is call-by-name. As in the case of game
semantics~\cite{abramsky97cbv}, directly handling call-by-value
---without specific encoding, such as
CPS~\cite{wadler2003call-by-value}--- typically requires
side-effects~\cite{schopp2014call-by-value,hoshino2014memoryful}.

The strict sequentiality is a problem in the context of additional
nodes reflecting operations on atomic types. Suppose for instance that
one of the type $\alpha$ stands for the natural numbers $\mathbb{N}$:
we can add to token states a register holding a natural number,
and have a special node ``$+$'' for addition. If $M$ and $N$
are terms of type $\Nx$, the term $\vdash M+N : \Nx$ then corresponds to
some net
\[
  \includegraphics[scale=.9,page=8]{fig/mll-pf-nets.pdf}
\]
Computing with an IAM-style machine requires to start from the
conclusion; but what should we do when we reach the $+$-node? Should
we go left, right?
Traditional solutions involve making an arbitrary choice on which
premise the token should explore first. However, despite the fact that
this requires a state to store the intermediate result, in languages
such as the quantum lambda calculus, this is not always possible as
some operators act on non-atomic types ---for instance, 2-qubit
unitary gates act on $\qbit{\tensor}\qbit$. This makes it difficult to
adapt to the single-token IAM, and requires a novel approach.

\subsection{Multi-Token Geometry of Interaction}
\label{sec:smll}

In order to answer the sequentiality problem listed in
Section~\ref{sec:goi-limits}, dal Lago \textit{et
  al.}~\cite{dallago2014geometry} offers an alternative approach for a
Geometry of Interaction: Instead of starting from conclusion to fetch
values, values ``flow'' on their own from inputs \emph{towards}
conclusions ---in a call-by-value spirit. Instead of one single
token, the GoI machine of \cite{dallago2014geometry} fires one token
per potential value. Tokens are then emitted from negative conclusions
and, if any, from nodes introducing atomic types.
This solves the problem discussed in
Section~\ref{sec:goi-limits}: each one of the premises of the $+$-node
eventually meets with a value-token. The problem
\cite{dallago2014geometry} addresses ---in the very restricted case of
MLL--- is the synchronization issue: for the $+$-node to fire, it
needs \emph{both} of its argument-tokens to have arrived, as shown in
this example
\[
  \includegraphics[scale=.7,page=9]{fig/mll-pf-nets.pdf}
\]
This might however lead to deadlock, as illustrated in the following (bogus) run
\[
  \includegraphics[scale=.7,page=10]{fig/mll-pf-nets.pdf}
\]
In the last panel, the +-box is unable to fire anything before the
arrival of a token on its right input. But this token will never come
since it would be resulting from the output of the same +-box.

Formally, dal Lago \etal \cite{dallago2014geometry} introduce
proof-nets for SMLL, an extension of MLL with synchronization points.
The authors describe a correctness criterion for ruling out deadlocks,
and they present a token-based Geometry of Interaction where tokens
flow from values to conclusions. They then sketch how this can be used
to model a \emph{strictly linear} quantum lambda calculus.

\subsection{Towards a Quantum Geometry of Interaction}
\label{sec:cbv-goi}

\begin{mylife}
  I joined the research project at the time of the publication
  of~\cite{dallago2014geometry}, and my participation led to two
  publications \cite{lago2015parallelism,lago2017geometry}.
  In this section,
  I summarize the corresponding contributions.
\end{mylife}

The paper \cite{lago2015parallelism} presents a generalization
of SMLL nets and their multi-token GoI to support exponential
modalities and recursive behavior. The resulting nets ---called
SMEYLL\footnote{We thought of using the name ``SMELLY'' but it was
  ruled out despite my heavy lobbying}--- therefore add to MELL-nets
synchronization points and two additional boxes: the $\bot$-box of
SMLL, to encode a primitive conditional, and the $Y$-box, to represent
fixpoints.  The presence of fixpoints forces us to consider a
restricted notion of reduction, namely closed surface reduction (i.e.,
reduction never takes place inside a box). Cuts cannot be eliminated
(in general) from SMEYLL proofs, as one expects in a system with
fixpoints. Reduction, however, is proved to be deadlock-free, i.e.,
normal forms cannot contain surface cuts.

\begin{figure}
  {\begin{subfigure}[t]{2.1in}
    {\includegraphics[scale=.7,page=6]{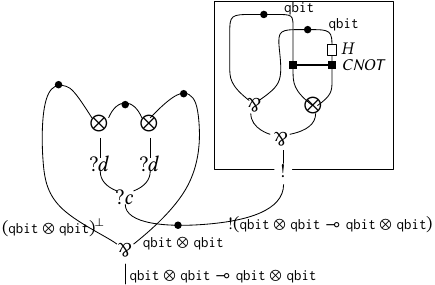}}
    \caption{Original}\label{fig:ex-prog-pf-net-q-1}
  \end{subfigure}}
  \hfill
  {\begin{subfigure}[t]{1.66in}
    {\includegraphics[scale=.7,page=5]{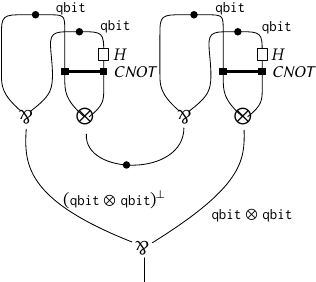}}
    \caption{Intermediary state}\label{fig:ex-prog-pf-net-q-2}
  \end{subfigure}}
  \hfill
  {\begin{subfigure}[t]{1.05in}
    {\includegraphics[scale=.6,page=7]{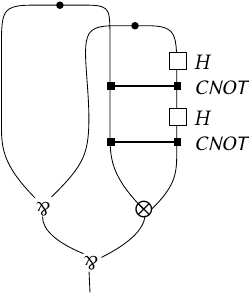}}
    \caption{Final Unfolding}\label{fig:ex-prog-pf-net-q-3}
  \end{subfigure}}
  \caption{Instantiating and unrolling the net of Fig.~\ref{fig:ex-prog-pf-net}.}
  \label{fig:ex-prog-pf-net-q}
\end{figure}

If we invite the reader to read the full paper for details
\cite{lago2015parallelism}, we present here a small example to
illustrate the setting. Let us instantiate the term of
Section~\ref{sec:net-lc} to
\[
(\lambda f.\lambda x.f(f\,x))(\lambda w.
\lettermx{\tuple{x,y}}{w}{\CNOT\tuple{x,H(y)}})
\]
The type $A$ is $\qbit\tensor\qbit$, and the net is presented in
Fig.~\ref{fig:ex-prog-pf-net-q}: Fig.~\ref{fig:ex-prog-pf-net-q-1}
and~\ref{fig:ex-prog-pf-net-q-2} recall the original state and the
result of the partial unfolding. Fig.~\ref{fig:ex-prog-pf-net-q-3}
shows the final result, and highlights an informal result: SMEYLL nets
describe ``folded'' quantum circuits that the rewriting unfolds.

SMEYLL nets are seen as interactive objects through a synchronous
interactive abstract machine (SIAM in the following). As for SMLL,
multiple tokens circulate around the net simultaneously, and
synchronize at sync nodes. In \cite{lago2015parallelism} however,
SMEYLL nets and SIAM tokens do not support probabilistic
behavior and can only carry very simple additional states such as
natural numbers or Boolean values.

The follow-up paper \cite{lago2017geometry} extends the setting to
support quantum information and probabilistic side-effects. The
resulting model then supports all of the structures needed to model
the quantum lambda calculus. 
The model addresses two problems: the handling of entanglement, and
the probability behavior.

The problem of entanglement can be exemplified by the following
example: in the state $\frac1{\sqrt2}(\ket{00}+\ket{11})$, we want to
be able to manipulate both qubits independently. In previous
approaches \cite{hasuo2017semantics, delbecque2008quantum,
  delbecque2008game-store, delbecque2006quantum}, either this was not
possible, or the expressive power of the system was too weak, lacking
recursion and duplication \cite{lago2014wave-style,
  dallago2014geometry}.

In the literature, probabilistic behavior is usually handled through
sequentialization. This can be done with the help of a reduction
strategy, as in the probabilistic lambda calculus
\cite{ehrhard2014probabilistic}, or with the use of polarity
\cite{danos02probabilistic,hoshino2014memoryful}. The proposal in
\cite{lago2017geometry} defines instead a confluent probabilistic
transition system that is both infinitary and parallel.

Unlike the original proposal of \cite{selinger2006lambda}, the new
framework proposed in \cite{lago2017geometry} models memory and
choice effects in a parametric way, via a \emphidx{memory
  structure}. The memory structure is presented in an algebraic manner
and relies on nominal sets \cite{pitts2013nominal}. This can be 
regarded as a generalization of Staton's equational system
\cite{staton2015algebraic}, specific to quantum computation.
Compared to purely categorical
presentations of quantum GoI \cite{hasuo2017semantics}, this semantics
offers a concrete, executable model where tokens follow the folded
quantum circuit hidden inside a quantum program.

\section{A Categorical Semantics for Circuit-Description}
\label{sec:red-ql}

Up until this point, in the presentation there have been two
distinct lines of research. On one hand, Chapter~\ref{ch:compil} 
discusses how quantum programming languages are all about
circuit-description languages. On the other hand, this chapter has
only been presenting semantics of quantum lambda calculi based on a
very simple, QRAM operational semantics. 

This section is devoted to closing the gap: we discuss a denotational
semantics for a circuit-description lambda calculus. The challenge is
that circuits are syntactic description: one cannot only rely on (some
extension) of CPM. The usual suspect for sketching an answer is to
rely on pure categorical constructs to capture all of the
required structures of the language.

Section~\ref{sec:cat-circ} introduces the existing attempts at
formalizing circuit-description languages with a focus on
\protoquipper: a formal, core subset of \quipper expressed as an
lambda calculus extension for circuit manipulation.
Section~\ref{sec:cat-circ-opalg} sketches the proposal of Rennela and
Staton \cite{rennela2017classical} for extending a concrete model of
first-order quantum computation based on $C^*$ algebras to express
quantum circuit manipulation. However, because circuits are identified
with the operations they represent, the semantics fails at
encompassing syntactic circuit operations.
Section~\ref{sec:proto-quipper-M} discusses a purely categorical, more
general semantics proposed by Selinger and Rios
\cite{rios2017categorical}. It is still limited in the sense that it
only supports a limited form of measurement. Finally,
Section~\ref{sec:dong-ho} presents our proposal, answering this
problem \cite{lee2022concrete}.

\begin{mylife}
  The work presented in Section~\ref{sec:dong-ho} has been
  the result of the Ph.D thesis of 이동호 (Dongho Lee)
  \cite{lee2022phd} who I co-supervised with Valentin Perrelle
  (CEA-LIST/LSL).
\end{mylife}

\subsection{Formalizing Circuit-Description Languages}
\label{sec:cat-circ}

The formalization of circuit description languages started with two
research threads. Arguably the first one is the formal language
\protoquipper \cite{ross2015algebraic}, aimed at describing the core
computational behavior of \quipper. The other approach has been the
language QWIRE embedded in Coq \cite{paykin2017qwire}, aiming at
proving properties of quantum programs. Both works follow a similar
approach to quantum programming: the circuit is a data-structure that
is being constructed in a dynamic manner by a classical program. In
the following I will concentrate on \protoquipper, as its structure is
closer to the quantum lambda calculus already discussed in
Section~\ref{sec:qlc-ll}. Moreover, \protoquipper has been the seminal
work for many more improvements on the semantics side of quantum
description languages \cite{rios2017categorical,
  lindenhovius2018enriching,
  fu2020linear,
  fu2020tutorial,
  lee2022concrete,
  colledan2022dynamic,
  fu2022biset-enriched,
  fu2022proto-quipper}.

\protoquipper can be regarded as an extension of the quantum
lambda calculus. Instead of sending gates to the QRAM one at a time,
the language features a constructor \tbox for turning functions into
circuits---i.e. buffering gates into an
circuit-object that can then be manipulated as any other object. This
\tbox-operation can be regarded as a kind of thunk
\cite{ingerman1961thunks,hatcliff1996thunksw} with partial evaluation
\cite{consel1993tutorial}: a term $\tbox\,M$ will become a
circuit-object, for instance a list of gates, but the gates will not
be sent to the QRAM. In order to do so, another construct \tunbox aims
at ``running'' the circuit, effectively sending the gates
downstream. Circuits are modeled in the language using a special
arrow-type $\tcirc$: a circuit with input $A$ and output $B$ is typed
with $\tcirc(A,B)$. We can therefore give the following type to \tbox
and \tunbox.
\begin{align*}
  \tbox &{}: {!}(A\loli B) \loli \tcirc(A,B),
  \\
  \tunbox &{}: \tcirc(A,B)\loli !(A\loli B).
\end{align*}
The constant \tbox makes a circuit out of the (partial evaluation) of
a function, while \tbox turns a circuit into a function.

In the original \protoquipper of Neil Ross \cite{ross2015algebraic},
the language would not support probabilistic behavior. So measurement
is only allowed as a circuit gate sending a wire of type qubit to a
wire of type bit. The possibility to turn a bit into a regular
Boolean value on which to run the if-then-else construct of
lambda calculus ---the \emphidx{dynamic lifting} feature--- is not
part of the formalism. It is then for instance not possible to realize
dynamic circuits such as the one sketched in Figure~\ref{fig:wf-dyn}.

Following Ross's work \cite{ross2015algebraic}, the team at Dalhousie
has developed a categorical semantics for circuit-description
languages \cite{rios2017categorical}, based on the variant
\protoquipperx{M}. This language and its categorical semantics has been
the seminal work on which most of the later works step up:
\cite{lindenhovius2018enriching} discusses (classical) recursion,
\cite{fu2020linear, fu2020tutorial} generalizes the model to support
dependent-types, while \cite{fu2022biset-enriched,
  fu2022proto-quipper} (with \protoquipperx{Dyn}) and
\cite{lee2022concrete} (with \protoquipperx{L}) study the addition of
dynamic lifting to the language. The former approach \cite{fu2022biset-enriched,
  fu2022proto-quipper} describes the set of axioms
required for the categorical semantics to be sound, whereas the latter
\cite{lee2022concrete} constructs a concrete category based on quantum
channels, and shows how the branching stemming from measurements can be
seen as a Kleisli category in this framework (see
Section~\ref{sec:dong-ho} for a more detailed discussion).

\subsection{Semantics based on Operator Algebras}
\label{sec:cat-circ-opalg}

The semantics of regular, first-order quantum computation---with both
unitaries and mea\-surements---have been studied for a long time. If
one trend of research focuses on mathematical, concrete models
extending the original semantics of trace-non-increasing completely
positive maps
\cite{selinger04quantum,westerbaan2016quantum,westerbaan2019category},
other works follows a more axiomatic approach.  Sam Staton
\cite{staton2015algebraic} in particular proposes a complete
equational theory of first-order quantum computation, characterized by
unitary applications and measurements. The equational theory is
complete and comes with nine axioms, relying on $C^*$ algebras:
positive elements of $C^*$ algebras can be regarded as observables in
quantum theory.  With Matthys Rennela
\cite{rennela2017classical,rennela2020classical}, they later explore
how to build a linear-non-linear category à la Benton
\cite{benton94mixed-tech}. The model is very general and models any
interacting computation involving a notion of circuit. To recover
quantum computation (with measurement), they instantiate the model on
Staton's equational theory \cite{staton2015algebraic} (and
$C^*$-algebras). As presented, the model is therefore very intentional:
in its $C^*$ instantiation, one can for instance hardly count the
number of gates of a circuit within the model.

\subsection{Semantics based on Category Theory}
\label{sec:proto-quipper-M}

\newcommand\doverline[1]{%
\tikz[baseline=(nodeAnchor.base)]{
    \node[inner sep=0] (nodeAnchor) {$#1$}; 
    \draw[line width=0.1ex,line cap=round] 
        ($(nodeAnchor.north west)+(0.0em,0.2ex)$) 
            --
        ($(nodeAnchor.north east)+(0.0em,0.2ex)$) 
        ($(nodeAnchor.north west)+(0.0em,0.5ex)$) 
            --
        ($(nodeAnchor.north east)+(0.0em,0.5ex)$) 
    ;
}}

Following Ross's formalization of \protoquipper
\cite{ross2015algebraic}, Rios and Selinger \cite{rios2017categorical}
offer a categorical semantics of a related circuit-description
language dubbed \protoquipper{M}. The semantics accounts for the \tbox
and \tunbox operations, as well as---when correctly
instantiated---classical operators on circuits such as gate-count. The
model is built from the following.
\begin{itemize}
\item A symmetric monoidal category $M$. Objects corresponds to
  bunches of wires and morphisms to circuits.
\item A symmetric monoidal closed category
  $\overline{M}$ with arbitrary products encapsulating $M$. This category is a technical
  vessel for the category $\doverline{M}$ defined next.
\item A category $\doverline{M}$ aiming at modeling
  parameterized circuits:
  An object of $\doverline{M}$ is a pair
  $(X,\{A_x\}_{x\in X})$ where $X$ is a set
    and $A_x$ an object of $\overline{M}$.
    A morphism $(X,\{A_x\}_{x\in X})\to(Y,\{B_y\}_{y\in Y})$ is a
    pair $(f_0,\{f_x\}_{x\in X})$ where $f_0:X\to Y$ is a set-function
    and where for all $x\in X$, $f_x:A_x\to B_{f_0(x)}$ is a morphism
    of $\overline{M}$.
\end{itemize}
One can canonically construct the embedding
$p:\textsf{Set}\to\doverline{M}$, and $p$ features an
adjoint functor $\flat:\doverline{M}\to\textsf{Set}$. This
adjunction describes a linear-non-linear category and turns
$\doverline{M}$ into a model of linear
logic \cite{benton94mixed}.
Moreover, it makes it possible to model boxing and unboxing in
a natural way: the homset $M(A,B)$ ---the representation of a circuit
between $A$ and $B$--- can be regarded as a map $A\loli B$ in
$\doverline{M}$.

If Rios and Selinger's construction can be extended to support
recursive datatypes and fixpoints \cite{lindenhovius2018enriching}, it
is \textit{a priori} not expressive enough to support measurements as
such. Indeed, if syntactic circuits can feature wires of type
bit, this bit cannot be \emph{lifted} to a Boolean value in the category
$\doverline{M}$ of regular, classical computations.

\subsection{Semantics for Circuits with Measurements}
\label{sec:dong-ho}

\begin{figure}[tb]
  \centering
  \includegraphics[scale=.9,page=1]{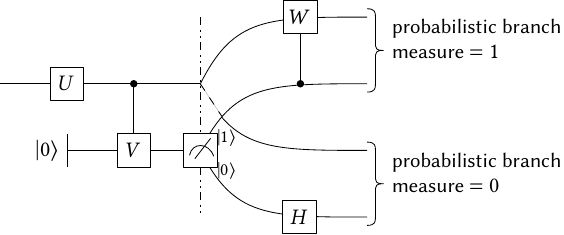}
  \caption{Example of quantum channel.}
  \label{fig:q-chan}
\end{figure}

In \cite{lee2022concrete}, we propose the language \protoquipper{L},
extending \protoquipper with dynamic lifting. The \tbox operation now
not only captures unitary operations but also measurements, so that
one can for instance box the function
\[
  \vdash\lambda x.\lettermx{z}{H\,(\qinit\,\ffalse)}{\iftermx{\meas\,z}{U\,x}{V\,x}} : \qbit\to\qbit
\]
(written in the language of Section~\ref{sec:qlc-ll}). Circuits are
therefore now not only lists of gates, but \emph{branching trees}\index{branching tree}
accounting for the choice to make for continuing a circuit after a
measure. Such generalized circuits are called \emph{quantum
  channels}\index{quantum channel}. A typical quantum channel is presented in
Figure~\ref{fig:q-chan}: the measurement spawns two independent
branches, one for each result of the measure.
This follows the intuition drawn by the data-structure {\tt CircIO}
underlying \quipper's {\tt Circ} monad presented in
Section~\ref{sec:quipper-desc}.

We design a companion concrete category $M$ of syntactic quantum
channels that can account for (first-order) quantum computation with
both unitaries and measurements, in the similar spirit as what was
proposed by Ross Duncan \cite{duncan2009generalized}. In particular,
the category $M$ is already monoidal closed and features products, so
that we can identify $\overline{M}$ and $M$.
We show how in this situation, the category $\doverline{M}$
features a monad capturing branching side-effects coming from the
measure. This branching monad is the categorical interpretation of the
\texttt{Circ} monad of \quipper.
Quantum computations with dynamic lifting can then naturally
be represented in the corresponding Kleisli category.

\begin{table}[p]
  \centering
  \scalebox{.8}{\begin{minipage}{1.25\textwidth}
        \begin{mylife*}
          \begin{compactdesc}
            \itempaper{selinger2009quantum} \myciteselingerquantum
            \itempaper{valiron2010orthogonality} \mycitevalironorthogonality
            \itempaper{valiron2012quantum} \mycitevalironquantumtuto
            \itempaper{valiron2013quantum} \mycitevalironquantumtutobis
            \itempaper{pagani2014applying} \mycitepaganiapplying
            \itempaper{valiron2014finite} \mycitevalironfinite
            \itempaper{valiron2014modeling} \mycitevalironmodeling
            \itempaper{lago2015parallelism} \mycitelagoparallelism
            \itempaper{lago2017geometry} \mycitelagogeometry
            \itempaper{diaz-caro2019realizability} \mycitediazcarorealizability
            \itempaper{xu2021reasoning} \mycitexureasoning
            \itempaper{lee2022concrete} \myciteleeconcrete
          \end{compactdesc}
          \caption{Personal publications related to Chapter~\ref{ch:sem}.}
          \label{tab:publis-sem}
       \end{mylife*}
     \end{minipage}}
\end{table}


\clearpage{\thispagestyle{empty}\cleardoublepage}

\chapter[Quantum Control and Reversible Computation]{%
  Quantum Control and\\*Reversible Computation%
}
\label{ch:qcont}
\rhead{Quantum Control}

A typical execution in the coprocessor model consists of elementary
gates applied to the quantum memory. The memory state consists
of a superposition of basis elements: the gates are applied
indistinctly on all basis elements at once. This model can be
summarized by the slogan ``quantum data, classical control''
\cite{selinger04quantum}.  In this scheme, a quantum program is merely
a classical program with classical control-flow, manipulating a
quantum memory. The only thing in superposition in this model is the
data.

However, a particular circuit combinator is hinting at a possibly finer
execution model: the \emph{control} of an operator. This combinator
makes it possible to filter out the state space and only act on a
subspace. A computational model of \emph{purely quantum} executions
generalizes the notion of controlled gate with quantum
\emphidx{superposition of executions} instead of only superposition of
data \cite{nielsen1997computable}.
One shifts from a model of \emphidx{classical
  control}-\emph{flow} of programs to a \emphidx{quantum
  control}-\emph{flow}, where program (or circuits) can be superposed,
yielding an alternative slogan:
\begin{quote}
  ``Quantum data, quantum control''.
\end{quote}
This non-standard model of computation raises several challenges: this
chapter discusses two of them. The first one is whether this model is
realistic and can bring anything new compared to the
standard circuit model. Another challenge of interest to us is the
design of a suitable language to express superpositions of
programs. In particular, the difficulty is to \emph{preserve unitarity}.
\begin{itemize} 
\item Section~\ref{sec:qcont-intro} discusses the problem with the
  concrete implementation of quantum control. It discusses the
  literature on the subject and focuses on one of our main
  contributions: the quantum SWITCH \cite{chiribella2013quantum}.
  This small protocol highlights how
  quantum control is not reducible to quantum circuits even though
  it was shown to be physically implementable. The section
  concludes by discussing the several approaches followed in the
  literature to define a syntax for superpositions of executions. In particular,
  the
  section discusses the notion of tests, loops, and
  recursion in a purely quantum context. It describes the problems
  that one must overcome while designing a syntax for quantum
  control.
\item Section~\ref{sec:veclc} presents one of our contributions on the
  design of a syntax accounting for superpositions of programs. We
  focus on an extension of lambda calculus featuring terms in
  superposition, and we discuss the design of two possible type systems
  accounting for superpositions of terms and unitarity
  \cite{arrighi2017vectorial,diaz-caro2019realizability}. In this
  approach, the language supports arbitrary linear combinations of
  terms. The type systems aims at determining which terms are ``valid'',
  i.e., make sense as quantum superpositions.
\item Section~\ref{sec:qcont-patt} presents our other contribution for
  a syntax of quantum control \cite{sabry2018symmetric}.
  Leaving the realm of pure
  lambda calculus, we propose a language based on pattern
  matching. The approach is dual to the one of
  Section~\ref{sec:veclc}: Instead of allowing any linear combination
  of any programs, the syntax enforces valid quantum programs from
  first principles. We discuss how the language handles naturally both
  (some form of) recursion and unitarity. We also discuss how the
  corresponding type system agrees with an extension of linear logic:
  the logic {\mumall} \cite{chardonnet2022curry-howard}.
\end{itemize}

\section{Implementing Quantum Control}
\label{sec:qcont-intro}

Deciding to turn controlled gates into a general superposition of
execution raises several questions: does it make sense in general? If
yes, how does it differ from the regular model of quantum circuits?
And, last but not least, how to \emph{program} in this model?
Each subsection addresses one question. Section~\ref{sec:qcont-phys}
discusses the debates pertaining to the physicality of quantum
control. Section~\ref{sec:qcont-qswitch} presents our seminal
contribution on the topic: the quantum SWITCH. It consists in a
minimal protocol exhibiting quantum superposition of execution. We
show how the quantum SWITCH cannot be realized with quantum
circuits. Section~\ref{sec:review-qcont} finally reviews the attempts
at capturing quantum control within a syntax, and highlights the
problems that occur.

\subsection{Physicality of Quantum Control}
\label{sec:qcont-phys}

This is part of a larger
problem: the \emph{physical Church-Turing thesis} whose scope is
to describe what computation means within the constraints imposed by
the laws of physics.

Citing Gandy \cite{gandy1980churchs}, the standard Church-Turing
thesis\footnote{According to Copeland
  \cite{copeland2020church-turing}, the term ``Church-Turing thesis''
  was coined by Kleene \cite{kleene1967mathematical}} states that
``Every effectively calculable function is a computable function''.
In the 1930s, on one hand, two main computational models were
developed: ``purely mechanical devices'', the soon-to-become Turing
machines \cite{turing1936computable}, and Church and Kleene's
$\lambda$-definable functions
\cite{kleene35theory1,kleene35theory2,church36unsolvable}. On the
other hand, as described by Turing \cite{turing1938phd}, the notion of
\emph{effectively calculable} ``refers to the intuitive idea without
particular identification with any one of these definitions''.  Turing
showed how these two definitions turn in fact out to be equal
\cite{turing1936computable}, yielding the aforementioned thesis.

With its \emph{physical Church-Turing thesis}, the problem unearthed
by Gandy \cite{gandy1980churchs} can be summarized by asking what
physical process can be regarded as a valid ``purely mechanical
device''. Unlike approaches attempting to describe the notion of
computability from an axiomatic standpoint
\cite{dershowitz2008natural}, Gandy derives a few physical principles
entailing computational constraints on the behavior of any
\emph{reasonable} physical machine. His thesis ``M'' then states that
``what can be calculated by a machine is computable''
\cite{gandy1980churchs}.

Gandy was only considering the context of classical machines, leaving
open the case of quantum computation. For the latter, Deutsch
\cite{deutsch1985quantum} discussed its computational power and
derived that it is no more powerful than classical computation: is the
case closed?  The question is not so clear. For instance, Nielsen
\cite{nielsen1997computable} discusses a paradox, with an
(infinite-dimensional) unitary operator solving the halting problem:
unitaries being the core elementary constructions for quantum
computation, how to reconciliate the paradox with Deutsch's thesis
\cite{james2012embracing, arrighi2012physical}?
According to Arrighi\,\&\,Dowek \cite{arrighi2012physical}, the main
problem lies with the infinite dimensionality that needs to be tamed:
they provide a few physical principles, quantum equivalent to the one
proposed by Gandy and ruling out Nielsen's paradox.
This analysis sheds a new light on the physicality of
non-standard models of quantum computation, such as quantum automata
\cite{dowek2012around,arrighi2019overview}, and generally models based
on \emph{indefinite causal orders}: quantum causal graphs
\cite{arrighi2017quantum}; causal boxes \cite{portmann2017causal};
routed quantum circuits \cite{vanrietvelde2021routed}; quantum
switches \cite{chiribella2013quantum,wilson2020diagrammatic};
supermaps \cite{chiribella2008transforming,wechs2021quantum};
extended circuit diagrams \cite{lorenz2021causal}.

\subsection{A Minimal Quantum Control: the Quantum SWITCH}
\label{sec:qcont-qswitch}

One of the seminal works on \emphidx{quantum control} and superposition
of
causal orders is \cite{chiribella2013quantum}, presenting
the simplest example of \emphidx{non-causal gate}
ordering: the so-called \emphidx{quantum SWITCH}. A presentation
proceeds as follows. Suppose that you are given one \emph{single} copy
of a gate $U$ and a gate $V$, and that you are asked to realize the
operation $\qswitch(U,V)$ acting on two qubits $A$ and $B$:
\begin{equation}\label{eq:qsw-sem}
  \begin{array}{l@{\quad\mapsto\quad}l}
    \ket{0}_A\tensor\ket{y}_B & \ket{0}_A\tensor(UV\ket{y}_B),
    \\
    \ket{1}_A\tensor\ket{y}_B & \ket{1}_A\tensor(VU\ket{y}_B).
  \end{array}
\end{equation}
Provided that $U$ and $V$ are unitary, it is easy to check that this
2-qubit operator is unitary. Depending on the state of the first
qubit, the action on the second qubit is either $UV$ or $VU$. But if
the first qubit is in superposition, the action on the second qubit is
\emph{non-causal}.

Provided that $U$ and $V$ are known, fixed operators, $\qswitch(U,V)$
can be synthesized as a circuit without problem. The difficulty
appears when $U$ and $V$ are unknown: we are therefore looking for a
``higher-order'' circuit with two holes such as

\begin{center}
  \includegraphics[scale=.8]{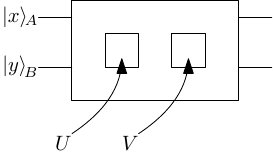}
\end{center}

\noindent
and whose behavior would be the one described in
Eq.~\eqref{eq:qsw-sem}. As we showed in \cite{chiribella2013quantum},
such a ``circuit with a hole'', also known as \emph{quantum comb}
\cite{chiribella2008quantum,chiribella2009theoretical}, cannot
possibly implement the behavior of Eq.~\eqref{eq:qsw-sem}. Of course,
the quantum SWITCH can be realized if you ever had \emph{two} copies
of $U$ and $V$ ---and if you were allowed to control unknown
gates \cite{friis2014implementing}--- as follows:
\[
  \xymatrix@=2ex{
    \ar@{-}[r] & 
    *{\bullet}\ar@{-}[r]\ar@{-}[d] & 
    *{\bullet}\ar@{-}[r]\ar@{-}[d] & 
    *{\circ}\ar@{-}[r]\ar@{-}[d] & 
    *{\circ}\ar@{-}[r]\ar@{-}[d] & 
    \\
    \ar@{-}[r] & 
    *++[F]{U}\ar@{-}[r] & 
    *++[F]{V}\ar@{-}[r] & 
    *++[F]{V}\ar@{-}[r] & 
    *++[F]{U}\ar@{-}[r] & 
  }
\]
But with only one copy of each, this is not possible.

Despite this impossibility within the circuit model, not only the
quantum SWITCH has been shown to be physically realizable
\cite{procopio2015experimental,taddei2021computational} but it has
also been proven to bring a computational advantage
\cite{araujo2014computational,taddei2021computational} to be relevant
in the context of quantum metrology \cite{zhao2020quantum} and
thermodynamics \cite{capela2023reassessing}.

\subsection{Syntactic Approaches for Quantum Control}
\label{sec:review-qcont}

In order to analyze quantum superpositions of execution paths, another
approach followed in the literature 
consists in focusing on the programming language constructs able to
yield such a behavior. The focus is moved towards the control flow of
a quantum computation happening \emph{inside} the coprocessor: a
\emphidx{quantum control} flow instead of the (standard) \emph{classical}
control flow.\index{classical control}

In conventional models of computation with algebraic effects such as
non-deterministic or probabilistic computation, a successful approach
has been to adapt the versatile lambda calculus to the new
paradigm. The mainstream technique consists in encapsulating the
side-effect inside a monad, following Moggi's proposal
\cite{moggi89computational}: the effect is \emph{outside} of the
calculus, only appearing as an epiphenomenon along the execution. A
generic \emph{computational} lambda calculus can be used for any
side-effects representable with a monad. Another approach instead
takes the algebraic effect as a \emph{part of the computation}: the
language is augmented with the corresponding algebraic structure. In
this paradigm, a term non-deterministically reducing to $M$ or to $N$
is typically represented with $M+N$
\cite{deliguoro1995nondet}. Similarly, one can equip the
lambda calculus with a probabilistic sum
\cite{lago2012probabilistic,leventis2016probabilistic}, and more
general cases can be handled by furthermore adding scalar
multiplications \cite{vaux2009algebraic}.

\subsubsection{Van Tonder's Quantum Lambda Calculus}
To make superpositions part of the computation, a natural solution
therefore consists in considering \emph{superpositions of
  lambda-terms}. The first author to try it out was van Tonder
\cite{tonder04lambda} in 2004. His proposal is very intuitive: instead
of coding lambda-terms on a regular memory, let us encode them on the
\emph{quantum} memory; the superposition therefore comes for free. In
this language, one can for instance write
\(
(\lambda x.H\,x)\frac1{\sqrt2}(0+1)
\),
represented in the memory as
\[\textstyle
\ket{(}\otimes\ket{\lambda}\otimes\ket{x}\otimes\ket{.}\otimes\ket{H}\otimes\ket{x}\otimes\ket{)}\otimes\frac1{\sqrt2}(\ket{0}+\ket{1}),
\]
and meant to reduce to $\ket{0}$.
The beta-reduction has to be encoded with quantum operations:
typically with with a unitary map. In his paper, van Tonder describes
the three main problems occurring along the way, and proposes solutions
to them. The first problem is the fact that beta-reduction is not
reversible: As for the case of reversible abstract machines
\cite{kluge1999reversible}, this can be countered by keeping track of
previous moves. The second problem is concerned with implicit
weakening.  Consider as an illustration the term
$\ket{\lambda x.0}\otimes\frac1{\sqrt2}(\ket0+\ket1)$: a naive
interpretation yields $\frac2{\sqrt2}\ket0$, whose norm is not $1$. The
problem comes from the non-linearity of the lambda-term.  Finally, the
third problem occurs with non-trivial terms in
superpositions. Consider for instance
$\frac1{\sqrt2}(\ket{(\lambda xy.x)\,1\,0}+\ket{(\lambda
  xy.y)\,0\,1})$: although this state is arguably of norm $1$, it
should reduce to a state of norm different from $1$. Van Tonder then
describes his solution, which is to restrict the language to a system
where terms in superposition have to be equal, apart for $0$'s and
$1$'s: this essentially amounts to only having classical control, as
this corresponds precisely to the quantum lambda calculus of
\cite{selinger2006lambda}. In other words, this simple, naive approach
fails to capture any quantum control.

\subsubsection{QML}
The first successful attempt at quantum control can be traced back to
QML
\cite{grattage05functional,altenkirch2005qml,altenkirch2005algebra}:
in this line of
work, the authors present the first example of a programmable
\emph{quantum test}, together with a compiler to circuits. In QML, it
is possible to give a formal meaning to the intuitive program
\[
  x:\qbit,y:\qbit\vdash
  \iftermq{x}{{\tuple{x,U\,y}}}{{\tuple{x,V\,y}}}:\qbit\tensor\qbit,
\]
which inputs two qubit wires $x,y$ and compiles down to the circuit
\[
  \xymatrix@=2ex{
    x\ar@{-}[r] & 
    *{\bullet}\ar@{-}[r]\ar@{-}[d] & 
    *{\circ}\ar@{-}[r]\ar@{-}[d] & 
    \\
    y\ar@{-}[r] & 
    *++[F]{U}\ar@{-}[r] & 
    *++[F]{V}\ar@{-}[r] & 
  }
\]
The $\texttt{if}^\circ$-construct consists in a \emph{quantum} test:
unlike the if-then-else construct in the quantum lambda calculus
presented in Section~\ref{sec:qlc-ll}, the qubit $x$ is not measured, and
both branches happen in parallel. For this to make sense, we however
need both branches to somehow yield orthogonal states. For instance,
the term
\[
  x:\qbit,y:\qbit\vdash
  \iftermq{x}{x}{\NOT\,x}:\qbit
\]
is not valid, since it maps $\frac1{\sqrt2}(\ket0+\ket1)$ to
$\frac2{\sqrt2}\ket1$, therefore not preserving the norm.

Altenkirch\,\&\,Grattage proposes a small, first-order language with a
simple type system of tensors of qubits. The system comes with a
syntactic notion of orthogonality, but, partly because of the limited
expressiveness of the type system it is very constrained.
Despite its limitations, it does compile to quantum circuits: it is
therefore ``fully quantum''.


\subsubsection{Linear Algebraic Lambda Calculus}\label{sec:lineal}
QML answers one of the problems of van Tonder's quantum
lambda calculus: superposing distinct execution flows, but at the
expense of expressiveness. An alternative to gain expressive power is to
lift part of the restrictions imposed by the encoding onto the quantum
memory.

Instead of requiring a strict unitarity of the beta-reduction
while asking for a norm condition on terms, \emphidx{Lineal}, the
\emph{linear, algebraic lambda calculus} of Arrighi\,\&\,Dowek
\cite{arrighi2008linear-algebraic,arrighi2017lineal} support
\emph{any} linear combination of terms:
\[
  M,N \quad{:}{:}{=}\quad
  x\bor\lambda x.M\bor MN\bor \alpha\cdot M\bor M+N\bor \vec{0}
\]
This line of work questions the fundamental notion of computation:
what does it mean to \emph{compute in a vector space}
\cite{arrighi2005computational}?

The operational semantics of Lineal formalizes the idea of
``terms-as-operators'': a (pure) term is seen as a basis
vector. The
application is then distributive over sum and scalar multiplication:
\begin{multline}\label{eq:vec-red}
  (\alpha_1\cdot M_1 + \alpha_2\cdot M_2)
  (\beta_1\cdot N_1 + \beta_2\cdot N_2) \to^*\\
  \alpha_1\beta_1\cdot(M_1N_1) + \alpha_1\beta_2\cdot(M_1N_2)
 + \alpha_2\beta_1\cdot(M_2N_1) + \alpha_2\beta_2\cdot(M_2N_2),
\end{multline}
while the $\lambda$-constructor is not, acting like a \emph{thunk}
\cite{ingerman1961thunks,hatcliff1996thunksw}. Formally, the
beta-reduction is extended with a set of rules for manipulating sum
and scalar multiple of terms, such that Eq.~\eqref{eq:vec-red} can be
deduced. Several possibilities exists: are terms considered modulo
associativity and commutativity? modulo the algebraic equational
theory?
Each of these possibilities provide sensible ---and
related--- models of
computation~\cite{diaz-caro2010equivalence,assaf2014call-by-value}:
as shown in \cite{arrighi2005computational}, the equational theory for vector
spaces can be made into a confluent rewrite system.

Lineal follows a call-by-value reduction strategy. Or, more precisely,
a \emphidx{call-by-base} reduction strategy: $(\lambda x.M)V$
only evaluates whenever $V$ is a \emph{pure} term: a term that is not
a distribution.

The underlying idea is that terms are regarded as generalized
operators. In particular, it is then possible to encode matrices:
\[
  U \quad\triangleq\quad
  \left(
    \begin{array}{cc}
      \alpha&\beta\\
      \delta&\gamma
    \end{array}
  \right)
\]
can be regarded as a map acting on the space of terms generated by
$\ttrue \triangleq \lambda xy.x$ and $\ffalse \triangleq \lambda
xy.y$. The matrix $U$ can be modeled with
\[
  M_U\quad\triangleq\quad
  \lambda b.b\,
  (\lambda z.(\alpha\cdot\ttrue+\delta\cdot\ffalse))
  (\lambda z.(\beta\cdot\ttrue+\gamma\cdot\ffalse))\lambda z.z.
\]
The term $M_U\,(a\cdot\ttrue+b\cdot\ffalse)$ then reduces to
\(
(a\alpha+b\beta)\cdot\ttrue+(a\delta+b\gamma)\cdot\ffalse,
\)
corresponding to the operation
\[
  \left(
    \begin{array}{cc}
      \alpha&\beta\\
      \delta&\gamma
    \end{array}
  \right)
  \left(
    \begin{array}{c}
      a\\
      b
    \end{array}
  \right).
\]
As Arrighi\,\&\,Dowek discuss, in Lineal one can encode matrices, vectors,
but also tensor products, and therefore simulate quantum circuits.

If this extension of lambda calculus looks promising, it is however to
take with care. Indeed, consider the following term
\begin{equation}\label{eq:yb}
  Y_M\quad\triangleq\quad (\lambda x.(xx+M))(\lambda x.(xx+M)).
\end{equation}
The term $Y_M$ reduces to $Y_M + M$. But then
\(
Y_M-Y_M
\)
is beta-equivalent to both $M$ and $0$, the null vector: all terms
collapse to zero.

Arrighi\,\&\,Dowek address the problem by by enforcing a rewriting strategy
disallowing the reduction of terms such as $Y_M-Y_M$. This consistency problem
however appears in many algebraic extensions of lambda calculi, and
several approaches to deal with the problem
have been proposed in the literature
\cite{vaux2009algebraic, valiron2010semantics, valiron2013typed}.

\subsubsection{Other Algebraic Extensions of Lambda Calculus}\label{sec:alg-lc}
Lineal fits within the large class of \emph{algebraic extensions of
  lambda calculus}. The origin of the study can be traced back to
Breazu-Tannen, discussing code optimization at compilation time
\cite{tannen1987computable}, and how replacing \texttt{x - x} with
\texttt{0} can be problematic within an untyped setting. This seminal
paper yielded a line of works showing how type discipline can help
\cite{tannen1988combining}, and how the consistency of the system is
related to the confluence of the underlying algebraic rewrite system,
whether in a typed \cite{breazu-tannen1989polymorphic,
  tannen1991polymorphic, tannen1994polymorphic} or in an untyped
setting \cite{dougherty1992adding}.

Algebraic lambda calculi in the style of Lineal ---that is, where
linear combinations of terms are themselves terms--- were introduced
concurrently to Arrighi\,\&\,Dowek \cite{arrighi2008linear-algebraic}
within the context of the differential lambda calculus
\cite{ehrhard2003differential}, stemming from an analysis of
quantitative models of linear logic \cite{ehrhard2002kothe}.  The
interaction of the algebraic structure and the lambda calculus in this
context has then been studied by Vaux \cite{vaux2009algebraic}, who
rediscovered the problems discussed by Breazu-Tannen 20 years earlier
\cite{tannen1987computable}. With linear combinations, the algebraic
structure of terms is very rigid, and Vaux discusses several ways to
recover consistency: with a type system enforcing strong normalization
of terms, or with positive scalars (thus ruling out terms such as the
one of Eq.~\eqref{eq:yb}), or with finitely splittable scalars.

Compared to Arrighi\,\&\,Dowek approach, Vaux's algebraic lambda calculus
\cite{vaux2009algebraic} is \emph{call-by-name}: application is not
distributive on the right, so
\[
  (\lambda x.M)(\alpha\cdot N_1 + \beta\cdot N_2)
  \to
  M[x := \alpha\cdot N_1 + \beta\cdot N_2]
\]
while
\[
  M(\alpha\cdot N_1 + \beta\cdot N_2)
  \quad
  \not=
  \quad
  \alpha\cdot MN_1 + \beta\cdot MN_2.
\]
The two are incompatible, as it is already the case, say, in
probabilistic computation: tossing a coin and duplicating the result
is not the same thing as tossing twice the coin.

\section{Typing Superpositions of Lambda-Terms}
\label{sec:veclc}

This section presents our work on the development of a type system for
the lambda calculus presented in Section~\ref{sec:lineal}. Extended
with linear combinations of terms, this lambda calculus aims at
modeling quantum superposition of programs.

The challenge addressed in this section concerns the validity of a
lambda-term in superposition. How can we decide whether such a program
indeed represents a physical, quantum operation? We want for instance
a program to correspond to a unitary operation.

One of the formal tools to separate ``valid'' programs
from ``invalid'' ones is the use of a type system. It consists in a
formal term annotation, stable under composition, and characterizing a
property we want ``valid'' programs to satisfy. Typical use-cases for a
type system are termination and error-freeness.

This section presents two type systems for a linear algebraic
lambda calculus. The type system of Section~\ref{sec:vectype} comes as
a set of sophisticated, compositional definitions. Proving properties
of well-typed terms then requires complex proofs. The type system
presented in Section~\ref{sec:realiz} is instead defined organically
using the operational semantics: a type is a set of terms with
suitable properties. The compositionality of the type system is
derived as a corollary. We discuss how the system we obtain is more
fine-grained.

\begin{mylife}
  Both of the works presented here are collaborations with Alejandro
  Díaz Caro. The one discussed in Section~\ref{sec:vectype} started
  while Alejandro was doing his Ph.D---we are still collaborators
  nowadays. In Section~\ref{sec:realiz}, I present a work Alejandro
  and I realized later on, on a collaboration with the logic group in
  Montevideo (Uruguay).
\end{mylife}

\subsection{An Axiomatic Type System: Vectorial System-F}
\label{sec:vectype}

As discussed in Section~\ref{sec:alg-lc}, the simplest strategy to recover
consistency for an algebraic lambda calculus is to add a type system
enforcing termination.

Vaux's simple type system is very natural: it consists in typing
$\sum_i\alpha_i\cdot M_i$ with $A$ as long as each $M_i$ can be typed
with $A$. This approach is akin to the approach one can follow in the
context of probabilistic or non-deterministic behavior: terms ``in
superpositions'' should share the same type $A$, and the overall
``computation'' is then given the type $A$.

Instead, the approach we followed in \cite{arrighi2017vectorial}
is to allow terms with
distinct types to be summed. This section describes the approach.

\subsubsection{Simply-Typed Vectorial Lambda Calculus}
\label{sec:st-vec}

The grammar for the vectorial lambda calculus is the same as the one
of Lineal,
presented in Section~\ref{sec:lineal}. A simple type system is as follows:
\[
  A,B\quad{:}{:}{=}\quad
  X \bor A\tto B.
\]
Following the Lineal approach, the rewrite-system of the vectorial
lambda calculus is \emphidx{call-by-base}.

\paragraph{Coding Qubits.}
In the regular lambda calculus the Boolean values $\ttrue$ and $\ffalse$ can
be coded with $\lambda xy.x$ and $\lambda xy.y$. These terms can both
be typed with $X\tto X\tto X$. Within the vectorial
lambda calculus, it is possible to write any linear combination
\[
  \alpha\cdot\lambda xy.x + \beta\cdot\lambda xy.y,
\]
and the typing rules can give to all of these terms the type
$X\tto X\tto X$.  If scalars range over the complex field,
and if we impose $|\alpha|^2+|\beta|^2=1$, we can claim to have
embedded the state of quantum bits in the vectorial lambda calculus.

\paragraph{Quantum If.}
With the Boolean values coded as $\lambda xy.x$ and $\lambda xy.y$, in
the regular lambda calculus, the if-then-else construct
$\iftermx{M}{N}{P}$ can simply be written with $(MN)P$. As we are in a
call-by-value setting, we want to forbid the branches of the test to
evaluate: we can use \emphidxalt{thunks}{thunk}
\cite{ingerman1961thunks,hatcliff1996thunksw} to ``freeze'' the
computations in the branches, as follows:
\begin{equation}\label{eq:defifvec}
  \iftermx{M}{N}{P}
  \quad\triangleq\quad
  ((M(\lambda z.N))(\lambda z.P))\,\ast
\end{equation}
where $z$ is a fresh variable and $\ast$ is any closed normal form,
for instance $\lambda x.x$.

The vectorial lambda calculus being call-by-value, we can rely on the
encoding of Eq.~\eqref{eq:defifvec} to emulate the behavior of a
``quantum test'' as in QML. As the language does not enforce any
unitary constraint, we can in fact encode \emph{any} matrix. Consider
for instance the map $U$ sending $\ttrue$ to
$\alpha\cdot\ttrue+\beta\cdot\ffalse$ and $\ffalse$ to
$\gamma\cdot\ttrue+\delta\cdot\ffalse$. The operator $U$ can be
emulated with the term
\begin{equation}\label{eq:opAvec}
  U
  \quad\triangleq\quad
  \lambda x.\iftermx{x}{
    (\alpha\cdot\ttrue+\beta\cdot\ffalse)
  }{
    (\gamma\cdot\ttrue+\delta\cdot\ffalse)
  }
\end{equation}
using the encoding of Eq.~\eqref{eq:defifvec}.

Typing the operator $U$ with the simple type system is akin to typing
if-then-else in the regular simply-typed lambda calculus: to get a
portable solution the type system misses universal quantifiers. Such
quantifiers can be \emph{à priori} easily added to the vectorial
lambda calculus, with the two standard typing rules
\[
  \infer{\Delta\vdash M:A[X:=B]}{
    \Delta\vdash M:\forall X.A
  }
  \qquad
  \infer{\Delta\vdash M:\forall X.A}{
    \Delta\vdash M:A
    &
    X\not\in\text{FV}(\Delta)
  }
\]
The Boolean values (and their linear combinations)
can then be typed with $\forall X.X\tto X\tto X$, and the operator $U$
with
\[
  (\forall X.X\tto X\tto X)\tto(\forall X.X\tto X\tto X).
\]
However, as expressive as it is this type system is unable to capture
algebraic properties of terms. Several studies have in particular been
performed by Arrighi\,\&\,Díaz-Caro \cite{arrighi2009scalar,
  diaz-caro2017typing,
  arrighi2012system,arrighi2011subject,arrighi2017vectorial}, with the
addition of scalars or more generally a vectorial structure to types.

\subsubsection{Quantifiers: Vectorial System-F}
\label{sec:vecsysF}

The objective of this section is to present the work initiated in
\cite{arrighi2011subject} and achieved in
\cite{arrighi2017vectorial}. Its aim is to capture some algebraic
properties of the vectorial lambda calculus within a type
system. Schematically, if $M:A$ and $N:B$, we aim at a meaningful way
of saying that $\alpha\cdot M+\beta\cdot N$ is of type
$\alpha\cdot A + \beta\cdot B$. The type-system should also be
expressive enough to be able to give a parametric type to the operator
$U$ presented in Eq~\eqref{eq:opAvec}. This was conceived as a first
step towards a type system enforcing unitarity constraints.

The language of Section~\ref{sec:st-vec}
is now equipped with the type grammar
\[
  \begin{array}{llll}
    \text{Types}
    & T,R,S
    & {:}{:}{=}
    & U \bor \alpha\cdot T \bor T+R \bor \mathbb{X},
    \\
    \text{Unit Types}\qquad
    & U,V,W
    & {:}{:}{=}
    & \mathcal{X} \bor U\tto T \bor \forall\mathcal{X}.U \bor \forall\mathbb{X}.U.
  \end{array}
\]
The type system acknowledges the fact that any term is first and
foremost a linear combination of \emphidxalt{base terms}{base term}: a
general type is therefore a linear combination of \emph{unit types},
where pure types are meant to type \emph{base terms}. A base type
cannot be a linear combination: it is therefore either an arrow-type
or a quantified type.
The type system features two kinds of type variables: type variables
$\mathbb{X}$ standing for general types, and type variables $\mathcal{X}$
standing for unit types. 
Arrow types reflect the fact that the language is call-by-base: the
domain of an arrow type is a unit-type. Indeed, consider
$\lambda x.M$: although $M$ can be any term, the term variable $x$ can
only be replaced by a base term, and base terms are meant to be typed
with unit types.

The types come equipped with an equivalence relation $\equiv$, defined
as follows:
\[
  \begin{array}{r@{~\equiv~}l@{\qquad}r@{~\equiv~}l}
    1\cdot T & T
    & \alpha\cdot T + \beta\cdot T &(\alpha+\beta)\cdot T
    \\
    \alpha\cdot(\beta\cdot T) & (\alpha\beta)\cdot T
    & T + R & R + T
    \\
    \alpha\cdot T + \alpha\cdot R & \alpha\cdot(T+R)
    & T + (R + S) & (T + R) + S
  \end{array}
\]
This relation is used in the typing rules for the algebraic aspect of
the language, as follows:
\[
  \infer[(\alpha_I)]{
    \Gamma\vdash \alpha\cdot M : \alpha\cdot T
  }{
    \Gamma\vdash M : T
  }
  \quad
  \infer[(+_I)]{
    \Gamma\vdash M + N : R + T
  }{
    \Gamma\vdash M : R
    &
    \Gamma\vdash N : T
  }
  \quad
  \infer[(\equiv)]{
    \Gamma\vdash M : T
  }{
    \Gamma\vdash M : R
    &
    R \equiv T
  }
\]
Finally, the term $0$ can be given any \emph{inhabitated} type, as
follows:
\[
  \infer[(0_I)]{
    \Gamma\vdash 0 : 0\cdot T
  }{
    \Gamma\vdash M : T
  }
\]
This rules out the possibility to introduce bogus, empty types inside
a linear combination. It stems from the fact that the only possible
way to introduce a $0$-term is through the rewrite rule
$0\cdot M\to 0$.

One peculiar thing to note in the type system is the absence of a
$0$-type: there is no notion of ``empty'' linear combination of
types. One of the reasons is consistency: With a $0$-type, it would
make sense to ask for $0\cdot T \equiv 0$, thus rendering all
$0$-scalared types equal, including empty ones.

\paragraph{Dealing with functions.}
The typing rule for the lambda-abstraction follows the intuition given
while describing the type system:
\[
  \infer[(\tto_I)]{
    \Gamma\vdash \lambda x.M : U\tto R
  }{
    \Gamma, x:U \vdash M : R
  }
\]
The rule for the application is more involved, as we want to be able
to handle for instance terms of the form $(M_1+M_2)(N_1+N_2)$. For the
purpose of the discussion, we only give an example
and we invite the reader to consult the full paper
\cite{arrighi2017vectorial} for details.

Because quantifiers can only happen at the level of unit-types, the
typing rule for application contains both an arrow-elimination and a
quantifier-elimination. In order to illustrate what we mean, assume
that we have defined pairs and projections in the usual way, using the
second-order.
Now, the term $(\pi_1+\pi_2)(\tuple{U_1,U_2} + \tuple{V_1,V_2})$ reduces to
$U_1+U_2+V_1+V_2$: assuming that the $U_i$'s and $V_i$'s are
well-typed, this should also be well-typed. Following the way the
rewrite procedure distributes the application over the sum, the rule can be
stated as
\[
  \infer{
    \Gamma\vdash MN : U_1 + U_2 + V_1 + V_2.
  }{
    \Gamma\vdash M : \forall
    \mathcal{X}\mathcal{Y}.(\mathcal{X}\times\mathcal{Y}\tto\mathcal{X})
    +
    \forall
    \mathcal{X}\mathcal{Y}.(\mathcal{X}\times\mathcal{Y}\tto\mathcal{Y})
    &
    \Gamma\vdash N : (U_1\times U_2) + (V_1\times V_2)
  }
\]
Both of the types $U_1\times U_2$ and $V_1\times V_2$ are matched
against the domain $\mathcal{X}\times\mathcal{Y}$ of the function, and
in each case each summand of the type of the function yields its part,
therefore producing all of the $U_i$'s and $V_j$'s.

This instance of the application rule features all of the subtleties
of the general version presented in \cite{arrighi2017vectorial}.

\paragraph{Discussion.}
The vectorial lambda calculus can emulate (finite-dimensional) linear
operations: one can encode quantum circuits in the
language. The proposed vectorial System-F is then expressive
enough to correctly type this encoding.
Moreover, the type system enforces consistency: all typed terms are strongly
normalizing.
However, the standard property of subject reduction, stating that if $M\to N$
and $M:A$ then $N:A$ does not quite hold in our system. The problem
comes from the mismatch between the equivalence on types that does not
equate $T$ and $T + 0\cdot R$ and the rewrite system, sending all
terms of the form $0\cdot M$ to $0$. The language however features a
weakened version, based on a relation $\sqsupseteq$ satisfying several
rules, among which
\(
  T \sqsupseteq T + 0\cdot R,
\)
capturing the ``problem'' with the zero-term.

\subsection{A Type System Based on Realizability}
\label{sec:realiz}

Two strategies can be followed in order to design a type system. On
one hand, one can define a formal grammar of types, and types can then
be attached to terms using an (external) set of axiom rules.
This is what has been done in Section~\ref{sec:vectype}
A type system usually aims at capturing various properties of typed terms:
These properties are then proven from the typing rules, as
corollaries. On the other hand, one can follow the route of
\emphidx{realizability} \cite{kleene1945interpretation} and instead
define types inductively, as \emph{sets of terms}. The properties to
enforce on typed terms can be added as constraints on the definition
of the types. In this situation, typing rules becomes lemmas to be
proven, instead of primitive axioms. The fact that a well-typed term
verifies one of desired property is obtained ``by definition''.

\subsubsection{Example based on simply-typed lambda calculus}

The archetypal example is the proof of strong normalization of a
typed lambda calculus. On one hand, one can set up the type up front
with an abstract grammar:
\[
  \begin{array}{lll}
    M,M & {:}{:}{=} & x \bor \lambda x.M \bor MN,
    \\
    A,B & {:}{:}{=} & \sigma \bor A \tto B,
  \end{array}
\]
set up a notion of typing judgment $x_1:A_1,\ldots,x_n:A_n\vdash M:B$
and define what it means to be a valid typing judgment with a series
of typing rules, posed as axioms:
\begin{equation}\label{eq:st-lc}
  \infer{
    \Delta,x:A\vdash x:A
  }{
  }
  \qquad  
  \infer{
    \Delta\vdash\lambda x.M:A\tto B
  }{
    \Delta,x:A\vdash M:B
  }
  \qquad
  \infer{
    \Delta\vdash MN:B
  }{
    \Delta\vdash M:A\tto B
    &
    \Delta\vdash N:A
  }
\end{equation}
The rewrite system of lambda calculus is based on beta-reduction:
$M\to N$ is defined as the smallest congruent relation satisfying
\(
  (\lambda x.M)N\to M[x\mapsto N].
\)
It is well-known that although the untyped lambda calculus is not
strongly normalizing, the simply-typed fragment is. A proof can be
designed using \emphidx{reducibility candidates}
\cite{tait1967intensional,girard89proofs}.

The intuition behind reducibility candidates consists in defining for
each type $A$ a set of terms $\RED{A}$, called \emph{reducibility
  candidates of type $A$}. They are defined by induction, following
the ``structure'' of the types. For instance, $M\in\RED{A\tto B}$
whenever for all $N\in\RED{A}$, we have $MN\in\RED{B}$. For the base
case $\RED{\sigma}$, we enforce the desired property:
$M\in\RED{\sigma}$ whenever it is strongly normalizing and of type
$\sigma$. One then derives various properties of these sets of terms,
from which one can conclude strong normalization of well-typed terms.

An alternative approach to typing consists in directly starting from the
computational behavior given by the beta-reduction. In this setting,
we start from an untyped lambda calculus, and we define types in a
semantic manner as \emph{closed normal forms}.
A term $M$ is a \emphidx{realizer} for a type $A$,
denoted with $M\Vdash A$, if $M$ reduces to a term in $A$.
We then define the arrow $(\tto)$ as an operator on sets of terms as
follows:
\begin{equation}\label{eq:realiz-func}
  X\tto Y \quad\triangleq\quad
  \left\{~~
    \lambda x.M \text{ closed term}
    ~~\middle|~~
    \forall N \in X, ~ M[x\mapsto N] \Vdash Y
  ~~\right\}.
\end{equation}
Note how realizers of $X\tto Y$ are strongly normalizing when the
realizers of $X$ and $Y$ are strongly normalizing.
Typing judgments of the form $x_1:A_1,\ldots,x_n:A_n\vdash M:B$ are
then defined as a shortcut notation for
\[
  \forall N_1\in A_1,\ldots N_n\in A_n,~~M[x_1\mapsto
  N_1,\ldots,x_n\mapsto N_n]\Vdash B.
\]
The 3 typing rules of Eq.~\eqref{eq:st-lc} then become
lemmas: we essentially have to choose as base type $\sigma$ a set of
strongly normalizing closed terms\ldots

The system is very versatile: it can be applied to extended
lambda calculi, and other type operators can be constructed from their
computational interpretation: products, co-products, lists,
quantifiers, \etc \cite{lepigre2016semantics}.

\subsubsection{Weak Vector Spaces}

A realizability model heavily relies on the \emph{computational
  behavior} of an \emph{untyped} language. As discussed in Section~\ref{sec:lineal},
without types the consistency of the vectorial lambda calculus is not
guaranteed, seemingly jeopardizing any meaningful notion of normal
form and, therefore, computation.

In a research thread \cite{valiron2010semantics,valiron2013typed}, we
propose a solution to this problem. The idea is akin to what was
proposed for the algebraic structure of the vectorial system-F:
disconnect the $0$-vector and vectors scaled to $0$: the latter
register what information got zero-ed out while in the former case
everything is discarded. When the information is consistent one can
indeed identify both approaches --- but this is not possible anymore
if the information is inconsistent, as for instance with the term
$Y_M$ of Eq.~\eqref{eq:yb}.

The solution we propose for retaining consistency in a vectorial
lambda calculus where arbitrary fixpoints are allowed is to
\emph{weaken} the equations of module by disallowing the rule equating
$0\cdot M$ and $\vec{0}$: if we still have $Y_M-Y_M=0\cdot Y_M$, one
cannot anymore get to $\vec{0}$.
Formally, if $(\mathcal{A},1,\star,0,+)$ is a ring, an
\emphidx{weak $\mathcal{A}$-module} $(M,+,\vec{0},\cdot)$ is the data
consisting of a commutative monoid $(M,+,\vec{0})$ and an operation
$(\cdot):\mathcal{A}\times M\to M$ such that for all
$a,b\in\mathcal{A}$ and for all $x,y\in M$,
\[
  \begin{array}{r@{~=~}l@{\quad}r@{~=~}l}
    a\cdot(x+y) & a\cdot x + a\cdot y,
    &  a\cdot (b\cdot x) & (a\star b) \cdot x,
    \\
    (a+b)\cdot x & a\cdot x + b\cdot x,
    & 1\cdot x & x.
  \end{array}
\]
In particular, we do \emph{not} impose $0\cdot x = \vec{0}$, meaning
that $\vec{0}$ is not the same as $x + (-1)\cdot x$.

By turning the vectorial lambda calculus into a \emph{weak} module,
the term $Y_M-Y_M$ still exists but cannot be used to collapse the
equational theory anymore: it does not equate $\vec{0}$ anymore.  This
has been formalized in \cite{valiron2013typed}, where I show how a
typed vectorial lambda calculus with fixpoint based on an equational
theory of weak module admits a non-trivial model.

\subsubsection{Realizability Model Capturing Unitary}

Based on \cite{valiron2013typed}, it is possible to give a sound
computational interpretation of a vectorial lambda calculus (based on
weak modules). With such a computational interpretation, one can then
design a realizability model. Interestingly enough, it is even
possible to capture \emph{within a type system} a notion of unitarity:
this is the topic of a collaboration with Alejandro Díaz Caro,
Mauricio Guillermo and Alexandre Miquel
\cite{diaz-caro2019realizability}.

The language we consider is a (vectorial) lambda calculus extended
with constructs to deal with a unit term $\star$, pairing and (binary)
injections. For the purpose of the discussion, we note the pairing of
$M$ and $N$ as $\tuple{M,N}$, and we consider Boolean values $\ttrue$
and $\ffalse$ defined in the usual way using injections. The pairing
is bilinear with respect to the weak module structure, so for instance
we have
\[
  \tuple{M_1+M_2,\alpha\cdot N}
  =
  \alpha\cdot\tuple{M_1,N} + \alpha\cdot\tuple{M_2,N}.
\]
The operational semantics is ``call-by-base'': one does not reduce
under lambda-abstractions.

\begin{notation}
  For this section, we use the following notation: $V,W$ stands for
  pure values: lambda-abstractions, pairs of values, or injections
  of values ; $\vec{V},\vec{W}$ stands for linear combinations of pure
  values, and $M,N$ stands for general terms.
\end{notation}

Since the ring is the field of complex numbers, we show in the paper
that some of the notion of Hilbert spaces can be defined in this
weakened context. In particular, one can define the naive notion of
scalar product and $\ell^2$-norm as follows:
\[
  \begin{array}{l@{~=~}l}
    \left\langle~
    \sum_i\alpha_i\cdot V_i
    ~\middle|~
    \sum_j\beta_j\cdot W_j
    ~\right\rangle
    &
      \sum_{i,j}\overline{\alpha_i}\beta_j\cdot\delta_{V_i,W_j}
    \\
    \left|\left|~
    \sum_i\alpha_i\cdot V_i
    ~\right|\right|
    &
      \sqrt{\sum_i|\alpha_i|^2}
  \end{array}
\]
where the $V_i$'s and the $W_j$'s are pure values and $\delta_{V,W}=1$
if $V=W$ and $0$ otherwise\footnote{There is some subtlety in term
  equality. Please refer to the paper
  \cite{diaz-caro2019realizability} for details}. This gives a notion
of orthogonality, making for instance $\ttrue$ and $\ffalse$
orthogonal. We can define the \emph{span} $\text{span}(X)$ and the
\emph{basis} $\flat X$ of a set of terms $X$
\[
  \begin{array}{l@{~=~}l}
    \text{span}(X) & \left\{~\sum_{i=1}^n\alpha_i\cdot
                     M_i~~\middle|~~n\in\mathbb{N},~\forall i, M_i\in
                     X~\right\},
    \\
    \flat X & \left\{~ V~~\middle|~~\alpha\cdot V + M \in X~\right\}.
  \end{array}
\]
We can also define the \emphidx{unit sphere} of values as
\[
  S_1 = \left\{~\vec{V}~~\middle|~~||\vec{V}|| = 1~\right\}.
\]
This gives a canonical notion of ``normalized vector'' in the ``vector
space'' of linear combinations of terms.

In this model, the type of Boolean values can then be defined as
$\mathbb{B} \triangleq \{\,\ttrue,\ffalse\,\}$, and the type of
quantum bits as
\(
\mathbb{Q}~\triangleq~\text{span}(\mathbb{B})\cap S_1.
\)
A qubit is then literally a superposition of Boolean values.

In general, we define a \emphidx{unitary type} as a subset of $S_1$:
In \cite{diaz-caro2019realizability}, on top of $\flat$ we
define several unitary type operators, among which:
\begin{align*}
  A\times B &= \left\{~\tuple{\vec{V},\vec{W}}~\middle|~\vec{V}\in A,
              \vec{W}\in B~\right\},
  &
  A\otimes B &= \text{span}(A\times B)\cup S_1.
\end{align*}
Note how $\mathbb{Q}\otimes\mathbb{Q}$ consists in normalized linear
distributions of pairs of Boolean values.

We can define function types as discussed for
Eq.~\eqref{eq:realiz-func}, and from the system we can derive that the
function type $\mathbb{Q}\to\mathbb{Q}$ corresponds to unitary
operators on the Hilbert space $\mathbb{C}^2$. We then derive a set of typing rules
\cite[Tab.~6]{diaz-caro2019realizability}, providing a compositional
way to construct terms that enforces unitarity ``by construction''.

\paragraph{Discussion.}
In order to assess the expressiveness (and the versatility) of the
language, we show how to embed the quantum lambda calculus (albeit
without measurement), similar to the one discussed in
Section~\ref{sec:qlc-ll}. Moreover, because of vectorial structure of
the language, we show that we can also define a control operator, and,
in general define a form of quantum SWITCH (discussed in
Chapter~\ref{ch:qcont}), thus offering a framework for both classical
and quantum control.

However, although this paper offers a solution to the long-standing
question of seeing vectorial lambda calculus as a medium for
representing quantum computation, it is still unsatisfactory.
The main limitation stands in the discrepancy between the
\emph{semantic} interpretation of unitary terms in the realizability
model and their concrete, syntactic structure. One interesting
research direction consists in developing a compilation framework to
turn valid terms into executable quantum circuits.

\section{Reversible and Quantum Pattern-Matching}
\label{sec:qcont-patt}

The vectorial lambda calculus is based on a inherently irreversible
model of computation. Instead of trying to adjust its defects to
purely quantum computation, an alternative approach consists in
changing the paradigm and moving towards \emphidx{reversible
  computation}.

This section is devoted to a model of computation alternative to
lambda calculus, reversible and based on pattern-matching
\cite{sabry2018symmetric}. This computational model shares links with
the vectorial lambda calculus, yet it allows a finer control over
problematic aspects: there is no need for weak
modules, yet the system supports a notion of recursive behavior.

Section~\ref{sec:back-rev-lang} presents the concept of reversible
language, while Section~\ref{sec:rev-pat} discusses how reversible
pattern-matching can be seen as a primitive design for a reversible
language.  Section~\ref{sec:rev-cat} sketches its categorical
interpretation, while Section~\ref{sec:revpat-ind} provides an
strategy for introducing inductive types and
recursion. Section~\ref{sec:rev-qcont} discusses how to extend the
language to quantum control, and Section~\ref{rev:mumall} reviews a
Curry-Howards interpretation based on the logic {\mumall}.

\begin{mylife}
  If this line of work started from an epistolary discussion with Amr
  Sabry and Juliana Vizzotto, it has become one of my main research
  vehicles in recent years. In particular, reversible pattern matching
  has been the subject of the Ph.D of my students Kostia Chardonnet
  \cite{chardonnet2023phd}, co-supervised with Alexis Saurin (IRIF)
  and Louis Lemonnier \cite{lemonnier2024phd}, co-supervised with
  Vladimir Zamdzhiev.
\end{mylife}

\subsection{Background on Reversible Language}
\label{sec:back-rev-lang}

Discussed in Section~\ref{sec:oracle-gen}, the subject of reversible
computation has spurred an avenue of research in programming
languages and type systems.

On one hand, following the trend of research oriented towards
reversible architectures \cite{frank2020special}, a line of research
aims at designing imperative programming languages targeting
reversible computation. The ancestor of such languages is arguably
Janus \cite{lutz1986janus}, an imperative flow-chart language
rediscovered in \cite{yokoyama2007reversible} and studied in details
in \cite{yokoyama2016fundamentals}. Reversible computation is linked
to linearity, and \cite{baker1992nreversal,matos2003linear} explore
the subject with respectively $\Psi$-lisp and SRL, akin to a linear
Janus language.

In the realm of syntactic functional languages, lambda calculus is not
well-suited --- although there has been a proposal for a reversible
combinatory algebra \cite{pierro2006reversible}. In terms of syntax, a
seminal proposal for a reversible, functional language is the untyped
Rfun \cite{yokoyama2011reversible} --- although Thomsen concurrently
aimed at a proposal \cite{thomsen2012functional}. From this seminal
Rfun several extensions were developed: heap manipulation and
algebraic datatypes in \cite{axelsen2013reversible}, the ability to
manipulate non-linear objects in \cite{mogensen2014reference}, the
addition of garbage collection in \cite{mogensen2018reversible}. On
the side of concrete use of the language, a series of examples of code
have been proposed \cite{thomsen2015interpretation}, while on the
formal side a core fragment of the language (CoreFun) has been
analyzed in \cite{jacobsen2018corefun}.

A last trend consists in the study of point-free languages. The first
approach is the language INV \cite{mu2004injective}, followed by
$\Pi$, presented by James and Sabry \cite{james2012information} with a
sound presentation of isomorphism between types and a discussion on
irreversibility as side effect. A (reversible) compiler for the
language has also been designed \cite{james2012isomorphic}. A specific
discussion on type systems for reversible programming as semirings
can be found in \cite{carette2016computing}.

To conclude, an approach linking point-free languages and regular,
typed functional language is Theseus
\cite{james2014theseus}. Initiated in \cite{sabry2018symmetric}, the
line of work I follow can be seen as the study of a formalization of a
core of Theseus.

\subsection{Reversible Pattern-Matching}
\label{sec:rev-pat}

In \cite{sabry2018symmetric}, we propose a simple yet extensible model
of reversible computation. Consider a grammar of typed patterns\index{pattern}
defined as follows:
\[
  \begin{array}{rcl}
    v,w & ~{:}{:}{=}~ & x \bor \star \bor \tuple{v,w} \bor \inl(v) \bor
                        \inr(v),
    \\
    a,b & ~{:}{:}{=}~ & 1 \bor a\otimes b \bor a\oplus b,
  \end{array}
\]
where $x$ ranges over a set of (typed) variables, $\star$ is a unit
term of unit type $1$, $\tuple{v,w}:a\otimes b$ is a pair of a value
$v:a$ and a value $w:b$, and provided that $v:a$, we have
$\inl(v):a\oplus b$ and $\inr(v):b\oplus a$.
A pattern $v$ is valid if each variable appears at most once in it. We
denote with $\FV(v)$ the set of variables appearing in $v$.
A closed pattern is called a \emphidx{value}.

In a functional setting, a reversible computation between type $a$ and
type $b$ is a bijection between values of the respective types. Aiming
at an applicative language on structured types, in
\cite{sabry2018symmetric} we follow the standard strategy
\cite{landin1966next,burstall1969proving,turner1979implementation}
and propose a syntax of isomorphisms based on
pattern-matching.

An \emphidx{iso} consists of a set of clauses
\[
  \left\{
    \begin{array}{rcl}
      v_1 & \iso & v'_1 \\ & \vdots & \\ v_n & \iso & v'_n
    \end{array}
  \right\},
\]
where for each $i$, we have $\FV(v_i)=\FV(v'_i)$. The iso is well-type
of type $a\iso b$ whenever each $v_i$ is of type $a$ and each $v'_j$
is of type $b$. For the iso to be well-defined, the patterns on the
left of each clause should be \emphidx{non-overlapping}. For the iso
to be injective, the patterns should furthermore be
\emphidx{exhaustive}: all values of type $a$ should have a matching
pattern in the iso. To enforce bijectivity, the same two constraints
should also be set on the \emph{right}-hand-side of the clauses of the
iso.

In other word, an iso defines a bijection between values of type $a$
and values of type $b$ if and only if the patterns are exhaustive and
non-overlapping on the left \emph{and} on the right. For instance, the
(bijective) iso $\omega$ of type
$a\otimes(b\oplus c)\iso(a\otimes b)\oplus (a\otimes c)$ can be
defined
\[
  \left\{
    \begin{array}{rcl}
      \tuple{x,\inl\,y} & \iso & \inl\,\tuple{x,y}
      \\
      \tuple{x,\inr\,y} & \iso & \inr\,\tuple{x,y}
    \end{array}
  \right\}.
\]
One can check that the left hand-side of the iso satisfies the two
required properties: Any value of type $a\otimes(b\oplus c)$ is either
of the form $\tuple{x,\inl\,y}$ or of the form $\tuple{x,\inr\,y}$.

\subsection{A Categorical Interpretation}
\label{sec:rev-cat}

\begin{mylife}
  Together with Kostia Chardonnet and Louis Lemonnier, we studied the
  categorical structure underlying the pattern-matching presented in
  Section \ref{sec:rev-pat}. This work yielded a publication
  \cite{lemonnier2021categorical} presented in this section. The text
  is taken from the introduction of the paper.
\end{mylife}

The categorical analysis of partial injective maps have been thoroughly
analyzed since 1979, first by Kastl \cite{kastl1979inverse}, and then
by Cockett and Lack
\cite{cockett2002restriction-I,cockett2003restriction-II,cockett2007restriction-III}. This
led to the development of inverse category: a category equipped with
an inverse operator in which all morphisms have partial inverses and
are therefore reversible. The main aspect of this line of research is
that partiality can have a purely algebraic description: one can
introduce a restriction operator on morphisms, associating to a
morphism a partial identity on its domain.  This categorical framework
has recently been put to use to develop the semantics of specific
reversible programming constructs and concrete reversible languages:
analysis of recursion in the context of reversibility
\cite{axelsen2016join,kaarsgaard2019inversion,kaarsgaard2019engarde},
formalization of reversible flowchart languages [12, 22], analysis of
side-effects \cite{heunen2015reversible,heunen2018reversible},
etc. Interestingly enough however, the adequacy of the developed
categorical constructs with reversible functional programming
languages has been seldom studied. For instance, if Kaarsgaard et
al. \cite{kaarsgaard2017join} mention Theseus as a potential use-case,
they do not discuss it in details. So far, the semantics of functional
and applicative reversible languages has always been done in concrete
categories of partial isomorphisms
\cite{kaarsgaard2019engarde,kaarsgaard2021join}.

In particular, one important aspect that has not been addressed yet in
detail is the categorical interpretation of pattern-matching. If
pattern-matching can be added to reversible imperative languages
\cite{gluck2019reversible}, it is particularly relevant in the context
of functional languages where it is one of the core construct needed
for manipulating structured data. This is for instance emphasized by
the several existing languages making use of it
\cite{yokoyama2011reversible,thomsen2015interpretation,
  james2014theseus,jacobsen2018corefun,sabry2018symmetric,
  chardonnet2020curry-howard,chardonnet2022curry-howard}.
In the literature, pattern-matching has
either been considered in the context of a Set-based semantics
\cite{gluck2019reversible}, or more generally in categorical models
making heavy use of rig structures \cite{carette2016computing} or
co-products \cite{kaarsgaard2021join,kaarsgaard2019engarde} to
represent it. If such rich structures are clearly enough to capture
pattern-matching, we show in \cite{lemonnier2021categorical} that they
are too coarse, and that a weaker structure is enough for
characterizing pattern-matching.

\subsection{Inductive Types, Fixpoints and termination}
\label{sec:revpat-ind}

As it stands, the language is very limited: it is for instance not
possible to express natural numbers, or lists. One simple solution
consists in extending the language with \emphidx{inductive types} and
\emphidx{recursion}.

Through the Curry-Howard isomorphism, inductive types are the twin siblings of
inductively defined predicates. Arguably, the formalization of
induction takes its root on one hand \cite{backhouse1989diy} from the
theory of inductive definitions
\cite{feferman1970formal,buchholz1981iterated,aczel1977introduction}
and Martin Löf's type theory
\cite{martin-lof1971hauptsatz,martin-lof1984intuitionistic}, and on
the other hand from de Bakker and Scott's $\mu$-calculus
\cite{scott1969theory,bakker1971recursive,pratt1981decidable,kozen1983results}. The
main design choice consists in whether to use an equational
presentation with named constructors
\cite{coquand1988inductively,dybjer1991inductive,dybjer1994inductive}
or an \emph{anonymous} presentation using $\mu$-abstractions
\cite{bakker1972calculus,deroever1974recursive,mendler1988phd,matthes1998phd},
in the same way $\lambda$-abstractions can be used to express
(unnamed) functionals.

If in \cite{sabry2018symmetric} we use named constructors for the
dedicated type constructor $[a]$ for lists, in later works
\cite{chardonnet2020curry-howard,chardonnet2022curry-howard}
we work with the more generic anonymous inductive types, using a
$\mu$-construction, as follows.
\[
  a,b \quad{:}{:}{=}\quad X \bor 1 \bor a\otimes b \bor a\oplus b
  \bor \mu X.a.
\]
The additional type constructor $\mu X.a$ comes with a pattern/value
constructor $\tfold$. The typing rules enforces the equivalence
\[
  \tfold\, v : \mu X.a \quad\sim\quad v : a[X \mapsto \mu X.a],
\]
encapsulating de Bakker and Scott's induction principle
\cite{scott1969theory}.
A type constructor $[a]$ for lists can for instance be defined as
$[a]\triangleq \mu X.1\oplus(a\otimes X)$. The list constructor
$\tnil$ and $v_1\tcons v_2$, standing respectively for the empty list
and the cons operation, can be defined as
$\tnil\triangleq \tfold\,\inl\,\star$ and
$v_1\tcons v_2\triangleq \tfold\,\inr\,\tuple{v1,v2}$. 

An inductive type might have arbitrarily large values: in a functional
setting, the traditional method \cite{jones1987implementation}
consists in using \emphidx{fixpoints}. To do so, one missing feature
of the iso language is the capability to manipulate variables
representing isos. In \cite{sabry2018symmetric} we extend the language
by adding iso variables, application, lambda-abstraction over
iso-variables, and a fixpoint operator $\tfix$.
Assuming $\tfix\,f.\omega$ has the behavior
\[
  \tfix\,f.\omega \to \omega[f \mapsto \tfix\,f.\omega]
\]
and assuming a suitable syntax extension for the right-hand-side of
isos, we can then define the (higher-order) map operation of type
$(a\iso b)\tto([a]\iso[b])$ as follows:
\[
  \texttt{map}\quad\triangleq\quad
  \lambda g.\tfix\,f.
    \left\{
    \begin{array}{rcl}
      \tnil & \iso & \tnil
      \\
      h\tcons t & \iso & (g\,h)\tcons(f\,h)
    \end{array}
  \right\}.
\]
Assuming the iso $g$ is of type $a\iso b$, the operation
$\texttt{map}\,g$ sends $\tnil$ to $\tnil$ and otherwise applies $g$
to the head of the list and itself to the tail.

Introducing fixpoints pose the question of the termination of
programs: without termination, the isos describes injective maps
between sets of values. Bijections are only obtained in the case of
\emph{terminating} isos. One of the result of
\cite{sabry2018symmetric} is to formalize this fact.

\subsection{Pattern-Matching for Quantum Control}
\label{sec:rev-qcont}

The reversible iso-language described in Section~\ref{sec:revpat-ind}
is amenable to the same algebraic extension as the vectorial
lambda calculus presented in Section~\ref{sec:lineal}: values are
extended to \emph{linear combinations}. Therefore, assuming
$\ttrue\triangleq\inl\star$ and $\ffalse\triangleq\inr\star$ are the
two (standard) values of type $1\oplus 1$, one can now consider values
of the form $\alpha\cdot\ttrue+\beta\cdot\ffalse$, where
$\alpha,\beta$ are scalars: the type $1\oplus 1$ is a $2$-dimensional
module. In \cite{sabry2018symmetric} we formalize an algebraic
extension of the iso-language supporting the encoding of
\emph{unitary} maps (when scalars ranges over complex numbers). For
instance, the Hadamard gate can be represented with the iso
\[
  \left\{
    \begin{array}{r@{~~\iso~~}l}
      \ffalse
      &
        \frac{\sqrt2}2(\ffalse+\ttrue)
      \\
      \ttrue
      &
        \frac{\sqrt2}2(\ffalse-\ttrue)
    \end{array}
  \right\}
\]
of type $(1\oplus1)\iso(1\oplus1)$. In the paper, we discuss how this
idea can be extended to types of infinite dimension such as lists, and
we provide a (restricted) formal setting for the support of lists and
linear combinations. In particular, we provide a compositional
interpretation of isos as unitaries in $\ell^2$-spaces.

\paragraph{The quantum SWITCH.}
In the context of the algebraic extension of the iso-language, the
quantum SWITCH is simple to write. Indeed, suppose that $u:a\iso b$ and
$v:b\iso a$ are two isos, then one can write the iso
\[
  \left\{
    \begin{array}{r@{~~\iso~~}l}
      \inl\,x
      &
        \inl (v\,(u\,x))
      \\
      \inr\,x
      &
        \inr (u\,(v\,x))
    \end{array}
  \right\}
\]
of type $(a\oplus b)\iso(a\oplus b)$. Depending on the branch (left or
right), either $u\,v$ or $v\,u$ is applied.
The remaining question is whether $u$ and $v$ are indeed ``used'' only
once.

\subsection{Relationship with the Logic \texorpdfstring{$\mu$MALL}{muMALL}}
\label{rev:mumall}

\begin{mylife}
  This section summarizes a series of papers obtained with Kostia
  Chardonnet, a Ph.D student I co-supervised with Alexis Saurin
  \cite{chardonnet2020curry-howard, chardonnet2021curry-howard,
    chardonnet2022curry-howard, chardonnet2023phd}.
\end{mylife}

\noindent
The extended type system presented in Section~\ref{sec:revpat-ind} is
very reminiscent of the logic {\mumall}, an extension of \simpleidx{MALL} with
inductive formulae\index{inductive formula}.
Introduced by Baelde, \cite{baelde2007least,baelde2008linear} the
logic {\mumall} is linear and features tensors and
coproducts. Inductive formulae are dealt with infinite proof
structures. The problem in this context is then to characterize
whether a given infinite derivation is indeed correct: there exist
several validity criteria \cite{baelde2012least,
  baelde2016infinitary, doumane2017infinitary, nollet2018local}.
The strict linearity of the logic together its ability to express
inductive formulae makes {\mumall} a good candidate logic to serve as
a Curry-Howard correspondence with the iso-language 
sketched in Section~\ref{sec:revpat-ind}.

As mentioned earlier, an iso $\omega : A\iso B$
corresponds to both a computation sending a value of type $A$ to a
result of type $B$ and a computation sending a value of type $B$ to a
result of type $A$. The logical counterpart consists in two proofs: a
proof $\pi$ of $A\vdash B$ and a proof $\pi^\bot$ of
$B\vdash A$. These proofs describes an isomorphism if the two possible
cuts between them reduce to an identity proof.

The case of \emph{structurally recursive} isos is considered in
\cite{chardonnet2022curry-howard}.
We show that the resulting
language can encode any primitive recursive function
\cite{rogers1987theory}, and we give an interpretation of well-typed
isos as proofs of isomorphisms in {\mumall}. 
The work described in \cite{chardonnet2022curry-howard} discusses how
the syntactical constraints of structural recursion are linked to the
validity of infinite proofs corresponding to programs.

The example discussed in the paper is the proof $\denot{\texttt{map}\,g}$
corresponding to
the term $\texttt{map}\,g$ with $g:A\iso B$ being
\[
  \includegraphics[page=1,scale=.7]{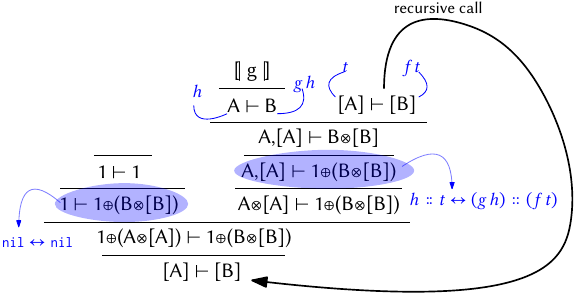}
\]
and where $\denot{g}$ is the proof corresponding to $g$. Although the proof is
folded in a cyclic manner, it corresponds to an infinite proof: the
back-edge is really an infinite branch consisting of
``copy-and-paste'' of the proof structure:
\[
  \includegraphics[page=2,scale=.5]{fig/map-mumall.pdf}
\]
The fact that proof is valid comes from the infinite number of
left-unfolding of the list.

\begin{table}[p]
  \centering
  \scalebox{.8}{\begin{minipage}{1.25\textwidth}
        \begin{mylife*}
          \begin{compactdesc}
            \itempaper{valiron2010semantics} \mycitevalironsemantics
            \itempaper{arrighi2011subject} \mycitearrighisubject
            \itempaper{chiribella2013quantum} \mycitechiribellaquantum
            \itempaper{valiron2013typed} \mycitevalirontyped
            \itempaper{assaf2014call-by-value} \myciteassafcall
            \itempaper{valiron2014finite} \mycitevalironfinite
            \itempaper{valiron2014modeling} \mycitevalironmodeling
            \itempaper{arrighi2017vectorial} \mycitearrighivectorial
            \itempaper{sabry2018symmetric} \mycitesabrysymmetric
            \itempaper{diaz-caro2019realizability} \mycitediazcarorealizability
            \itempaper{chardonnet2020curry-howard} \mycitechardonnetcurryrc
            \itempaper{lemonnier2021categorical} \mycitelemonniercategorical
            \itempaper{chardonnet2022curry-howard} \mycitechardonnetcurry
          \end{compactdesc}
          \caption{Personal publications related to Chapter~\ref{ch:qcont}.}
          \label{tab:publis-qcont}
       \end{mylife*}
     \end{minipage}}
\end{table}


\clearpage{\thispagestyle{empty}\cleardoublepage}

\chapter[Opening]{%
  Opening%
}
\label{ch:opening}
\rhead{Opening}

Over the past dozen years, the field of quantum programming languages
has experienced substantial maturation. Fifteen years ago, programming
languages were restricted to toy languages, independent from hardware,
and only relied on small mathematical specifications such as Knill's
QRAM model \cite{knill96conventions}.  What happened below this layer
was left ``to the physicists''. Nowadays, quantum programming
languages address large, industrial-scale problem instances and target
concrete hardware \cite{mckinsey2021}.
It has become clear that many low-level aspects
require scrutiny and are of interest to the programmer: Timing,
coprocessor expressivity power, quantum superposition of execution,
etc. All of this needs to be taken into account at once, and it is not
clear whether this can be compartmented into an intermediate
representation, with the compiler responsible for the translation, or
if some or all of the constraints have to be lifted to the high-level
language. Both points of view are interesting: leveraging intermediate
representations for quantum computation to this extended paradigm but
also having a full-fledged high-level language for describing such
processes.

Our seminal work on \quipper is arguably a milestone in the design of
quantum programming languages \cite{green2013quipper}.  \quipper can
indeed be seen as an experiment in the design of a
scalable language with sound principles. The central axiom is that
programming quantum algorithms is, first of all, the description of
the construction of a circuit: any quantum programming language
should, therefore, be a classical programming language providing a
series of specialized constructs to realize circuits as efficiently as
possible. Several design principles ensue. For instance, one such
design is the capability to box functions---i.e., considering a
function from qubit to qubit as a circuit--- and unbox circuits---that
is, considering a circuit as a function acting on a qubit. Another
principle is the ability to build and apply higher-order circuit
combinators: the programming language should make it easy to control a
circuit, use it in the context of local, ancilla wires, and otherwise
perform arbitrary transformations on it. A last principle worth
mentioning is the ability to generate a circuit from a classical
description. These principles have been laid out in the context of
\quipper: they are still state-of-the-art in the design of current
quantum programming languages.

The field is, however, moving fast, and as NISQ
\cite{preskill2018nisq} reaches the level of industrialization, new
paradigms are needed. On the one hand, the so-called realm of
\emph{hybrid computation} requires more than circuit-description
languages: a distributed programming model where the classical
processor and the quantum coprocessor speak on equal
grounds. Programming in this context opens novel questions. For
instance, what is the classical expressive power of the quantum
coprocessor compared to that of the classical processor? What are the
programming features needed within the quantum coprocessor, i.e.,
what are the reasonable capabilities of the coprocessor in terms of
memory, clock, and timing related in the interactions?
The large-scale
development of quantum hardware also opens the door to the question of
quantum compilation. Quantum circuits can no longer be considered
low-level targets, and a quantum programming environment cannot be
reduced to circuit description and evaluation
\cite{quingo,pennylane,heurtel2023perceval,qiskit,meijer2024advances}.

From a theoretical standpoint, in the last 15 years, we have proposed
a novel computational construct: quantum SWITCH
\cite{chiribella2013quantum}.  This construct can be summarized by
asking whether it is conceivable to realize \emph{in superposition} the
sequence ``$U$ then $V$'' and the sequence ``$V$ then $U$'', given
only one single copy of each. Said otherwise, instead of only having
data in superposition, is it possible to also have executions in
superposition? If $U$ and $V$ are unitary, the overall operation is
unitary, so this should not be a problem. The issue's crux is the
impossibility of building a circuit with only two holes--one for $U$
and one for $V$--when realizing this procedure.

Although this ``quantum test'' makes sense from a computational
perspective and is mathematically meaningful, it lies outside the
circuit model. Nonetheless, it has been realized by concrete physical
experiment \cite{procopio2015experimental,taddei2021computational}.
This exemplifies the cross-fertilization happening between physics and
computer science: The scientific community has
derived several exciting results from the realization of quantum
superposition of executions, including speedup in communication and
finer analysis of quantum metrology
\cite{araujo2014computational,taddei2021computational,zhao2020quantum}.
From the quantum programming language side, the promise of quantum
control lies in the \emph{programming} of quantum unitaries. Instead
of constructing circuits as lists of opaque low-level gates, quantum
control opens the door to a complete programming environment
manipulating classical and quantum operations within a unified
paradigm.

Since 2008, in parallel with the development of quantum programming
languages, semantics have seen a significant development
\cite{rennela2017classical,lindenhovius2018enriching,westerbaan2019category}.
Indeed, semantics is essential to unearth the structures underlying quantum
computation and to shed light on the suitable structures for
manipulating quantum computation soundly and consistently. The
development of semantics for a programming language is, in particular,
what makes it possible to obtain language-supporting techniques to
express and prove the properties of programs.

At the dawn of 2010, the semantics of quantum programming languages
were either very abstract or very close to physics textbooks: based on
superoperators and completely positive maps, following the standard
models of quantum information theory, or, on the opposite, following
purely categorical constructions
\cite{selinger06fully,valiron2008categorical,malherbe2010categorical}. Since
then, the models have tremendously evolved, capturing more
fine-grained concrete, useful programmatic paradigms. An example is
the interplay between quantum circuits and measurements in the context
of circuit-description languages: we now have mathematical
representations of dynamic circuits, where the circuit's shape might
depend on the result of previous measurements. Another example is the
range of sophisticated type systems based on linear logic. Such type
systems can now enforce the non-duplicability of quantum data and, at
the same time, provide refined logical properties by relying on
dependent types \cite{fu2020linear}. Finally, a last example concerns
quantum control and superposition of executions: we now have several
proposals of formal languages equipped with rewrite systems able to
formalize what it means to feature quantum superposition of
executions.

\section*{A Few Current Trends of Research}

The rest of the chapter broadens the focus to a set of ongoing
trends of research.

\paragraph{Rise of Graphical Languages.}
In the field of quantum programming languages, one crucial event in
recent years has been the advent of graphical languages targeted
toward quantum computation. Arguably, the first one is the ZX
calculus .
Abstracting away from quantum circuits, the language is a
formal graph-based language akin to what is used in the context of
tensor networks but specialized for the specificities of quantum
computation. In ZX, wires correspond to qubit states, and the
available nodes in the graph stand for linear maps related to
elementary operations such as rotations along $X$ and $Z$
basis. Unlike quantum circuits, it is equipped with sound semantics
and a complete equational theory, making it a strong candidate for
reasoning and processing quantum computation.

Over time, the ZX calculus has proven more versatile than quantum
circuits when considering several classes of problems such as circuit
optimization, qubit layout, post-selection, or reasoning over
error-correcting schemes \cite{zx-publications}.
The language is also easily extensible. It
has, for instance, been extended to support features such as
measurement and trace, or arrays of qubits. Compared to syntactic,
more conventional approaches such as \qasm implementing plain quantum
circuits, the ZX calculus has shown to be a credible approach for
serving as an intermediate representation in the context of a quantum
compilation toolchain \cite{meijer2024advances}.

The success of the ZX calculus has spurred a line of research in the
design of graphical languages with a focus on specific
backends. Indeed, the ZX calculus is particularly well-suited for
hardware based on the gate-set Clifford+T but possibly less fitted for
other models. For instance, for Clifford+Toffoli---the canonical
gate-set for cat-qubits---the ZH calculus is better suited
\cite{vilmart2018zh}. Similarly,
the ZW calculus \cite{hadzihasanovic2018zw}
is considered adequate for reasoning on quantum
computation with Rydberg atoms. Recently, one can also cite languages
such as the \lov-calculus developed for linear optical circuits
\cite{clement2022lov}.

These trends emphasize connecting theoretical computer science
concepts with physical implementations. Such a cross-disciplinary
integration has already been shown to be effective: we have recently
shown a completeness result for quantum circuits based on the
development of the \lov-calculus \cite{clement2023complete}.

The story is still ongoing, and exciting questions await us. For
instance, existing languages currently only model quantum computation
in finite-dimensional spaces; the ability of graphical languages to
handle infinite-dimensional objects is still work in
progress \cite{felice2022qath,shaikh2024fockedup}.
Another
question is the interoperability of these languages. In particular, if
the Kronecker product is the canonical operation to join systems
together in ZX-based languages, linear-optical languages use the
product. Finally---and maybe more generally---these languages are
still far from many considerations closer to the hardware: they do not
handle (yet) timing, noise, nor hybrid computation.

\paragraph{Unification of Quantum and Classical Control.}
Hybrid quantum computation is a model of computation where the
interaction with the quantum coprocessor can depend on the results of
intermediate measurements. In the standard model, the coprocessor is
considered a closed box that can be manipulated using an interface
given once for all. In particular, the elementary operations available
in the quantum coprocessor are not programmable.

The model of quantum control instead considers the case where the
programmer can describe a superposition of executions instead of a
simple list of gates. In this model, one can express the purely
quantum part with native, quantum-specific programming constructs such
as the quantum SWITCH \cite{chiribella2013quantum}.

One missing aspect of quantum control is the interaction with the
classical machine. Indeed, current descriptions of quantum control
only focus on the purely quantum part, and the model does not
encompass a hybrid system where, for instance, quantum information
could be measured, let alone used to drive quantum
evolution. Therefore, finding a model unifying both classical and
quantum control in the same framework is an open question.

Another more foundational issue is how to handle quantum, recursive
datatypes. Consider, for instance, the type of lists of qubits and the
element consisting of a list of size $2$ and a list of size $5$. What
does it mean to iterate over this superposition of lists? The question
can be generalized: What kind of recursion is allowed in a purely
quantum context? This question lies in the more general problem
of the expressive power of quantum control. Indeed, if superpositions
of programs makes a powerful computational paradigm, the constraint of
unitarity is a subtle condition to enforce at the syntactic level.

A last large research avenue open for quantum control is the problem
of turning the description of a superposition of executions into
something that can be physically executed on a quantum backend. The
question is twofold. On the one hand, several non-standard models of
quantum computation feature some notion of quantum control, such as
AQG or routed circuits. Compiling on these (formal) backends is
already a stimulating question. On the other hand, we do have concrete
hardware candidates for (physical) quantum computation. A natural
question is to study how much quantum control these hardware
candidates can handle and devise a suitable compilation scheme for
them.

\paragraph{Quantum Compilation Toolchain.}
Although the field of quantum programming languages is now reaching a
mature state, turning a quantum program into a realistic set of
low-level operations executable on a quantum coprocessor is still a
work in progress. Currently, each vendor offers a specific solution,
usually with a particular quantum dialect and a compilation framework
specifically tailored for one particular hardware. Existing quantum
compilers are also currently very limited. For once, they provide
little parametrization and do not scale well: they mainly target
(small) NISQ devices, and LSQ is still an open field of
research. Another issue is the ability to handle hybrid computation;
static circuits remain the norm of what is possible to compile.

A crucial open question in this realm consists in devising tools and
techniques to effectively compile quantum programs down to low-level,
executable physical operation. Following what has been done in the
classical setting, it would be natural to consider one or several
intermediate representations specific to quantum computation, such as
graphical languages. In any case, problems such as timing and
parallelism of quantum operations and topological constraints of the
hardware (or of the quantum error-correcting layer) need to be
addressed in a consistent manner. Because of the very distinct kinds
of hardware, it is now admitted that there will not be a one-fit-all
solution. Nonetheless, many problems are cross-platform, yielding
similar answers. Although a generic hardware-independent rigid
compiler might not be doable, devising a common framework for building
and composing compiling tools and modules is an active research area.

A complementary problem in quantum compilation concerns optimizing and
estimating the resources needed to run a given quantum
program. Although this program is written in a hardware-agnostic,
higher-order language, the target backend is one specific
coprocessor. The problem is akin to what happens for critical systems:
This processor has limited memory. Error correction---if any---is
costly: we want the code to be as optimized and as parallel as
possible. This tension calls for developing optimization schemes to
minimize resources and static analysis tools to evaluate these
resources.

\paragraph{Static Analysis for Quantum Programs.}
Verification techniques for quantum programs are currently in their
early stages of development. So far, various attempts have been
followed with little of a unified approach \cite{chareton2021formal}.
Among these, one can
mention the use of proof assistants such as Coq and Isabelle/HOL to
prove properties of quantum programs, but also deductive verification
techniques, either standalone and based on a crafted Hoare logic or
embedded in existing tools such as Why3. Although other novel
SMT-based tools have recently emerged, the field still needs to be
more structured, making it challenging to compare these methods and
ascertain the overall direction of the discipline.

The main problem with static analysis of quantum programs is the
nature of the thing we want to analyze. Many aspects can be
considered. If the program describes a static circuit, as discussed in
the trend related to quantum compilation, one can be interested in the
circuit's size, shape, or depth. Another critical aspect is ensuring
that the circuit generated by the program implements the correct
unitary map. In the context of a probabilistic algorithm relying on
measurements, one should also guarantee the probability of success.

In the context of a quantum compilation toolchain, one can also be
interested in certifying other layers of the compilation stack, such
as the optimization schemes, the qubit layout process, and, in
general, circuit transformations and translations to dedicated
graphical, intermediate representations.

Finally, an untouched aspect of the static analysis of quantum
programs is quantum control. What can be asserted in this context, how
to do it, and how to verify it is still a completely open area of
research.

The questions discussed above rely on sound, expressive semantics of
both quantum programs and layers in the quantum compilation
toolchain. Unlike classical computation, where models are discrete
structures, in quantum computation, the mix of discrete and continuous
structure, duplicable and non-duplicable objects, linear algebra, and
operator theory renders the development of powerful analysis tools
challenging without a fine-grain understanding of the structures at
stake. The development of semantics for quantum programming languages
is a continuous dialog between three actors: physics, hinting at the
underlying constraints; computer science, discovering unknown
computational structures and techniques; and semantics, formalizing
them and providing a firm, sound framework upon which to build and
reason.

\section*{Conclusion}

\begin{mylife}
  This thesis has presented how I understand the evolution of
  quantum programming language within the past fifteen years. In this
  time frame, the field of quantum programming languages has shifted
  from toy examples to a more mature state. A dialog between physics
  and computer science has fostered unforeseen discoveries along this
  path. The story is, however, not over, and the field remains open,
  presenting exciting questions and opportunities for future
  developments.
    
  I am already involved in some of the research paths presented in
  this section, in particular with currently ongoing Ph.D
  students. With Nicolas Heurtel, Ph.D funded by a CIFRE with
  Quandela, we are studying graphical languages for quantum linear
  optics---this yielded for instance the LOv calculus
  \cite{clement2022lov}. With Julien Lamiroy, co-supervised with
  Renaud Vilmart, we are investigating graphical languages for quantum
  control. With Jérome Ricciardi, Ph.D funded by CEA and co-supervised
  with Christophe Chareton, we are studying static analysis methods
  for quantum programs with measurements.
\end{mylife}

\begin{table}[p]
  \centering
  \scalebox{.8}{\begin{minipage}{1.25\textwidth}
        \begin{mylife*}
          \begin{compactdesc}
            \itempaper{qpl2020} \myciteqpl
            \itempaper{chardonnet2021geometry} \mycitechardonnetgeometry
            \itempaper{brugiere2021reducing} \mycitebrugierereducing
            \itempaper{goubault2021gaussian} \mycitegoubaultgaussian
            \itempaper{valiron2022semantics} \mycitevalironsemanticsjlamp
            \itempaper{brugiere2022decoding} \mycitebrugieredecoding
            \itempaper{clement2022lov} \myciteclementlov
            \itempaper{chardonnet2022many-worlds} \mycitechardonnetmanyworlds
            \itempaper{chapuis-chkaiban2023pagerank} \mycitechapuispagerank
            \itempaper{arrighi2023addressable} \mycitearrighiaddressable
            \itempaper{heurtel2023perceval} \myciteheurtelperceval
            \itempaper{heurtel2023strong} \myciteheurtelstrong
            \itempaper{clement2023complete} \myciteclementcomplete
            \itempaper{clement2023complete-tqc} \myciteclementcompletetqc
            \itempaper{qpl2023} \myciteqplbis
          \end{compactdesc}
          \caption{Personal publications since $\sim$2020 not yet mentioned.}
          \label{tab:publis-opening}
       \end{mylife*}
     \end{minipage}}
\end{table}


\clearpage{\thispagestyle{empty}\cleardoublepage}
\lhead{Bibliography}
\rhead{}

\emergencystretch=1em

\clearpage{\thispagestyle{empty}\cleardoublepage}
\lhead{Index}
\rhead{}

\addcontentsline{toc}{chapter}{Index}

\printindex


\addcontentsline{toc}{chapter}{Bibliography}

{\footnotesize
\begin{thebibliography}{99}
%
\bibitem[ABGV18]{allouche2018reuse}
C. Allouche, M. Baboulin, T. Goubault de Brugière, and B. Valiron. ``Reuse method for quantum circuit
synthesis''. In: \emph{Recent Advances in Mathematical and Statistical Methods, post-proceedings of the IV
AMMCS International Conference on Applied Mathematics, Modeling and Computational Science,
Waterloo, Canada, August 20 -- 25, 2017}. Ed. by D. Marc Kilgour, Herb Kunze, Roman Makarov,
Roderick Melnik, and Xu Wang. Springer International Publishing, 2018, pp. 3--12. \textsc{isbn}:
978-3-319-99719-3. \textsc{doi}: \href {https://doi.org/10.1007/978-3-319-99719-3_1} {\nolinkurl
{10.1007/978-3-319-99719-3_1}}. \textsc{hal}: \href {https://hal.archives-ouvertes.fr/hal-01711378} {\nolinkurl
{hal-01711378}}.
%
\bibitem[ABIMBK19]{arrazola2019machine}
Juan Miguel Arrazola, Thomas R Bromley, Josh Izaac, Casey R Myers, Kamil Brádler, and Nathan Killoran.
``Machine learning method for state preparation and gate synthesis on photonic quantum computers''. In:
\emph{Quantum Science and Technology} 4.2 (Jan. 2019), p. 024004. \textsc{doi}: \href
{https://doi.org/10.1088/2058-9565/aaf59e} {\nolinkurl {10.1088/2058-9565/aaf59e}}.
%
\bibitem[Abr05]{abramsky2005structural}
Samson Abramsky. ``A structural approach to reversible computation''. In: \emph{Theoretical Computer
Science} 247.3 (2005), pp. 441--464. \textsc{doi}: \href {https://doi.org/10.1016/j.tcs.2005.07.002} {\nolinkurl
{10.1016/j.tcs.2005.07.002}}. \textsc{arXiv}: \href {https://www.arxiv.org/abs/1111.7154} {\nolinkurl {1111.7154}}.
%
\bibitem[Abr07]{abramsky2007temperleylieb}
Samson Abramsky. ``Temperley-Lieb algebra: from knot theory to logic and computation via quantum
mechanics''. In: \emph{Mathematics of Quantum Computation and Quantum Technology}. Ed. by
Goong Chen, Louis Kauffman, and Samuel J. Lomonaco. New York: Chapman, Hall/CRC, Taylor, and
Francis, 2007. Chap. 15, pp. 415--458. \textsc{isbn}: 978-1-58488-900-7. \textsc{doi}: \href
{https://doi.org/10.1201/9781584889007} {\nolinkurl {10.1201/9781584889007}}. \textsc{arXiv}: \href
{https://www.arxiv.org/abs/0910.2737} {\nolinkurl {0910.2737}}.
%
\bibitem[Abr93]{abramsky93computational}
Samson Abramsky. ``Computational interpretations of linear logic''. In: \emph{Theoretical Computer
Science} 111.1-2 (1993), pp. 3--57. \textsc{doi}: \href {https://doi.org/10.1016/0304-3975(93)90181-R} {\nolinkurl
{10.1016/0304-3975(93)90181-R}}.
%
\bibitem[AC04]{abramsky04categorical}
Samson Abramsky and Bob Coecke. ``A categorical semantics of quantum protocols''. In: \cite{lics04},
pp. 415--425. \textsc{doi}: \href {https://doi.org/10.1109/LICS.2004.1319636} {\nolinkurl
{10.1109/LICS.2004.1319636}}. \textsc{arXiv}: \href {https://www.arxiv.org/abs/quant-ph/0402130} {\nolinkurl
{quant-ph/0402130}}.
%
\bibitem[ACB14]{araujo2014computational}
Mateus Araújo, Fabio Costa, and Časlav Brukner. ``Computational advantage from quantum-controlled
ordering of gates''. In: \emph{Physical Review Letters} 113 (25 2014), p. 250402. \textsc{doi}: \href
{https://doi.org/10.1103/PhysRevLett.113.250402} {\nolinkurl {10.1103/PhysRevLett.113.250402}}. \textsc{arXiv}:
\href {https://www.arxiv.org/abs/1401.8127} {\nolinkurl {1401.8127}}.

%
\bibitem[Acc15]{accattoli2015proof}
Beniamino Accattoli. ``Proof Nets and the Call-by-Value Lambda-Calculus''. In: \emph{Theoretical Computer
Science} 606 (2015), pp. 2--24. \textsc{doi}: \href {https://doi.org/10.1016/j.tcs.2015.08.006} {\nolinkurl
{10.1016/j.tcs.2015.08.006}}.
%
\bibitem[ACCRV23]{arrighi2023addressable}
Pablo Arrighi, Christopher Cedzich, Marin Costes, Ulysse Rémond, and Benoît Valiron. ``Addressable
quantum gates''. In: \emph{ACM Transactions on Quantum Computing} 4.3 (2023), pp. 1--41. \textsc{doi}:
\href {https://doi.org/10.1145/3581760} {\nolinkurl {10.1145/3581760}}. \textsc{hal}: \href
{https://hal.archives-ouvertes.fr/hal-03936367} {\nolinkurl {hal-03936367}}. \textsc{arXiv}: \href
{https://www.arxiv.org/abs/2109.08050} {\nolinkurl {2109.08050}}.
%
\bibitem[ACRŠZ10]{ambainis2010andor}
Andris Ambainis, Andrew M. Childs, Ben Reichardt, Robert Špalek, and Shengyu Zhang. ``Any AND-OR
formula of size $N$ can be evaluated in time $N^{1/2+o(1)}$
on a quantum computer''. In: \emph{SIAM Journal on
Computing} 39.6 (2010), pp. 2513--2530. \textsc{doi}: \href {https://doi.org/10.1137/080712167} {\nolinkurl
{10.1137/080712167}}.
%
\bibitem[Acz77]{aczel1977introduction}
Peter Aczel. ``An introduction to inductive definitions''. In: \emph{Handbook of Mathematical Logic}. Ed. by
John Barwise. Vol. 90. Studies in Logic and the Foundations of Mathematics. North Holland, 1977,
pp. 739--782. \textsc{doi}: \href {https://doi.org/10.1016/S0049-237X(08)71120-0} {\nolinkurl
{10.1016/S0049-237X(08)71120-0}}.
%
\bibitem[AD05]{arrighi2005computational}
Pablo Arrighi and Gilles Dowek. ``A computational definition of the notion of vectorial space''. In:
\emph{Proceedings of the Fifth International Workshop on Rewriting Logic and its Applications (WRLA'04)}.
Vol. 117. Electronic Notes in Theoretical Computer Science. 2005, pp. 249--261. \textsc{doi}: \href
{https://doi.org/10.1016/j.entcs.2004.06.013} {\nolinkurl {10.1016/j.entcs.2004.06.013}}.
%
\bibitem[AD08]{arrighi2008linear-algebraic}
Pablo Arrighi and Gilles Dowek. ``Linear-algebraic lambda-calculus: higher-order, encodings, and
confluence.'' In: \emph{Proceedings of the 19th International Conference on Rewriting Techniques and
Applications, RTA'08} (Hagenberg, Austria). Ed. by Andrei Voronkov. Vol. 5117. Lecture Notes in Computer
Science. Springer, 2008, pp. 17--31. \textsc{isbn}: 978-3-540-70588-8. \textsc{doi}: \href
{https://doi.org/10.1007/978-3-540-70590-1_2} {\nolinkurl {10.1007/978-3-540-70590-1_2}}.
%
\bibitem[AD09]{arrighi2009scalar}
Pablo Arrighi and Alejandro Díaz-Caro. ``Scalar system F for linear-algebraic lambda-calculus: towards a
quantum physical logic''. In: \emph{Proceedings of the 6th International Workshop on Quantum Physics and
Logic, QPL'09} (Oxford, UK.). Ed. by B. Coecke, P. Panangaden, and P. Selinger. Vol. 270-2. Electronic Notes
in Theoretical Computer Science. 2009, pp. 219--229. \textsc{doi}: \href
{https://doi.org/10.1016/j.entcs.2011.01.033} {\nolinkurl {10.1016/j.entcs.2011.01.033}}.
%
\bibitem[AD12a]{arrighi2012system}
Pablo Arrighi and Alejandro Díaz-Caro. ``A system F accounting for scalars''. In: \emph{Logical Methods in
Computer Science} 8.1 (2012). See also extended abstract \cite{arrighi2009scalar}. \textsc{doi}: \href
{https://doi.org/10.2168/LMCS-8(1:11)2012} {\nolinkurl {10.2168/LMCS-8(1:11)2012}}.
%
\bibitem[AD12b]{arrighi2012physical}
Pablo Arrighi and Gilles Dowek. ``The physical Church-Turing thesis and the principles of quantum theory''.
In: \emph{International Journal of Foundations of Computer Science} 23.5 (2012), pp. 1131--1146. \textsc{doi}:
\href {https://doi.org/10.1142/S0129054112500153} {\nolinkurl {10.1142/S0129054112500153}}. \textsc{arXiv}:
\href {https://www.arxiv.org/abs/1102.1612} {\nolinkurl {1102.1612}}.
%
\bibitem[AD17]{arrighi2017lineal}
Pablo Arrighi and Gilles Dowek. ``Lineal: a linear-algebraic lambda-calculus''. In: \emph{Logical Methods in
Computer Science} 13.1 (2017). \textsc{doi}: \href {https://doi.org/10.23638/LMCS-13(1:8)2017} {\nolinkurl
{10.23638/LMCS-13(1:8)2017}}.
%
\bibitem[Ada02]{adamatzky2002collision-based}
Andrew Adamatzky, ed. \emph{Collision-Based Computing}. Springer-Verlag, 2002. \textsc{url}: \url{http://www.cems.uwe.ac.uk/~aadamatz/compiled.htm}.
%
\bibitem[ADPTV14]{assaf2014call-by-value}
Ali Assaf, Alejandro Díaz-Caro, Simon Perdrix, Christine Tasson, and Benoît Valiron. ``Call-by-value,
call-by-name and the vectorial behaviour of the algebraic lambda-calculus''. In: \emph{Logical Methods in
Computer Science} 10.4 (2014). \textsc{doi}: \href {https://doi.org/10.2168/LMCS-10(4:8)2014} {\nolinkurl
{10.2168/LMCS-10(4:8)2014}}. \textsc{arXiv}: \href {https://www.arxiv.org/abs/1005.2897v7} {\nolinkurl
{1005.2897v7}}.
%
\bibitem[ADV11]{arrighi2011subject}
Pablo Arrighi, Alejandro Díaz-Caro, and Benoît Valiron. ``Subject reduction in a curry-style polymorphic
type system with a vectorial structure''. In: \emph{Proceedings of the 7th International Workshop on
Developments of Computational Methods, DCM 2011} (Zurich, Switzerland, July 3, 2021). Ed. by
Elham Kashefi, Jean Krivine, and Femke van Raamsdonk. Vol. 88. Electronic Proceedings in Theoretical
Computer Science. Preliminary work to the journal paper \cite{arrighi2017vectorial}. 2011, pp. 1--15. \textsc{doi}: \href
{https://doi.org/10.4204/EPTCS.88.1} {\nolinkurl {10.4204/EPTCS.88.1}}. \textsc{hal}: \href
{https://hal.archives-ouvertes.fr/hal-00924926} {\nolinkurl {hal-00924926}}. \textsc{arXiv}: \href
{https://www.arxiv.org/abs/1012.4032} {\nolinkurl {1012.4032}}.
%
\bibitem[ADV17]{arrighi2017vectorial}
Pablo Arrighi, Alejandro Díaz-Caro, and Benoît Valiron. ``The vectorial lambda-calculus''. In:
\emph{Information and Computation} 254 (2017), pp. 105--139. \textsc{doi}: \href
{https://doi.org/10.1016/j.ic.2017.04.001} {\nolinkurl {10.1016/j.ic.2017.04.001}}. \textsc{hal}: \href
{https://hal.archives-ouvertes.fr/hal-00921087} {\nolinkurl {hal-00921087}}.
%
\bibitem[AG05a]{grattage05functional}
Thorsten Altenkirch and Jonathan Grattage. ``A functional quantum programming language''. In:
\emph{Proceedings of the 20th Symposium on Logic in Computer Science, LICS'05} (Chicago, Illinois, US.).
Ed. by Prakash Panangaden. IEEE. IEEE Computer Society Press, 2005, pp. 249--258. \textsc{doi}: \href
{https://doi.org/10.1109/LICS.2005.1} {\nolinkurl {10.1109/LICS.2005.1}}. \textsc{arXiv}: \href
{https://www.arxiv.org/abs/quant-ph/0409065} {\nolinkurl {quant-ph/0409065}}.
%
\bibitem[AG05b]{altenkirch2005qml}
Thorsten Altenkirch and Jonathan Grattage. ``QML: Quantum data and control''. Draft, extended version of
the LICS publication \cite{grattage05functional}. 2005.
%
\bibitem[AG09]{green2009qio}
Thorsten Altenkirch and Alexander S Green. ``The quantum IO monad''. In: \cite{gay2009semantic}, pp. 173--205.
%
\bibitem[AG13]{axelsen2013reversible}
Holger Bock Axelsen and Robert Glück. ``Reversible representation and manipulation of constructor terms in
the heap''. In: \cite{rc2013}, pp. 96--109. \textsc{doi}: \href {https://doi.org/10.1007/978-3-642-38986-3_9} {\nolinkurl
{10.1007/978-3-642-38986-3_9}}.
%
\bibitem[AGVS05]{altenkirch2005algebra}
Thorsten Altenkirch, Jonathan Grattage, Juliana Kaizer Vizzotto, and Amr Sabry. ``An algebra of pure
quantum programming''. In: \cite{qpl2005}, pp. 23--47. \textsc{doi}: \href
{https://doi.org/10.1016/j.entcs.2006.12.010} {\nolinkurl {10.1016/j.entcs.2006.12.010}}.
%
\bibitem[AK16]{axelsen2016join}
Holger Bock Axelsen and Robin Kaarsgaard. ``Join inverse categories as models of reversible recursion''. In:
\emph{Proceedings of the 19th International Conference on Foundations of Software Science and
Computation Structures, FoSSaCS'16} (Eindhoven, The Netherlands). Ed. by Bart Jacobs and
Christof Löding. Vol. 9634. Lecture Notes in Computer Science. Springer, 2016, pp. 73--90. \textsc{doi}: \href
{https://doi.org/10.1007/978-3-662-49630-5_5} {\nolinkurl {10.1007/978-3-662-49630-5_5}}.
%
\bibitem[AM17]{arrighi2017quantum}
Pablo Arrighi and Simon Martiel. ``Quantum causal graph dynamics''. In: \emph{Physical Review D} 96 (2
2017), p. 024026. \textsc{doi}: \href {https://doi.org/10.1103/PhysRevD.96.024026} {\nolinkurl
{10.1103/PhysRevD.96.024026}}. \textsc{arXiv}: \href {https://www.arxiv.org/abs/1607.06700} {\nolinkurl
{1607.06700}}.
%
\bibitem[AM97]{abramsky97cbv}
Samson Abramsky and Guy McCusker. ``Call-by-value games''. In: \emph{Computer Science Logic, 11th
International Workshop, CSL'97 Annual Conference of the EACSL} (Aarhus, Denmark). Ed. by
Mogens Nielsen and Wolfgang Thomas. Vol. 1414. Lecture Notes in Computer Science. European Association
for Computer Science Logic. Springer Verlag, Aug. 1997, pp. 1--17. \textsc{doi}: \href
{https://doi.org/10.1007/BFb0028004} {\nolinkurl {10.1007/BFb0028004}}.
%
\bibitem[Amb12]{ambainis2012variable}
Andris Ambainis. ``Variable time amplitude amplification and quantum algorithms for linear algebra
problems''. In: \emph{Proceedings of the 29th International Symposium on Theoretical Aspects of Computer
Science, STACS 2012} (Paris, France, Feb. 29--Mar. 3, 2012). Ed. by Christoph Dürr and Thomas Wilke. Vol. 14.
LIPIcs. Schloss Dagstuhl - Leibniz-Zentrum für Informatik, 2012, pp. 636--647. \textsc{isbn}:
978-3-939897-35-4. \textsc{doi}: \href {https://doi.org/10.4230/LIPIcs.STACS.2012.636} {\nolinkurl
{10.4230/LIPIcs.STACS.2012.636}}.
%
\bibitem[AMM14]{amy2014polynomial-time}
Matthew Amy, Dmitri Maslov, and Michele Mosca. ``Polynomial-time T-depth optimization of Clifford+t
circuits via matroid partitioning''. In: \emph{IEEE Transactions on Computer-Aided Design of Integrated
Circuits and Systems} 33.10 (2014), pp. 1476--1489. \textsc{doi}: \href
{https://doi.org/10.1109/TCAD.2014.2341953} {\nolinkurl {10.1109/TCAD.2014.2341953}}.
%
\bibitem[AMMR13]{amy2013meet-in-the-middle}
Matthew Amy, Dmitri Maslov, Michele Mosca, and Martin Roetteler. ``A meet-in-the-middle algorithm for
fast synthesis of depth-optimal quantum circuits''. In: \emph{IEEE Transactions on Computer-Aided Design
of Integrated Circuits and Systems} 32.6 (2013), pp. 818--830. \textsc{doi}: \href
{https://doi.org/10.1109/TCAD.2013.2244643} {\nolinkurl {10.1109/TCAD.2013.2244643}}.
%
\bibitem[Amy13]{amy2013algorithms}
Matthew Amy. ``Algorithms for the Optimizations of Quantum Circuits''. MA thesis. University of Waterloo,
2013.
%
\bibitem[Amy18]{amy2018large-scale}
Matthew Amy. ``Towards large-scale functional verification of universal quantum circuits''. In: \cite{qpl2018},
pp. 1--21. \textsc{doi}: \href {https://doi.org/10.4204/EPTCS.287.1} {\nolinkurl {10.4204/EPTCS.287.1}}.
%
\bibitem[Amy19]{amy2019formal}
Matthew Amy. ``Formal Methods in Quantum Circuit Design''. PhD thesis. University of Waterloo, Ontario,
Canada, 2019. \textsc{url}: \url {http://hdl.handle.net/10012/14480}.
%
\bibitem[AR02]{asperti2002intuitionistic}
Andrea Asperti and Luca Roversi. ``Intuitionistic light affine logic''. In: \emph{ACM Transactions on
Computational Logic} 3.1 (2002), pp. 137--175. \textsc{doi}: \href {https://doi.org/10.1145/504077.504081}
{\nolinkurl {10.1145/504077.504081}}.
%
\bibitem[Arr19]{arrighi2019overview}
Pablo Arrighi. ``An overview of quantum cellular automata''. In: \emph{Natural Computing} 18.4 (2019),
pp. 885--899. \textsc{doi}: \href {https://doi.org/10.1007/s11047-019-09762-6} {\nolinkurl
{10.1007/s11047-019-09762-6}}. \textsc{arXiv}: \href {https://www.arxiv.org/abs/1904.12956} {\nolinkurl
{1904.12956}}.
%
\bibitem[ARS17]{amy2017verified}
Matthew Amy, Martin Roetteler, and Krysta M. Svore. ``Verified compilation of space-efficient reversible
circuits''. In: \emph{Computer Aided Verification - 29th International Conference, CAV 2017, Heidelberg,
Germany, July 24-28, 2017, Proceedings, Part II}. Ed. by Rupak Majumdar and Viktor Kuncak. Vol. 10427.
Lecture Notes in Computer Science. Springer, 2017, pp. 3--21. \textsc{isbn}: 978-3-319-63389-3. \textsc{doi}:
\href {https://doi.org/10.1007/978-3-319-63390-9_1} {\nolinkurl {10.1007/978-3-319-63390-9_1}}. \textsc{arXiv}:
\href {https://www.arxiv.org/abs/1603.01635} {\nolinkurl {1603.01635}}.
%
\bibitem[Ato22]{aqasm-github}
Atos. \emph{MyQLM Documentation: The AQASM Format}. 2022. \textsc{url}: \url
{https://myqlm.github.io/aqasm.html} (visited on Sept. 1, 2022).
%
\bibitem[BACS07]{berry2007efficient}
Dominic W. Berry, Graeme Ahokas, Richard Cleve, and Barry C. Sanders. ``Efficient quantum algorithms for
simulating sparse Hamiltonians''. In: \emph{Communications in Mathematical Physics} 270 (2 2007),
pp. 359--371. \textsc{doi}: \href {https://doi.org/10.1007/s00220-006-0150-x} {\nolinkurl
{10.1007/s00220-006-0150-x}}.
%
\bibitem[Bae08]{baelde2008linear}
David Baelde. ``A Linear Approach to the Proof-Theory of Least and Greatest Fixed Points''. PhD thesis.
École Polytechnique, Palaiseau, France, 2008.
%
\bibitem[Bae12]{baelde2012least}
David Baelde. ``Least and greatest fixed points in linear logic''. In: \emph{ACM Transactions on
Computational Logic} 13.1 (2012), 2:1--2:44. \textsc{doi}: \href {https://doi.org/10.1145/2071368.2071370}
{\nolinkurl {10.1145/2071368.2071370}}. \textsc{url}: \url {http://doi.acm.org/10.1145/2071368.2071370}.
%
\bibitem[Bak71]{bakker1971recursive}
J. W. Bakker. \emph{Recursive Procedures}. Vol. 24. Mathematical Centre Tracts. Mathematisch Centrum
Amsterdam, 1971.
%
\bibitem[Bak92]{baker1992nreversal}
Henry G. Baker. ``NREVERSAL of fortune - the thermodynamics of garbage collection''. In:
\emph{International Workshop on Memory Management, IWMM 92} (St. Malo, France, Sept. 17--19, 1992).
Ed. by Yves Bekkers and Jacques Cohen. Vol. 637. Lecture Notes in Computer Science. Springer, 1992,
pp. 507--524. \textsc{isbn}: 3-540-55940-X. \textsc{doi}: \href {https://doi.org/10.1007/BFb0017210} {\nolinkurl
{10.1007/BFb0017210}}.
%
\bibitem[Bar10]{baratella2010quantum}
Stefano Baratella. ``Quantum coherent spaces and linear logic''. In: \emph{RAIRO Theoretical Informatics
Applications} 44.4 (2010), pp. 419--441. \textsc{doi}: \href {https://doi.org/10.1051/ita/2010021} {\nolinkurl
{10.1051/ita/2010021}}.
%
\bibitem[Bar84]{barendregt84lambda}
Henk P. Barendregt. \emph{The Lambda-Calculus, its Syntax and Semantics}. 2nd ed. Vol. 103. Studies in
Logic and the Foundation of Mathematics. North Holland, 1984. \textsc{isbn}: 0-444-86748-1.
%
\bibitem[BB17]{biamonte2017tensor}
Jacob Biamonte and Ville Bergholm. ``Tensor Networks in a Nutshell''. To appear in Contemporary Physics.
Draft available as arXiv:1708.00006. 2017. \textsc{arXiv}: \href {https://www.arxiv.org/abs/1708.00006v1}
{\nolinkurl {1708.00006v1}}.
%
\bibitem[BB83]{beckenbach1983inequalities}
Edwin F. Beckenbach and Richard Bellman. \emph{Inequalities}. Fourth. Vol. 30. Ergebnisse der Mathematik
und ihrer Grenzgebiete. Springer-Verlag, 1983.
%
\bibitem[BBCD+95]{barenco1995elementary}
Adriano Barenco, Charles H. Bennett, Richard Cleve, David P. DiVincenzo, Norman Margolus, Peter Shor,
Tycho Sleator, John A. Smolin, and Harald Weinfurter. ``Elementary gates for quantum computation''. In:
\emph{Physical Review A} 52 (5 Nov. 1995), pp. 3457--3467. \textsc{doi}: \href
{https://doi.org/10.1103/PhysRevA.52.3457} {\nolinkurl {10.1103/PhysRevA.52.3457}}.
%
\bibitem[BBGV20]{silq}
Benjamin Bichsel, Maximilian Baader, Timon Gehr, and Martin T. Vechev. ``Silq: a high-level quantum
language with safe uncomputation and intuitive semantics''. In: \emph{Proceedings of the 41st ACM
SIGPLAN International Conference on Programming Language Design and Implementation, PLDI'20}
(London, UK). Ed. by Alastair F. Donaldson and Emina Torlak. ACM, 2020, pp. 286--300. \textsc{doi}: \href
{https://doi.org/10.1145/3385412.3386007} {\nolinkurl {10.1145/3385412.3386007}}.
%
\bibitem[BBHP92]{benton92linear}
Nick Benton, Gavin M. Bierman, Martin Hyland, and Valeria C. V. de Paiva. ``Linear lambda-calculus and
categorical models revisited''. In: \emph{Computer Science Logic, Sixth International Workshop, CSL'92,
Selected Papers} (San Miniato, Italy). Ed. by E. Börger, G. Jäger, H. Kleine Büning, S. Martini, and
M. M. Richter. Vol. 702. Lecture Notes in Computer Science. European Association for Computer Science
Logic. Springer Verlag, Sept. 1992. \textsc{isbn}: 978-3-540-56992-3. \textsc{doi}: \href
{https://doi.org/10.1007/3-540-56992-8_6} {\nolinkurl {10.1007/3-540-56992-8_6}}.
%
\bibitem[BBPH93]{benton93term}
Nick Benton, Gavin M. Bierman, Valeria C. V. de Paiva, and Martin Hyland. ``A term calculus for
intuitionistic linear logic''. In: \emph{Proceedings of the International Conference on Typed Lambda Calculi
and Applications, TLCA'93} (Ultrech, Netherlands). Ed. by Marc Bezem and Jan Friso Groote. Vol. 664.
Lecture Notes in Computer Science. Springer Verlag, Mar. 1993, pp. 75--90. \textsc{isbn}: 3-540-56517-5.
\textsc{doi}: \href {https://doi.org/10.1007/BFb0037099} {\nolinkurl {10.1007/BFb0037099}}.
%
\bibitem[BBVA19]{brugiere2019synthesizing}
Timothée Goubault de Brugière, Marc Baboulin, Benoît Valiron, and Cyril Allouche. ``Synthesizing quantum
circuits via numerical optimization''. In: \emph{Proceedings of the 19th International Conference on
Computational Science, ICCS 2019, Part II} (Faro, Portugal, June 12--14, 2019). Ed. by João M. F. Rodrigues,
Pedro J. S. Cardoso, Jânio M. Monteiro, Roberto Lam, Valeria V. Krzhizhanovskaya, Michael Harold Lees,
Jack J. Dongarra, and Peter M. A. Sloot. Vol. 11537. Lecture Notes in Computer Science. Springer, 2019,
pp. 3--16. \textsc{isbn}: 978-3-030-22740-1. \textsc{doi}: \href {https://doi.org/10.1007/978-3-030-22741-8_1}
{\nolinkurl {10.1007/978-3-030-22741-8_1}}. \textsc{hal}: \href {https://hal.archives-ouvertes.fr/hal-02174967}
{\nolinkurl {hal-02174967}}. \textsc{arXiv}: \href {https://www.arxiv.org/abs/2004.07714} {\nolinkurl
{2004.07714}}.
%
\bibitem[BBVA20]{brugiere2020householder}
Timothée Goubault de Brugière, Marc Baboulin, Benoît Valiron, and Cyril Allouche. ``Quantum circuits
synthesis using Householder transformations''. In: \emph{Computer Physics Communications} 248 (2020),
p. 107001. \textsc{doi}: \href {https://doi.org/10.1016/j.cpc.2019.107001} {\nolinkurl
{10.1016/j.cpc.2019.107001}}. \textsc{hal}: \href {https://hal.archives-ouvertes.fr/hal-02545123} {\nolinkurl
{hal-02545123}}. \textsc{arXiv}: \href {https://www.arxiv.org/abs/2004.07710} {\nolinkurl {2004.07710}}.
%
\bibitem[BBVMA20]{brugiere2020quantum}
Timothée Goubault de Brugière, Marc Baboulin, Benoît Valiron, Simon Martiel, and Cyril Allouche.
``Quantum CNOT circuits synthesis for NISQ architectures using the syndrome decoding problem''. In:
[RC20], pp. 189--205. \textsc{doi}: \href {https://doi.org/10.1007/978-3-030-52482-1_11} {\nolinkurl
{10.1007/978-3-030-52482-1_11}}.
%
\bibitem[BBVMA21]{brugiere2021reducing}
Timothee Goubault De Brugiere, Marc Baboulin, Benoît Valiron, Simon Martiel, and Cyril Allouche.
``Reducing the depth of linear reversible quantum circuits''. In: \emph{IEEE Transactions on Quantum
Engineering} 2 (2021), p. 3102422. \textsc{doi}: \href {https://doi.org/10.1109/TQE.2021.3091648} {\nolinkurl
{10.1109/TQE.2021.3091648}}. \textsc{hal}: \href {https://hal.archives-ouvertes.fr/hal-03553916} {\nolinkurl
{hal-03553916}}.
%
\bibitem[BBVMA22]{brugiere2022decoding}
Timothée Goubault de Brugière, Marc Baboulin, Benoît Valiron, Simon Martiel, and Cyril Allouche.
``Decoding techniques applied to the compilation of CNOT circuits for NISQ architectures''. In:
\emph{Science of Computer Programming} 214 (2022), p. 102726. \textsc{doi}: \href
{https://doi.org/10.1016/J.SCICO.2021.102726} {\nolinkurl {10.1016/J.SCICO.2021.102726}}. \textsc{hal}: \href
{https://hal.archives-ouvertes.fr/hal-03547113} {\nolinkurl {hal-03547113}}.
%
\bibitem[BC13]{braibant2013formal}
Thomas Braibant and Adam Chlipala. ``Formal verification of hardware synthesis''. In: \emph{Proceedings of
the 25th International Conference on Computer Aided Verification, CAV 2013} (Saint Petersburg, Russia,
July 13--19, 2013). Ed. by Natasha Sharygina and Helmut Veith. Vol. 8044. Lecture Notes in Computer
Science. Springer, 2013, pp. 213--228. \textsc{isbn}: 978-3-642-39798-1. \textsc{doi}: \href
{https://doi.org/10.1007/978-3-642-39799-8_14} {\nolinkurl {10.1007/978-3-642-39799-8_14}}. \textsc{arXiv}:
\href {https://www.arxiv.org/abs/1301.4779} {\nolinkurl {1301.4779}}.
%
\bibitem[BCCKS14]{berry2014exponential}
Dominic W. Berry, Andrew M. Childs, Richard Cleve, Robin Kothari, and Rolando D. Somma. ``Exponential
improvement in precision for simulating sparse Hamiltonians''. In: \emph{Proceedings of the Symposium on
Theory of Computing, STOC 2014} (New York, NY, USA, May 31--June 3, 2014). Ed. by David B. Shmoys.
ACM, 2014, pp. 283--292. \textsc{isbn}: 978-1-4503-2710-7. \textsc{doi}: \href
{https://doi.org/10.1145/2591796.2591854} {\nolinkurl {10.1145/2591796.2591854}}. \textsc{arXiv}: \href
{https://www.arxiv.org/abs/1312.1414} {\nolinkurl {1312.1414}}.
%
\bibitem[BCK15]{berry2015hamiltonian}
Dominic W. Berry, Andrew M. Childs, and Robin Kothari. ``Hamiltonian simulation with nearly optimal
dependence on all parameters''. In: \emph{Proceedings of the IEEE 56th Annual Symposium on Foundations
of Computer Science, FOCS 2015} (Berkeley, CA, USA, Oct. 17, 2014--Oct. 20, 2015). Ed. by
Venkatesan Guruswami. IEEE Computer Society, 2015, pp. 792--809. \textsc{doi}: \href
{https://doi.org/10.1109/FOCS.2015.54} {\nolinkurl {10.1109/FOCS.2015.54}}. \textsc{arXiv}: \href
{https://www.arxiv.org/abs/1501.01715} {\nolinkurl {1501.01715}}.
%
\bibitem[BCKH+22]{bharti2022noisy}
Kishor Bharti, Alba Cervera-Lierta, Thi Ha Kyaw, Tobias Haug, Sumner Alperin-Lea, Abhinav Anand,
Matthias Degroote, Hermanni Heimonen, Jakob S. Kottmann, Tim Menke, Wai-Keong Mok, Sukin Sim,
Leong-Chuan Kwek, and Alán Aspuru-Guzik. ``Noisy intermediate-scale quantum algorithms''. In:
\emph{Reviews of Modern Physics} 94 (1 Feb. 2022), p. 015004. \textsc{doi}: \href
{https://doi.org/10.1103/RevModPhys.94.015004} {\nolinkurl {10.1103/RevModPhys.94.015004}}.
%
\bibitem[BCMS89]{backhouse1989diy}
Roland Backhouse, Paul Chisholm, Grant Malcolm, and Erik Saaman. ``Do-it-yourself type theory''. In:
\emph{Formal Aspects of Computing} 1.1 (1989), pp. 19--84. \textsc{doi}: \href
{https://doi.org/10.1007/BF01887198} {\nolinkurl {10.1007/BF01887198}}.
%
\bibitem[BCS03]{bettelli03architecture}
Stefano Bettelli, Tommaso Calarco, and Luciano Serafini. ``Toward an architecture for quantum
programming''. In: \emph{The European Physical Journal D - Atomic, Molecular and Optical Physics} 25.2
(Aug. 2003), pp. 181--200. \textsc{doi}: \href {https://doi.org/10.1140/epjd/e2003-00242-2} {\nolinkurl
{10.1140/epjd/e2003-00242-2}}. \textsc{arXiv}: \href {https://www.arxiv.org/abs/cs/0103009} {\nolinkurl
{cs/0103009}}.
%
\bibitem[BDHP19]{beaudrap2019pauli}
Niel de Beaudrap, Ross Duncan, Dominic Horsman, and Simon Perdrix. ``Pauli fusion: a computational
model to realise quantum transformations from ZX terms''. In: \cite{qpl2019}, pp. 85--105. \textsc{doi}: \href
{https://doi.org/10.4204/EPTCS.318.6} {\nolinkurl {10.4204/EPTCS.318.6}}.
%
\bibitem[BDS13]{barendregt2013lambda}
Henk Barendregt, Wil Dekkers, and Richard Statman. \emph{Lambda Calculus with Types}. Perspective in
Logic. Cambridge University Press and ASL, 2013.
%
\bibitem[BDS16]{baelde2016infinitary}
David Baelde, Amina Doumane, and Alexis Saurin. ``Infinitary proof theory: the multiplicative additive case''.
In: \emph{Proceedings of the 25th EACSL Annual Conference on Computer Science Logic (CSL'16)}
(Marseille, France). Ed. by Jean-Marc Talbot and Laurent Regnier. Vol. 62. LIPIcs. Schloss Dagstuhl Leibniz-Zentrum fuer Informatik, 2016, 42:1--42:17. \textsc{isbn}: 978-3-95977-022-4. \textsc{doi}: \href
{https://doi.org/10.4230/LIPIcs.CSL.2016.42} {\nolinkurl {10.4230/LIPIcs.CSL.2016.42}}. \textsc{url}: \url
{http://www.dagstuhl.de/dagpub/978-3-95977-022-4}.
%
\bibitem[Bea03]{beauregard2003circuit}
Stephane Beauregard. ``Circuit for Shor's algorithm using $2n + 3$ qubits''. In: \emph{Quantum Information
and Computation} 3.2 (2003), pp. 175--185. \textsc{doi}: \href {https://doi.org/10.26421/QIC3.2-8} {\nolinkurl
{10.26421/QIC3.2-8}}.
%
\bibitem[Ben00]{bennett2000notes}
Charles H. Bennett. ``Notes on the history of reversible computation''. In: \emph{IBM Journal of Research
and Development} 44.1 (2000), pp. 270--278. \textsc{doi}: \href {https://doi.org/10.1147/rd.441.0270} {\nolinkurl
{10.1147/rd.441.0270}}.
%
\bibitem[Ben73]{bennett1973logical}
Charles H Bennett. ``Logical reversibility of computation''. In: \emph{IBM Journal of Research and
Development} 17.6 (1973), pp. 525--532. \textsc{doi}: \href {https://doi.org/10.1147/rd.176.0525} {\nolinkurl
{10.1147/rd.176.0525}}.
%
\bibitem[Ben89]{bennett1989timespace}
Charles H. Bennett. ``Time/space trade-offs for reversible computation''. In: \emph{SIAM Journal on
Computing} 18.4 (1989), pp. 766--776. \textsc{doi}: \href {https://doi.org/10.1137/0218053} {\nolinkurl
{10.1137/0218053}}.
%
\bibitem[Ben94a]{benton94mixed-tech}
Nick Benton. \emph{A Mixed Linear and Non-Linear Logic: Proofs, Terms and Models}. Tech. rep.
UCAM-CL-TR-352. 65 pages. Computer Science department, Cambridge University, 1994.
%
\bibitem[Ben94b]{benton94mixed}
Nick Benton. ``A mixed linear and non-linear logic: proofs, terms and models (extended abstract)''. In:
\emph{Computer Science Logic, Eighth International Workshop, CSL'94, Selected Papers} (Kazimierz,
Poland). Ed. by Leszek Pacholski and Jerzy Tiuryn. Vol. 933. Lecture Notes in Computer Science. European
Association for Computer Science Logic. Springer Verlag, Sept. 1994, pp. 121--135. \textsc{isbn}:
3-540-60017-5. \textsc{doi}: \href {https://doi.org/10.1007/BFb0022251} {\nolinkurl {10.1007/BFb0022251}}.
%
\bibitem[Ber00]{berry2000foundations}
Gérard Berry. ``The foundations of Esterel''. In: \emph{Proof, Language, and Interaction, Essays in Honour of
Robin Milner}. Ed. by Gordon D. Plotkin, Colin Stirling, and Mads Tofte. The MIT Press, 2000, pp. 425--454.
\textsc{isbn}: 978-0-262-16188-6. \textsc{doi}: \href {https://doi.org/10.7551/mitpress/5641.003.0021} {\nolinkurl
{10.7551/mitpress/5641.003.0021}}.
%
\bibitem[BFMP11]{bobot2011why3}
François Bobot, Jean-Christophe Filliâtre, Claude Marché, and Andrei Paskevich. ``Why3: shepherd your
herd of provers''. In: \emph{Proceedings of Boogie 2011: First International Workshop on Intermediate
Verification Languages} (Wroclaw, Poland). 2011, pp. 53--64. \textsc{hal}: \href
{https://hal.archives-ouvertes.fr/hal-00790310} {\nolinkurl {hal-00790310}}.
%
\bibitem[BFPS81]{buchholz1981iterated}
Wilfried Buchholz, Solomon Feferman, Wolfram Pohlers, and Wilfried Sieg. \emph{Iterated Inductive
Definitions and Subsystems of Analysis: Recent Proof-Theoretical Studies}. Vol. 897. Lecture Notes in
Mathematics. Springer-Verlag, 1981.
%
\bibitem[BG89]{breazu-tannen1989polymorphic}
Val Breazu-Tannen and Jean H. Gallier. ``Polymorphic rewriting conserves algebraic strong normalization
and confluence''. In: \emph{Proceedings of the 16th International Colloquium on Automata, Languages and
Programming, ICALP'89} (Stresa, Italy, July 11--15, 1989). Ed. by Giorgio Ausiello,
Mariangiola Dezani-Ciancaglini, and Simona Ronchi Della Rocca. Vol. 372. Lecture Notes in Computer
Science. Springer, 1989, pp. 137--150. \textsc{isbn}: 3-540-51371-X. \textsc{doi}: \href
{https://doi.org/10.1007/BFb0035757} {\nolinkurl {10.1007/BFb0035757}}.
%
\bibitem[BG91]{tannen1991polymorphic}
Val Breazu-Tannen and Jean H. Gallier. ``Polymorphic rewriting conserves algebraic strong normalization''.
In: \emph{Theoretical Computer Science} 83.1 (1991), pp. 3--28. \textsc{doi}: \href
{https://doi.org/10.1016/0304-3975(91)90037-3} {\nolinkurl {10.1016/0304-3975(91)90037-3}}.
%
\bibitem[BG94]{tannen1994polymorphic}
Val Breazu-Tannen and Jean H. Gallier. ``Polymorphic rewriting conserves algebraic confluence''. In:
\emph{Information and Computation} 114.1 (1994). Available as U. Penn. Tech. Report MS-CIS-90-37.,
pp. 1--29. \textsc{doi}: \href {https://doi.org/10.1006/inco.1994.1078} {\nolinkurl {10.1006/inco.1994.1078}}.
%
\bibitem[BH20]{beaudrap2017zx}
Niel de Beaudrap and Dominic Horsman. ``The ZX calculus is a language for surface code lattice surgery''. In:
\emph{Quantum} 4 (2020), p. 218. \textsc{arXiv}: \href {https://www.arxiv.org/abs/1704.08670} {\nolinkurl
{1704.08670}}.
%
\bibitem[BHYYZ20]{barthe2020relational}
Gilles Barthe, Justin Hsu, Mingsheng Ying, Nengkun Yu, and Li Zhou. ``Relational proofs for quantum
programs''. In: \emph{Proceedings of the ACM on Programming Languages} 4.POPL (2020), 21:1--21:29.
\textsc{doi}: \href {https://doi.org/10.1145/3371089} {\nolinkurl {10.1145/3371089}}.
%
\bibitem[Bie93]{bierman93intuitionistic}
Gavin M. Bierman. ``On Intuitionistic Linear Logic''. Available as Technical Report 346, August 1994.
PhD thesis. England, UK.: Computer Science department, Cambridge University, 1993.
%
\bibitem[BISG+20]{pennylane}
Ville Bergholm, Josh Izaac, Maria Schuld, Christian Gogolin, M. Sohaib Alam, Shahnawaz Ahmed,
Juan Miguel Arrazola, Carsten Blank, Alain Delgado, Soran Jahangiri, Keri McKiernan,
Johannes Jakob Meyer, Zeyue Niu, Antal Száva, and Nathan Killoran. ``PennyLane: Automatic differentiation
of hybrid quantum-classical computations''. 2020. \textsc{arXiv}: \href {https://www.arxiv.org/abs/1811.04968}
{\nolinkurl {1811.04968}}.
%
\bibitem[BK18]{backens2018zh}
Miriam Backens and Aleks Kissinger. ``ZH: a complete graphical calculus for quantum computations
involving classical non-linearity''. In: \cite{qpl2018}, pp. 23--42. \textsc{doi}: \href
{https://doi.org/10.4204/EPTCS.287.2} {\nolinkurl {10.4204/EPTCS.287.2}}.
%
\bibitem[BKN15]{boender2015formalization}
Jaap Boender, Florian Kammüller, and Rajagopal Nagarajan. ``Formalization of quantum protocols using
coq''. In: \emph{Proceedings of the 12th International Workshop on Quantum Physics and Logic, QPL 2015}
(Oxford, UK). Ed. by Chris Heunen, Peter Selinger, and Jamie Vicary. Vol. 195. Electronic Proceedings in
Theoretical Computer Science. 2015, pp. 71--83. \textsc{doi}: \href {https://doi.org/10.4204/EPTCS.195.6}
{\nolinkurl {10.4204/EPTCS.195.6}}.
%
\bibitem[Blu96]{blute96hopf}
Richard F. Blute. ``Hopf algebras and linear logic''. In: \emph{Mathematical Structures in Computer Science}
6.2 (1996), pp. 189--212. \textsc{doi}: \href {https://doi.org/10.1017/S0960129500000943} {\nolinkurl
{10.1017/S0960129500000943}}.
%
\bibitem[BM04]{bullock2004asymptotically}
Stephen S. Bullock and Igor L. Markov. ``Asymptotically optimal circuits for arbitrary n-qubit diagonal
computations''. In: \emph{Quantum Information and Computation} 4.1 (2004), pp. 27--47.
%
\bibitem[BM07]{baelde2007least}
David Baelde and Dale Miller. ``Least and greatest fixed points in linear logic''. In: \emph{Proceedings of the
14th International Conference on Logic for Programming, Artificial Intelligence, LPAR'07} (Yerevan,
Armenia). Ed. by Nachum Dershowitz and Andrei Voronkov. Vol. 4790. Lecture Notes in Computer Science.
Springer, 2007, pp. 92--106. \textsc{isbn}: 978-3-540-75558-6. \textsc{doi}: \href
{https://doi.org/10.1007/978-3-540-75560-9_9} {\nolinkurl {10.1007/978-3-540-75560-9_9}}.
%
\bibitem[BM87]{tannen1987computable}
Val Breazu-Tannen and Albert R. Meyer. ``Computable values can be classical''. In: \emph{Proceedings of the
14th ACM SIGPLAN-SIGACT Symposium on Principles of Programming Languages, POPL'87} (Munich,
West Germany). ACM, 1987, pp. 238--245. \textsc{isbn}: 0-89791-215-2. \textsc{doi}: \href
{https://doi.org/10.1145/41625.41646} {\nolinkurl {10.1145/41625.41646}}.
%
\bibitem[BN36]{birkhoff1936logic}
Garrett Birkhoff and John Von Neumann. ``The logic of quantum mechanics''. In: \emph{Annals of
Mathematics} 37.4 (1936), pp. 823--843. \textsc{doi}: \href {https://doi.org/10.2307/1968621} {\nolinkurl
{10.2307/1968621}}.
%
\bibitem[BPS93a]{blute94fock}
Richard F. Blute, Prakash Panangaden, and Robert A.G. Seely. ``Fock Space: A Model of Linear Exponential
Types''. Manuscript, revised version of \cite{blute93holomorphic}. 1993.
%
\bibitem[BPS93b]{blute93holomorphic}
Richard F. Blute, Prakash Panangaden, and Robert A.G. Seely. ``Holomorphic models of exponential types in
linear logic''. In: \cite{mfps93}, pp. 474--512.
%
\bibitem[BPV21]{borgna2021hybrid}
Agustín Borgna, Simon Perdrix, and Benoît Valiron. ``Hybrid quantum-classical circuit simplification with
the ZX-calculus''. In: \emph{Proceedings of the 19th Asian Symposium on Programming Languages and
Systems, APLAS 2021} (Chicago, IL, USA (Online Conference), Oct. 17--18, 2021). Ed. by Hakjoo Oh.
Vol. 13008. Lecture Notes in Computer Science. Springer, 2021, pp. 121--139. \textsc{doi}: \href
{https://doi.org/10.1007/978-3-030-89051-3_8} {\nolinkurl {10.1007/978-3-030-89051-3_8}}. \textsc{arXiv}: \href
{https://www.arxiv.org/abs/2109.06071} {\nolinkurl {2109.06071}}.
%
\bibitem[BR72]{bakker1972calculus}
J. W. de Bakker and Willem P. de Roever. ``A calculus for recursive program schemes''. In: \emph{Automata,
Languages and Programming: Proceedings of a Symposium Organized by IRIA} (Rocquencourt, France,
1972). Ed. by Maurice Nivat. Also found as a Technical Report of \emph{Stichting Mathematisch Centrum},
CWI. North-Holland, Amsterdam, 1972, pp. 167--196. \textsc{isbn}: 0-7204-2074-1. \textsc{url}: \url
{https://ir.cwi.nl/pub/9145/} (visited on Aug. 11, 2022).
%
\bibitem[Bre88]{tannen1988combining}
Val Breazu-Tannen. ``Combining algebra and higher-order types''. In: \emph{Proceedings of the Third Annual
Symposium on Logic in Computer Science, LICS '88} (Edinburgh, Scotland, UK, July 5--8, 1988). IEEE
Computer Society, 1988, pp. 82--90. \textsc{isbn}: 0-8186-0853-6. \textsc{doi}: \href
{https://doi.org/10.1109/LICS.1988.5103} {\nolinkurl {10.1109/LICS.1988.5103}}.
%
\bibitem[Bro70]{broyden1970convergence}
C. G. Broyden. ``The convergence of a class of double-rank minimization algorithms''. In: \emph{IMA Journal
of Applied Mathematics} 6.1 (1970), pp. 76--90.
%
\bibitem[Bru20]{brugiere2020phd}
Timothée Goubault de Brugière. ``Methods for optimizing the synthesis of quantum circuits''. Thèse de
Doctorat. Université Paris-Saclay, 2020. \textsc{hal}: \href {https://hal.archives-ouvertes.fr/tel-03127089}
{\nolinkurl {tel-03127089}}.
%
\bibitem[BSDCM19]{bhattacharjee2019reversible}
Debjyoti Bhattacharjee, Mathias Soeken, Srijit Dutta, Anupam Chattopadhyay, and Giovanni De Micheli.
``Reversible pebble games for reducing qubits in hierarchical quantum circuit synthesis''. In:
\emph{Proceedings of the 49th International Symposium on Multiple-Valued Logic, ISMVL 2019}
(Fredericton, NB, Canada, May 21--23, 2019). IEEE, 2019, pp. 102--107. \textsc{isbn}: 978-1-7281-0092-0.
\textsc{doi}: \href {https://doi.org/10.1109/ISMVL.2019.00026} {\nolinkurl {10.1109/ISMVL.2019.00026}}.
%
\bibitem[BT04]{baillot04light}
Patrick Baillot and Kazushige Terui. ``Light types for polynomial time computation in lambda-calculus''. In:
[LICS04], pp. 266--275. \textsc{doi}: \href {https://doi.org/10.1109/LICS.2004.1319621} {\nolinkurl
{10.1109/LICS.2004.1319621}}. \textsc{hal}: \href {https://hal.archives-ouvertes.fr/hal-00003468} {\nolinkurl
{hal-00003468}}. \textsc{arXiv}: \href {https://www.arxiv.org/abs/cs/0402059} {\nolinkurl {cs/0402059}}.
%
\bibitem[Bur69]{burstall1969proving}
Rod M. Burstall. ``Proving properties of programs by structural induction''. In: \emph{The Computer Journal}
12.1 (1969), pp. 41--48. \textsc{doi}: \href {https://doi.org/10.1093/comjnl/12.1.41} {\nolinkurl
{10.1093/comjnl/12.1.41}}.
%
\bibitem[Bur74]{burstall1974program}
Rod M. Burstall. ``Program proving as hand simulation with a little induction''. In: \emph{Proceedings of the
6th IFIP Congress on Information Processing} (Stockholm, Sweden, Aug. 5--10, 1974). Ed. by
Jack L. Rosenfeld. North-Holland, 1974, pp. 308--312. \textsc{isbn}: 0-7204-2803-3.
%
\bibitem[BW15]{bian2015graphical}
Xiaoning Bian and Quanlong Wang. ``Graphical calculus for qutrit systems''. In: \emph{IEICE Transactions
on Fundamentals of Electronics, Communications and Computer Sciences} 98-A.1 (2015), pp. 391--399.
\textsc{doi}: \href {https://doi.org/10.1587/transfun.E98.A.391} {\nolinkurl {10.1587/transfun.E98.A.391}}.
%
\bibitem[CABB+21]{cerezo2021variational}
M. Cerezo, Andrew Arrasmith, Ryan Babbush, Simon C. Benjamin, Suguru Endo, Keisuke Fujii,
Jarrod R. McClean, Kosuke Mitarai, Xiao Yuan, Lukasz Cincio, and Patrick J. Coles. ``Variational quantum
algorithms''. In: \emph{Nature Reviews Physics} 3.9 (2021), pp. 625--644. \textsc{doi}: \href
{https://doi.org/10.1038/s42254-021-00348-9} {\nolinkurl {10.1038/s42254-021-00348-9}}.
%
\bibitem[CBBPV21]{chareton2021automated}
Christophe Chareton, Sébastien Bardin, François Bobot, Valentin Perrelle, and Benoît Valiron. ``An
automated deductive verification framework for circuit-building quantum programs''. In: \emph{Proceedings
of the 30th European Symposium on Programming Languages and Systems, ESOP 2021} (Luxembourg City,
Luxembourg, Mar. 27--Apr. 1, 2021). Ed. by Nobuko Yoshida. Vol. 12648. Lecture Notes in Computer Science.
Springer, 2021, pp. 148--177. \textsc{isbn}: 978-3-030-72018-6. \textsc{doi}: \href
{https://doi.org/10.1007/978-3-030-72019-3_6} {\nolinkurl {10.1007/978-3-030-72019-3_6}}. \textsc{arXiv}: \href
{https://www.arxiv.org/abs/2003.05841} {\nolinkurl {2003.05841}}.
%
\bibitem[CBLVVX21]{chareton2021formal}
Christophe Chareton, Sébastien Bardin, Dongho Lee, Benoît Valiron, Renaud Vilmart, and Zhaowei Xu.
``Formal Methods for Quantum Programs: A Survey''. Draft, to appear as a book chapter. 2021. \textsc{arXiv}:
\href {https://www.arxiv.org/abs/2109.06493} {\nolinkurl {2109.06493}}.
%
\bibitem[CBSNG19]{cross2019validating}
Andrew W. Cross, Lev S. Bishop, Sarah Sheldon, Paul D. Nation, and Jay M. Gambetta. ``Validating quantum
computers using randomized model circuits''. In: \emph{Physical Review A} 100 (3 Sept. 2019), p. 032328.
\textsc{doi}: \href {https://doi.org/10.1103/PhysRevA.100.032328} {\nolinkurl {10.1103/PhysRevA.100.032328}}.
\textsc{arXiv}: \href {https://www.arxiv.org/abs/1811.12926} {\nolinkurl {1811.12926}}.
%
\bibitem[CCDFGS03]{childs2003exponential}
Andrew M. Childs, Richard Cleve, Enrico Deotto, Edward Farhi, Sam Gutmann, and Daniel A. Spielman.
``Exponential algorithmic speedup by a quantum walk''. In: \emph{Proceedings of the 35th Annual ACM
Symposium on Theory of Computing, STOC'03} (San Diego, CA, USA, June 2003). Ed. by
Lawrence L. Larmore and Michel X. Goemans. ACM, 2003, pp. 59--68. \textsc{isbn}: 1-58113-674-9.
\textsc{doi}: \href {https://doi.org/10.1145/780542.780552} {\nolinkurl {10.1145/780542.780552}}.
%
\bibitem[CD07]{coecke2007graphical}
Bob Coecke and Ross Duncan. ``A graphical calculus for quantum observables''. Historical note from Bob
Coecke: First paper containing ZX-diagrams. Rejected from QIP with reports like: ``nice pictures, so what?''.
2007. \textsc{url}: \url {http://www.cs.ox.ac.uk/people/bob.coecke/GreenRed.pdf} (visited on Aug. 24, 2022).
%
\bibitem[CD08]{coecke2008interacting}
Bob Coecke and Ross Duncan. ``Interacting quantum observables''. In: \emph{Proceedings of the 35th
International Colloquium on Automata, Languages and Programming (ICALP 2008), Part II} (Reykjavik,
Iceland, July 7--11, 2008). Ed. by Luca Aceto, Ivan Damgård, Leslie Ann Goldberg, Magnús M. Halldórsson,
Anna Ingólfsdóttir, and Igor Walukiewicz. Vol. 5126. Lecture Notes in Computer Science. Springer, 2008,
pp. 298--310. \textsc{doi}: \href {https://doi.org/10.1007/978-3-540-70583-3\_25} {\nolinkurl
{10.1007/978-3-540-70583-3\_25}}.
%
\bibitem[CD11]{coecke2011interacting}
Bob Coecke and Ross Duncan. ``Interacting quantum observables: categorical algebra and diagrammatics''.
In: \emph{New Journal of Physics} 13.4 (Apr. 2011), p. 043016. \textsc{doi}: \href
{https://doi.org/10.1088/1367-2630/13/4/043016} {\nolinkurl {10.1088/1367-2630/13/4/043016}}. \textsc{arXiv}:
\href {https://www.arxiv.org/abs/0906.4725} {\nolinkurl {0906.4725}}.
%
\bibitem[CD22]{colledan2022dynamic}
Andrea Colledan and Ugo Dal Lago. ``On Dynamic Lifting and Effect Typing in Circuit Description
Languages (Extended Version)''. Presented at TYPES 2022. 2022. \textsc{arXiv}: \href
{https://www.arxiv.org/abs/2202.07636} {\nolinkurl {2202.07636}}.
%
\bibitem[CD93]{consel1993tutorial}
Charles Consel and Olivier Danvy. ``Tutorial notes on partial evaluation''. In: \emph{Conference Record of
the Twentieth Annual ACM SIGPLAN-SIGACT Symposium on Principles of Programming Languages,
POPL'93} (Charleston, South Carolina, USA). Ed. by Mary S. Van Deusen and Bernard Lang. ACM Press,
1993, pp. 493--501. \textsc{isbn}: 0-89791-560-7. \textsc{doi}: \href {https://doi.org/10.1145/158511.158707}
{\nolinkurl {10.1145/158511.158707}}.
%
\bibitem[CDP08a]{chiribella2008quantum}
G. Chiribella, G. M. D'Ariano, and P. Perinotti. ``Quantum circuit architecture''. In: \emph{Physical Review
Letters} 101 (6 Aug. 2008), p. 060401. \textsc{doi}: \href {https://doi.org/10.1103/PhysRevLett.101.060401}
{\nolinkurl {10.1103/PhysRevLett.101.060401}}. \textsc{arXiv}: \href {https://www.arxiv.org/abs/0712.1325}
{\nolinkurl {0712.1325}}.
%
\bibitem[CDP08b]{chiribella2008transforming}
G. Chiribella, G. M. D'Ariano, and P. Perinotti. ``Transforming quantum operations: quantum supermaps''.
In: \emph{EPL (Europhysics Letters)} 83.3 (2008), p. 30004. \textsc{doi}: \href
{https://doi.org/10.1209/0295-5075/83/30004} {\nolinkurl {10.1209/0295-5075/83/30004}}. \textsc{arXiv}: \href
{https://www.arxiv.org/abs/0804.0180} {\nolinkurl {0804.0180}}.
%
\bibitem[CDP09]{chiribella2009theoretical}
Giulio Chiribella, Giacomo Mauro D'Ariano, and Paolo Perinotti. ``Theoretical framework for quantum
networks''. In: \emph{Physical Review A} 80 (2 Aug. 2009), p. 022339. \textsc{doi}: \href
{https://doi.org/10.1103/PhysRevA.80.022339} {\nolinkurl {10.1103/PhysRevA.80.022339}}.
%
\bibitem[CDPV13]{chiribella2013quantum}
G. Chiribella, G. M. D'Ariano, P. Perinotti, and B. Valiron. ``Quantum computations without definite causal
structure''. In: \emph{Physical Review A} 88 (2013), p. 022318. \textsc{doi}: \href
{https://doi.org/10.1103/PhysRevA.88.022318} {\nolinkurl {10.1103/PhysRevA.88.022318}}. \textsc{arXiv}: \href
{https://www.arxiv.org/abs/0912.0195} {\nolinkurl {0912.0195}}.
%
\bibitem[CFC58]{curry1958combinatory}
Haskell H. Curry, Robert Feys, and William Craig. \emph{Combinatory Logic}. Vol. 22. Studies in Logic and
the Foundations of Mathematics. North-Holland, 1958.
%
\bibitem[CFM17]{chong2017programming}
Frederic T. Chong, Diana Franklin, and Margaret Martonosi. ``Programming languages and compiler design
for realistic quantum hardware''. In: \emph{Nature} 549 (7671 2017), pp. 180--187. \textsc{doi}: \href
{https://doi.org/10.1038/nature23459} {\nolinkurl {10.1038/nature23459}}.
%
\bibitem[Cha23]{chardonnet2023phd}
Kostia Chardonnet. ``Vers une correspondance de Curry-Howard pour le calcul quantique''. Thèse de
Doctorat. Université Paris-Saclay, 2023.
%
\bibitem[CHMPV22]{clement2022lov}
Alexandre Clément, Nicolas Heurtel, Shane Mansfield, Simon Perdrix, and Benoît Valiron. ``LOv-calculus: a
graphical language for linear optical quantum circuits''. In: \emph{47th International Symposium on
Mathematical Foundations of Computer Science, MFCS 2022, August 22-26, 2022, Vienna, Austria}. Ed. by
Stefan Szeider, Robert Ganian, and Alexandra Silva. Vol. 241. LIPIcs. 2022, 35:1--35:16. \textsc{doi}: \href
{https://doi.org/10.4230/LIPICS.MFCS.2022.35} {\nolinkurl {10.4230/LIPICS.MFCS.2022.35}}. \textsc{url}: \url
{https://doi.org/10.4230/LIPIcs.MFCS.2022.35}.
%
\bibitem[CHMPV23a]{clement2023complete}
Alexandre Clément, Nicolas Heurtel, Shane Mansfield, Simon Perdrix, and Benoît Valiron. ``A complete
equational theory for quantum circuits''. In: \emph{38th Annual ACM/IEEE Symposium on Logic in
Computer Science, LICS 2023, Boston, MA, USA, June 26-29, 2023}. IEEE, 2023, pp. 1--13. \textsc{doi}: \href
{https://doi.org/10.1109/LICS56636.2023.10175801} {\nolinkurl {10.1109/LICS56636.2023.10175801}}.
\textsc{hal}: \href {https://hal.archives-ouvertes.fr/hal-03926757} {\nolinkurl {hal-03926757}}.
%
\bibitem[CHMPV23b]{clement2023complete-tqc}
Alexandre Clément, Nicolas Heurtel, Shane Mansfield, Simon Perdrix, and Benoît Valiron. ``A Complete
Equational Theory for Quantum Circuits''. Presentation accepted at the 18th Conference on the Theory of
Quantum Computation, Communication and Cryptography (TQC 2023), in Aveiro, Portugal. 2023.
\textsc{hal}: \href {https://hal.archives-ouvertes.fr/hal-04318291v1} {\nolinkurl {hal-04318291v1}}.
%
\bibitem[Cho75]{choi1975completely}
Man-Duen Choi. ``Completely positive linear maps on complex matrices''. In: \emph{Linear Algebra and its
Applications} 10.3 (1975), pp. 285--290.
%
\bibitem[CHP19]{carette2019szx}
Titouan Carette, Dominic Horsman, and Simon Perdrix. ``SZX-calculus: scalable graphical quantum
reasoning''. In: \emph{Proceedings of the 44th International Symposium on Mathematical Foundations of
Computer Science, MFCS 2019} (Aachen, Germany, Aug. 26--30, 2019). Ed. by Peter Rossmanith,
Pinar Heggernes, and Joost-Pieter Katoen. Vol. 138. LIPIcs. Schloss Dagstuhl - Leibniz-Zentrum für
Informatik, 2019, 55:1--55:15. \textsc{isbn}: 978-3-95977-117-7. \textsc{doi}: \href
{https://doi.org/10.4230/LIPIcs.MFCS.2019.55} {\nolinkurl {10.4230/LIPIcs.MFCS.2019.55}}.
%
\bibitem[Chu36]{church36unsolvable}
Alonzo Church. ``An unsolvable problem of elementary number theory''. In: \emph{American Journal of
Mathematics} 58.2 (1936), pp. 345--363. \textsc{doi}: \href {https://doi.org/10.2307/2371045} {\nolinkurl
{10.2307/2371045}}.
%
\bibitem[CJ20]{carette2020recipe}
Titouan Carette and Emmanuel Jeandel. ``A recipe for quantum graphical languages''. In: \emph{Proceedings
of the 47th International Colloquium on Automata, Languages, and Programming, ICALP 2020}
(Saarbrücken, Germany (Virtual Conference), July 8, 2022--July 11, 2020). Ed. by Artur Czumaj, Anuj Dawar,
and Emanuela Merelli. Vol. 168. LIPIcs. Schloss Dagstuhl - Leibniz-Zentrum für Informatik, 2020,
118:1--118:17. \textsc{doi}: \href {https://doi.org/10.4230/LIPIcs.ICALP.2020.118} {\nolinkurl
{10.4230/LIPIcs.ICALP.2020.118}}.
%
\bibitem[CJPV19]{carette2019completeness}
Titouan Carette, Emmanuel Jeandel, Simon Perdrix, and Renaud Vilmart. ``Completeness of graphical
languages for mixed states quantum mechanics''. In: \emph{Proceedings of the 46th International
Colloquium on Automata, Languages, and Programming, ICALP 2019} (Patras, Greece, July 9--12, 2019).
Ed. by Christel Baier, Ioannis Chatzigiannakis, Paola Flocchini, and Stefano Leonardi. Vol. 132. LIPIcs.
Schloss Dagstuhl - Leibniz-Zentrum für Informatik, 2019, 108:1--108:15. \textsc{isbn}: 978-3-95977-109-2.
\textsc{doi}: \href {https://doi.org/10.4230/LIPIcs.ICALP.2019.108} {\nolinkurl {10.4230/LIPIcs.ICALP.2019.108}}.
\textsc{url}: \url {http://www.dagstuhl.de/dagpub/978-3-95977-109-2}.
%
\bibitem[CJS13]{clader2013preconditioned}
B. D. Clader, B. C. Jacobs, and C. R. Sprouse. ``Preconditioned quantum linear system algorithm''. In:
\emph{Physical Review Letters} 110 (25 2013), p. 250504. \textsc{doi}: \href
{https://doi.org/10.1103/PhysRevLett.110.250504} {\nolinkurl {10.1103/PhysRevLett.110.250504}}. \textsc{arXiv}:
\href {https://www.arxiv.org/abs/1301.2340} {\nolinkurl {1301.2340}}.
%
\bibitem[CJV93]{chatterjee1993edge-based}
A. Chatterjee, J. M. Jin, and J. L. Volakis. ``Edge-based finite elements and vector abc's applied to 3D
scattering''. In: \emph{IEEE Transactions on Antennas and Propagation} 41.2 (1993).
%
\bibitem[CK17]{coecke2017picturing}
Bob Coecke and Aleks Kissinger. \emph{Picturing Quantum Processes: A First Course in Quantum Theory
and Diagrammatic Reasoning}. Cambridge University Press, 2017. \textsc{isbn}: 9781316219317. \textsc{doi}:
\href {https://doi.org/10.1017/9781316219317} {\nolinkurl {10.1017/9781316219317}}.
%
\bibitem[CK97]{cosmo1997strong}
Roberto Di Cosmo and Delia Kesner. ``Strong normalization of explicit substitutions via cut elimination in
proof nets (extended abstract)''. In: \emph{Proceedings of the Twelfth Annual IEEE Symposium on Logic in
Computer Science, LICS'97} (Warsaw, Poland, June 29--July 2, 1997). IEEE Computer Society, 1997, pp. 35--46.
\textsc{isbn}: 0-8186-7925-5. \textsc{doi}: \href {https://doi.org/10.1109/LICS.1997.614927} {\nolinkurl
{10.1109/LICS.1997.614927}}.
%
\bibitem[CKRZH18]{chancellor2018graphical}
Nicholas Chancellor, Aleks Kissinger, Joschka Roffe, Stefan Zohren, and Dominic Horsman. ``Graphical
Structures for Design and Verification of Quantum Error Correction''. Draft. 2018.
%
\bibitem[CL02]{cockett2002restriction-I}
J. Robin B. Cockett and Stephen Lack. ``Restriction categories I: categories of partial maps''. In:
\emph{Theoretical Computer Science} 270.1 (2002), pp. 223--259. \textsc{doi}: \href
{https://doi.org/10.1016/S0304-3975(00)00382-0} {\nolinkurl {10.1016/S0304-3975(00)00382-0}}.
%
\bibitem[CL03]{cockett2003restriction-II}
J. Robin B. Cockett and Stephen Lack. ``Restriction categories ii: partial map classification''. In:
\emph{Theoretical Computer Science} 294.1 (2003), pp. 61--102. \textsc{doi}: \href
{https://doi.org/10.1016/S0304-3975(01)00245-6} {\nolinkurl {10.1016/S0304-3975(01)00245-6}}.
%
\bibitem[CL07]{cockett2007restriction-III}
Robin Cockett and Stephen Lack. ``Restriction categories III: colimits, partial limits and extensivity''. In:
\emph{Mathematical Structures in Computer Science} 17.4 (2007), pp. 775--817. \textsc{doi}: \href
{https://doi.org/10.1017/S0960129507006056} {\nolinkurl {10.1017/S0960129507006056}}.
%
\bibitem[Cla01]{claessen2001embedded}
Koen Claessen. ``Embedded Languages for Describing and Verifying Hardware''. Doktorsavhandlingar.
Chalmers University of Technology, Gothenburg, Sweden, 2001. \textsc{isbn}: 91-7291-014-3.
%
\bibitem[CLV21]{lemonnier2021categorical}
Kostia Chardonnet, Louis Lemonnier, and Benoît Valiron. ``Categorical semantics of reversible
pattern-matching''. In: \cite{mfps2021}, pp. 18--33. \textsc{doi}: \href {https://doi.org/10.4204/EPTCS.351.2}
{\nolinkurl {10.4204/EPTCS.351.2}}.
%
\bibitem[Coe04]{coecke04informationflow}
Bob Coecke. ``Quantum information-flow, concretely, abstractly''. In: \cite{qpl04}, pp. 57--73.
%
\bibitem[Cop17]{copeland2020church-turing}
B. Jack Copeland. ``The Church-Turing thesis''. In: \emph{The Stanford Encyclopedia of Philosophy}. Ed. by
Edward N. Zalta. Summer 2020 Edition. 2017. \textsc{url}: \url
{https://plato.stanford.edu/archives/sum2020/entries/church-turing/} (visited on Sept. 6, 2021).
%
\bibitem[CP90]{coquand1988inductively}
T. Coquand and C. Paulin. ``Inductively defined types''. In: \cite{colog1988}, pp. 50--66.
%
\bibitem[CPV13]{coecke2013description}
Bob Coecke, Dusko Pavlovic, and Jamie Vicary. ``A new description of orthogonal bases''. In:
\emph{Mathematical Structures in Computer Science} 23.3 (2013), pp. 555--567. \textsc{doi}: \href
{https://doi.org/10.1017/S0960129512000047} {\nolinkurl {10.1017/S0960129512000047}}.
%
\bibitem[CS16]{carette2016computing}
Jacques Carette and Amr Sabry. ``Computing with semirings and weak rig groupoids''. In:
\emph{Proceedings of the 25th European Symposium on Programming Languages and Systems (ESOP'16)}
(Eindhoven, The Netherlands). Ed. by Peter Thiemann. Vol. 9632. Lecture Notes in Computer Science.
Springer, Apr. 2016, pp. 123--148. \textsc{doi}: \href {https://doi.org/10.1007/978-3-662-49498-1_6} {\nolinkurl
{10.1007/978-3-662-49498-1_6}}.
%
\bibitem[CSV20]{chardonnet2020curry-howard}
Kostia Chardonnet, Alexis Saurin, and Benoît Valiron. ``Toward a Curry-Howard equivalence for linear,
reversible computation - work-in-progress''. In: \cite{rc2020}, pp. 144--152. \textsc{doi}: \href
{https://doi.org/10.1007/978-3-030-52482-1_8} {\nolinkurl {10.1007/978-3-030-52482-1_8}}. \textsc{hal}: \href
{https://hal.archives-ouvertes.fr/hal-03103455} {\nolinkurl {hal-03103455}}.
%
\bibitem[CSV21]{chardonnet2021curry-howard}
Kostia Chardonnet, Alexis Saurin, and Benoît Valiron. ``Towards a Curry-Howard correspondence for linear,
reversible computation''. In: \emph{Proceedings of the 5th International Workshop on Trends in Linear Logic
and Applications (TLLA 2021)} (Rome (virtual), Italy). 2021. \textsc{hal}: \href
{https://hal.archives-ouvertes.fr/lirmm-03271484} {\nolinkurl {lirmm-03271484}}.
%
\bibitem[CSV23]{chardonnet2022curry-howard}
Kostia Chardonnet, Alexis Saurin, and Benoît Valiron. ``A curry-howard correspondence for linear, reversible
computation''. In: \emph{31st EACSL Annual Conference on Computer Science Logic, CSL 2023} (Warsaw,
Poland, Feb. 13--16, 2023). Ed. by Bartek Klin and Elaine Pimentel. Vol. 252. LIPIcs. 2023, 13:1--13:18.
\textsc{doi}: \href {https://doi.org/10.4230/LIPICS.CSL.2023.13} {\nolinkurl {10.4230/LIPICS.CSL.2023.13}}.
%
\bibitem[CTV23]{chapuis-chkaiban2023pagerank}
Théodore Chapuis-Chkaiban, Zeno Toffano, and Benoît Valiron. ``On new pagerank computation methods
using quantum computing''. In: \emph{Quantum Information Processing} 22.3 (2023), p. 138. \textsc{doi}:
\href {https://doi.org/10.1007/S11128-023-03856-Y} {\nolinkurl {10.1007/S11128-023-03856-Y}}. \textsc{hal}:
\href {https://hal.archives-ouvertes.fr/hal-04056045} {\nolinkurl {hal-04056045}}.
%
\bibitem[CV20]{clairambault2020full}
Pierre Clairambault and Marc de Visme. ``Full abstraction for the quantum lambda-calculus''. In:
\emph{Proceedings of the ACM on Programming Languages} 4.POPL (2020), 63:1--63:28. \textsc{doi}: \href
{https://doi.org/10.1145/3371131} {\nolinkurl {10.1145/3371131}}.
%
\bibitem[CVCC23]{capela2023reassessing}
Matheus Capela, Harshit Verma, Fabio Costa, and Lucas C. Céleri. ``Reassessing thermodynamic advantage
from indefinite causal order''. In: \emph{Physical Review A} 107 (6 2023), p. 062208. \textsc{doi}: \href
{https://doi.org/10.1103/PhysRevA.107.062208} {\nolinkurl {10.1103/PhysRevA.107.062208}}.
%
\bibitem[CVV21]{chardonnet2021geometry}
Kostia Chardonnet, Benoît Valiron, and Renaud Vilmart. ``Geometry of interaction for ZX diagrams''. In:
\emph{Proceedings of the 46th International Symposium on Mathematical Foundations of Computer
Science, MFCS 2021} (Tallinn, Estonia). Ed. by Filippo Bonchi and Simon J. Puglisi. Vol. 202. LIPIcs. Schloss
Dagstuhl - Leibniz-Zentrum fuer Informatik, 2021, 30:1--30:16. \textsc{isbn}: 978-3-95977-201-3. \textsc{doi}:
\href {https://doi.org/10.4230/LIPIcs.MFCS.2021.30} {\nolinkurl {10.4230/LIPIcs.MFCS.2021.30}}.
%
\bibitem[CVVV22]{chardonnet2022many-worlds}
Kostia Chardonnet, Marc de Visme, Benoît Valiron, and Renaud Vilmart. ``The Many-Worlds Calculus:
Representing Quantum Control''. 2022.
%
\bibitem[CW96]{clarke1996formal}
Edmund M. Clarke and Jeannette M. Wing. ``Formal methods: state of the art and future directions''. In:
\emph{ACM Computing Surveys} 28.4 (1996), pp. 626--643. \textsc{doi}: \href
{https://doi.org/10.1145/242223.242257} {\nolinkurl {10.1145/242223.242257}}.
%
\bibitem[Cyb01]{cybenko2001reducing}
George Cybenko. ``Reducing quantum computations to elementary unitary operations''. In:
\emph{Computing in Science \& Engineering} 3.2 (2001), pp. 27--32. \textsc{doi}: \href
{https://doi.org/10.1109/5992.908999} {\nolinkurl {10.1109/5992.908999}}.
%
\bibitem[Dan90]{danos1990phd}
Vincent Danos. ``La Logique Linéaire appliquée à l'étude de divers processus de normalisation
(principalement du Lambda-calcul)''. Thèse de Doctorat en Mathématiques. Université Paris 7, 1990.
%
\bibitem[DD17]{diaz-caro2017typing}
Alejandro Díaz-Caro and Gilles Dowek. ``Typing quantum superpositions and measurement''. In:
\emph{Proceedings of the 6th International Conference on the Theory and Practice of Natural Computing
(TPNC'17)} (Prague, Czech Republic). Ed. by Carlos Martín-Vide, Roman Neruda, and
Miguel A. Vega-Rodríguez. Vol. 10687. Lecture Notes in Computer Science. Springer, 2017, pp. 281--293.
\textsc{doi}: \href {https://doi.org/10.1007/978-3-319-71069-3_22} {\nolinkurl {10.1007/978-3-319-71069-3_22}}.
%
\bibitem[DE11]{danos2011probabilistic}
Vincent Danos and Thomas Ehrhard. ``Probabilistic coherence spaces as a model of higher-order
probabilistic computation''. In: \emph{Information and Computation} 209.6 (2011), pp. 966--991. \textsc{doi}:
\href {https://doi.org/10.1016/j.ic.2011.02.001} {\nolinkurl {10.1016/j.ic.2011.02.001}}.
%
\bibitem[Del08a]{delbecque2006quantum}
Yannick Delbecque. ``A quantum game semantics for the measurement calculus''. In: \cite{qpl06}, pp. 33--48.
%
\bibitem[Del08b]{delbecque2008quantum}
Yannick Delbecque. ``Quantum Games as Quantum Types''. PhD thesis. McGill University, 2008.
%
\bibitem[Del11]{delbecque2008game-data}
Yannick Delbecque. ``Game semantics for quantum data''. In: \cite{qpl2008}, pp. 41--57. \textsc{doi}: \href
{https://doi.org/10.1016/j.entcs.2011.01.005} {\nolinkurl {10.1016/j.entcs.2011.01.005}}.
%
\bibitem[Deu85]{deutsch1985quantum}
David Deutsch. ``Quantum theory, the Church-Turing principle and the universal quantum computer''. In:
\emph{Proceedings of the Royal Society of London A} 400.1818 (1985), pp. 97--117. \textsc{doi}: \href
{https://doi.org/10.1098/rspa.1985.0070} {\nolinkurl {10.1098/rspa.1985.0070}}.
%
\bibitem[DG02]{dalla2002quantum}
Maria Luisa Dalla Chiara and Roberto Giuntini. ``Quantum logics''. In: \emph{Handbook of Philosophical
Logic}. Ed. by Dov M. Gabbay and F. Guenthner. 2nd ed. Vol. 6. Springer, Dordrecht, 2002, pp. 129--228.
\textsc{isbn}: 978-1-4020-0583-1. \textsc{doi}: \href {https://doi.org/10.1007/978-94-017-0460-1_2} {\nolinkurl
{10.1007/978-94-017-0460-1_2}}.
%
\bibitem[DG08]{dershowitz2008natural}
Nachum Dershowitz and Yuri Gurevich. ``A natural axiomatization of computability and proof of Church's
thesis''. In: \emph{Bulletin of Symbolic Logic} 14.3 (2008), pp. 299--350. \textsc{doi}: \href
{https://doi.org/10.2178/bsl/1231081370} {\nolinkurl {10.2178/bsl/1231081370}}.
%
\bibitem[DGMV19]{diaz-caro2019realizability}
Alejandro Díaz-Caro, Mauricio Guillermo, Alexandre Miquel, and Benoît Valiron. ``Realizability in the unitary
sphere''. In: \cite{lics2019}, pp. 1--13. \textsc{doi}: \href {https://doi.org/10.1109/LICS.2019.8785834} {\nolinkurl
{10.1109/LICS.2019.8785834}}. \textsc{hal}: \href {https://hal.archives-ouvertes.fr/hal-02175168} {\nolinkurl
{hal-02175168}}. \textsc{arXiv}: \href {https://www.arxiv.org/abs/1904.08785} {\nolinkurl {1904.08785}}.
%
\bibitem[DH02]{danos02probabilistic}
Vincent Danos and Russel S. Harmer. ``Probabilistic game semantics''. In: \emph{ACM Transactions on
Computational Logic} 3.3 (2002), pp. 359--382. \textsc{doi}: \href {https://doi.org/10.1145/507382.507385}
{\nolinkurl {10.1145/507382.507385}}.
%
\bibitem[DHKK95]{deussen1995verification}
Peter Deussen, A. Hansmann, Thomas Käufl, and Stefan Klingenbeck. ``The verification system
Tatzelwurm''. In: \emph{KORSO - Methods, Languages, and Tools for the Construction of Correct Software}.
Ed. by Manfred Broy and Stefan Jähnichen. Vol. 1009. Lecture Notes in Computer Science. Springer, 1995,
pp. 285--298. \textsc{isbn}: 3-540-60589-4. \textsc{doi}: \href {https://doi.org/10.1007/BFb0015468} {\nolinkurl
{10.1007/BFb0015468}}.
%
\bibitem[Dij76]{dijkstra1976discipline}
Edsger W. Dijkstra. \emph{A Discipline of Programming}. Prentice-Hall, 1976. \textsc{isbn}: 013215871X.
\textsc{url}: \url {https://www.worldcat.org/oclc/01958445}.
%
\bibitem[DKPW20]{duncan2020graphtheoretic}
Ross Duncan, Aleks Kissinger, Simon Perdrix, and John van de Wetering. ``Graph-theoretic simplification of
quantum circuits with the ZX-calculus''. In: \emph{Quantum} 4 (2020), p. 279. \textsc{issn}: 2521-327X.
\textsc{doi}: \href {https://doi.org/10.22331/q-2020-06-04-279} {\nolinkurl {10.22331/q-2020-06-04-279}}.
%
\bibitem[DKRS06]{draper2006logarithmic-depth}
Thomas G. Draper, Samuel A. Kutin, Eric M. Rains, and Krysta M. Svore. ``A logarithmic-depth quantum
carry-lookahead adder''. In: \emph{Quantum Information and Computation} 6.4--5 (2006), pp. 351--369.
\textsc{arXiv}: \href {https://www.arxiv.org/abs/quant-ph/0406142} {\nolinkurl {quant-ph/0406142}}.
%
\bibitem[DL13]{duncan2013verifying}
Ross Duncan and Maxime Lucas. ``Verifying the Steane code with Quantomatic''. In: \emph{Proceedings of
the 10th International Workshop on Quantum Physics and Logic, QPL 2013} (Castelldefels (Barcelona),
Spain, July 17--19, 2013). Ed. by Bob Coecke and Matty J. Hoban. Vol. 171. EPTCS. 2013, pp. 33--49.
\textsc{doi}: \href {https://doi.org/10.4204/EPTCS.171.4} {\nolinkurl {10.4204/EPTCS.171.4}}.
%
\bibitem[Dou17]{doumane2017infinitary}
Amina Doumane. ``On the Infinitary Proof Theory of Logics with Fixed Points''. Thèse de doctorat.
Université Paris Diderot, 2017.
%
\bibitem[Dou92]{dougherty1992adding}
Daniel J. Dougherty. ``Adding algebraic rewriting to the untyped lambda calculus''. In: \emph{Information
and Computation} 101.2 (1992), pp. 251--267. \textsc{doi}: \href
{https://doi.org/10.1016/0890-5401(92)90064-M} {\nolinkurl {10.1016/0890-5401(92)90064-M}}.
%
\bibitem[Dow12]{dowek2012around}
Gilles Dowek. ``Around the physical Church-Turing thesis: cellular automata, formal languages, and the
principles of quantum theory''. In: \emph{Proceedingsof the 6th International Conference on Language and
Automata Theory and Applications, LATA 2012} (A Coruña, Spain, Mar. 5--9, 2012). Ed. by
Adrian-Horia Dediu and Carlos Martín-Vide. Vol. 7183. Lecture Notes in Computer Science. Springer, 2012,
pp. 21--37. \textsc{isbn}: 978-3-642-28331-4. \textsc{doi}: \href {https://doi.org/10.1007/978-3-642-28332-1\_3}
{\nolinkurl {10.1007/978-3-642-28332-1\_3}}.
%
\bibitem[dP04]{hondt04weakest}
Ellie d'Hondt and Prakash Panangaden. ``Quantum weakest preconditions''. In: \cite{qpl04}, pp. 75--90.
%
\bibitem[dP06]{dhondt2006quantum}
Ellie d'Hondt and Prakash Panangaden. ``Quantum weakest preconditions''. In: \emph{Mathematical
Structures in Computer Science} 16.3 (2006), pp. 429--451. \textsc{doi}: \href
{https://doi.org/10.1017/S0960129506005251} {\nolinkurl {10.1017/S0960129506005251}}. \textsc{arXiv}: \href
{https://www.arxiv.org/abs/quant-ph/0501157} {\nolinkurl {quant-ph/0501157}}.
%
\bibitem[DP08]{delbecque2008game-store}
Yannick Delbecque and Prakash Panangaden. ``Game semantics for quantum stores''. In: \emph{Proceedings
of the 24th Conference on the Mathematical Foundations of Programming Semantics, MFPS XXIV}
(Philadelphia, PA, USA). Ed. by A. Bauer and M. Mislove. Vol. 218. Electronic Notes in Theoretical Computer
Science. May 2008, pp. 153--170. \textsc{doi}: \href {https://doi.org/10.1016/j.entcs.2008.10.010} {\nolinkurl
{10.1016/j.entcs.2008.10.010}}.
%
\bibitem[DP10]{duncan2010rewriting}
Ross Duncan and Simon Perdrix. ``Rewriting measurement-based quantum computations with generalised
flow''. In: \emph{Proceedings of the 37th International Colloquium on Automata, Languages and
Programming, ICALP'10, Part II} (Bordeaux, France). Ed. by Samson Abramsky, Cyril Gavoille,
Claude Kirchner, Friedhelm Meyer auf der Heide, and Paul G. Spirakis. Vol. 6199. Lecture Notes in
Computer Science. Springer, 2010, pp. 285--296. \textsc{doi}: \href
{https://doi.org/10.1007/978-3-642-14162-1_24} {\nolinkurl {10.1007/978-3-642-14162-1_24}}.
%
\bibitem[dP95]{deliguoro1995nondet}
Ugo de'Liguoro and Adolfo Piperno. ``Non deterministic extensions of untyped lambda-calculus''. In:
\emph{Information and Computation} 122.2 (1995), pp. 149--177. \textsc{doi}: \href
{https://doi.org/10.1006/inco.1995.1145} {\nolinkurl {10.1006/inco.1995.1145}}.
%
\bibitem[DPTV10]{diaz-caro2010equivalence}
Alejandro Díaz-Caro, Simon Perdrix, Christine Tasson, and Benoît Valiron. ``Equivalence of algebraic
lambda-calculi (work in progress)''. In: \emph{Pre-Proceedings of the 5th International Workshop on
Higher-Order Rewriting (HOR'10), Edinburgh, 14 juillet 2010}. This work has been finalized in \cite{assaf2014call-by-value}.
2010, pp. 6--11.
%
\bibitem[DR89]{danos1989structure}
Vincent Danos and Laurent Regnier. ``The structure of multiplicatives''. In: \emph{Archive for Mathematical
Logic} 28.3 (1989), pp. 181--203. \textsc{doi}: \href {https://doi.org/10.1007/BF01622878} {\nolinkurl
{10.1007/BF01622878}}.
%
\bibitem[DR99]{danos1999reversible}
Vincent Danos and Laurent Regnier. ``Reversible, irreversible and optimal lambda-machines''. In:
\emph{Theoretical Computer Science} 227.1-2 (1999), pp. 79--97. \textsc{doi}: \href
{https://doi.org/10.1016/S0304-3975(99)00049-3} {\nolinkurl {10.1016/S0304-3975(99)00049-3}}.
%
\bibitem[Dun09]{duncan2009generalized}
Ross Duncan. ``Generalized proof-nets for compact categories with biproducts''. In: \cite{gay2009semantic}. Chap. 3,
pp. 70--134. \textsc{doi}: \href {https://doi.org/10.1017/CBO9781139193313.004} {\nolinkurl
{10.1017/CBO9781139193313.004}}. \textsc{arXiv}: \href {https://www.arxiv.org/abs/0903.5154} {\nolinkurl
{0903.5154}}.
%
\bibitem[Dun13]{duncan2013graphical}
Ross Duncan. ``A graphical approach to measurement-based quantum computing''. In: \emph{Quantum
Physics and Linguistics: A Compositional, Diagrammatic Discourse}. Ed. by Chris Heunen,
Mehrnoosh Sadrzadeh, and Edward Grefenstette. Oxford University Press, 2013. Chap. 3. \textsc{isbn}:
9780199646296. \textsc{doi}: \href {https://doi.org/10.1093/acprof:oso/9780199646296.003.0003} {\nolinkurl
{10.1093/acprof:oso/9780199646296.003.0003}}. \textsc{arXiv}: \href {https://www.arxiv.org/abs/1203.6242}
{\nolinkurl {1203.6242}}.
%
\bibitem[Dur02]{durand-lose2002computing}
Jérome Durand-Lose. ``Computing inside the billiard ball model''. In: \cite{adamatzky2002collision-based}, pp. 135--160. \textsc{url}: \url
{http://www.univ-orleans.fr/lifo/Members/Jerome.Durand-Lose/Recherche/Publications/2002_BBM_book.pdf}.
%
\bibitem[Dyb91]{dybjer1991inductive}
Peter Dybjer. ``Inductive sets and families in Martin-Löf's type theory and their set-theoretic semantics''. In:
ed. by Gérard Huet and Gordon D. Plotkin. Cambridge University Press, 1991, pp. 280--306. \textsc{isbn}:
9780511569807. \textsc{doi}: \href {https://doi.org/10.1017/CBO9780511569807.012} {\nolinkurl
{10.1017/CBO9780511569807.012}}.
%
\bibitem[Dyb94]{dybjer1994inductive}
Peter Dybjer. ``Inductive families''. In: \emph{Formal Aspects of Computing} 6.4 (1994), pp. 440--465.
\textsc{doi}: \href {https://doi.org/10.1007/BF01211308} {\nolinkurl {10.1007/BF01211308}}.
%
\bibitem[Ehr02]{ehrhard2002kothe}
Thomas Ehrhard. ``On köthe sequence spaces and linear logic''. In: \emph{Mathematical Structures in
Computer Science} 12.5 (2002), pp. 579--623. \textsc{doi}: \href {https://doi.org/10.1017/S0960129502003729}
{\nolinkurl {10.1017/S0960129502003729}}.
%
\bibitem[Ehr05]{ehrhard2005finiteness}
Thomas Ehrhard. ``Finiteness spaces''. In: \emph{Mathematical Structures in Computer Science} 15.4 (2005),
pp. 615--646. \textsc{doi}: \href {https://doi.org/10.1017/S0960129504004645} {\nolinkurl
{10.1017/S0960129504004645}}.
%
\bibitem[EPT11]{ehrhard2011computational}
Thomas Ehrhard, Michele Pagani, and Christine Tasson. ``The computational meaning of probabilistic
coherence spaces''. In: \cite{lics2011}, pp. 87--96. \textsc{doi}: \href {https://doi.org/10.1109/LICS.2011.29}
{\nolinkurl {10.1109/LICS.2011.29}}. \textsc{hal}: \href {https://hal.archives-ouvertes.fr/hal-00627490}
{\nolinkurl {hal-00627490}}.
%
\bibitem[ER03]{ehrhard2003differential}
Thomas Ehrhard and Laurent Regnier. ``The differential lambda-calculus''. In: \emph{Theoretical Computer
Science} 309.1-3 (2003), pp. 1--41. \textsc{doi}: \href {https://doi.org/10.1016/S0304-3975(03)00392-X}
{\nolinkurl {10.1016/S0304-3975(03)00392-X}}.
%
\bibitem[ER06]{ehrhard2006differential}
Thomas Ehrhard and Laurent Regnier. ``Differential interaction nets''. In: \emph{Theoretical Computer
Science} 364.2 (2006), pp. 166--195. \textsc{doi}: \href {https://doi.org/10.1016/j.tcs.2006.08.003} {\nolinkurl
{10.1016/j.tcs.2006.08.003}}. \textsc{hal}: \href {https://hal.archives-ouvertes.fr/hal-00150274} {\nolinkurl
{hal-00150274}}.
%
\bibitem[ETP14]{ehrhard2014probabilistic}
Thomas Ehrhard, Christine Tasson, and Michele Pagani. ``Probabilistic coherence spaces are fully abstract
for probabilistic PCF''. In: \cite{popl2014}, pp. 309--320. \textsc{doi}: \href {https://doi.org/10.1145/2535838.2535865}
{\nolinkurl {10.1145/2535838.2535865}}.
%
\bibitem[FBCH+20]{frank2020special}
Michael P. Frank, Robert W. Brocato, Thomas M. Conte, Alexander H. Hsia, Anirudh Jain, Nancy A. Missert,
Karpur Shukla, and Brian D. Tierney. ``Special session: exploring the ultimate limits of adiabatic circuits''. In:
\emph{38th IEEE International Conference on Computer Design (ICCD)}. 2020, pp. 21--24. \textsc{doi}: \href
{https://doi.org/10.1109/ICCD50377.2020.00018} {\nolinkurl {10.1109/ICCD50377.2020.00018}}.
%
\bibitem[FC22]{felice2022qath}
Giovanni de Felice and Bob Coecke. ``Quantum linear optics via string diagrams''. In: \cite{qpl2022}, pp. 83--100.
\textsc{doi}: \href {https://doi.org/10.4204/EPTCS.394.6} {\nolinkurl {10.4204/EPTCS.394.6}}.
%
\bibitem[FD19]{fagan2018optimising}
Andrew Fagan and Ross Duncan. ``Optimising Clifford circuits with Quantomatic''. In: \cite{qpl2018}, pp. 85--105.
\textsc{doi}: \href {https://doi.org/10.4204/EPTCS.287.5} {\nolinkurl {10.4204/EPTCS.287.5}}.
%
\bibitem[FDDB14]{friis2014implementing}
Nicolai Friis, Vedran Dunjko, Wolfgang Dür, and Hans J. Briegel. ``Implementing quantum control for
unknown subroutines''. In: \emph{Physical Review A} 89 (3 2014), p. 030303. \textsc{doi}: \href
{https://doi.org/10.1103/PhysRevA.89.030303} {\nolinkurl {10.1103/PhysRevA.89.030303}}. \textsc{arXiv}: \href
{https://www.arxiv.org/abs/1401.8128} {\nolinkurl {1401.8128}}.
%
\bibitem[Fef70]{feferman1970formal}
Solomon Feferman. ``Formal theories for transfinite iterations of generalized inductive definitions and some
subsystems of analysis''. In: \emph{Intuitionism and Proof Theory: Proceedings of the Summer Conference
at Buffalo N.Y. 1968}. Ed. by A. Kino, J. Myhill, and R. E. Vesley. Vol. 60. Studies in Logic and the Foundations
of Mathematics. North-Holland, 1970, pp. 303--326. \textsc{doi}: \href
{https://doi.org/10.1016/S0049-237X(08)70761-4} {\nolinkurl {10.1016/S0049-237X(08)70761-4}}.
%
\bibitem[Fey82]{feynman1982simulating}
Richard P. Feynman. ``Simulating physics with computers''. In: \emph{International Journal of Theoretical
Physics} 21.7-8 (1982), pp. 467--488.
%
\bibitem[FGG14]{farhi2014quantum}
Edward Farhi, Jeffrey Goldstone, and Sam Gutmann. \emph{A Quantum Approximate Optimization
Algorithm}. Tech. rep. MIT-CTP/4610. MIT, 2014.
%
\bibitem[FGGLLP01]{farhi2001quantum}
Edward Farhi, Jeffrey Goldstone, Sam Gutmann, Joshua Lapan, Andrew Lundgren, and Daniel Preda. ``A
quantum adiabatic evolution algorithm applied to random instances of an NP-complete problem''. In:
\emph{Science} 292.5516 (2001), pp. 472--475. \textsc{doi}: \href {https://doi.org/10.1126/science.1057726}
{\nolinkurl {10.1126/science.1057726}}. \textsc{arXiv}: \href {https://www.arxiv.org/abs/quant-ph/0104129}
{\nolinkurl {quant-ph/0104129}}.
%
\bibitem[FGGS00]{farhi2000quantum}
Edward Farhi, Jeffrey Goldstone, Sam Gutmann, and Michael Sipser. \emph{Quantum Computation by
Adiabatic Evolution}. Tech. rep. MIT-CTP-2936. MIT, 2000.
%
\bibitem[FH65]{feynman1965quantum}
Richard P. Feynman and A. R. Hibbs. \emph{Quantum Mechanics and Path Integrals}. McGraw-Hill
Publishing Company, 1965. \textsc{isbn}: 0-07-020650-3.
%
\bibitem[FHTZ15]{feng2015qpmc}
Yuan Feng, Ernst Moritz Hahn, Andrea Turrini, and Lijun Zhang. ``QPMC: a model checker for quantum
programs and protocols''. In: \emph{Proceedings of the 20th International Symposium on Formal Methods
(FM 2015)} (Oslo, Norway). Ed. by Nikolaj Bjørner and Frank S. de Boer. Vol. 9109. Lecture Notes in
Computer Science. Springer, 2015, pp. 265--272. \textsc{isbn}: 978-3-319-19248-2. \textsc{doi}: \href
{https://doi.org/10.1007/978-3-319-19249-9_17} {\nolinkurl {10.1007/978-3-319-19249-9_17}}.
%
\bibitem[FKRS20]{fu2020tutorial}
Peng Fu, Kohei Kishida, Neil J. Ross, and Peter Selinger. ``A tutorial introduction to quantum circuit
programming in dependently typed proto-quipper''. In: \cite{rc2020}, pp. 153--168. \textsc{doi}: \href
{https://doi.org/10.1007/978-3-030-52482-1_9} {\nolinkurl {10.1007/978-3-030-52482-1_9}}. \textsc{arXiv}: \href
{https://www.arxiv.org/abs/2005.08396} {\nolinkurl {2005.08396}}.
%
\bibitem[FKRS22a]{fu2022biset-enriched}
Peng Fu, Kohei Kishida, Neil J. Ross, and Peter Selinger. ``A biset-enriched categorical model for
Proto-Quipper with dynamic lifting''. In: \cite{qpl2022}. \textsc{doi}: \href {https://doi.org/10.4204/EPTCS.394.16}
{\nolinkurl {10.4204/EPTCS.394.16}}. \textsc{arXiv}: \href {https://www.arxiv.org/abs/2204.13039} {\nolinkurl
{2204.13039}}.
%
\bibitem[FKRS22b]{fu2022proto-quipper}
Peng Fu, Kohei Kishida, Neil J. Ross, and Peter Selinger. ``Proto-Quipper with dynamic lifting''. See also the
companion paper \cite{fu2022biset-enriched}. 2022. \textsc{arXiv}: \href {https://www.arxiv.org/abs/2204.13041} {\nolinkurl
{2204.13041}}.
%
\bibitem[FKS20]{fu2020linear}
Peng Fu, Kohei Kishida, and Peter Selinger. ``Linear dependent type theory for quantum programming
languages: extended abstract''. In: \emph{Proceedings of the 35th Annual ACM/IEEE Symposium on Logic in
Computer Science, LICS 2020} (Saarbrücken, Germany). Ed. by Holger Hermanns, Lijun Zhang,
Naoki Kobayashi, and Dale Miller. ACM, July 2020, pp. 440--453. \textsc{isbn}: 978-1-4503-7104-9. \textsc{doi}:
\href {https://doi.org/10.1145/3373718.3394765} {\nolinkurl {10.1145/3373718.3394765}}. \textsc{arXiv}: \href
{https://www.arxiv.org/abs/2004.13472} {\nolinkurl {2004.13472}}.
%
\bibitem[Fle70]{fletcher1970approach}
R. Fletcher. ``A new approach to variable metric algorithms''. In: \emph{The Computer Journal} 13.3 (1970),
pp. 317--322. \textsc{doi}: \href {https://doi.org/10.1093/comjnl/13.3.317} {\nolinkurl {10.1093/comjnl/13.3.317}}.
%
\bibitem[Flo67]{floyd1967assigning}
Robert W. Floyd. ``Assigning meanings to programs''. In: \emph{Mathematical Aspects of Computer Science}.
Ed. by J. T. Schwartz. Vol. 19. Proceedings of Symposia in Applied Mathematics. AMS, 1967, pp. 19--32.
\textsc{doi}: \href {https://doi.org/10.1090/psapm/019} {\nolinkurl {10.1090/psapm/019}}.
%
\bibitem[FLY22]{feng2022verification}
Yuan Feng, Sanjiang Li, and Mingsheng Ying. ``Verification of distributed quantum programs''. In:
\emph{ACM Transactions in Computational Logic} 23.3 (2022), 19:1--19:40. \textsc{doi}: \href
{https://doi.org/10.1145/3517145} {\nolinkurl {10.1145/3517145}}.
%
\bibitem[FM07]{filliatre2007whykrakatoacaduceus}
Jean-Christophe Filliâtre and Claude Marché. ``The Why/Krakatoa/Caduceus platform for deductive
program verification''. In: \emph{Proceedings of the 19th International Conference on Computer Aided
Verification, CAV 2007} (Berlin, Germany, July 3--7, 2007). Ed. by Werner Damm and Holger Hermanns.
Vol. 4590. Lecture Notes in Computer Science. Springer, 2007, pp. 173--177. \textsc{isbn}: 978-3-540-73367-6.
\textsc{doi}: \href {https://doi.org/10.1007/978-3-540-73368-3_21} {\nolinkurl {10.1007/978-3-540-73368-3_21}}.
\textsc{hal}: \href {https://hal.archives-ouvertes.fr/inria-00270820} {\nolinkurl {inria-00270820}}.
%
\bibitem[FP13]{filliatre2013why3}
Jean-Christophe Filliâtre and Andrei Paskevich. ``Why3 -- where programs meet provers''. In:
\emph{Proceedings of the 22nd European Symposium on Programming Languages and Systems, ESOP 2013}
(Rome, Italy, Mar. 16--24, 2013). Ed. by Matthias Felleisen and Philippa Gardner. Vol. 7792. Lecture Notes in
Computer Science. Springer, 2013, pp. 125--128. \textsc{isbn}: 978-3-642-37035-9. \textsc{doi}: \href
{https://doi.org/10.1007/978-3-642-37036-6\_8} {\nolinkurl {10.1007/978-3-642-37036-6\_8}}. \textsc{hal}: \href
{https://hal.archives-ouvertes.fr/hal-00789533} {\nolinkurl {hal-00789533}}.
%
\bibitem[Fra99]{frank1999reversibility}
Michael Patrick Frank. ``Reversibility for Efficient Computing''. PhD thesis. MIT, 1999.
%
\bibitem[FT82]{fredkin1982conservative}
Edward Fredkin and Tommaso Toffoli. ``Conservative logic''. In: \emph{International Journal of Theoretical
Physics} 21 (3 1982), pp. 219--253. \textsc{doi}: \href {https://doi.org/10.1007/BF01857727} {\nolinkurl
{10.1007/BF01857727}}.
%
\bibitem[FTR07]{fazel2007esop-based}
K. Fazel, M. A. Thornton, and J. E. Rice. ``ESOP-based Toffoli gate cascade generation''. In: \emph{Proceedings
of the 2007 IEEE Pacific Rim Conference on Communications, Computers and Signal Processing, PACRIM
2007} (Victoria, BC, Canada, Aug. 22--24, 2007). IEEE, 2007. \textsc{isbn}: 978-1-4244-1189-4. \textsc{doi}: \href
{https://doi.org/10.1109/PACRIM.2007.4313212} {\nolinkurl {10.1109/PACRIM.2007.4313212}}.
%
\bibitem[GAJ06]{gupta2006algorithm}
Pallav Gupta, Abhinav Agrawal, and Niraj K. Jha. ``An algorithm for synthesis of reversible logic circuits''. In:
\emph{IEEE Transactions on Computer-Aided Design of Integrated Circuits and Systems} 25.11 (2006),
pp. 2317--2330. \textsc{doi}: \href {https://doi.org/10.1109/TCAD.2006.871622} {\nolinkurl
{10.1109/TCAD.2006.871622}}.
%
\bibitem[Gan80]{gandy1980churchs}
Robin Gandy. ``Church's thesis and principles for mechanisms''. In: \emph{The Kleene Symposium}
(Madison, Wisconsin, USA, June 18--24, 1978). Ed. by Jon Barwise, H. Jerome Keisler, and Kenneth Kunen.
Vol. 101. Studies in Logic and the Foundations of Mathematics. North-Holland Publishing Company, 1980,
pp. 123--148. \textsc{isbn}: 978-0-444-85345-5. \textsc{doi}: \href
{https://doi.org/10.1016/S0049-237X(08)71257-6} {\nolinkurl {10.1016/S0049-237X(08)71257-6}}.
%
\bibitem[Gay06]{gay2006quantum}
Simon J. Gay. ``Quantum programming languages: survey and bibliography''. In: \emph{Mathematical
Structures in Computer Science} 16.4 (2006), pp. 581--600. \textsc{doi}: \href
{https://doi.org/10.1017/S0960129506005378} {\nolinkurl {10.1017/S0960129506005378}}.
%
\bibitem[GBVMA21]{goubault2021gaussian}
Timothée Goubault de Brugière, Marc Baboulin, Benoît Valiron, Simon Martiel, and Cyril Allouche.
``Gaussian elimination versus greedy methods for the synthesis of linear reversible circuits''. In: \emph{ACM
Transactions on Quantum Computing} 2.3 (2021), p. 11. \textsc{doi}: \href {https://doi.org/10.1145/3474226}
{\nolinkurl {10.1145/3474226}}. \textsc{hal}: \href {https://hal.archives-ouvertes.fr/hal-03547117} {\nolinkurl
{hal-03547117}}.
%
\bibitem[GD17]{garvie2017verifying}
Liam Garvie and Ross Duncan. ``Verifying the smallest interesting colour code with Quantomatic''. In:
[QPL18], pp. 147--163. \textsc{doi}: \href {https://doi.org/10.4204/EPTCS.266.10} {\nolinkurl
{10.4204/EPTCS.266.10}}.
%
\bibitem[GE21]{gidney2021howtofactor}
Craig Gidney and Martin Ekerå. ``How to factor 2048 bit RSA integers in 8 hours using 20 million noisy
qubits''. In: \emph{Quantum} 5 (2021), p. 433. \textsc{doi}: \href {https://doi.org/10.22331/q-2021-04-15-433}
{\nolinkurl {10.22331/q-2021-04-15-433}}. \textsc{arXiv}: \href {https://www.arxiv.org/abs/1905.09749}
{\nolinkurl {1905.09749}}.
%
\bibitem[GHHNP13]{gaboardi2013linear}
Marco Gaboardi, Andreas Haeberlen, Justin Hsu, Arjun Narayan, and Benjamin C. Pierce. ``Linear dependent
types for differential privacy''. In: \emph{Proceedings of the 40th Annual ACM SIGPLAN-SIGACT
Symposium on Principles of Programming Languages, POPL'13} (Rome, Italy). Ed. by Roberto Giacobazzi
and Radhia Cousot. ACM, 2013, pp. 357--370. \textsc{isbn}: 978-1-4503-1832-7. \textsc{doi}: \href
{https://doi.org/10.1145/2429069.2429113} {\nolinkurl {10.1145/2429069.2429113}}.
%
\bibitem[Ghi07]{ghica2007geometry}
Dan R. Ghica. ``Geometry of synthesis: a structured approach to VLSI design''. In: \emph{Proceedings of the
34th ACM SIGPLAN-SIGACT Symposium on Principles of Programming Languages, POPL 2007} (Nice,
France). Ed. by Martin Hofmann and Matthias Felleisen. ACM, Jan. 2007, pp. 363--375. \textsc{isbn}:
1-59593-575-4. \textsc{doi}: \href {https://doi.org/10.1145/1190216.1190269} {\nolinkurl
{10.1145/1190216.1190269}}. \textsc{url}: \url {https://doi.org/10.1145/1190216.1190269}.
%
\bibitem[Ghi12]{ghica2012geometry}
Dan R. Ghica. ``The geometry of synthesis -- how to make hardware out of software''. In: \emph{Proceedings
of the 11th International Conference on Mathematics of Program Construction, MPC 2012} (Madrid, Spain).
Ed. by Jeremy Gibbons and Pablo Nogueira. Vol. 7342. Lecture Notes in Computer Science. Abstract of
Invited Talk. Springer, June 2012, pp. 23--24. \textsc{isbn}: 978-3-642-31112-3. \textsc{doi}: \href
{https://doi.org/10.1007/978-3-642-31113-0_3} {\nolinkurl {10.1007/978-3-642-31113-0_3}}.
%
\bibitem[GHKLMS03]{gierz2003continuous}
G. Gierz, K. H. Hofmann, K. Keimel, J. D. Lawson, M. Mislove, and D. S. Scott. \emph{Continuous Lattices
and Domains}. Vol. 93. Encyclopedia of Mathematics and its Applications. Cambridge University Press, 2003.
\textsc{isbn}: 0-521-80338-1.
%
\bibitem[Gir03]{girard2003goiIV}
Jean-Yves Girard. ``Geometry of interaction IV: the feedback equation''. In: \emph{Logic Colloquium '03.
Proceeedings of the Annual European Summer Meeting of the Association for Symbolic Logic, held in
Helsinki, Finland, August 14--20, 2003}. Ed. by Viggo Stoltenberg-Hansen and Jouko Väänänen. Vol. 24.
Lecture Notes in Logic. ASL, 2003, pp. 76--217.
%
\bibitem[Gir04]{girard2004between}
Jean-Yves Girard. ``Between logic and quantic -- a tract''. In: ed. by Thomas Ehrhard, Jean-Yves Girard,
Paul Ruet, and Philip Scott. Vol. 316. London Mathematical Society Lecture Notes Series. Cambridge
University Press, 2004. Chap. 10, pp. 346--381. \textsc{isbn}: 0-521-60857-0. \textsc{doi}: \href
{https://doi.org/10.1017/CBO9780511550850.011} {\nolinkurl {10.1017/CBO9780511550850.011}}.
%
\bibitem[Gir11]{girard2011goiV}
Jean-Yves Girard. ``Geometry of interaction V: logic in the hyperfinite factor''. In: \emph{Theoretical
Computer Science} 412.20 (2011), pp. 1860--1883. \textsc{doi}: \href {https://doi.org/10.1016/j.tcs.2010.12.016}
{\nolinkurl {10.1016/j.tcs.2010.12.016}}.
%
\bibitem[Gir86]{girard1986system}
Jean-Yves Girard. ``The system F of variable types, fifteen years later''. In: \emph{Theor. Comput. Sci.} 45.2
(1986), pp. 159--192. \textsc{doi}: \href {https://doi.org/10.1016/0304-3975(86)90044-7} {\nolinkurl
{10.1016/0304-3975(86)90044-7}}.
%
\bibitem[Gir87]{girard87linear}
Jean-Yves Girard. ``Linear logic''. In: \emph{Theoretical Computer Science} 50.1 (1987), pp. 1--101. \textsc{doi}:
\href {https://doi.org/10.1016/0304-3975(87)90045-4} {\nolinkurl {10.1016/0304-3975(87)90045-4}}.
%
\bibitem[Gir88]{girard1988normal}
Jean-Yves Girard. ``Normal functors, power series and lambda-calculus''. In: \emph{Annals of Pure and
Applied Logic} 37.2 (1988), pp. 129--177. \textsc{doi}: \href {https://doi.org/10.1016/0168-0072(88)90025-5}
{\nolinkurl {10.1016/0168-0072(88)90025-5}}.
%
\bibitem[Gir89]{girard1988goiI}
Jean-Yves Girard. ``Geometry of interaction I: interpretation of system F''. In: \emph{Logic Colloquium '88.
Proceedings of the Colloquium Held in Padova, Italy, August 22--31, 1988}. Ed. by R. Ferro, C. Bonotto,
S. Valentini, and A. Zanardo. Vol. 127. Studies in Logic and the Foundations of Mathematics. North-Holland,
1989, pp. 221--260. \textsc{doi}: \href {https://doi.org/10.1016/S0049-237X(08)70271-4} {\nolinkurl
{10.1016/S0049-237X(08)70271-4}}.
%
\bibitem[Gir90]{girard1988goiII}
Jean-Yves Girard. ``Geometry of interaction II: deadlock-free algorithms''. In: \cite{colog1988}, pp. 76--93. \textsc{doi}:
\href {https://doi.org/10.1007/3-540-52335-9_49} {\nolinkurl {10.1007/3-540-52335-9_49}}.
%
\bibitem[Gir95a]{girard1995goiIII}
Jean-Yves Girard. ``Geometry of interaction III: accommodating the additives''. In: \cite{girard1995advances}, pp. 329--389.
%
\bibitem[Gir95b]{girard95machine}
Jean-Yves Girard, ed. \emph{La Machine de Turing}. Vol. 131. Points Sciences. Contains \cite{turing1936computable} and \cite{turing50computing}
in integrality, with comments. Editions du Seuil, 1995. \textsc{isbn}: 2-02-036928-1.
%
\bibitem[Gir98]{girard1998light}
Jean-Yves Girard. ``Light linear logic''. In: \emph{Information and Computation} 143.2 (1998), pp. 175--204.
\textsc{doi}: \href {https://doi.org/10.1006/inco.1998.2700} {\nolinkurl {10.1006/inco.1998.2700}}.
%
\bibitem[Gir99]{girard1999coherent}
Jean-Yves Girard. ``Coherent Banach spaces: a continuous denotational semantics''. In: \emph{Theoretical
Computer Science} 227.1-2 (1999), pp. 275--297. \textsc{doi}: \href
{https://doi.org/10.1016/S0304-3975(99)00056-0} {\nolinkurl {10.1016/S0304-3975(99)00056-0}}.
%
\bibitem[GKY19]{gluck2019reversible}
Robert Glück, Robin Kaarsgaard, and Tetsuo Yokoyama. ``Reversible programs have reversible semantics''. In:
\emph{Formal Methods. FM 2019 International Workshops - Porto, Portugal, October 7-11, 2019, Revised
Selected Papers, Part II}. Ed. by Emil Sekerinski, Nelma Moreira, José N. Oliveira, Daniel Ratiu,
Riccardo Guidotti, Marie Farrell, Matt Luckcuck, Diego Marmsoler, José Campos, Troy Astarte,
Laure Gonnord, Antonio Cerone, Luis Couto, Brijesh Dongol, Martin Kutrib, Pedro Monteiro, and
David Delmas. Vol. 12233. Lecture Notes in Computer Science. Springer, 2019, pp. 413--427. \textsc{isbn}:
978-3-030-54996-1. \textsc{doi}: \href {https://doi.org/10.1007/978-3-030-54997-8_26} {\nolinkurl
{10.1007/978-3-030-54997-8_26}}.
%
\bibitem[GLR95]{girard1995advances}
Jean-Yves Girard, Yves Lafont, and Laurent Regnier, eds. \emph{Advances in Linear Logic}. Vol. 222. London
Mathematical Society Lecture Note Series. Cambridge University Press, 1995. \textsc{isbn}: 0-521-55961-8.
%
\bibitem[GLRSV12]{green2012report}
Alexander S. Green, Peter L. Lumsdaine, Neil J. Ross, Peter Selinger, and Benoî Valiron. \emph{Report on the
Quipper language, version 0.3, with GFI algorithm implementations (Updated for revision 0.3-4)}. Report to
IARPA, for official use only. 2012.
%
\bibitem[GLRSV13a]{green2013introduction}
Alexander S. Green, Peter LeFanu Lumsdaine, Neil J. Ross, Peter Selinger, and Benoît Valiron. ``An
introduction to quantum programming in quipper''. In: \cite{rc2013}, pp. 110--124. \textsc{doi}: \href
{https://doi.org/10.1007/978-3-642-38986-3_10} {\nolinkurl {10.1007/978-3-642-38986-3_10}}. \textsc{arXiv}:
\href {https://www.arxiv.org/abs/1304.5485} {\nolinkurl {1304.5485}}.
%
\bibitem[GLRSV13b]{green2013quipper}
Alexander S. Green, Peter LeFanu Lumsdaine, Neil J. Ross, Peter Selinger, and Benoît Valiron. ``Quipper: a
scalable quantum programming language''. In: \emph{Proceedings of the ACM SIGPLAN Conference on
Programming Language Design and Implementation, PLDI'13} (Seattle, WA, USA). Ed. by
Hans-Juergen Boehm and Cormac Flanagan. ACM, 2013, pp. 333--342. \textsc{isbn}: 978-1-4503-2014-6.
\textsc{doi}: \href {https://doi.org/10.1145/2491956.2462177} {\nolinkurl {10.1145/2491956.2462177}}.
\textsc{arXiv}: \href {https://www.arxiv.org/abs/1304.3390} {\nolinkurl {1304.3390}}.
%
\bibitem[GLT90]{girard89proofs}
Jean-Yves Girard, Yves Lafont, and Paul Taylor. \emph{Proofs and Types}. 2nd ed. Vol. 7. Cambridge Tracts In
Theoretical Computer Science. Cambridge University Press, 1990. \textsc{isbn}: 0-521-37181-3. \textsc{url}:
\url {http://www.cs.man.ac.uk/~pt/stable/Proofs+Types.html}.
%
\bibitem[GM01]{guerrini2001parsing}
Stefano Guerrini and Andrea Masini. ``Parsing MELL proof nets''. In: \emph{Theoretical Computer Science}
254.1-2 (2001), pp. 317--335. \textsc{doi}: \href {https://doi.org/10.1016/S0304-3975(99)00299-6} {\nolinkurl
{10.1016/S0304-3975(99)00299-6}}.
%
\bibitem[GM09]{gay2009semantic}
Simon Gay and Ian Mackie, eds. \emph{Semantic Techniques in Quantum Computation}. Cambridge
University Press, 2009. \textsc{isbn}: 978-0-521-51374-6.
%
\bibitem[GNP08]{gay2008qmc}
Simon J. Gay, Rajagopal Nagarajan, and Nikolaos Papanikolaou. ``QMC: a model checker for quantum
systems''. In: \emph{Proceeding of the 20th International Conference on Computer Aided Verification (CAV
2008)} (Princeton, NJ, USA). Ed. by Aarti Gupta and Sharad Malik. Vol. 5123. Lecture Notes in Computer
Science. Springer, 2008, pp. 543--547. \textsc{isbn}: 978-3-540-70543-7. \textsc{doi}: \href
{https://doi.org/10.1007/978-3-540-70545-1_51} {\nolinkurl {10.1007/978-3-540-70545-1_51}}.
%
\bibitem[Gol70]{goldfarb1970family}
Donald Goldfarb. ``A family of variable-metric methods derived by variational means''. In:
\emph{Mathematics of Computation} 24 (1970), pp. 23--26.
%
\bibitem[Gra88]{graham1987closure}
Steven K. Graham. ``Closure properties of a probabilistic domain construction''. In: \emph{Proceedings of the
3rd Workshop on Mathematical Foundations of Programming Language Semantics, MFPS'87} (Tulane
University, New Orleans, Louisiana, USA). Ed. by Michael G. Main, Austin Melton, Michael W. Mislove, and
David A. Schmidt. Vol. 298. Lecture Notes in Computer Science. Springer, 1988, pp. 213--233. \textsc{isbn}:
3-540-19020-1. \textsc{doi}: \href {https://doi.org/10.1007/3-540-19020-1} {\nolinkurl {10.1007/3-540-19020-1}}.
%
\bibitem[Gro96]{grover1996fast}
Lov K. Grover. ``A fast quantum mechanical algorithm for database search''. In: \emph{Proceedings of the
Twenty-Eighth Annual ACM Symposium on the Theory of Computing, STOC'96} (Philadelphia,
Pennsylvania, USA, May 22--June 24, 1996). Ed. by Gary L. Miller. ACM, 1996, pp. 212--219. \textsc{isbn}:
0-89791-785-5. \textsc{doi}: \href {https://doi.org/10.1145/237814.237866} {\nolinkurl {10.1145/237814.237866}}.
%
\bibitem[GS10]{ghica2010gos-II}
Dan R. Ghica and Alex I. Smith. ``Geometry of synthesis II: from games to delay-insensitive circuits''. In:
\emph{Proceedings of the 26th Conference on the Mathematical Foundations of Programming Semantics,
MFPS 2010, Ottawa, Ontario, Canada, May 6-10, 2010} (Ottawa, Ontario, Canada). Ed. by
Michael W. Mislove and Peter Selinger. Vol. 265. Electronic Notes in Theoretical Computer Science. Elsevier,
May 2010, pp. 301--324. \textsc{doi}: \href {https://doi.org/10.1016/j.entcs.2010.08.018} {\nolinkurl
{10.1016/j.entcs.2010.08.018}}.
%
\bibitem[GS11]{ghica2011gos-III}
Dan R. Ghica and Alex I. Smith. ``Geometry of synthesis III: resource management through type inference''.
In: \emph{Proceedings of the 38th ACM SIGPLAN-SIGACT Symposium on Principles of Programming
Languages, POPL 2011, Austin, TX, USA, January 26-28, 2011} (Austin, TX, USA). Ed. by Thomas Ball and
Mooly Sagiv. ACM, Jan. 2011, pp. 345--356. \textsc{isbn}: 978-1-4503-0490-0. \textsc{doi}: \href
{https://doi.org/10.1145/1926385.1926425} {\nolinkurl {10.1145/1926385.1926425}}.
%
\bibitem[GSS11]{ghica2011gos-IV}
Dan R. Ghica, Alex Smith, and Satnam Singh. ``Geometry of synthesis IV''. In: \cite{icfp2011}, pp. 221--233.
\textsc{doi}: \href {https://doi.org/10.1145/2034773.2034805]]} {\nolinkurl {10.1145/2034773.2034805]]}}.
%
\bibitem[GSS92]{girard1992bounded}
Jean-Yves Girard, Andre Scedrov, and Philip J. Scott. ``Bounded linear logic: a modular approach to
polynomial-time computability''. In: \emph{Theoretical Computer Science} 97.1 (1992), pp. 1--66. \textsc{doi}:
\href {https://doi.org/10.1016/0304-3975(92)90386-T} {\nolinkurl {10.1016/0304-3975(92)90386-T}}.
%
\bibitem[Had15]{hadzihasanovic2015diagrammatic}
Amar Hadzihasanovic. ``A diagrammatic axiomatisation for qubit entanglement''. In: \cite{lics2015}. \textsc{doi}:
\href {https://doi.org/10.1109/LICS.2015.59} {\nolinkurl {10.1109/LICS.2015.59}}.
%
\bibitem[Had17]{hadzihasanovic2017algebra}
Amar Hadzihasanovic. ``The algebra of entanglement and the geometry of composition''. PhD thesis. Oxford
Univesity, 2017.
%
\bibitem[HAGH16]{huisman2016formal}
Marieke Huisman, Wolfgang Ahrendt, Daniel Grahl, and Martin Hentschel. ``Formal specification with the
java modeling language''. In: \emph{Deductive Software Verification - The KeY Book - From Theory to
Practice}. Ed. by Wolfgang Ahrendt, Bernhard Beckert, Richard Bubel, Reiner Hähnle, Peter H. Schmitt, and
Mattias Ulbrich. Vol. 10001. Lecture Notes in Computer Science. Springer, 2016. Chap. 7, pp. 193--241.
\textsc{isbn}: 978-3-319-49811-9. \textsc{doi}: \href {https://doi.org/10.1007/978-3-319-49812-6_7} {\nolinkurl
{10.1007/978-3-319-49812-6_7}}.
%
\bibitem[Hal07]{hallgren2007polynomial-time}
Sean Hallgren. ``Polynomial-time quantum algorithms for Pell's equation and the principal ideal problem''.
In: \emph{Journal of the ACM} 54.1 (2007), 4:1--4:19. \textsc{doi}: \href
{https://doi.org/10.1145/1206035.1206039} {\nolinkurl {10.1145/1206035.1206039}}.
%
\bibitem[Hal92]{hall1992electroid}
J.S. Hall. ``An electroid switching model for reversible computer architectures''. In: \emph{Workshop on
Physics and Computation}. 1992, pp. 237--247. \textsc{doi}: \href
{https://doi.org/10.1109/PHYCMP.1992.615549} {\nolinkurl {10.1109/PHYCMP.1992.615549}}.
%
\bibitem[Har87]{rogers1987theory}
Rogers Hartley Jr. \emph{Theory of Recursive Functions and Effective Computability}. MIT Press, 1987.
\textsc{isbn}: 978-0-262-68052-3.
%
\bibitem[HC18]{heyfron2018efficient}
Luke E Heyfron and Earl T Campbell. ``An efficient quantum compiler that reduces T count''. In:
\emph{Quantum Science and Technology} 4.1 (Sept. 2018), p. 015004. \textsc{doi}: \href
{https://doi.org/10.1088/2058-9565/aad604} {\nolinkurl {10.1088/2058-9565/aad604}}.
%
\bibitem[HD96]{hatcliff1996thunksw}
John Hatcliff and Olivier Danvy. \emph{Thunks and the Lambda-Calculus}. Tech. rep. RS-96-19. BRICS,
University of Aahrus, 1996.
%
\bibitem[HFGB+23]{heurtel2023perceval}
Nicolas Heurtel, Andreas Fyrillas, Grégoire de Gliniasty, Raphaël Le Bihan, Sébastien Malherbe,
Marceau Pailhas, Eric Bertasi, Boris Bourdoncle, Pierre-Emmanuel Emeriau, Rawad Mezher, Luka Music,
Nadia Belabas, Benoît Valiron, Pascale Senellart, Shane Mansfield, and Jean Senellart. ``Perceval: a software
platform for discrete variable photonic quantum computing''. In: \emph{Quantum} 7 (2023), p. 931.
\textsc{doi}: \href {https://doi.org/10.22331/Q-2023-02-21-931} {\nolinkurl {10.22331/Q-2023-02-21-931}}.
\textsc{hal}: \href {https://hal.archives-ouvertes.fr/hal-03874624} {\nolinkurl {hal-03874624}}.
%
\bibitem[HH11]{hasuo2011semantics}
Ichiro Hasuo and Naohiko Hoshino. ``Semantics of higher-order quantum computation via geometry of
interaction''. In: \cite{lics2011}, pp. 237--246. \textsc{doi}: \href {https://doi.org/10.1109/LICS.2011.26} {\nolinkurl
{10.1109/LICS.2011.26}}.
%
\bibitem[HH17]{hasuo2017semantics}
Ichiro Hasuo and Naohiko Hoshino. ``Semantics of higher-order quantum computation via geometry of
interaction''. In: \emph{Annals of Pure and Applied Logic} 168.2 (2017). Long version of LICS paper \cite{hasuo2011semantics}.,
pp. 404--469. \textsc{doi}: \href {https://doi.org/10.1016/j.apal.2016.10.010} {\nolinkurl
{10.1016/j.apal.2016.10.010}}.
%
\bibitem[HH19]{hahnle2019deductive}
Reiner Hähnle and Marieke Huisman. ``Deductive software verification: from pen-and-paper proofs to
industrial tools''. In: \emph{Computing and Software Science - State of the Art and Perspectives}. Ed. by
Bernhard Steffen and Gerhard J. Woeginger. Vol. 10000. Lecture Notes in Computer Science. Springer, 2019,
pp. 345--373. \textsc{doi}: \href {https://doi.org/10.1007/978-3-319-91908-9\_18} {\nolinkurl
{10.1007/978-3-319-91908-9\_18}}.
%
\bibitem[HHL09]{harrow2009quantum}
Aram W. Harrow, Avinatan Hassidim, and Seth Lloyd. ``Quantum algorithm for linear systems of equations''.
In: \emph{Physical Review Letters} 103 (15 2009), p. 150502. \textsc{doi}: \href
{https://doi.org/10.1103/PhysRevLett.103.150502} {\nolinkurl {10.1103/PhysRevLett.103.150502}}. \textsc{arXiv}:
\href {https://www.arxiv.org/abs/0811.3171} {\nolinkurl {0811.3171}}.
%
\bibitem[HHZYHW19]{hung2019quantitative}
Shih-Han Hung, Kesha Hietala, Shaopeng Zhu, Mingsheng Ying, Michael Hicks, and Xiaodi Wu.
``Quantitative robustness analysis of quantum programs''. In: \emph{Proceedings of the ACM on
Programming Languages} 3.POPL (2019), 31:1--31:29. \textsc{doi}: \href {https://doi.org/10.1145/3290344}
{\nolinkurl {10.1145/3290344}}.
%
\bibitem[HK15]{heunen2015reversible}
Chris Heunen and Martti Karvonen. ``Reversible monadic computing''. In: \emph{Proceedings of the 31st
Conference on the Mathematical Foundations of Programming Semantics, MFPS XXXI} (Nijmegen, The
Netherlands). Ed. by Dan Ghica. Vol. 319. Electronic Notes in Theoretical Computer Science. 2015,
pp. 217--237. \textsc{doi}: \href {https://doi.org/10.1016/j.entcs.2015.12.014} {\nolinkurl
{10.1016/j.entcs.2015.12.014}}. \textsc{arXiv}: \href {https://www.arxiv.org/abs/1505.04330} {\nolinkurl
{1505.04330}}.
%
\bibitem[HKK18]{heunen2018reversible}
Chris Heunen, Robin Kaarsgaard, and Martti Karvonen. ``Reversible effects as inverse arrows''. In:
\emph{Proceedings of the 34th Conference on the Mathematical Foundations of Programming Semantics,
MFPS XXXIV} (Dalhousie University, Halifax, Canada). Ed. by Sam Staton. Vol. 341. Electronic Notes in
Theoretical Computer Science. Elsevier, 2018, pp. 179--199. \textsc{doi}: \href
{https://doi.org/10.1016/j.entcs.2018.11.009} {\nolinkurl {10.1016/j.entcs.2018.11.009}}.
%
\bibitem[HM19a]{huang2018qdb}
Yipeng Huang and Margaret Martonosi. ``QDB: from quantum algorithms towards correct quantum
programs''. In: \emph{Proceedings of the 9th Workshop on Evaluation and Usability of Programming
Languages and Tools, PLATEAU@SPLASH 2018} (Boston, Massachusetts, USA, Nov. 5, 2018). Ed. by
Titus Barik, Joshua Sunshine, and Sarah E. Chasins. Vol. 67. OASICS. Schloss Dagstuhl - Leibniz-Zentrum
für Informatik, 2019, 4:1--4:14. \textsc{isbn}: 978-3-95977-091-0. \textsc{doi}: \href
{https://doi.org/10.4230/OASIcs.PLATEAU.2018.4} {\nolinkurl {10.4230/OASIcs.PLATEAU.2018.4}}. \textsc{url}:
\url {http://www.dagstuhl.de/dagpub/978-3-95977-091-0}.
%
\bibitem[HM19b]{huang2019statistical}
Yipeng Huang and Margaret Martonosi. ``Statistical assertions for validating patterns and finding bugs in
quantum programs''. In: \emph{Proceedings of the 46th International Symposium on Computer
Architecture, ISCA 2019} (Phoenix, AZ, USA, June 22--26, 2019). Ed. by Srilatha Bobbie Manne,
Hillery C. Hunter, and Erik R. Altman. ACM, 2019, pp. 541--553. \textsc{isbn}: 978-1-4503-6669-4. \textsc{doi}:
\href {https://doi.org/10.1145/3307650.3322213} {\nolinkurl {10.1145/3307650.3322213}}.
%
\bibitem[HMH14]{hoshino2014memoryful}
Naohiko Hoshino, Koko Muroya, and Ichiro Hasuo. ``Memoryful geometry of interaction: from coalgebraic
components to algebraic effects''. In: \cite{lics2014}, 52:1--52:10. \textsc{doi}: \href
{https://doi.org/10.1145/2603088.2603124} {\nolinkurl {10.1145/2603088.2603124}}.
%
\bibitem[HMSM18]{haaswijk2018sat}
Winston Haaswijk, Alan Mishchenko, Mathias Soeken, and Giovanni De Micheli. ``SAT based exact
synthesis using DAG topology families''. In: \emph{Proceedings of the 55th Annual Design Automation
Conference, DAC 2018} (San Francisco, CA, USA). ACM, 2018, 53:1--53:6. \textsc{isbn}: 978-1-5386-4114-9.
\textsc{doi}: \href {https://doi.org/10.1145/3195970.3196111} {\nolinkurl {10.1145/3195970.3196111}}.
%
\bibitem[HMSV23]{heurtel2023strong}
Nicolas Heurtel, Shane Mansfield, Jean Senellart, and Benoît Valiron. ``Strong simulation of linear optical
processes''. In: \emph{Computer Physics Communications} 291 (2023), p. 108848. \textsc{doi}: \href
{https://doi.org/10.1016/J.CPC.2023.108848} {\nolinkurl {10.1016/J.CPC.2023.108848}}. \textsc{hal}: \href
{https://hal.archives-ouvertes.fr/hal-03874624v1} {\nolinkurl {hal-03874624v1}}.
%
\bibitem[HNW18]{hadzihasanovic2018zw}
Amar Hadzihasanovic, Kang Feng Ng, and Quanlong Wang. ``Two complete axiomatisations of pure-state
qubit quantum computing''. In: \cite{lics2018}, pp. 502--511. \textsc{doi}: \href
{https://doi.org/10.1145/3209108.3209128} {\nolinkurl {10.1145/3209108.3209128}}.
%
\bibitem[Hoa69]{hoare1969axiomatic}
C. A. R. Hoare. ``An axiomatic basis for computer programming''. In: \emph{Communications of the ACM}
12.10 (1969), pp. 576--580. \textsc{doi}: \href {https://doi.org/10.1145/363235.363259} {\nolinkurl
{10.1145/363235.363259}}.
%
\bibitem[Hor11]{horsman2011quantum}
Dominic Horsman. ``Quantum picturalism for topological cluster-state computing''. In: \emph{New Journal
of Physics} 13.9 (Sept. 2011), p. 095011. \textsc{doi}: \href {https://doi.org/10.1088/1367-2630/13/9/095011}
{\nolinkurl {10.1088/1367-2630/13/9/095011}}. \textsc{arXiv}: \href {https://www.arxiv.org/abs/1101.4722}
{\nolinkurl {1101.4722}}.
%
\bibitem[How80]{howard1980formulae}
W. A. Howard. ``The formulae-as-type notion of construction''. In: \emph{To H.B. Curry : Essays on
Combinatory Logic, Lambda Calculus and Formalism}. Ed. by Jonathan Paul Seldin and
James Roger Hindley. Academic Press, 1980.
%
\bibitem[HRHLH21]{hietala2021proving}
Kesha Hietala, Robert Rand, Shih-Han Hung, Liyi Li, and Michael Hicks. ``Proving quantum programs
correct''. In: \emph{Proceedings of the 12th International Conference on Interactive Theorem Proving, ITP
2021} (Rome, Italy (Virtual Conference), June 29--July 1, 2021). Ed. by Liron Cohen and Cezary Kaliszyk.
Vol. 193. LIPIcs. Schloss Dagstuhl - Leibniz-Zentrum für Informatik, 2021, 21:1--21:19. \textsc{isbn}:
978-3-95977-188-7. \textsc{doi}: \href {https://doi.org/10.4230/LIPIcs.ITP.2021.21} {\nolinkurl
{10.4230/LIPIcs.ITP.2021.21}}.
%
\bibitem[HRHWH21]{hietala2021verified}
Kesha Hietala, Robert Rand, Shih-Han Hung, Xiaodi Wu, and Michael Hicks. ``A verified optimizer for
quantum circuits''. In: \emph{Proceedings of the ACM on Programming Languages} 5.POPL (2021),
37:1--37:29. \textsc{doi}: \href {https://doi.org/10.1145/3434318} {\nolinkurl {10.1145/3434318}}.
%
\bibitem[HRS18]{haner2018optimizing}
Thomas Häner, Martin Roetteler, and Krysta M. Svore. ``Optimizing Quantum Circuits for Arithmetic''.
Draft. 2018. \textsc{arXiv}: \href {https://www.arxiv.org/abs/1805.12445} {\nolinkurl {1805.12445}}.
%
\bibitem[HSRS18]{haner2018quantum}
Thomas Häner, Mathias Soeken, Martin Roetteler, and Krysta M. Svore. ``Quantum circuits for
floating-point arithmetic''. In: \cite{rc2018}, pp. 162--174. \textsc{doi}: \href
{https://doi.org/10.1007/978-3-319-99498-7_11} {\nolinkurl {10.1007/978-3-319-99498-7_11}}.
%
\bibitem[IC1405]{cost-rev}
\emph{Official Webpage of the COST Action IC1405 on Reversible Computation}. \textsc{url}: \url
{http://www.revcomp.eu/} (visited on Aug. 27, 2021).
%
\bibitem[ICFP11]{icfp2011}
Manuel M. T. Chakravarty, Zhenjiang Hu, and Olivier Danvy, eds. \emph{Proceeding of the 16th ACM
SIGPLAN international conference on Functional Programming, ICFP 2011} (Tokyo, Japan, Sept. 19--21,
2011). ACM, 2011. \textsc{isbn}: 978-1-4503-0865-6.
%
\bibitem[Ing61]{ingerman1961thunks}
Peter Z. Ingerman. ``Thunks: a way of compiling procedure statements with some comments on procedure
declarations''. In: \emph{Communications of the ACM} 4.1 (1961), pp. 55--58.
%
\bibitem[JKT18]{jacobsen2018corefun}
Petur Andrias Højgaard Jacobsen, Robin Kaarsgaard, and Michael Kirkedal Thomsen. ``CoreFun : a typed
functional reversible core language''. In: \cite{rc2018}, pp. 304--321. \textsc{doi}: \href
{https://doi.org/10.1007/978-3-319-99498-7_21} {\nolinkurl {10.1007/978-3-319-99498-7_21}}.
%
\bibitem[Jon87]{jones1987implementation}
Simon Peyton Jones. \emph{The Implementation of Functional Programming Languages}. Prentice Hall,
1987.
%
\bibitem[Jon90]{jones1990phd}
Claire Jones. ``Probabilistic Non-Determinism''. PhD thesis. University of Edinburgh, 1990.
%
\bibitem[JPKH+15]{javadiabhari2015scaffcc}
Ali JavadiAbhari, Shruti Patil, Daniel Kudrow, Jeff Heckey, Alexey Lvov, Frederic T. Chong, and
Margaret Martonosi. ``ScaffCC: scalable compilation and analysis of quantum programs''. In: \emph{Parallel
Computing} 45 (2015), pp. 2--17. \textsc{doi}: \href {https://doi.org/10.1016/j.parco.2014.12.001} {\nolinkurl
{10.1016/j.parco.2014.12.001}}.
%
\bibitem[JPV18]{jeandel2017complete}
Emmanuel Jeandel, Simon Perdrix, and Renaud Vilmart. ``A complete axiomatisation of the ZX-calculus for
Clifford+T quantum mechanics''. In: \cite{lics2018}, pp. 559--568. \textsc{doi}: \href
{https://doi.org/10.1145/3209108.3209131} {\nolinkurl {10.1145/3209108.3209131}}. \textsc{arXiv}: \href
{https://www.arxiv.org/abs/1705.11151} {\nolinkurl {1705.11151}}.
%
\bibitem[Jr08]{brooks1995mythical}
Frederick P. Brooks Jr. \emph{The Mythical Man-Month - Essays on Software Engineering}. Anniversary
Edition. Addison-Wesley, 2008. \textsc{isbn}: 0-201-83595-9.
%
\bibitem[JS12a]{james2012embracing}
Roshan P. James and Amr Sabry. \emph{Embracing the Laws of Physics}. Presented at OBT'12. 2012.
%
\bibitem[JS12b]{james2012information}
Roshan P. James and Amr Sabry. ``Information effects''. In: \emph{Proceedings of the 39th ACM
SIGPLAN-SIGACT Symposium on Principles of Programming Languages, POPL'12} (Philadelphia,
Pennsylvania, USA). Ed. by John Field and Michael Hicks. ACM, 2012, pp. 73--84. \textsc{isbn}:
978-1-4503-1083-3. \textsc{doi}: \href {https://doi.org/10.1145/2103656.2103667} {\nolinkurl
{10.1145/2103656.2103667}}. \textsc{url}: \url {http://dl.acm.org/citation.cfm?id=2103656}.
%
\bibitem[JS12c]{james2012isomorphic}
Roshan P. James and Amr Sabry. ``Isomorphic interpreters from logically reversible abstract machines''. In:
\emph{Post-Proceedings of the 4th International Workshop on Reversible Computation, RC'12}
(Copenhagen, Denmark). Ed. by Robert Glück and Tetsuo Yokoyama. Vol. 7581. Lecture Notes in Computer
Science. Springer, 2012, pp. 57--71. \textsc{doi}: \href {https://doi.org/10.1007/978-3-642-36315-3_5} {\nolinkurl
{10.1007/978-3-642-36315-3_5}}.
%
\bibitem[JS14]{james2014theseus}
Rosham P. James and Amr Sabry. ``Theseus: A High-Level Language for Reversible Computing''. Booklet of
work-in-progress and short reports for RC 2014. 2014.
%
\bibitem[JT98]{jung1998troublesome}
Achim Jung and Regina Tix. ``The troublesome probabilistic powerdomain''. In: \emph{Comprox III, Third
Workshop on Computation and Approximation} (Birmingham, England, Sept. 11--13, 1997). Ed. by
Abbas Edalat, Achim Jung, Klaus Keimel, and Marta Kwiatkowska. Vol. 13. Electronic Notes in Theoretical
Computer Science. 1998, pp. 70--91. \textsc{doi}: \href {https://doi.org/10.1016/S1571-0661(05)80216-6}
{\nolinkurl {10.1016/S1571-0661(05)80216-6}}.
%
\bibitem[Kaa19]{kaarsgaard2019inversion}
Robin Kaarsgaard. ``Inversion, iteration, and the art of dual wielding''. In: \cite{rc2019}, pp. 34--50. \textsc{doi}:
\href {https://doi.org/10.1007/978-3-030-21500-2_3} {\nolinkurl {10.1007/978-3-030-21500-2_3}}. \textsc{arXiv}:
\href {https://www.arxiv.org/abs/1904.01679} {\nolinkurl {1904.01679}}.
%
\bibitem[KAG17]{kaarsgaard2017join}
Robin Kaarsgaard, Holger Bock Axelsen, and Robert Glück. ``Join inverse categories and reversible
recursion''. In: \emph{Journal of Logical and Algebraic Methods in Programming} 87 (2017), pp. 33--50.
\textsc{issn}: 2352-2208. \textsc{doi}: \href {https://doi.org/10.1016/j.jlamp.2016.08.003} {\nolinkurl
{10.1016/j.jlamp.2016.08.003}}.
%
\bibitem[Kas79]{kastl1979inverse}
J. Kastl. ``Inverse categories''. In: \emph{Algebraische Modelle, Kategorien und Gruppoide}. Studien zur
Algebra und ihre Anwendungen, Band 7. Berlin, Akademie-Verlag, 1979, pp. 51--60.
%
\bibitem[KIQBAW19]{strawberry-fields}
Nathan Killoran, Josh Izaac, Nicolás Quesada, Ville Bergholm, Matthew Amy, and Christian Weedbrook.
``Strawberry Fields: a software platform for photonic quantum computing''. In: \emph{Quantum} 3 (2019),
p. 129. \textsc{doi}: \href {https://doi.org/10.22331/q-2019-03-11-129} {\nolinkurl {10.22331/q-2019-03-11-129}}.
\textsc{arXiv}: \href {https://www.arxiv.org/abs/1804.03159} {\nolinkurl {1804.03159}}.
%
\bibitem[Kit95]{kitaev1995quantum}
A Yu Kitaev. ``Quantum measurements and the Abelian stabilizer problem''. 1995. \textsc{arXiv}: \href
{https://www.arxiv.org/abs/quant-ph/9511026} {\nolinkurl {quant-ph/9511026}}.
%
\bibitem[KJS10]{kiselyov2010type-families}
Oleg Kiselyov, Simon Peyton Jones, and Chung-Chieh Shan. ``Fun with type functions''. In:
\emph{Reflections on the Work of C.A.R. Hoare}. Springer, 2010. Chap. 14, pp. 301--331. \textsc{doi}: \href
{https://doi.org/10.1007/978-1-84882-912-1_14} {\nolinkurl {10.1007/978-1-84882-912-1_14}}.
%
\bibitem[KKPSY15]{kirchner2015frama-c}
Florent Kirchner, Nikolai Kosmatov, Virgile Prevosto, Julien Signoles, and Boris Yakobowski. ``Frama-C: a
software analysis perspective''. In: \emph{Formal Aspects of Computing} 27.3 (2015), pp. 573--609.
\textsc{doi}: \href {https://doi.org/10.1007/s00165-014-0326-7} {\nolinkurl {10.1007/s00165-014-0326-7}}.
%
\bibitem[Kle35a]{kleene35theory1}
Stephen C. Kleene. ``A theory of positive integers in formal logic, part I''. In: \emph{American Journal of
Mathematics} 57.1 (1935), pp. 153--173. \textsc{doi}: \href {https://doi.org/10.2307/2372027} {\nolinkurl
{10.2307/2372027}}.
%
\bibitem[Kle35b]{kleene35theory2}
Stephen C. Kleene. ``A theory of positive integers in formal logic, part II''. In: \emph{American Journal of
Mathematics} 57.2 (1935), pp. 219--244. \textsc{doi}: \href {https://doi.org/10.2307/2371199} {\nolinkurl
{10.2307/2371199}}.
%
\bibitem[Kle45]{kleene1945interpretation}
Stephen Cole Kleene. ``On the interpretation of intuitionistic number theory''. In: \emph{Journal of Symbolic
Logic} 10.4 (1945), pp. 109--124. \textsc{doi}: \href {https://doi.org/10.2307/2269016} {\nolinkurl
{10.2307/2269016}}.
%
\bibitem[Kle67]{kleene1967mathematical}
Stephen Cole Kleene. \emph{Mathematical Logic}. Dover Publication Inc., 1967.
%
\bibitem[Klu99]{kluge1999reversible}
Werner E. Kluge. ``A reversible SE(M)CD machine''. In: \emph{Selected Papers of the 11th International
Workshop on the Implementation of Functional Languages, IFL'99} (Lochem, The Netherlands, Sept. 7--10,
1999). Ed. by Pieter W. M. Koopman and Chris Clack. Vol. 1868. Lecture Notes in Computer Science.
Springer, 1999, pp. 95--113. \textsc{isbn}: 3-540-67864-6. \textsc{doi}: \href {https://doi.org/10.1007/10722298_6}
{\nolinkurl {10.1007/10722298_6}}.
%
\bibitem[KMR89]{liber-amicorum:1989}
Jan Willem Klop, J. J. C. Meijer, and Jan J. M. M. Rutten, eds. \emph{J.W. de Bakker, 25 Jaar Semantiek: Liber
Amicorum}. CWI, 1989. \textsc{url}: \url {https://ir.cwi.nl/pub/20371/} (visited on Aug. 11, 2022).
%
\bibitem[Kni95]{knill1995approximation}
E. Knill. \emph{Approximation by Quantum Circuits}. Tech. rep. LANL report LAUR-95-2225. Los Alamos
National Laboratory, 1995.
%
\bibitem[Kni96]{knill96conventions}
Emanuel H. Knill. \emph{Conventions for Quantum Pseudocode}. Tech. rep. LAUR-96-2724. Los Alamos,
New Mexico, US.: Los Alamos National Laboratory, 1996.
%
\bibitem[Koz83]{kozen1983results}
Dexter Kozen. ``Results on the propositional $\mu$-calculus''. In: \emph{Theoretical Computer Science} 27 (1983),
pp. 333--354. \textsc{doi}: \href {https://doi.org/10.1016/0304-3975(82)90125-6} {\nolinkurl
{10.1016/0304-3975(82)90125-6}}.
%
\bibitem[KR21]{kaarsgaard2021join}
Robin Kaarsgaard and Mathys Rennela. ``Join inverse rig categories for reversible functional programming,
and beyond''. In: \cite{mfps2021}, pp. 152--167. \textsc{doi}: \href {https://doi.org/10.4204/EPTCS.351.10} {\nolinkurl
{10.4204/EPTCS.351.10}}.
%
\bibitem[Kri07]{krivine2007call-by-name}
Jean-Louis Krivine. ``A call-by-name lambda-calculus machine''. In: \emph{Higher-Order and Symbolic
Computation} 20.3 (2007), pp. 199--207. \textsc{doi}: \href {https://doi.org/10.1007/s10990-007-9018-9}
{\nolinkurl {10.1007/s10990-007-9018-9}}.
%
\bibitem[KV19]{kaarsgaard2019engarde}
Robin Kaarsgaard and Niccolò Veltri. ``En garde! unguarded iteration for reversible computation in the delay
monad''. In: \emph{Proceedings of the 13th International Conference on Mathematics of Program
Construction, MPC 2019} (Porto, Portugal). Ed. by Graham Hutton. Vol. 11825. Lecture Notes in Computer
Science. Springer Verlag, Oct. 2019, pp. 366--384. \textsc{isbn}: 978-3-030-33635-6. \textsc{doi}: \href
{https://doi.org/10.1007/978-3-030-33636-3_13} {\nolinkurl {10.1007/978-3-030-33636-3_13}}.
%
\bibitem[KW20]{kissinger2020reducing}
Aleks Kissinger and John van de Wetering. ``Reducing the number of non-Clifford gates in quantum
circuits''. In: \emph{Physical Review A} 102 (2 2020), p. 022406. \textsc{doi}: \href
{https://doi.org/10.1103/PhysRevA.102.022406} {\nolinkurl {10.1103/PhysRevA.102.022406}}. \textsc{arXiv}:
\href {https://www.arxiv.org/abs/1903.10477} {\nolinkurl {1903.10477}}.
%
\bibitem[KZ15]{kissinger2015quantomatic}
Aleks Kissinger and Vladimir Zamdzhiev. ``Quantomatic: a proof assistant for diagrammatic reasoning''. In:
\emph{Proceedings of the 25th International Conference on Automated Deduction, CADE-25} (Berlin,
Germany, Aug. 1--7, 2015). Ed. by Amy P. Felty and Aart Middeldorp. Vol. 9195. Lecture Notes in Computer
Science. Springer, 2015, pp. 326--336. \textsc{isbn}: 978-3-319-21400-9. \textsc{doi}: \href
{https://doi.org/10.1007/978-3-319-21401-6_22} {\nolinkurl {10.1007/978-3-319-21401-6_22}}. \textsc{arXiv}:
\href {https://www.arxiv.org/abs/1503.01034} {\nolinkurl {1503.01034}}.
%
\bibitem[Laf04]{lafont04soft}
Yves Lafont. ``Soft linear logic and polynomial time''. In: \emph{Theoretical Computer Science} 318.1--2
(2004), pp. 163--180. \textsc{doi}: \href {https://doi.org/10.1016/j.tcs.2003.10.018} {\nolinkurl
{10.1016/j.tcs.2003.10.018}}.
%
\bibitem[Laf90]{lafont1990interaction}
Yves Lafont. ``Interaction nets''. In: \emph{Proceedings of the 17th ACM SIGPLAN-SIGACT Symposium on
Principles of Programming Languages, POPL'90} (San Francisco, California, USA). Ed. by Frances E. Allen.
ACM, 1990, pp. 95--108. \textsc{isbn}: 0-89791-343-4. \textsc{doi}: \href {https://doi.org/10.1145/96709.96718}
{\nolinkurl {10.1145/96709.96718}}.
%
\bibitem[Laf95]{lafont1995proofnets}
Yves Lafont. ``From proof-nets to interaction nets''. In: \cite{girard1995advances}, pp. 225--247.
%
\bibitem[Lan02]{lang2002algebra}
Serge Lang. \emph{Algebra}. Revised Third Edition. Vol. 211. Graduate Texts in Mathematics. Springer, 2002.
\textsc{isbn}: 0-387-95385-X.
%
\bibitem[Lan61]{landauer1961irreversibility}
Rolf Landauer. ``Irreversibility and heat generation in the computing process''. In: \emph{IBM Journal of
Research and Development} 5.3 (1961), pp. 183--191. \textsc{doi}: \href {https://doi.org/10.1147/rd.53.0183}
{\nolinkurl {10.1147/rd.53.0183}}.
%
\bibitem[Lan66]{landin1966next}
Peter J. Landin. ``The next 700 programming languages''. In: \emph{Communications of the ACM} 9.3 (1966),
pp. 157--166. \textsc{doi}: \href {https://doi.org/10.1145/365230.365257} {\nolinkurl {10.1145/365230.365257}}.
\textsc{url}: \url {https://doi.org/10.1145/365230.365257}.
%
\bibitem[Lau13]{laurent2013introduction}
Olivier Laurent. ``An Introduction to Proof-Nets''. Notes. 2013.
%
\bibitem[LB21]{lorenz2021causal}
Robin Lorenz and Jonathan Barrett. ``Causal and compositional structure of unitary transformations''. In:
\emph{Quantum} 5 (2021), p. 511. \textsc{doi}: \href {https://doi.org/10.22331/q-2021-07-28-511} {\nolinkurl
{10.22331/q-2021-07-28-511}}. \textsc{arXiv}: \href {https://www.arxiv.org/abs/2001.07774} {\nolinkurl
{2001.07774}}.
%
\bibitem[LBK05]{lim2005repeat-until-success}
Yuan Liang Lim, Almut Beige, and Leong Chuan Kwek. ``Repeat-Until-Success linear optics distributed
quantum computing''. In: \emph{Physical Review Letters} 95 (3 2005), p. 030505. \textsc{doi}: \href
{https://doi.org/10.1103/PhysRevLett.95.030505} {\nolinkurl {10.1103/PhysRevLett.95.030505}}.
%
\bibitem[Lee22]{lee2022phd}
Dongho Lee. ``Formal Methods for Quantum Programming Languages''. Thèse de doctorat. Université
Paris-Saclay, 2022.
%
\bibitem[Lee90]{leeuwen1990formal}
Jan van Leeuwen, ed. \emph{Formal Models and Semantics}. Vol. B. Handbook of Theoretical Computer
Science. Elsevier, 1990. \textsc{isbn}: 0-444-88074-7.
%
\bibitem[Lem24]{lemonnier2024phd}
Louis Lemonnier. ``The Semantics of Effects : Centrality, Quantum Control and Reversible Recursion''.
PhD thesis. Université Paris Saclay, 2024.
%
\bibitem[Lep16]{lepigre2016semantics}
Rodolphe Lepigre. ``Semantics and Implementation of an Extension of ML for Proving Programs''. Thèse de
Doctorat. Université de Grenoble, 2016.
%
\bibitem[Lev16]{leventis2016probabilistic}
Thomas Leventis. ``Probabilistic $\lambda$-Theories''. Thèse de Doctorat. Aix-Marseille Université, 2016. \textsc{hal}:
\href {https://hal.archives-ouvertes.fr/tel-01427279} {\nolinkurl {tel-01427279}}.
%
\bibitem[LFFP11]{leuschel2011automated}
Michael Leuschel, Jérôme Falampin, Fabian Fritz, and Daniel Plagge. ``Automated property verification for
large scale B models with ProB''. In: \emph{Formal Aspects of Computing} 23.6 (2011), pp. 683--709.
\textsc{doi}: \href {https://doi.org/10.1007/s00165-010-0172-1} {\nolinkurl {10.1007/s00165-010-0172-1}}.
%
\bibitem[LFHY14]{dallago2014geometry}
Ugo Dal Lago, Claudia Faggian, Ichiro Hasuo, and Akira Yoshimizu. ``The geometry of synchronization''. In:
[LICS14], 35:1--35:10. \textsc{doi}: \href {https://doi.org/10.1145/2603088.2603154} {\nolinkurl
{10.1145/2603088.2603154}}. \textsc{arXiv}: \href {https://www.arxiv.org/abs/1405.3427} {\nolinkurl {1405.3427}}.
%
\bibitem[LFSMC20]{litteken2020updated}
Andrew Litteken, Yung-Ching Fan, Devina Singh, Margaret Martonosi, and Frederic T Chong. ``An updated
LLVM-based quantum research compiler with further OpenQASM support''. In: \emph{Quantum Science
and Technology} 5.3 (2020), p. 034013.
%
\bibitem[LFVY15]{lago2015parallelism}
Ugo Dal Lago, Claudia Faggian, Benoît Valiron, and Akira Yoshimizu. ``Parallelism and synchronization in an
infinitary context''. In: \cite{lics2015}, pp. 559--572. \textsc{doi}: \href {https://doi.org/10.1109/LICS.2015.58}
{\nolinkurl {10.1109/LICS.2015.58}}. \textsc{hal}: \href {https://hal.archives-ouvertes.fr/hal-01231831}
{\nolinkurl {hal-01231831}}. \textsc{arXiv}: \href {https://www.arxiv.org/abs/1505.03635} {\nolinkurl
{1505.03635}}.
%
\bibitem[LFVY17]{lago2017geometry}
Ugo Dal Lago, Claudia Faggian, Benoît Valiron, and Akira Yoshimizu. ``The geometry of parallelism:
classical, probabilistic, and quantum effects''. In: \cite{popl2017}, pp. 833--845. \textsc{doi}: \href
{https://doi.org/10.1145/3009837.3009859} {\nolinkurl {10.1145/3009837.3009859}}. \textsc{hal}: \href
{https://hal.archives-ouvertes.fr/hal-01474620} {\nolinkurl {hal-01474620}}. \textsc{arXiv}: \href
{https://www.arxiv.org/abs/1610.09629} {\nolinkurl {1610.09629}}.
%
\bibitem[LH09]{lago2009bounded}
Ugo Dal Lago and Martin Hofmann. ``Bounded linear logic, revisited''. In: \emph{Proceedings of the 9th
International Conference on Typed Lambda Calculi and Applications (TLCA'09)} (Brasilia, Brazil). Ed. by
Pierre-Louis Curien. Vol. 5608. Lecture Notes in Computer Science. See also the journal's version \cite{lago2010bounded}.
Springer, 2009. \textsc{isbn}: 978-3-642-02272-2. \textsc{doi}: \href
{https://doi.org/10.1007/978-3-642-02273-9_8} {\nolinkurl {10.1007/978-3-642-02273-9_8}}.
%
\bibitem[LH10]{lago2010bounded}
Ugo Dal Lago and Martin Hofmann. ``Bounded linear logic, revisited''. In: \emph{Logical Methods in
Computer Science} 6.4 (2010). Long version of the TLCA'09 publication \cite{lago2009bounded}. \textsc{doi}: \href
{https://doi.org/10.2168/LMCS-6(4:7)2010} {\nolinkurl {10.2168/LMCS-6(4:7)2010}}. \textsc{arXiv}: \href
{https://www.arxiv.org/abs/0904.2675} {\nolinkurl {0904.2675}}.
%
\bibitem[LH85]{luckham1985overview}
David C. Luckham and Friedrich W. von Henke. ``An overview of Anna, a specification language for Ada''. In:
\emph{IEEE Software} 2.2 (1985), pp. 9--22. \textsc{doi}: \href {https://doi.org/10.1109/MS.1985.230345}
{\nolinkurl {10.1109/MS.1985.230345}}.
%
\bibitem[LICS04]{lics04}
\emph{Proceedings of the 19th Symposium on Logic in Computer Science, LICS'04} (Turku, Finland,
July 14--17, 2004). IEEE. IEEE Computer Society Press, July 2004. \textsc{isbn}: 0-7695-2192-4.
%
\bibitem[LICS11]{lics2011}
\emph{Proceedings of the 26th Annual IEEE Symposium on Logic in Computer Science, LICS 2011} (Toronto,
Ontario, Canada, June 21--24, 2011). IEEE Computer Society, 2011. \textsc{isbn}: 978-0-7695-4412-0.
%
\bibitem[LICS14]{lics2014}
Thomas A. Henzinger and Dale Miller, eds. \emph{Proceedings of the Joint Meeting of the Twenty-Third
EACSL Annual Conference on Computer Science Logic and the Twenty-Ninth Annual ACM/IEEE
Symposium on Logic in Computer Science, CSL-LICS'14} (Vienna, Austria). ACM, 2014. \textsc{isbn}:
978-1-4503-2886-9.
%
\bibitem[LICS15]{lics2015}
\emph{Proceedings of the 30th Annual ACM/IEEE Symposium on Logic in Computer Science, LICS'15}
(Kyoto, Japan). IEEE Computer Society, 2015. \textsc{isbn}: 978-1-4799-8875-4.
%
\bibitem[LICS18]{lics2018}
Anuj Dawar and Erich Grädel, eds. \emph{Proceedings of the 33rd Annual ACM/IEEE Symposium on Logic
in Computer Science, LICS'2018} (Oxford, UK). ACM, 2018. \textsc{doi}: \href
{https://doi.org/10.1145/3209108} {\nolinkurl {10.1145/3209108}}.
%
\bibitem[LICS19]{lics2019}
\emph{Proceedings of the 34th Annual ACM/IEEE Symposium on Logic in Computer Science, LICS'19}
(Vancouver, BC, Canada, June 24--27, 2019). IEEE, 2019. \textsc{isbn}: 978-1-7281-3608-0.
%
\bibitem[LMMP13]{laird2013weighted}
Jim Laird, Giulio Manzonetto, Guy McCusker, and Michele Pagani. ``Weighted relational models of typed
lambda-calculi''. In: \emph{Proceedings of the 28th Annual ACM/IEEE Symposium on Logic in Computer
Science, LICS 2013} (New Orleans, LA, USA, June 25--28, 2013). IEEE Computer Society, 2013, pp. 301--310.
\textsc{isbn}: 978-1-4799-0413-6. \textsc{doi}: \href {https://doi.org/10.1109/LICS.2013.36} {\nolinkurl
{10.1109/LICS.2013.36}}.
%
\bibitem[LMZ10]{lago2010quantum}
Ugo Dal Lago, Andrea Masini, and Margherita Zorzi. ``Quantum implicit computational complexity''. In:
\emph{Theoretical Computer Science} 411.2 (2010), pp. 377--409. \textsc{doi}: \href
{https://doi.org/10.1016/j.tcs.2009.07.045} {\nolinkurl {10.1016/j.tcs.2009.07.045}}.
%
\bibitem[LMZ18]{lindenhovius2018enriching}
Bert Lindenhovius, Michael W. Mislove, and Vladimir Zamdzhiev. ``Enriching a linear/non-linear lambda
calculus: A programming language for string diagrams''. In: \cite{lics2018}, pp. 659--668. \textsc{doi}: \href
{https://doi.org/10.1145/3209108.3209196} {\nolinkurl {10.1145/3209108.3209196}}. \textsc{hal}: \href
{https://hal.archives-ouvertes.fr/hal-03018477} {\nolinkurl {hal-03018477}}. \textsc{arXiv}: \href
{https://www.arxiv.org/abs/1804.09822} {\nolinkurl {1804.09822}}.
%
\bibitem[LN98]{leino1998extended}
K. Rustan M. Leino and Greg Nelson. ``An extended static checker for Modular-3''. In: \emph{Proceedings of
the 7th International Conference on Compiler Construction, CC'98} (Lisbon, Portugal, Mar. 28--Apr. 4, 1998).
Ed. by Kai Koskimies. Vol. 1383. Lecture Notes in Computer Science. Springer, 1998, pp. 302--305.
\textsc{isbn}: 3-540-64304-4. \textsc{doi}: \href {https://doi.org/10.1007/BFb0026441} {\nolinkurl
{10.1007/BFb0026441}}.
%
\bibitem[Loe50]{loewner1950some}
Charles Loewner. ``Some classes of functions defined by difference or differential inequalities''. In:
\emph{Bulletin of the American Mathematical Society} 56.6 (1950), pp. 308--319.
%
\bibitem[Low19]{low2019hamiltonian}
Guang Hao Low. ``Hamiltonian simulation with nearly optimal dependence on spectral norm''. In:
\emph{Proceedings of the 51st Annual ACM SIGACT Symposium on Theory of Computing, STOC 2019}
(Phoenix, AZ, USA, June 23--26, 2019). Ed. by Moses Charikar and Edith Cohen. ACM, 2019, pp. 491--502.
\textsc{doi}: \href {https://doi.org/10.1145/3313276.3316386} {\nolinkurl {10.1145/3313276.3316386}}.
\textsc{arXiv}: \href {https://www.arxiv.org/abs/1807.03967} {\nolinkurl {1807.03967}}.
%
\bibitem[Löw34]{lowner1934order}
Karl Löwner. ``Über monotone Matrixfunktionen''. In: \emph{Mathematische Zeitschrift} 38.1 (1934),
pp. 177--216. \textsc{doi}: \href {https://doi.org/10.1007/BF01170633} {\nolinkurl {10.1007/BF01170633}}.
%
\bibitem[LPVX21]{lee2022concrete}
Dongho Lee, Valentin Perrelle, Benoît Valiron, and Zhaowei Xu. ``Concrete categorical model of a quantum
circuit description language with measurement''. In: \emph{Proceedings of the 41st IARCS Annual
Conference on Foundations of Software Technology and Theoretical Computer Science, FSTTCS 2021}.
Ed. by Mikolaj Bojanczyk and Chandra Chekuri. Vol. 213. LIPIcs. 2021, 51:1--51:20. \textsc{doi}: \href
{https://doi.org/10.4230/LIPIcs.FSTTCS.2021.51} {\nolinkurl {10.4230/LIPIcs.FSTTCS.2021.51}}.
%
\bibitem[LS89]{lambek89introduction}
Joachim Lambek and Philip Scott. \emph{Introduction to Higher Order Categorical Logic}. 2nd ed. Vol. 7.
Cambridge studies in advanced mathematics. Cambridge University Press, 1989. \textsc{isbn}: 0-521-35653-9.
%
\bibitem[LS90]{levin1990note}
Robert Y. Levin and Alan T. Sherman. ``A note on Bennett's time-space tradeoff for reversible computation''.
In: \emph{SIAM Journal on Computing} 19.4 (1990), pp. 673--677. \textsc{doi}: \href
{https://doi.org/10.1137/0219046} {\nolinkurl {10.1137/0219046}}.
%
\bibitem[Lut86]{lutz1986janus}
Christopher Lutz. ``Janus: a time-reversible language''. Letter to R. Landauer, posted online by Tetsuo
Yokoyama on \texttt{http://www.tetsuo.jp/ref/janus.html}. 1986.
%
\bibitem[LWZG+18]{liu2018qsi}
Shusen Liu, Xin Wang, Li Zhou, Ji Guan, Yinan Li, Yang He, Runyao Duan, and Mingsheng Ying. ``Q|SI⟩ : a
quantum programming environment''. In: \emph{Symposium on Real-Time and Hybrid Systems - Essays
Dedicated to Professor Chaochen Zhou on the Occasion of His 80th Birthday}. Ed. by Cliff B. Jones, Ji Wang,
and Naijun Zhan. Vol. 11180. Lecture Notes in Computer Science. 2018, pp. 133--164. \textsc{doi}: \href
{https://doi.org/10.1007/978-3-030-01461-2\_8} {\nolinkurl {10.1007/978-3-030-01461-2\_8}}.
%
\bibitem[LY18]{li2018algorithmic}
Yangjia Li and Mingsheng Ying. ``Algorithmic analysis of termination problems for quantum programs''. In:
\emph{Proceedings of the ACM on Programming Languages} 2.POPL (2018), 35:1--35:29. \textsc{doi}: \href
{https://doi.org/10.1145/3158123} {\nolinkurl {10.1145/3158123}}.
%
\bibitem[LYY14]{li2014termination}
Yangjia Li, Nengkun Yu, and Mingsheng Ying. ``Termination of nondeterministic quantum programs''. In:
\emph{Acta Informatica} 51.1 (2014), pp. 1--24. \textsc{doi}: \href {https://doi.org/10.1007/s00236-013-0185-3}
{\nolinkurl {10.1007/s00236-013-0185-3}}.
%
\bibitem[LZ12]{lago2012probabilistic}
Ugo Dal Lago and Margherita Zorzi. ``Probabilistic operational semantics for the lambda calculus''. In:
\emph{RAIRO Theor. Informatics Appl.} 46.3 (2012), pp. 413--450. \textsc{doi}: \href
{https://doi.org/10.1051/ita/2012012} {\nolinkurl {10.1051/ita/2012012}}. \textsc{arXiv}: \href
{https://www.arxiv.org/abs/1104.0195} {\nolinkurl {1104.0195}}.
%
\bibitem[LZ14]{lago2014wave-style}
Ugo Dal Lago and Margherita Zorzi. ``Wave-style token machines and quantum lambda calculi''. In:
\emph{Proceedings Third International Workshop on Linearity, LINEARITY 2014} (Vienna, Austria, July 13,
2014). Ed. by Sandra Alves and Iliano Cervesato. Vol. 176. Electronic Proceedings in Theoretical Computer
Science. 2014, pp. 64--78. \textsc{doi}: \href {https://doi.org/10.4204/EPTCS.176.6} {\nolinkurl
{10.4204/EPTCS.176.6}}.
%
\bibitem[LZBY22]{liu2022quantum}
Junyi Liu, Li Zhou, Gilles Barthe, and Mingsheng Ying. ``Quantum weakest preconditions for reasoning
about expected runtimes of quantum programs''. In: \emph{Proceedings of the 37th Annual ACM/IEEE
Symposium on Logic in Computer Science, LICS '22} (Haifa, Israel, Aug. 2--5, 2022). Ed. by Christel Baier and
Dana Fisman. ACM, 2022, 4:1--4:13. \textsc{isbn}: 978-1-4503-9351-5. \textsc{doi}: \href
{https://doi.org/10.1145/3531130.3533327} {\nolinkurl {10.1145/3531130.3533327}}.
%
\bibitem[LZWY+19a]{liu2019formal}
Junyi Liu, Bohua Zhan, Shuling Wang, Shenggang Ying, Tao Liu, Yangjia Li, Mingsheng Ying, and
Naijun Zhan. ``Formal verification of quantum algorithms using quantum Hoare logic''. In:
\emph{Proceedings of the 31st International Conference on Computer Aided Verification, CAV 2019, Part II}
(New York City, NY, USA, July 15--18, 2019). Ed. by Isil Dillig and Serdar Tasiran. Vol. 11562. Lecture Notes in
Computer Science. Springer, 2019, pp. 187--207. \textsc{isbn}: 978-3-030-25542-8. \textsc{doi}: \href
{https://doi.org/10.1007/978-3-030-25543-5\_12} {\nolinkurl {10.1007/978-3-030-25543-5\_12}}. \textsc{arXiv}:
\href {https://www.arxiv.org/abs/1601.03835} {\nolinkurl {1601.03835}}.
%
\bibitem[LZWY+19b]{liu2019quantum}
Junyi Liu, Bohua Zhan, Shuling Wang, Shenggang Ying, Tao Liu, Yangjia Li, Mingsheng Ying, and
Naijun Zhan. ``Quantum Hoare logic''. In: \emph{Archive of Formal Proofs} 2019 (2019). \textsc{url}: \url
{https://www.isa-afp.org/entries/QHLProver.html} (visited on Aug. 15, 2022).
%
\bibitem[LZYDYX20]{li2020projection-based}
Gushu Li, Li Zhou, Nengkun Yu, Yufei Ding, Mingsheng Ying, and Yuan Xie. ``Projection-based runtime
assertions for testing and debugging quantum programs''. In: \emph{Proceedings of the ACM on
Programming Languages} 4.OOPSLA (2020), 150:1--150:29. \textsc{doi}: \href {https://doi.org/10.1145/3428218}
{\nolinkurl {10.1145/3428218}}. \textsc{arXiv}: \href {https://www.arxiv.org/abs/1911.12855} {\nolinkurl
{1911.12855}}.
%
\bibitem[Mac94]{mackie1994geometry}
Ian Mackie. ``The Geometry of Implementation : Applications of the Geometry of Interaction to Language
Implementation''. PhD thesis. University of London, 1994.
%
\bibitem[Mac95]{mackie1995geometry}
Ian Mackie. ``The geometry of interaction machine''. In: \emph{Proceedings of the 22nd ACM
SIGPLAN-SIGACT Symposium on Principles of Programming Languages, POPL'95} (San Francisco,
California, US.). ACM. ACM Press, 1995, pp. 198--208. \textsc{doi}: \href
{https://doi.org/10.1145/199448.199483} {\nolinkurl {10.1145/199448.199483}}.
%
\bibitem[Mal10]{malherbe2010categorical}
Octavio Malherbe. ``Categorical models of computation: partially traced categories and presheaf models of
quantum computation''. PhD thesis. University of Ottawa, 2010. \textsc{arXiv}: \href
{https://www.arxiv.org/abs/1301.5087} {\nolinkurl {1301.5087}}.
%
\bibitem[Mar71]{martin-lof1971hauptsatz}
Per Martin-Löf. ``Hauptsatz for the intuitionistic theory of iterated inductive definitions''. In:
\emph{Proceedings of the Second Scandinavian Logic Symposium}. Ed. by J. E. Fenstad. Vol. 63. Studies in
Logic and the Foundations of Mathematics. North-Holland, 1971, pp. 179--216. \textsc{doi}: \href
{https://doi.org/10.1016/S0049-237X(08)70847-4} {\nolinkurl {10.1016/S0049-237X(08)70847-4}}.
%
\bibitem[Mar84]{martin-lof1984intuitionistic}
Per Martin-Löf. \emph{Intuitionistic Type Theory}. Studies in Proof Theories. Napoli, Italy: Bibliopolis, 1984.
%
\bibitem[Mat03]{matos2003linear}
Armando B. Matos. ``Linear programs in a simple reversible language''. In: \emph{Theoretival Computer
Science} 290.3 (2003), pp. 2063--2074. \textsc{doi}: \href {https://doi.org/10.1016/S0304-3975(02)00486-3}
{\nolinkurl {10.1016/S0304-3975(02)00486-3}}.
%
\bibitem[MAT14]{mahmoud2014formal}
Mohamed Yousri Mahmoud, Vincent Aravantinos, and Sofiène Tahar. ``Formal verification of optical
quantum flip gate''. In: \emph{Proceedings of the 5th International Conference on Interactive Theorem
Proving, ITP'14} (Vienna, Austria). Ed. by Gerwin Klein and Ruben Gamboa. Vol. 8558. Lecture Notes in
Computer Science. Springer, 2014, pp. 358--373. \textsc{doi}: \href
{https://doi.org/10.1007/978-3-319-08970-6_23} {\nolinkurl {10.1007/978-3-319-08970-6_23}}.
%
\bibitem[Mat98]{matthes1998phd}
Ralph Matthes. ``Extensions of System F by Iteration and Primitive Recursion on Monotone Inductive
Types''. PhD thesis. Ludwig-Maximilians-Universität, München, Germany, 1998.
%
\bibitem[Maz06]{mazza2006phd}
Damiano Mazza. ``Interaction Nets : Semantics and Concurrent Extensions''. Thèse de Doctorat. Université
de la Méditerranée/Università degli Studi Roma Tre, 2006.
%
\bibitem[McKinsey21]{mckinsey2021}
Matteo Biondi, Anna Heid, Ivan Ostojic, Nicolaus Henke, Lorenzo Pautasso, Niko Mohr, Linde Wester, and
Rodney Zemmel. \emph{Quantum computing: An emerging ecosystem and industry use cases}. Report.
McKinsey \& Company, 2021.
%
\bibitem[MDM03]{maslov2003fredkintoffoli}
Dmitri Maslov, Gerhard W. Dueck, and D. Michael Miller. ``Fredkin/Toffoli templates for reversible logic
synthesis''. In: \emph{Proceedings of the International Conference on Computer-Aided Design, ICCAD'03}
(San Jose, CA, USA). IEEE Computer Society / ACM, 2003, pp. 256--261. \textsc{isbn}: 1-58113-762-1.
\textsc{doi}: \href {https://doi.org/10.1109/ICCAD.2003.1257667} {\nolinkurl {10.1109/ICCAD.2003.1257667}}.
%
\bibitem[MDM05]{maslov2005toffoli}
Dmitri Maslov, Gerhard W. Dueck, and D. Michael Miller. ``Toffoli network synthesis with templates''. In:
\emph{IEEE Transactions on Computer-Aided Design of Integrated Circuits and Systems} 24.6 (2005),
pp. 807--817. \textsc{doi}: \href {https://doi.org/10.1109/TCAD.2005.847911} {\nolinkurl
{10.1109/TCAD.2005.847911}}.
%
\bibitem[Mei24]{meijer2024advances}
Arianne Meijer--van de Griend. ``Advances in Quantum Compilation in the NISQ Era''. PhD thesis.
University of Helsinki, 2024.
%
\bibitem[Men88]{mendler1988phd}
Paul Mendler. ``Inductive Definition in Type Theory''. PhD thesis. Cornell University, USA, 1988. \textsc{url}:
\url {https://hdl.handle.net/1813/6710} (visited on Aug. 3, 2022).
%
\bibitem[Mey92]{meyer1992applying}
Bertrand Meyer. ``Applying ``design by contract''''. In: \emph{IEEE Computer} 25.10 (1992), pp. 40--51.
\textsc{doi}: \href {https://doi.org/10.1109/2.161279} {\nolinkurl {10.1109/2.161279}}.
%
\bibitem[MFPS21]{mfps2021}
Ana Sokolova, ed. \emph{Proceedings 37th Conference on Mathematical Foundations of Programming
Semantics (MFPS 2021)} (Hybrid: Salzburg, Austria and Online, Aug. 30--Sept. 2, 2021). Vol. 351. EPTCS. 2021.
\textsc{doi}: \href {https://doi.org/10.4204/EPTCS.351} {\nolinkurl {10.4204/EPTCS.351}}.
%
\bibitem[MFPS93]{mfps93}
Stephen Brookes, Michael Main, Austin Melton, Michael Mislove, and David Schmidt, eds.
\emph{Mathematical Foundations of Programming Semantics: Ninth International Conference, MFPS IX}
(New Orleans, Louisiana, US.). Vol. 802. Lecture Notes in Computer Science. Springer Verlag, Apr. 1993.
\textsc{isbn}: 3-540-58027-1. \textsc{doi}: \href {https://doi.org/10.1007/3-540-58027-1} {\nolinkurl
{10.1007/3-540-58027-1}}.
%
\bibitem[MHT04]{mu2004injective}
Shin-Cheng Mu, Zhenjiang Hu, and Masato Takeichi. ``An injective language for reversible computation''. In:
\emph{Proceedings of the 7th International Conference on Mathematics of Program Construction, MPC
2004} (Stirling, Scotland, UK). Ed. by Dexter Kozen. Vol. 3125. Lecture Notes in Computer Science. Springer
Verlag, July 2004, pp. 289--313. \textsc{isbn}: 978-3-540-22380-1. \textsc{doi}: \href
{https://doi.org/10.1007/978-3-540-27764-4_16} {\nolinkurl {10.1007/978-3-540-27764-4_16}}.
%
\bibitem[Mit78]{mittelstaedt1978quantum}
Peter Mittelstaedt. \emph{Quantum Logic}. Vol. 126. Synthese Library. Dordrecht, Holland: D. Reidel
Publishing Company, 1978. \textsc{isbn}: 978-94-009-9873-5. \textsc{doi}: \href
{https://doi.org/10.1007/978-94-009-9871-1} {\nolinkurl {10.1007/978-94-009-9871-1}}.
%
\bibitem[MM16]{matteo2016parallelizing}
Olivia Di Matteo and Michele Mosca. ``Parallelizing quantum circuit synthesis''. In: \emph{Quantum Science
and Technology} 1.1 (Mar. 2016), p. 015003. \textsc{doi}: \href {https://doi.org/10.1088/2058-9565/1/1/015003}
{\nolinkurl {10.1088/2058-9565/1/1/015003}}.
%
\bibitem[MM90]{colog1988}
Per Martin-Löf and Grigori Mints, eds. \emph{Proceedings of the International Conference on Computer
Logic (COLOG-88)} (Tallinn, USSR). Vol. 417. Lecture Notes in Computer Science. Springer, 1990.
%
\bibitem[MMNSB16]{martinez2016compiling}
Esteban A. Martinez, Thomas Monz, Daniel Nigg, Philipp Schindler, and Rainer Blatt. ``Compiling quantum
algorithms for architectures with multi-qubit gates''. In: \emph{New Journal of Physics} 18 (2016), p. 063029.
%
\bibitem[Mog14]{mogensen2014reference}
Torben Ægidius Mogensen. ``Reference counting for reversible languages''. In: \emph{Proceedings of hte 6th
International Conference on Reversible Computation, RC 2014} (Kyoto, Japan, July 10--11, 2014). Ed. by
Shigeru Yamashita and Shin-ichi Minato. Vol. 8507. Lecture Notes in Computer Science. Springer, 2014,
pp. 82--94. \textsc{isbn}: 978-3-319-08493-0. \textsc{doi}: \href {https://doi.org/10.1007/978-3-319-08494-7_7}
{\nolinkurl {10.1007/978-3-319-08494-7_7}}.
%
\bibitem[Mog18]{mogensen2018reversible}
Torben Ægidius Mogensen. ``Reversible garbage collection for reversible functional languages''. In:
\emph{New Generation Computing} 36.3 (2018), pp. 203--232. \textsc{doi}: \href
{https://doi.org/10.1007/s00354-018-0037-3} {\nolinkurl {10.1007/s00354-018-0037-3}}.
%
\bibitem[Mog19]{mogensen2019reversible}
Torben Ægidius Mogensen. ``Reversible in-place carry-lookahead addition with few ancillae''. In: \cite{rc2019},
pp. 224--237. \textsc{doi}: \href {https://doi.org/10.1007/978-3-030-21500-2\_14} {\nolinkurl
{10.1007/978-3-030-21500-2\_14}}.
%
\bibitem[Mog89]{moggi89computational}
Eugenio Moggi. ``Computational lambda-calculus and monads''. In: \emph{Proceedings of the Fourth
Symposium on Logic in Computer Science, LICS'89} (Pacific Grove, California, US.). IEEE. IEEE Computer
Society Press, June 1989, pp. 14--23. \textsc{isbn}: 0-8186-1954-6. \textsc{doi}: \href
{https://doi.org/10.1109/LICS.1989.39155} {\nolinkurl {10.1109/LICS.1989.39155}}. \textsc{url}: \url
{http://www.lfcs.inf.ed.ac.uk/reports/88/ECS-LFCS-88-66/}.
%
\bibitem[Moo62]{moore1962machine}
Edward F. Moore. ``Machine models of self-reproduction''. In: \emph{Mathematics Problems in Biological
Sciences} (New York City, US. Apr. 5--8, 1961). Ed. by R. E. Bellman. Vol. XIV. Proceedings of a Symposia in
Applied Mathematics. AMS, 1962, pp. 17--33.
%
\bibitem[MOTW95]{maraist1995call-by-name}
John Maraist, Martin Odersky, David N. Turner, and Philip Wadler. ``Call-by-name, call-by-value,
call-by-need and the linear lambda calculus''. In: \emph{Proceedings of the 11th Annual Conference on
Mathematical Foundations of Programming Semantics, MFPS XI} (New Orleans, Louisiana, USA). Vol. 1.
Electronic Notes in Theoretical Computer Science. See also journal version \cite{maraist1999call-by-name}. 1995, pp. 370--392.
\textsc{doi}: \href {https://doi.org/10.1016/S1571-0661(04)00022-2} {\nolinkurl
{10.1016/S1571-0661(04)00022-2}}.
%
\bibitem[MOTW99]{maraist1999call-by-name}
John Maraist, Martin Odersky, David N. Turner, and Philip Wadler. ``Call-by-name, call-by-value,
call-by-need and the linear lambda calculus''. In: \emph{Theoretical Computer Science} 228.1-2 (1999). Long
version of \cite{maraist1995call-by-name}., pp. 175--210. \textsc{doi}: \href {https://doi.org/10.1016/S0304-3975(98)00358-2}
{\nolinkurl {10.1016/S0304-3975(98)00358-2}}.
%
\bibitem[MRBA16]{mcclean2016theory}
Jarrod R McClean, Jonathan Romero, Ryan Babbush, and Alán Aspuru-Guzik. ``The theory of variational
hybrid quantum-classical algorithms''. In: \emph{New Journal of Physics} 18 (2016), p. 023023. \textsc{doi}:
\href {https://doi.org/10.1088/1367-2630/18/2/023023} {\nolinkurl {10.1088/1367-2630/18/2/023023}}.
\textsc{arXiv}: \href {https://www.arxiv.org/abs/1509.04279} {\nolinkurl {1509.04279}}.
%
\bibitem[MS99]{molmer1999multiparticle}
Klaus Mølmer and Anders Sørensen. ``Multiparticle entanglement of hot trapped ions''. In: \emph{Physical
Review Letters} 82 (9 1999), pp. 1835--1838. \textsc{doi}: \href {https://doi.org/10.1103/PhysRevLett.82.1835}
{\nolinkurl {10.1103/PhysRevLett.82.1835}}.
%
\bibitem[MSRH20]{meuli2020enabling}
Giulia Meuli, Mathias Soeken, Martin Roetteler, and Thomas Häner. ``Enabling accuracy-aware quantum
compilers using symbolic resource estimation''. In: \emph{Proceedings of the ACM on Programming
Languages} 4.OOPSLA (2020), 130:1--130:26. \textsc{doi}: \href {https://doi.org/10.1145/3428198} {\nolinkurl
{10.1145/3428198}}.
%
\bibitem[MSRM19]{meuli2019ros}
Giulia Meuli, Mathias Soeken, Martin Roetteler, and Giovanni De Micheli. ``ROS: resource constrained
oracle synthesis for quantum computers''. In: \cite{qpl2019}, pp. 119--130. \textsc{doi}: \href
{https://doi.org/10.4204/EPTCS.318.8} {\nolinkurl {10.4204/EPTCS.318.8}}.
%
\bibitem[MSS07]{magniez2007quantum}
Frédéric Magniez, Miklos Santha, and Mario Szegedy. ``Quantum algorithms for the triangle problem''. In:
\emph{SIAM Journal on Computing} 37.2 (2007), pp. 413--424. \textsc{doi}: \href
{https://doi.org/10.1137/050643684} {\nolinkurl {10.1137/050643684}}.
%
\bibitem[MSS13]{malherbe2013presheaf}
Octavio Malherbe, Philip Scott, and Peter Selinger. ``Presheaf models of quantum computation: an outline''.
In: \emph{Computation, Logic, Games, and Quantum Foundations. The Many Facets of Samson Abramsky --
Essays Dedicated to Samson Abramsky on the Occasion of His 60th Birthday}. Ed. by Bob Coecke, Luke Ong,
and Prakash Panangaden. Vol. 7860. Lecture Notes in Computer Science. Springer, 2013, pp. 178--194.
\textsc{doi}: \href {https://doi.org/10.1007/978-3-642-38164-5_13} {\nolinkurl {10.1007/978-3-642-38164-5_13}}.
\textsc{arXiv}: \href {https://www.arxiv.org/abs/1302.5652} {\nolinkurl {1302.5652}}.
%
\bibitem[MTT09]{mellies2009explicit}
Paul-André Melliès, Nicolas Tabareau, and Christine Tasson. ``An explicit formula for the free exponential
modality of linear logic''. In: \emph{Proceedings of the 36th Internatilonal Colloquium on Automata,
Languages and Programming, ICALP 2009, Part II} (Rhodes, Greece, July 5--12, 2009). Ed. by Susanne Albers,
Alberto Marchetti-Spaccamela, Yossi Matias, Sotiris E. Nikoletseas, and Wolfgang Thomas. Vol. 5556. Lecture
Notes in Computer Science. Springer, 2009, pp. 247--260. \textsc{isbn}: 978-3-642-02929-5. \textsc{doi}: \href
{https://doi.org/10.1007/978-3-642-02930-1_21} {\nolinkurl {10.1007/978-3-642-02930-1_21}}. \textsc{hal}: \href
{https://hal.archives-ouvertes.fr/hal-00391714} {\nolinkurl {hal-00391714}}.
%
\bibitem[MV06]{mottonen2006decompositions}
M. Mottonen and J. J. Vartiainen. ``Decompositions of general quantum gates''. In: \emph{Trends in Quantum
Computing Research}. Ed. by Susan Shannon. Nova Science Publishers, 2006. Chap. 7. \textsc{arXiv}: \href
{https://www.arxiv.org/abs/quant-ph/0504100} {\nolinkurl {quant-ph/0504100}}.
%
\bibitem[MWD09]{miller2009synthesizing}
D. Michael Miller, Robert Wille, and Gerhard W. Dueck. ``Synthesizing reversible circuits for irreversible
functions''. In: \emph{Proceedings of the 12th Euromicro Conference on Digital System Design,
Architectures, Methods and Tools, DSD 2009} (Patras, Greece, Aug. 27--29, 2009). Ed. by Antonio Núñez and
Pedro P. Carballo. IEEE Computer Society, 2009, pp. 749--756. \textsc{isbn}: 978-0-7695-3782-5. \textsc{doi}:
\href {https://doi.org/10.1109/DSD.2009.186} {\nolinkurl {10.1109/DSD.2009.186}}.
%
\bibitem[NC02]{nielsen02quantum}
Michael A. Nielsen and Isaac L. Chuang. \emph{Quantum Computation and Quantum Information}.
Cambridge University Press, 2002. \textsc{isbn}: 0-521-63503-9.
%
\bibitem[Nie97]{nielsen1997computable}
M. A. Nielsen. ``Computable functions, quantum measurements, and quantum dynamics''. In:
\emph{Physical Review Letters} 79 (15 1997), pp. 2915--2918. \textsc{doi}: \href
{https://doi.org/10.1103/PhysRevLett.79.2915} {\nolinkurl {10.1103/PhysRevLett.79.2915}}. \textsc{arXiv}: \href
{https://www.arxiv.org/abs/quant-ph/9706006} {\nolinkurl {quant-ph/9706006}}.
%
\bibitem[Nik04]{nikhil2004bluespec}
Rishiyur S. Nikhil. ``BlueSpec System Verilog: efficient, correct RTL from high level specifications''. In:
\emph{Proceedings of the 2nd ACM \& IEEE International Conference on Formal Methods and Models for
Co-Design, MEMOCODE 2004} (San Diego, California, USA, June 23--25, 2004). IEEE Computer Society,
2004, pp. 69--70. \textsc{isbn}: 0-7803-8509-8. \textsc{doi}: \href
{https://doi.org/10.1109/MEMCOD.2004.1459818} {\nolinkurl {10.1109/MEMCOD.2004.1459818}}.
%
\bibitem[NM07]{naurois2007correctness}
Paulin Jacobé de Naurois and Virgile Mogbil. ``Correctness of multiplicative (and exponential) proof
structures is \emph{NL} -complete''. In: \emph{Proceedings of the 21st International Workshop on Computer
Science Logic and 16th Annual Conference of the EACSL, CSL 2007} (Lausanne, Switzerland, Sept. 11--15,
2007). Ed. by Jacques Duparc and Thomas A. Henzinger. Vol. 4646. Lecture Notes in Computer Science.
Springer, 2007, pp. 435--450. \textsc{isbn}: 978-3-540-74914-1. \textsc{doi}: \href
{https://doi.org/10.1007/978-3-540-74915-8_33} {\nolinkurl {10.1007/978-3-540-74915-8_33}}. \textsc{hal}: \href
{https://hal.archives-ouvertes.fr/hal-00143926} {\nolinkurl {hal-00143926}}.
%
\bibitem[Nor98]{norrish1998c}
Michael Norrish. ``C Formalised in HOL''. Also: Technical Report UCAM-CL-TR-453. PhD thesis. University
of Cambridge, 1998.
%
\bibitem[NRSCM18]{nam2018automated}
Yunseong Nam, Neil J. Ross, Yuan Su, Andrew M. Childs, and Dmitri Maslov. ``Automated optimization of
large quantum circuits with continuous parameters''. In: \emph{npj Quantum Information} 4.1 (2018), p. 23.
\textsc{doi}: \href {https://doi.org/10.1038/s41534-018-0072-4} {\nolinkurl {10.1038/s41534-018-0072-4}}.
%
\bibitem[NST18]{nollet2018local}
Rémi Nollet, Alexis Saurin, and Christine Tasson. ``Local validity for circular proofs in linear logic with fixed
points''. In: \emph{27th EACSL Annual Conference on Computer Science Logic, CSL 2018} (Birmingham, UK,
Sept. 4--7, 2018). Ed. by Dan R. Ghica and Achim Jung. Vol. 119. LIPIcs. Schloss Dagstuhl - Leibniz-Zentrum
für Informatik, 2018, 35:1--35:23. \textsc{isbn}: 978-3-95977-088-0. \textsc{doi}: \href
{https://doi.org/10.4230/LIPIcs.CSL.2018.35} {\nolinkurl {10.4230/LIPIcs.CSL.2018.35}}. \textsc{hal}: \href
{https://hal.archives-ouvertes.fr/hal-01825477} {\nolinkurl {hal-01825477}}.
%
\bibitem[NTR11]{nachtigal2011design}
Michael Nachtigal, Himanshu Thapliyal, and Nagarajan Ranganathan. ``Design of a reversible floating-point
adder architecture''. In: \emph{Proceedings of the 11th IEEE International Conference on Nanotechnology}.
2011, pp. 451--456. \textsc{doi}: \href {https://doi.org/10.1109/NANO.2011.6144358} {\nolinkurl
{10.1109/NANO.2011.6144358}}.
%
\bibitem[NV14]{nguyen2013space-efficient}
Trung Duc Nguyen and Rodney Van Meter. ``A resource-efficient design for a reversible floating point adder
in quantum computing''. In: \emph{ACM Journal on Emerging Technologies in Computing Systems} 11.2
(2014), 13:1--13:18. \textsc{issn}: 1550-4832. \textsc{doi}: \href {https://doi.org/10.1145/2629525} {\nolinkurl
{10.1145/2629525}}. \textsc{arXiv}: \href {https://www.arxiv.org/abs/1306.3760} {\nolinkurl {1306.3760}}.
%
\bibitem[NW06]{nocedal2006numerical}
Jorge Nocedal and Stephen J. Wright. \emph{Numerical Optimization}. Springer Series in Operations
Research and Financial Engineering (ORFE). Springer, 2006.
%
\bibitem[Öme00]{omer2000quantum}
Berhnard Ömer. ``Quantum Programming in QCL''. PhD thesis. TU Wien, 2000.
%
\bibitem[Öme03]{omer2003structured}
Berhnard Ömer. ``Structured Quantum Programming''. PhD thesis. TU Wien, 2003.
%
\bibitem[Per08]{perdrix2008quantum}
Simon Perdrix. ``Quantum entanglement analysis based on abstract interpretation''. In: \emph{Proceedings
of the 15th International Symposium on Static Analysis (SAS'08)} (Valencia, Spain). Ed. by María Alpuente
and Germán Vidal. Vol. 5079. Lecture Notes in Computer Science. Springer, 2008, pp. 270--282. \textsc{doi}:
\href {https://doi.org/10.1007/978-3-540-69166-2_18} {\nolinkurl {10.1007/978-3-540-69166-2_18}}.
\textsc{arXiv}: \href {https://www.arxiv.org/abs/0801.4230} {\nolinkurl {0801.4230}}.
%
\bibitem[PHW06]{pierro2006reversible}
Alessandra Di Pierro, Chris Hankin, and Herbert Wiklicky. ``Reversible combinatory logic''. In:
\emph{Mathematical Structures in Computer Science} 16.4 (2006), pp. 621--637. \textsc{doi}: \href
{https://doi.org/10.1017/S0960129506005391} {\nolinkurl {10.1017/S0960129506005391}}.
%
\bibitem[Pie02]{pierce02types}
Benjamin C. Pierce. \emph{Types and Programming Languages}. MIT Press, 2002. \textsc{isbn}:
0-262-16209-1.
%
\bibitem[Pit13]{pitts2013nominal}
Andrew M. Pitts. \emph{Nominal Sets: Names and Symmetry in Computer Science}. Vol. 57. Cambridge
Tracts in Theoretical Computer Science. Cambridge University Press, 2013. \textsc{isbn}: 978-1-107-01778-8.
%
\bibitem[PKI08]{park2008functional}
Sungwoo Park, Jinha Kim, and Hyeonseung Im. ``Functional netlists''. In: \emph{Proceeding of the 13th ACM
SIGPLAN international conference on Functional programming, ICFP 2008} (Victoria, BC, Canada,
Sept. 20--28, 2008). Ed. by James Hook and Peter Thiemann. ACM, 2008, pp. 353--366. \textsc{isbn}:
978-1-59593-919-7. \textsc{doi}: \href {https://doi.org/10.1145/1411204.1411253} {\nolinkurl
{10.1145/1411204.1411253}}.
%
\bibitem[Plo75]{plotkin75callbyname}
Gordon D. Plotkin. ``Call-by-name, call-by-value and the lambda-calculus''. In: \emph{Theoretical Computer
Science} 1.2 (1975), pp. 125--159. \textsc{doi}: \href {https://doi.org/10.1016/0304-3975(75)90017-1} {\nolinkurl
{10.1016/0304-3975(75)90017-1}}.
%
\bibitem[Plo83]{plotkin83domains}
Gordon D. Plotkin. ``Domains, Pisa Notes''. Course notes on domain theory, available on the author's
website under the name ``Pisa Notes''. 1983.
%
\bibitem[PMAC+15]{procopio2015experimental}
Lorenzo M. Procopio, Amir Moqanaki, Mateus Araújo, Fabio Costa, Irati Alonso Calafell, Emma G. Dowd,
Deny R. Hamel, Lee A. Rozema, Časlav Brukner, and Philip Walther. ``Experimental superposition of orders
of quantum gates''. In: \emph{Nature Communications} 6.1 (2015), p. 7913. \textsc{doi}: \href
{https://doi.org/10.1038/ncomms8913} {\nolinkurl {10.1038/ncomms8913}}. \textsc{arXiv}: \href
{https://www.arxiv.org/abs/1412.4006} {\nolinkurl {1412.4006}}.
%
\bibitem[PMMRT17]{portmann2017causal}
Christopher Portmann, Christian Matt, Ueli Maurer, Renato Renner, and Björn Tackmann. ``Causal boxes:
quantum information-processing systems closed under composition''. In: \emph{IEEE Transactions on
Information Theory} 63.5 (2017), pp. 3277--3305. \textsc{doi}: \href {https://doi.org/10.1109/TIT.2017.2676805}
{\nolinkurl {10.1109/TIT.2017.2676805}}. \textsc{arXiv}: \href {https://www.arxiv.org/abs/1512.02240} {\nolinkurl
{1512.02240}}.
%
\bibitem[PMSY+14]{peruzzo2014variational}
Alberto Peruzzo, Jarrod McClean, Peter Shadbolt, Man-Hong Yung, Xiao-Qi Zhou, Peter J. Love,
Alán Aspuru-Guzik, and Jeremy L. O'Brien. ``A variational eigenvalue solver on a photonic quantum
processor''. In: \emph{Nature} 5 (2014), p. 4213. \textsc{doi}: \href {https://doi.org/10.1038/ncomms5213}
{\nolinkurl {10.1038/ncomms5213}}.
%
\bibitem[POPL14]{popl2014}
Suresh Jagannathan and Peter Sewell, eds. \emph{Proceedings of the 41st ACM SIGPLAN-SIGACT
Symposium on Principles of Programming Languages, POPL'14} (San Diego, California, USA). ACM, 2014.
\textsc{isbn}: 978-1-4503-2544-8.
%
\bibitem[POPL17]{popl2017}
Giuseppe Castagna and Andrew D. Gordon, eds. \emph{Proceedings of the 44th ACM SIGPLAN Symposium
on Principles of Programming Languages, POPL'17} (Paris, France). ACM, 2017. \textsc{isbn}:
978-1-4503-4660-3. \textsc{doi}: \href {https://doi.org/10.1145/3009837} {\nolinkurl {10.1145/3009837}}.
%
\bibitem[PPRZ20]{pechoux2020quantum}
Romain Péchoux, Simon Perdrix, Mathys Rennela, and Vladimir Zamdzhiev. ``Quantum programming with
inductive datatypes: causality and affine type theory''. In: \emph{Proceedings of the 23rd International
Conference on Foundations of Software Science and Computation Structures, FoSSaCS 2020} (Dublin,
Ireland, Apr. 25--30, 2020). Ed. by Jean Goubault-Larrecq and Barbara König. Vol. 12077. Lecture Notes in
Computer Science. Springer, 2020, pp. 562--581. \textsc{isbn}: 978-3-030-45230-8. \textsc{doi}: \href
{https://doi.org/10.1007/978-3-030-45231-5_29} {\nolinkurl {10.1007/978-3-030-45231-5_29}}. \textsc{hal}: \href
{https://hal.archives-ouvertes.fr/hal-03018513} {\nolinkurl {hal-03018513}}.
%
\bibitem[PPZ19]{paolini2019qpcf}
Luca Paolini, Mauro Piccolo, and Margherita Zorzi. ``qPCF: higher-order languages and quantum circuits''.
In: \emph{Journal of Automated Reasoning} 63.4 (2019). Extended version of a TAMC'17 paper \cite{paolini2017qpcf}.,
pp. 941--966. \textsc{doi}: \href {https://doi.org/10.1007/s10817-019-09518-y} {\nolinkurl
{10.1007/s10817-019-09518-y}}.
%
\bibitem[Pra81]{pratt1981decidable}
Vaughan R. Pratt. ``A decidable mu-calculus: preliminary report''. In: \emph{Proceedings of the 22nd Annual
Symposium on Foundations of Computer Science (FOCS'81)} (Nashville, Tennessee, USA). IEEE Computer
Society, 1981, pp. 421--427. \textsc{doi}: \href {https://doi.org/10.1109/SFCS.1981.4} {\nolinkurl
{10.1109/SFCS.1981.4}}.
%
\bibitem[Pre18]{preskill2018nisq}
John Preskill. ``Quantum computing in the NISQ era and beyond''. In: \emph{Quantum} 2 (2018), p. 79.
\textsc{doi}: \href {https://doi.org/10.22331/q-2018-08-06-79} {\nolinkurl {10.22331/q-2018-08-06-79}}.
\textsc{arXiv}: \href {https://www.arxiv.org/abs/1801.00862v3} {\nolinkurl {1801.00862v3}}.
%
\bibitem[PRS17]{parent2017revs}
Alex Parent, Martin Roetteler, and Krysta M. Svore. ``REVS: a tool for space-optimized reversible circuit
synthesis''. In: \cite{rc2017}, pp. 90--101. \textsc{doi}: \href {https://doi.org/10.1007/978-3-319-59936-6_7} {\nolinkurl
{10.1007/978-3-319-59936-6_7}}. \textsc{arXiv}: \href {https://www.arxiv.org/abs/1510.00377} {\nolinkurl
{1510.00377}}.
%
\bibitem[PRZ17]{paykin2017qwire}
Jennifer Paykin, Robert Rand, and Steve Zdancewic. ``QWIRE: a core language for quantum circuits''. In:
[POPL17], pp. 846--858. \textsc{doi}: \href {https://doi.org/10.1145/3009837.3009894} {\nolinkurl
{10.1145/3009837.3009894}}. \textsc{hal}: \href {https://hal.archives-ouvertes.fr/hal-01474620} {\nolinkurl
{hal-01474620}}. \textsc{arXiv}: \href {https://www.arxiv.org/abs/1610.09629} {\nolinkurl {1610.09629}}.
%
\bibitem[PS14]{paetznick2014repeat-until-success}
Adam Paetznick and Krysta M. Svore. ``Repeat-until-success: non-deterministic decomposition of
single-qubit unitaries''. In: \emph{Quantum Information and Computation} 14.15-016 (2014), pp. 1277--1301.
\textsc{doi}: \href {https://doi.org/10.26421/QIC14.15-16-2} {\nolinkurl {10.26421/QIC14.15-16-2}}.
\textsc{arXiv}: \href {https://www.arxiv.org/abs/1311.1074} {\nolinkurl {1311.1074}}.
%
\bibitem[PSV14]{pagani2014applying}
Michele Pagani, Peter Selinger, and Benoît Valiron. ``Applying quantitative semantics to higher-order
quantum computing''. In: \cite{popl2014}, pp. 647--658. \textsc{doi}: \href {https://doi.org/10.1145/2535838.2535879}
{\nolinkurl {10.1145/2535838.2535879}}. \textsc{arXiv}: \href {https://www.arxiv.org/abs/1311.2290} {\nolinkurl
{1311.2290}}.
%
\bibitem[pub22]{zx-publications}
ZX-calculus publications. \emph{East, Richard and van de Wetering, John}. 2022. \textsc{url}: \url
{https://zxcalculus.com/publications.html} (visited on Aug. 24, 2022).
%
\bibitem[PZ17]{paolini2017qpcf}
Luca Paolini and Margherita Zorzi. ``qPCF: a language for quantum circuit computations''. In:
\emph{Proceedings of the 14th Annual Conference on Theory and Applications of Models of Computation,
TAMC'17} (Bern, Switzerland). Ed. by T. V. Gopal, Gerhard Jäger, and Silvia Steila. Vol. 10185. Lecture Notes
in Computer Science. Extended journal version: \cite{paolini2019qpcf}. 2017, pp. 455--469. \textsc{isbn}: 978-3-319-55910-0.
\textsc{doi}: \href {https://doi.org/10.1007/978-3-319-55911-7_33} {\nolinkurl {10.1007/978-3-319-55911-7_33}}.
%
\bibitem[QCS]{iarpa-qcs}
\emph{Official webpage for the IARPA research program QCS}. \textsc{url}: \url
{https://www.iarpa.gov/research-programs/qcs} (visited on Aug. 30, 2024).
%
\bibitem[Qis]{qiskit}
Qiskit Development Team. \emph{Qiskit Documentation}. \textsc{url}: \url
{https://qiskit.org/documentation/} (visited on July 28, 2021).
%
\bibitem[QPL04]{qpl04}
Peter Selinger, ed. \emph{Proceedings of the Second International Workshop on Quantum Programming
Languages, QPL'04} (Turku, Finland). Vol. 33. TUCS General Publication. TUCS, 2004. \textsc{isbn}:
9-5212-1374-4. \textsc{url}: \url {http://urn.fi/URN:NBN:fi:bib:me:I00035039900}.
%
\bibitem[QPL07]{qpl2005}
Peter Selinger, ed. \emph{Proceedings of the 3rd International Workshop on Quantum Programming
Languages, QPL'05} (DePaul University, Chicago, USA). Vol. 170. Electronic Notes in Theoretical Computer
Science. 2007.
%
\bibitem[QPL08]{qpl06}
Peter Selinger, ed. \emph{Proceedings of the Fourth International Workshop on Quantum Programming
Languages, QPL'06} (Oxford, UK.). Vol. 210. Electronic Notes in Theoretical Computer Science. July 2008.
%
\bibitem[QPL11]{qpl2008}
B. Coecke, I. Mackie, P. Panangaden, and P. Selinger, eds. \emph{Proceedings of the Joint 5th International
Workshop on Quantum Physics and Logic and 4th Workshop on Developments in Computational Models,
QPL/DCM 2008} (Reykjavik, Iceland). Vol. 270-1. Electronic Notes in Theoretical Computer Science. 2011.
%
\bibitem[QPL14]{qpl2014}
Bob Coecke, Ichiro Hasuo, and Prakash Panangaden, eds. \emph{Proceedings of the 11th workshop on
Quantum Physics and Logic, QPL 2014} (Kyoto, Japan). Vol. 172. Electronic Proceedings in Theoretical
Computer Science. 2014. \textsc{doi}: \href {https://doi.org/10.4204/EPTCS.172} {\nolinkurl
{10.4204/EPTCS.172}}.
%
\bibitem[QPL18]{qpl2017}
Bob Coecke and Aleks Kissinger, eds. \emph{Proceedings 14th International Conference on Quantum
Physics and Logic, QPL 2017} (Nijmegen, The Netherlands). Vol. 266. Electronic Proceedings in Theoretical
Computer Science. 2018.
%
\bibitem[QPL19]{qpl2018}
Peter Selinger and Giulio Chiribella, eds. \emph{Proceedings 15th International Conference on Quantum
Physics and Logic, QPL 2018} (Halifax, Canada, June 3--7, 2018). Vol. 287. EPTCS. 2019.
%
\bibitem[QPL20a]{qpl2019}
Bob Coecke and Matthew Leifer, eds. \emph{Proceedings 16th International Conference on Quantum
Physics and Logic, QPL 2019} (Chapman University, Orange, CA, USA, June 10--14, 2019). Vol. 318. EPTCS.
2020.
%
\bibitem[QPL20b]{qpl2020}
Benoît Valiron, Shane Mansfield, Pablo Arrighi, and Prakash Panangaden, eds. \emph{Proceedings 17th
International Conference on Quantum Physics and Logic, QPL 2020} (Online (due to Covid), June 2--6, 2020).
Vol. 340. EPTCS. 2020.
%
\bibitem[QPL23a]{qpl2022}
Stefano Gogioso and Matty Hoban, eds. \emph{Proceedings 19th International Conference on Quantum
Physics and Logic, QPL 2022} (Wolfson College, Oxford, UK, June 27--July 1, 2022). Vol. 394. EPTCS. 2023.
%
\bibitem[QPL23b]{qpl2023}
Shane Mansfield, Benoît Valiron, and Vladimir Zamdzhiev, eds. \emph{Proceedings of the Twentieth
International Conference on Quantum Physics and Logic, QPL 2023} (Paris, France, July 17--21, 2023).
Vol. 384. EPTCS. 2023. \textsc{doi}: \href {https://doi.org/10.4204/EPTCS.384} {\nolinkurl {10.4204/EPTCS.384}}.
%
\bibitem[QTOOLS24]{list-tools}
GQI. \emph{List of Tools, by Quantum Computing Report}. 2024. \textsc{url}: \url
{https://quantumcomputingreport.com/tools/} (visited on June 30, 2024).
%
\bibitem[Qui20]{quingo}
Quingo Development Team. ``Quingo: A Programming Framework for Heterogeneous Quantum-Classical
Computing with NISQ Features''. 2020. \textsc{arXiv}: \href {https://www.arxiv.org/abs/2009.01686}
{\nolinkurl {2009.01686}}.
%
\bibitem[QZOO22]{quantum-zoo}
Stephen Jordan. \emph{Quantum Algorithm Zoo}. 2022. \textsc{url}: \url {https://quantumalgorithmzoo.org/}
(visited on Sept. 10, 2022).
%
\bibitem[Ran14]{ranchin2014depicting}
André Ranchin. ``Depicting qudit quantum mechanics and mutually unbiased qudit theories''. In: \cite{qpl2014},
pp. 68--91.
%
\bibitem[Ran18]{rand2018phd}
Robert Rand. ``Formally Verified Quantum Programming''. PhD thesis. University of Pennsylania, 2018.
\textsc{url}: \url {https://repository.upenn.edu/edissertations/3175/} (visited on Aug. 29, 2021).
%
\bibitem[RB01]{raussendorf2001one-way}
Robert Raussendorf and Hans J. Briegel. ``A one-way quantum computer''. In: \emph{Physical Review
Letters} 86 (22 2001), pp. 5188--5191. \textsc{doi}: \href {https://doi.org/10.1103/PhysRevLett.86.5188}
{\nolinkurl {10.1103/PhysRevLett.86.5188}}. \textsc{arXiv}: \href {https://www.arxiv.org/abs/quant-ph/0510135}
{\nolinkurl {quant-ph/0510135}}.
%
\bibitem[RBB03]{mraussendorf2003mbqc}
Robert Raussendorf, Daniel E. Browne, and Hans J. Briegel. ``Measurement-based quantum computation on
cluster states''. In: \emph{Physical Review A} 68 (2 2003), p. 022312. \textsc{doi}: \href
{https://doi.org/10.1103/PhysRevA.68.022312} {\nolinkurl {10.1103/PhysRevA.68.022312}}. \textsc{arXiv}: \href
{https://www.arxiv.org/abs/quant-ph/0301052} {\nolinkurl {quant-ph/0301052}}.
%
\bibitem[RC13]{rc2013}
Gerhard W. Dueck and D. Michael Miller, eds. \emph{Proceedings of the 5th International Conference on
Reversible Computation, RC'13} (Victoria, BC, Canada, July 4--5, 2013). Vol. 7948. Lecture Notes in Computer
Science. Springer, 2013. \textsc{isbn}: 978-3-642-38985-6. \textsc{doi}: \href
{https://doi.org/10.1007/978-3-642-38986-3} {\nolinkurl {10.1007/978-3-642-38986-3}}.
%
\bibitem[RC17]{rc2017}
Iain Phillips and Hafizur Rahaman, eds. \emph{Proceedings of the 9th International Conference on
Reversible Computation, RC'17} (Kolkata, India, July 6--7, 2017). Vol. 10301. Lecture Notes in Computer
Science. Springer, 2017. \textsc{isbn}: 978-3-319-59935-9. \textsc{doi}: \href
{https://doi.org/10.1007/978-3-319-59936-6} {\nolinkurl {10.1007/978-3-319-59936-6}}.
%
\bibitem[RC18]{rc2018}
Jarkko Kari and Irek Ulidowski, eds. \emph{Proceedings of the 10th International Conference on Reversible
Computation, RC 2018} (Leicester, UK). Vol. 11106. Lecture Notes in Computer Science. Springer, 2018.
\textsc{isbn}: 978-3-319-99497-0. \textsc{doi}: \href {https://doi.org/10.1007/978-3-319-99498-7} {\nolinkurl
{10.1007/978-3-319-99498-7}}.
%
\bibitem[RC19]{rc2019}
Michael Kirkedal Thomsen and Mathias Soeken, eds. \emph{Proceedings of the 11th International
Conference on Reversible Computation, RC 2019} (Lausanne, Switzerland). Vol. 11497. Lecture Notes in
Computer Science. Springer, 2019. \textsc{isbn}: 978-3-030-21499-9. \textsc{doi}: \href
{https://doi.org/10.1007/978-3-030-21500-2} {\nolinkurl {10.1007/978-3-030-21500-2}}.
%
\bibitem[RC20]{rc2020}
Ivan Lanese and Mariusz Rawski, eds. \emph{Proceedings of the 12th International Conference on
Reversible Computation, RC 2020}. Vol. 12227. Lecture Notes in Computer Science. Springer, 2020.
\textsc{isbn}: 978-3-030-52481-4. \textsc{doi}: \href {https://doi.org/10.1007/978-3-030-52482-1} {\nolinkurl
{10.1007/978-3-030-52482-1}}.
%
\bibitem[RC21]{rc2021}
Shigeru Yamashita and Tetsuo Yokoyama, eds. \emph{Proceedings of the 13th International Conference on
Reversible Computation, RC 2021} (Virtual Event, July 7--8, 2021). Vol. 12805. Lecture Notes in Computer
Science. Springer, 2021. \textsc{isbn}: 978-3-030-79836-9. \textsc{doi}: \href
{https://doi.org/10.1007/978-3-030-79837-6} {\nolinkurl {10.1007/978-3-030-79837-6}}.
%
\bibitem[Reg04]{regev2004quantum}
Oded Regev. ``Quantum computation and lattice problems''. In: \emph{SIAM Journal on Computing} 33.3
(2004), pp. 738--760. \textsc{doi}: \href {https://doi.org/10.1137/S0097539703440678} {\nolinkurl
{10.1137/S0097539703440678}}.
%
\bibitem[Reg92]{regnier1992phd}
Laurent Regnier. ``Lambda-Calcul et Réseaux''. Thèse de Doctorat. Université Paris 7, 1992.
%
\bibitem[RG17]{ruiz-perez2017quantum}
Lidia Ruiz-Perez and Juan Carlos García-Escartín. ``Quantum arithmetic with the quantum fourier
transform''. In: \emph{Quantum Information Processing} 16.6 (2017), p. 152. \textsc{doi}: \href
{https://doi.org/10.1007/s11128-017-1603-1} {\nolinkurl {10.1007/s11128-017-1603-1}}.
%
\bibitem[RLNS00]{esc-java-manual}
K. Rustan, M. Leino, Greg Nelson, and James B. Saxe. \emph{ESC/Java User's Manual}. SRC Technical Note
2000-002. Compaq Computer Corporation, 2000.
%
\bibitem[Rob97]{robinson1997b}
Ken Robinson. ``The B method and the B toolkit''. In: \emph{Proceedings of the 6th International Conference
on Algebraic Methodology and Software Technology, AMAST'97} (Sydney,Australia, Dec. 13--17, 1997). Ed. by
Michael Johnson. Vol. 1349. Lecture Notes in Computer Science. Springer, 1997, pp. 576--580. \textsc{isbn}:
3-540-63888-1. \textsc{doi}: \href {https://doi.org/10.1007/BFb0000503} {\nolinkurl {10.1007/BFb0000503}}.
%
\bibitem[Roe74]{deroever1974recursive}
W. P. de Roever. ``Recursive Program Schemes : Semantics and Proof Theory''. Proefschrift. Vrije Universiteit
te Amsterdam, 1974.
%
\bibitem[Ros15]{ross2015algebraic}
Neil J. Ross. ``Algebraic and Logical Methods in Quantum Computation''. PhD thesis. Dalhousie University,
2015. \textsc{arXiv}: \href {https://www.arxiv.org/abs/1510.02198} {\nolinkurl {1510.02198}}.
%
\bibitem[RP10]{reed2010distance}
Jason Reed and Benjamin C. Pierce. ``Distance makes the types grow stronger: a calculus for differential
privacy''. In: \emph{Proceeding of the 15th ACM SIGPLAN international conference on Functional
programming, ICFP 2010} (Baltimore, Maryland, USA, Sept. 27--29, 2010). Ed. by Paul Hudak and
Stephanie Weirich. ACM, 2010, pp. 157--168. \textsc{isbn}: 978-1-60558-794-3. \textsc{doi}: \href
{https://doi.org/10.1145/1863543.1863568} {\nolinkurl {10.1145/1863543.1863568}}.
%
\bibitem[RPLZ18]{rand2018reqwire}
Robert Rand, Jennifer Paykin, Dong-Ho Lee, and Steve Zdancewic. ``Reqwire: reasoning about reversible
quantum circuits''. In: \cite{qpl2018}, pp. 299--312. \textsc{doi}: \href {https://doi.org/10.4204/EPTCS.287.17}
{\nolinkurl {10.4204/EPTCS.287.17}}.
%
\bibitem[RS18a]{rennela2017classical}
Mathys Rennela and Sam Staton. ``Classical control and quantum circuits in enriched category theory''. In:
\emph{Proceedings of the Thirty-Fourth Conference on the Mathematical Foundations of Programming
Semantics, MFPS 2018} (Dalhousie University, Halifax, Canada, June 6--9, 2018). Ed. by Sam Staton. Vol. 341.
Electronic Notes in Theoretical Computer Science. See also the journal version \cite{rennela2020classical}. Elsevier, 2018,
pp. 257--279. \textsc{doi}: \href {https://doi.org/10.1016/j.entcs.2018.03.027} {\nolinkurl
{10.1016/j.entcs.2018.03.027}}. \textsc{arXiv}: \href {https://www.arxiv.org/abs/1711.05159} {\nolinkurl
{1711.05159}}.
%
\bibitem[RS18b]{rios2017categorical}
Francisco Rios and Peter Selinger. ``A categorical model for a quantum circuit description language''. In:
[QPL18], pp. 164--178. \textsc{doi}: \href {https://doi.org/10.4204/EPTCS.266.11} {\nolinkurl
{10.4204/EPTCS.266.11}}. \textsc{arXiv}: \href {https://www.arxiv.org/abs/1706.02630} {\nolinkurl {1706.02630}}.
%
\bibitem[RS20]{rennela2020classical}
Mathys Rennela and Sam Staton. ``Classical control, quantum circuits and linear logic in enriched category
theory''. In: \emph{Logical Methods in Computer Science} 16.1 (2020). Journal version of an MFPS
publication \cite{rennela2017classical}. \textsc{doi}: \href {https://doi.org/10.23638/LMCS-16(1:30)2020} {\nolinkurl
{10.23638/LMCS-16(1:30)2020}}.
%
\bibitem[RZBB94]{reck1994experimental}
Michael Reck, Anton Zeilinger, Herbert J. Bernstein, and Philip Bertani. ``Experimental realization of any
discrete unitary operator''. In: \emph{Physical Review Letters} 73 (1 July 1994), pp. 58--61. \textsc{doi}: \href
{https://doi.org/10.1103/PhysRevLett.73.58} {\nolinkurl {10.1103/PhysRevLett.73.58}}.
%
\bibitem[Saa03]{saad2003iterative}
Yousef Saad. \emph{Iterative Methods for Sparse Linear Systems}. 2nd ed. SIAM, 2003. \textsc{isbn}:
978-0-89871-534-7.
%
\bibitem[SB69]{scott1969theory}
Dana Scott and J. W. de Bakker. ``A Theory of Programs''. Unpublished, manuscript notes for an IBM
Seminar, Vienna, August 1969. Reprinted in \cite{liber-amicorum:1989}. 1969.
%
\bibitem[SBM06]{shende2006synthesis}
Vivek V. Shende, Stephen S. Bullock, and Igor L. Markov. ``Synthesis of quantum-logic circuits''. In:
\emph{IEEE Transactions on Computer-Aided Design of Integrated Circuits and Systems} 25.6 (2006),
pp. 1000--1010. \textsc{doi}: \href {https://doi.org/10.1109/TCAD.2005.855930} {\nolinkurl
{10.1109/TCAD.2005.855930}}. \textsc{arXiv}: \href {https://www.arxiv.org/abs/quant-ph/0406176} {\nolinkurl
{quant-ph/0406176}}.
%
\bibitem[Sch08]{schlosshauer2008decoherence}
Maximilian A. Schlosshauer. \emph{Decoherence and the Quantum-To-Classical Transition}. Springer, 2008.
%
\bibitem[Sch14]{schopp2014call-by-value}
Ulrich Schöpp. ``Call-by-value in a basic logic for interaction''. In: \emph{Proceedings of the 12th Asian
Symposium on Programming Languages and Systems,(APLAS 2014)} (Singapore). Ed. by Jacques Garrigue.
Vol. 8858. Lecture Notes in Computer Science. Springer, Nov. 2014, pp. 428--448. \textsc{isbn}:
978-3-319-12735-4. \textsc{doi}: \href {https://doi.org/10.1007/978-3-319-12736-1_23} {\nolinkurl
{10.1007/978-3-319-12736-1_23}}.
%
\bibitem[Sch86]{schmidt1986denotational}
David A. Schmidt. \emph{Denotational Semantics: A Methodology for Language Development}. Allyn and
Bacon, Inc, 1986.
%
\bibitem[SCZ17]{smith2017practical}
Robert S. Smith, Michael J. Curtis, and William J. Zeng. ``A Practical Quantum Instruction Set Architecture''.
White paper presenting the assembly language Quil used in pyQuil for the Rigetti Forest framework. 2017.
\textsc{arXiv}: \href {https://www.arxiv.org/abs/1608.03355v2} {\nolinkurl {1608.03355v2}}.
%
\bibitem[SDCSED20]{sivarajah2020tket}
Seyon Sivarajah, Silas Dilkes, Alexander Cowtan, Will Simmons, Alec Edgington, and Ross Duncan. ``T|ket⟩:
a retargetable compiler for NISQ devices''. In: \emph{Quantum Science and Technology} 6.1 (2020), p. 014003.
\textsc{doi}: \href {https://doi.org/10.1088/2058-9565/ab8e92} {\nolinkurl {10.1088/2058-9565/ab8e92}}.
%
\bibitem[Sel04a]{selinger04quantum}
Peter Selinger. ``Towards a quantum programming language''. In: \emph{Mathematical Structures in
Computer Science} 14.4 (2004), pp. 527--586. \textsc{doi}: \href {https://doi.org/10.1017/S0960129504004256}
{\nolinkurl {10.1017/S0960129504004256}}.
%
\bibitem[Sel04b]{selinger04semantics}
Peter Selinger. ``Towards a semantics for higher-order quantum computation''. In: \cite{qpl04}, pp. 127--143.
%
\bibitem[Sel07]{selinger2007dagger}
Peter Selinger. ``Dagger compact closed categories and completely positive maps (extended abstract)''. In:
[QPL07], pp. 139--163. \textsc{doi}: \href {https://doi.org/10.1016/j.entcs.2006.12.018} {\nolinkurl
{10.1016/j.entcs.2006.12.018}}.
%
\bibitem[SGLH11]{swamy2011lightweight}
Nikhil Swamy, Nataliya Guts, Daan Leijen, and Michael Hicks. ``Lightweight monadic programming in ML''.
In: \cite{icfp2011}, pp. 15--27. \textsc{doi}: \href {https://doi.org/10.1145/2034773.2034778} {\nolinkurl
{10.1145/2034773.2034778}}.
%
\bibitem[SGTA+18]{svore2018qsharp}
Krysta Svore, Alan Geller, Matthias Troyer, John Azariah, Christopher Granade, Bettina Heim,
Vadym Kliuchnikov, Mariia Mykhailova, Andres Paz, and Martin Roetteler. ``Q\#: enabling scalable quantum
computing and development with a high-level DSL''. In: \emph{Proceedings of the Real World Domain
Specific Languages Workshop 2018} (Vienna, Austria). ACM, 2018, 7:1--7:10. \textsc{isbn}: 978-1-4503-6355-6.
\textsc{doi}: \href {https://doi.org/10.1145/3183895.3183901} {\nolinkurl {10.1145/3183895.3183901}}.
\textsc{arXiv}: \href {https://www.arxiv.org/abs/1803.00652} {\nolinkurl {1803.00652}}.
%
\bibitem[Sha70]{shanno1970conditioning}
D. F. Shanno. ``Conditioning of quasi-Newton methods for function minimization''. In: \emph{Mathematics
of Computation} 24 (1970), pp. 647--656.
%
\bibitem[She94]{shewchuk1994introduction}
Jonathan R. Shewchuk. \emph{An Introduction to the Conjugate Gradient Method Without the Agonizing
Pain}. Tech. rep. CMU-CS-94-125. School of Computer Science, Carnegie Mellon University, 1994.
%
\bibitem[Sho94]{shor94algorithms}
Peter W. Shor. ``Algorithms for quantum computation: discrete log and factoring''. In: \emph{Proceedings of
the 35th Annual Symposium on Foundations of Computer Science (FOCS'94)} (Santa Fe, New Mexico, US.).
IEEE. IEEE Computer Society Press, Nov. 1994, pp. 124--134. \textsc{isbn}: 0-8186-6580-7. \textsc{doi}: \href
{https://doi.org/10.1109/SFCS.1994.365700} {\nolinkurl {10.1109/SFCS.1994.365700}}.
%
\bibitem[Sho97]{shor97polynomial}
Peter W. Shor. ``Polynomial-time algorithms for prime factorization and discrete logarithms on a quantum
computer''. In: \emph{SIAM Journal on Computing} 26.5 (1997), pp. 1484--1509. \textsc{doi}: \href
{https://doi.org/10.1137/S0097539795293172} {\nolinkurl {10.1137/S0097539795293172}}. \textsc{arXiv}: \href
{https://www.arxiv.org/abs/quant-ph/9508027} {\nolinkurl {quant-ph/9508027}}.
%
\bibitem[SHT18]{projectq}
Damian S. Steiger, Thomas Häner, and Matthias Troyer. ``ProjectQ: an open source software framework for
quantum computing''. In: \emph{Quantum} 2 (2018), p. 49. \textsc{doi}: \href
{https://doi.org/10.22331/q-2018-01-31-49} {\nolinkurl {10.22331/q-2018-01-31-49}}. \textsc{arXiv}: \href
{https://www.arxiv.org/abs/1612.08091v2} {\nolinkurl {1612.08091v2}}.
%
\bibitem[Sim05]{simpson2005reduction}
Alex K. Simpson. ``Reduction in a linear lambda-calculus with applications to operational semantics''. In:
\emph{Proceedings of the 16th International Conference on Term Rewriting and Applications, RTA'05}
(Nara, Japan). Ed. by Jürgen Giesl. Vol. 3467. Lecture Notes in Computer Science. Springer, 2005, pp. 219--234.
\textsc{isbn}: 3-540-25596-6. \textsc{doi}: \href {https://doi.org/10.1007/978-3-540-32033-3_17} {\nolinkurl
{10.1007/978-3-540-32033-3_17}}.
%
\bibitem[SJA17]{sander2017forsyde}
Ingo Sander, Axel Jantsch, and Seyed-Hosein Attarzadeh-Niaki. ``ForSyDe: system design using a functional
language and models of computation''. In: \emph{Handbook of Hardware/Software Codesign}. Ed. by
Soonhoi Ha and Jürgen Teich. Springer, 2017, pp. 99--140. \textsc{isbn}: 978-94-017-7266-2. \textsc{doi}: \href
{https://doi.org/10.1007/978-94-017-7267-9_5} {\nolinkurl {10.1007/978-94-017-7267-9_5}}.
%
\bibitem[SM13]{saeedi2013synthesis}
Mehdi Saeedi and Igor L. Markov. ``Synthesis and optimization of reversible circuits - a survey''. In:
\emph{ACM Computing Surveys} 45.2 (2013), 21:1--21:34. \textsc{doi}: \href
{https://doi.org/10.1145/2431211.2431220} {\nolinkurl {10.1145/2431211.2431220}}.
%
\bibitem[SMB04]{shende2004minimal}
Vivek V. Shende, Igor L. Markov, and Stephen S. Bullock. ``Minimal universal two-qubit
controlled-NOT-based circuits''. In: \emph{Physical Review A} 69 (6 2004), p. 062321. \textsc{doi}: \href
{https://doi.org/10.1103/PhysRevA.69.062321} {\nolinkurl {10.1103/PhysRevA.69.062321}}.
%
\bibitem[SPA11]{spark}
SPARK Team. \emph{SPARK -- The SPADE Ada Kernel (including RavenSPARK)}. 2011. \textsc{url}: \url
{https://docs.adacore.com/sparkdocs-docs/SPARK_LRM.htm} (visited on Sept. 10, 2022).
%
\bibitem[SRSV14]{smith2014quipper}
Jonathan M. Smith, Neil J. Ross, Peter Selinger, and Benoît Valiron. ``Quipper: concrete resource estimation
in quantum algorithms''. In: \emph{Informal Proceedings of QAPL'14, Grenoble, France}. 2014. \textsc{arXiv}:
\href {https://www.arxiv.org/abs/1412.0625} {\nolinkurl {1412.0625}}.
%
\bibitem[SRWD17]{soeken2017hierarchical}
Mathias Soeken, Martin Roetteler, Nathan Wiebe, and Giovanni De Micheli. ``Hierarchical reversible logic
synthesis using luts''. In: \emph{Proceedings of the 54th Annual Design Automation Conference (DAC'17)}
(Austin, TX, USA). ACM, 2017, 78:1--78:6. \textsc{isbn}: 978-1-4503-4927-7. \textsc{doi}: \href
{https://doi.org/10.1145/3061639.3062261} {\nolinkurl {10.1145/3061639.3062261}}.
%
\bibitem[Sta15]{staton2015algebraic}
Sam Staton. ``Algebraic effects, linearity, and quantum programming languages''. In: \emph{Proceedings of
the 42nd Annual ACM SIGPLAN-SIGACT Symposium on Principles of Programming Languages, POPL'15}
(Mumbai, India). Ed. by Sriram K. Rajamani and David Walker. ACM, 2015, pp. 395--406. \textsc{isbn}:
978-1-4503-3300-9. \textsc{doi}: \href {https://doi.org/10.1145/2676726.2676999} {\nolinkurl
{10.1145/2676726.2676999}}. \textsc{url}: \url {http://dl.acm.org/citation.cfm?id=2676726}.
%
\bibitem[Sto77]{stoy1977denotational}
Joseph E. Stoy. \emph{Denotational Semantics: The Scott-Strachey approach to Programming Language
Theory}. Vol. 1. MIT Press Series in Computer Science. MIT Press, 1977.
%
\bibitem[SV05]{selinger05lambda:conf}
Peter Selinger and Benoît Valiron. ``A lambda calculus for quantum computation with classical control''. In:
\emph{Proceedings of the Seventh International Conference on Typed Lambda Calculi and Applications,
TLCA'05} (Nara, Japan). Ed. by Pawel Urzyczyn. Vol. 3461. Lecture Notes in Computer Science. Journal
version appeared in MSCS \cite{selinger2006lambda}. Springer Verlag, Apr. 2005, pp. 354--368. \textsc{isbn}: 3-540-25593-1.
\textsc{doi}: \href {https://doi.org/10.1007/11417170_26} {\nolinkurl {10.1007/11417170_26}}. \textsc{hal}: \href
{https://hal.archives-ouvertes.fr/hal-00483924} {\nolinkurl {hal-00483924}}. \textsc{arXiv}: \href
{https://www.arxiv.org/abs/cs/0404056} {\nolinkurl {cs/0404056}}.
%
\bibitem[SV06]{selinger2006lambda}
Peter Selinger and Benoît Valiron. ``A lambda calculus for quantum computation with classical control''. In:
\emph{Mathematical Structures in Computer Science} 16 (3 2006), pp. 527--552. \textsc{doi}: \href
{https://doi.org/10.1017/S0960129506005238} {\nolinkurl {10.1017/S0960129506005238}}.
%
\bibitem[SV08a]{selinger06fully}
Peter Selinger and Benoît Valiron. ``On a fully abstract model for a quantum linear functional language''. In:
[QPL08], pp. 123--137. \textsc{doi}: \href {https://doi.org/10.1016/j.entcs.2008.04.022} {\nolinkurl
{10.1016/j.entcs.2008.04.022}}.
%
\bibitem[SV08b]{valiron2008categorical}
Peter Selinger and Benoît Valiron. ``A linear-non-linear model for a computational call-by-value lambda
calculus (extended abstract)''. In: \emph{Proceedings of the 11th International Conference on Foundations of
Software Science and Computational Structures, FoSSaCS'08} (Budapest, Hungary). Ed. by
Roberto M. Amadio. Vol. 4962. Lecture Notes in Computer Science. Springer, 2008, pp. 81--96. \textsc{doi}:
\href {https://doi.org/10.1007/978-3-540-78499-9_7} {\nolinkurl {10.1007/978-3-540-78499-9_7}}. \textsc{hal}:
\href {https://hal.archives-ouvertes.fr/hal-00483903} {\nolinkurl {hal-00483903}}.
%
\bibitem[SV09]{selinger2009quantum}
Peter Selinger and Benoît Valiron. ``Quantum lambda-calculus''. In: \cite{gay2009semantic}. Chap. 4, pp. 135--172.
%
\bibitem[SVMABC17]{scherer2017concrete}
Artur Scherer, Benoît Valiron, Siun-Chuon Mau, D. Scott Alexander, Eric van den Berg, and
Thomas E. Chapuran. ``Concrete resource analysis of the quantum linear-system algorithm used to compute
the electromagnetic scattering cross section of a 2D target''. In: \emph{Quantum Information Processing}
16.3 (2017), p. 60. \textsc{doi}: \href {https://doi.org/10.1007/s11128-016-1495-5} {\nolinkurl
{10.1007/s11128-016-1495-5}}. \textsc{hal}: \href {https://hal.archives-ouvertes.fr/hal-01474610} {\nolinkurl
{hal-01474610}}. \textsc{arXiv}: \href {https://www.arxiv.org/abs/1505.06552} {\nolinkurl {1505.06552}}.
%
\bibitem[SVV18]{sabry2018symmetric}

Amr Sabry, Benoît Valiron, and Juliana Kaizer Vizzotto. ``From symmetric pattern-matching to quantum
control''. In: \emph{Proceedings of the 21st International Conference on Foundations of Software Science
and Computation Structures, FoSSaCS 2018} (Thessaloniki, Greece). Ed. by Christel Baier and Ugo Dal Lago.
Vol. 10803. Lecture Notes in Computer Science. Springer, 2018, pp. 348--364. \textsc{doi}: \href
{https://doi.org/10.1007/978-3-319-89366-2_19} {\nolinkurl {10.1007/978-3-319-89366-2_19}}. \textsc{hal}: \href
{https://hal.archives-ouvertes.fr/hal-01763568} {\nolinkurl {hal-01763568}}. \textsc{arXiv}: \href
{https://www.arxiv.org/abs/1804.00952} {\nolinkurl {1804.00952}}.
%
\bibitem[SYG24]{shaikh2024fockedup}
Razin A. Shaikh, Lia Yeh, and Stefano Gogioso. ``The Focked-up ZX Calculus: Picturing Continuous-Variable
Quantum Computation''. Accepted for Communication at QPL 2024. 2024. \textsc{arXiv}: \href
{https://www.arxiv.org/abs/2406.02905} {\nolinkurl {2406.02905}}.
%
\bibitem[TA15]{thomsen2015interpretation}
Michael Kirkedal Thomsen and Holger Bock Axelsen. ``Interpretation and programming of the reversible
functional language RFUN''. In: \emph{Proceedings of the 27th Symposium on the Implementation and
Application of Functional Programming Languages, IFL 2015, Koblenz, Germany, September 14-16, 2015}.
Ed. by Ralf Lämmel. ACM, 2015, 8:1--8:13. \textsc{isbn}: 978-1-4503-4273-5. \textsc{doi}: \href
{https://doi.org/10.1145/2897336.2897345} {\nolinkurl {10.1145/2897336.2897345}}.
%
\bibitem[Tai67]{tait1967intensional}
W. W. Tait. ``Intensional interpretations of functionals of finite type''. In: \emph{Journal of Symbolic Logic}
32.2 (1967), pp. 198--212. \textsc{doi}: \href {https://doi.org/10.2307/2271658} {\nolinkurl {10.2307/2271658}}.
%
\bibitem[TCMG+21]{taddei2021computational}
Márcio M. Taddei, Jaime Cariñe, Daniel Martínez, Tania García, Nayda Guerrero, Alastair A. Abbott,
Mateus Araújo, Cyril Branciard, Esteban S. Gómez, Stephen P. Walborn, Leandro Aolita, and Gustavo Lima.
``Computational advantage from the quantum superposition of multiple temporal orders of photonic gates''.
In: \emph{PRX Quantum} 2 (1 2021), p. 010320. \textsc{doi}: \href
{https://doi.org/10.1103/PRXQuantum.2.010320} {\nolinkurl {10.1103/PRXQuantum.2.010320}}. \textsc{arXiv}:
\href {https://www.arxiv.org/abs/2002.07817} {\nolinkurl {2002.07817}}.
%
\bibitem[TempHask]{template-haskell}
\emph{Template Haskell Library}. \textsc{url}: \url {https://hackage.haskell.org/package/template-haskell}
(visited on Aug. 26, 2021).
%
\bibitem[Tho12]{thomsen2012functional}
Michael Kirkedal Thomsen. ``A functional language for describing reversible logic''. In: \emph{Proceeding of
the 2012 Forum on Specification and Design Languages (FDL'12)} (Vienna, Austria). IEEE, 2012, pp. 135--142.
\textsc{isbn}: 978-1-4673-1240-0. \textsc{url}: \url {http://ieeexplore.ieee.org/document/6336999/}.
%
\bibitem[TK05]{takahashi2005linear-size}
Yasuhiro Takahashi and Noboru Kunihiro. ``A linear-size quantum circuit for addition with no ancillary
qubits''. In: \emph{Quantum Information and Computation} 5.6 (2005), pp. 440--448.
%
\bibitem[TK08]{takahashi2008fast}
Yasuhiro Takahashi and Noboru Kunihiro. ``A fast quantum circuit for addition with few qubits''. In:
\emph{Quantum Information and Computation} 8.6--7 (2008), pp. 636--649.
%
\bibitem[TM22]{townsend2022simplification}
Alex Townsend-Teague and Konstantinos Meichanetzidis. ``Simplification Strategies for the Qutrit
ZX-Calculus''. Presentation at QPL 2022. 2022. \textsc{arXiv}: \href {https://www.arxiv.org/abs/2103.06914}
{\nolinkurl {2103.06914}}.
%
\bibitem[Tof77]{toffoli1977computation}
Tommaso Toffoli. ``Computation and construction universality of reversible cellular automata''. In:
\emph{Journal of Computer and System Sciences} 15.2 (1977), pp. 213--231. \textsc{doi}: \href
{https://doi.org/10.1016/S0022-0000(77)80007-X} {\nolinkurl {10.1016/S0022-0000(77)80007-X}}.
%
\bibitem[Tof80a]{toffoli1980reversible-tech}
Tommaso Toffoli. \emph{Reversible Computing}. Tech. rep. MIT/LCS/TM-151. See also the ICALP'80 paper
[Tof80b]. MIT, 1980.
%
\bibitem[Tof80b]{toffoli1980reversible}
Tommaso Toffoli. ``Reversible computing''. In: \emph{Automata, Languages and Programming, 7th
Colloquium, Noordweijkerhout, The Netherlands, July 14-18, 1980, Proceedings} (Noordweijkerhout, The
Netherlands, July 14--18, 1980). Ed. by J. W. de Bakker and Jan van Leeuwen. Vol. 85. Lecture Notes in
Computer Science. See also the corresponding technical report \cite{toffoli1980reversible}. Springer, 1980, pp. 632--644.
\textsc{isbn}: 3-540-10003-2. \textsc{doi}: \href {https://doi.org/10.1007/3-540-10003-2_104} {\nolinkurl
{10.1007/3-540-10003-2_104}}.
%
\bibitem[Ton04]{tonder04lambda}
André van Tonder. ``A lambda calculus for quantum computation''. In: \emph{SIAM Journal on Computing}
33.5 (2004), pp. 1109--1135. \textsc{doi}: \href {https://doi.org/10.1137/S0097539703432165} {\nolinkurl
{10.1137/S0097539703432165}}. \textsc{arXiv}: \href {https://www.arxiv.org/abs/quant-ph/0307150} {\nolinkurl
{quant-ph/0307150}}.
%
\bibitem[Tra11]{tranquilli2011intuitionistic}
Paolo Tranquilli. ``Intuitionistic differential nets and lambda-calculus''. In: \emph{Theoretical Computer
Science} 412.20 (2011), pp. 1979--1997. \textsc{doi}: \href {https://doi.org/10.1016/j.tcs.2010.12.022} {\nolinkurl
{10.1016/j.tcs.2010.12.022}}.
%
\bibitem[Tro92]{troelstra1992lectures}
Anne S. Troelstra. \emph{Lectures in Linear Logic}. Vol. 29. CSLI Lecture Notes. Stanford, California, US.:
Center for the Study of Language and Information, 1992. \textsc{isbn}: 0-937073-77-6.
%
\bibitem[Tur36]{turing1936computable}
Alan M. Turing. ``On computable numbers, with an application to the Entscheidungsproblem''. In:
\emph{Proceedings of the London Mathematical Society, Series 2} 42 (1936). Can be found integrally, and
commented, in \cite{girard95machine}, pp. 230--265.
%
\bibitem[Tur38]{turing1938phd}
Alan Turing. ``Systems of Logic Based on Ordinals''. PhD thesis. Princeton University, 1938.
%
\bibitem[Tur50]{turing50computing}
Alan M. Turing. ``Computing machinery and intelligence''. In: \emph{Journal of the Mind Association} 59.236
(1950). Can be found integrally, and commented, in \cite{girard95machine}, pp. 433--460.
%
\bibitem[Tur79]{turner1979implementation}
D. A. Turner. ``A new implementation technique for applicative languages''. In: \emph{Software -- Practice
and Experience} 9.1 (1979), pp. 31--49. \textsc{doi}: \href {https://doi.org/10.1002/spe.4380090105} {\nolinkurl
{10.1002/spe.4380090105}}.
%
\bibitem[Unr19a]{unruh2019qhl-ghost}
Dominique Unruh. ``Quantum Hoare logic with ghost variables''. In: \cite{lics2019}, pp. 1--13. \textsc{doi}: \href
{https://doi.org/10.1109/LICS.2019.8785779} {\nolinkurl {10.1109/LICS.2019.8785779}}.
%
\bibitem[Unr19b]{unruh2019rqhl}
Dominique Unruh. ``Quantum relational Hoare logic''. In: \emph{Proceedings of the ACM on Programming
Languages} 3.POPL (2019), 33:1--33:31. \textsc{doi}: \href {https://doi.org/10.1145/3290346} {\nolinkurl
{10.1145/3290346}}.
%
\bibitem[Val04]{valiron2004msc}
Benoît Valiron. ``A Functional Programming Language for Quantum Computation With Classical Control''.
Master thesis. University of Ottawa, 2004. \textsc{hal}: \href {https://hal.archives-ouvertes.fr/tel-00483944}
{\nolinkurl {tel-00483944}}.
%
\bibitem[Val08]{valiron2008phd}
Benoît Valiron. ``Semantics for a Higher Order Functional Programming Language for Quantum
Computation''. PhD thesis. University of Ottawa, 2008. \textsc{hal}: \href
{https://hal.archives-ouvertes.fr/tel-00483944} {\nolinkurl {tel-00483944}}.
%
\bibitem[Val10a]{valiron2010orthogonality}
Benoît Valiron. ``Orthogonality and algebraic lambda-calculus''. In: \emph{Proceedings of the 7th
International QPL Workshop Quantum Physics and Logic, QPL'10} (Oxford, UK). Ed. by Bob Coecke,
Prakash Panangaden, and Peter Selinger. 2010, pp. 169--175. \textsc{url}: \url
{http://www.cs.ox.ac.uk/people/bob.coecke/QPL_proceedings.html}.
%
\bibitem[Val10b]{valiron2010semantics}
Benoît Valiron. ``Semantics of a typed algebraic lambda-calculus''. In: \emph{Proceedings of the Sixth
Workshop on Developments in Computational Models: Causality, Computation, and Physics, DCM 2010}
(Edinburgh, Scotland, July 9--10, 2010). Ed. by S. Barry Cooper, Prakash Panangaden, and Elham Kashefi.
Vol. 26. Electronic Proceedings in Theoretical Computer Science. Preliminary work to the journal
paper \cite{valiron2013typed}. 2010, pp. 147--158. \textsc{doi}: \href {https://doi.org/10.4204/EPTCS.26.14} {\nolinkurl
{10.4204/EPTCS.26.14}}.
%
\bibitem[Val11]{valiron2008quantum}
Benoît Valiron. ``On quantum and probabilistic linear lambda-calculi (extended abstract)''. In: \cite{qpl2008},
pp. 121--128. \textsc{doi}: \href {https://doi.org/10.1016/j.entcs.2011.01.011} {\nolinkurl
{10.1016/j.entcs.2011.01.011}}.
%
\bibitem[Val12]{valiron2012quantum}
Benoît Valiron. ``Quantum computation: a tutorial''. In: \emph{New Generation Computing} 30.4 (2012),
pp. 271--296. \textsc{doi}: \href {https://doi.org/10.1007/s00354-012-0401-7} {\nolinkurl
{10.1007/s00354-012-0401-7}}.
%
\bibitem[Val13a]{valiron2013typed}
Benoît Valiron. ``A typed, algebraic, computational lambda-calculus''. In: \emph{Mathematical Structures in
Computer Science} 23.2 (2013). Journal, extended version of \cite{valiron2010semantics}., pp. 504--554. \textsc{doi}: \href
{https://doi.org/10.1017/S0960129512000205} {\nolinkurl {10.1017/S0960129512000205}}.
%
\bibitem[Val13b]{valiron2013quantum}
Benoît Valiron. ``Quantum computation: from a programmer's perspective''. In: \emph{New Generation
Computing} 31.1 (2013), pp. 1--26. \textsc{doi}: \href {https://doi.org/10.1007/s00354-012-0120-0} {\nolinkurl
{10.1007/s00354-012-0120-0}}.
%
\bibitem[Val16]{valiron2016generating}
Benoît Valiron. ``Generating reversible circuits from higher-order functional programs''. In:
\emph{Proceedings of the 8th International Conference on Reversible Computation, RC'16} (Bologna, Italy).
Ed. by Simon J. Devitt and Ivan Lanese. Vol. 9720. Lecture Notes in Computer Science. Springer, 2016,
pp. 289--306. \textsc{doi}: \href {https://doi.org/10.1007/978-3-319-40578-0_21} {\nolinkurl
{10.1007/978-3-319-40578-0_21}}. \textsc{hal}: \href {https://hal.archives-ouvertes.fr/hal-01474621} {\nolinkurl
{hal-01474621}}.
%
\bibitem[Val17]{valiron2017programmer}
Benoît Valiron. \emph{Programmer un ordinateur quantique}. Column in MathsInfos Hors-Série Numéro 3,
published by Fondation Mathématique de Paris. 2017. \textsc{hal}: \href
{https://hal.archives-ouvertes.fr/hal-01763585} {\nolinkurl {hal-01763585}}.
%
\bibitem[Val18]{valiron2018formal}
Benoît Valiron. ``A formal analysis of quantum algorithms''. In: \emph{ERCIM News} 112 (Jan. 2018),
pp. 23--24. \textsc{hal}: \href {https://hal.archives-ouvertes.fr/hal-01763602} {\nolinkurl {hal-01763602}}.
%
\bibitem[Val22]{valiron2022semantics}
Benoît Valiron. ``Semantics of quantum programming languages: classical control, quantum control''. In:
\emph{Journal of Logical and Algebraic Methods in Programming} 128 (2022), p. 100790. \textsc{doi}: \href
{https://doi.org/10.1016/J.JLAMP.2022.100790} {\nolinkurl {10.1016/J.JLAMP.2022.100790}}. \textsc{hal}: \href
{https://hal.archives-ouvertes.fr/hal-04038653} {\nolinkurl {hal-04038653}}.
%
\bibitem[VAS06]{vizzotto06structuring}
Juliana K. Vizzotto, Thorsten Altenkirch, and Amr Sabry. ``Structuring quantum effects: superoperators as
arrows''. In: \emph{Mathematical Structures in Computer Science} 16.3 (2006), pp. 453--468. \textsc{arXiv}:
\href {https://www.arxiv.org/abs/quant-ph/0501151} {\nolinkurl {quant-ph/0501151}}.
%
\bibitem[Vau09]{vaux2009algebraic}
Lionel Vaux. ``The algebraic lambda-calculus''. In: \emph{Mathematical Structures in Computer Science} 19.5
(2009), pp. 1029--1059. \textsc{doi}: \href {https://doi.org/10.1017/S0960129509990089} {\nolinkurl
{10.1017/S0960129509990089}}.
%
\bibitem[VBE96]{vedral1996quantum}
Vlatko Vedral, Adriano Barenco, and Artur Ekert. ``Quantum networks for elementary arithmetic
operations''. In: \emph{Physical Review A} 54.1 (1996), pp. 147--153. \textsc{doi}: \href
{https://doi.org/10.1103/PhysRevA.54.147} {\nolinkurl {10.1103/PhysRevA.54.147}}. \textsc{arXiv}: \href
{https://www.arxiv.org/abs/quant-ph/9511018} {\nolinkurl {quant-ph/9511018}}.
%
\bibitem[Vil18]{vilmart2018zh}
Renaud Vilmart. ``A zx-calculus with triangles for Toffoli-Hadamard, Clifford+T, and beyond''. In: \cite{qpl2018},
313--p344. \textsc{doi}: \href {https://doi.org/10.4204/EPTCS.287.18} {\nolinkurl {10.4204/EPTCS.287.18}}.
%
\bibitem[Vil21]{vilmart2021structure}
Renaud Vilmart. ``The structure of sum-over-paths, its consequences, and completeness for Clifford''. In:
\emph{Proceedings of the 24th International Conference on the Foundations of Software Science and
Computation Structures, FoSSaCS 2021} (Luxembourg City, Luxembourg, Mar. 27--Apr. 1, 2021). Ed. by
Stefan Kiefer and Christine Tasson. Vol. 12650. Lecture Notes in Computer Science. Springer, 2021,
pp. 531--550. \textsc{isbn}: 978-3-030-71994-4. \textsc{doi}: \href
{https://doi.org/10.1007/978-3-030-71995-1_27} {\nolinkurl {10.1007/978-3-030-71995-1_27}}. \textsc{hal}: \href
{https://hal.archives-ouvertes.fr/hal-02651473} {\nolinkurl {hal-02651473}}.
%
\bibitem[VKB21]{vanrietvelde2021routed}
Augustin Vanrietvelde, Hlér Kristjánsson, and Jonathan Barrett. ``Routed quantum circuits''. In:
\emph{Quantum} 5 (2021), p. 503. \textsc{doi}: \href {https://doi.org/10.22331/q-2021-07-13-503} {\nolinkurl
{10.22331/q-2021-07-13-503}}. \textsc{arXiv}: \href {https://www.arxiv.org/abs/2011.08120} {\nolinkurl
{2011.08120}}.
%
\bibitem[VRSAS15]{valiron2015programming}
Benoît Valiron, Neil J. Ross, Peter Selinger, Dana Scott Alexander, and Jonathan M. Smith. ``Programming
the quantum future''. In: \emph{Communications of the ACM} 58.8 (2015), pp. 52--61. \textsc{doi}: \href
{https://doi.org/10.1145/2699415} {\nolinkurl {10.1145/2699415}}. \textsc{url}: \url
{http://doi.acm.org/10.1145/2699415}. \textsc{hal}: \href {https://hal.archives-ouvertes.fr/hal-01194416}
{\nolinkurl {hal-01194416}}.
%
\bibitem[VZ14a]{valiron2014finite}
Benoît Valiron and Steve Zdancewic. ``Finite vector spaces as model of simply-typed lambda-calculi''. In:
\emph{Proceedings of the 11th International Colloquium on Theoretical Aspects of Computing, ICTAC 2014}
(Bucharest, Romania, Sept. 17--19, 2014). Ed. by Gabriel Ciobanu and Dominique Méry. Vol. 8687. Lecture
Notes in Computer Science. See \cite{valiron2014modeling} for the long version. Springer, 2014, pp. 442--459. \textsc{doi}: \href
{https://doi.org/10.1007/978-3-319-10882-7_26} {\nolinkurl {10.1007/978-3-319-10882-7_26}}.
%
\bibitem[VZ14b]{valiron2014modeling}
Benoît Valiron and Steve Zdancewic. ``Modeling simply-typed lambda calculi in the category of finite vector
spaces''. In: \emph{Scientific Annals of Computer Science} 24.2 (2014), pp. 325--368. \textsc{doi}: \href
{https://doi.org/10.7561/SACS.2014.2.325} {\nolinkurl {10.7561/SACS.2014.2.325}}.
%
\bibitem[Wad03]{wadler2003call-by-value}
Philip Wadler. ``Call-by-value is dual to call-by-name''. In: \emph{Proceedings of the Eighth ACM SIGPLAN
International Conference on Functional Programming, ICFP'03} (Uppsala, Sweden). Ed. by Colin Runciman
and Olin Shivers. ACM, 2003, pp. 189--201. \textsc{isbn}: 1-58113-756-7. \textsc{doi}: \href
{https://doi.org/10.1145/944705.944723} {\nolinkurl {10.1145/944705.944723}}.
%
\bibitem[Wad93]{wadler93syntax}
Philip Wadler. ``A syntax for linear logic''. In: \cite{mfps93}, pp. 513--529.
%
\bibitem[Wan17]{wang2017qutrit}
Quanlong Wang. ``Qutrit ZX-calculus is complete for stabilizer quantum mechanics''. In: \cite{qpl2017}, pp. 58--70.
\textsc{doi}: \href {https://doi.org/10.4204/EPTCS.266.3} {\nolinkurl {10.4204/EPTCS.266.3}}.
%
\bibitem[WB14]{wang2014qutrit}
Quanlong Wang and Xiaoning Bian. ``Qutrit dichromatic calculus and its universality''. In: \cite{qpl2014},
pp. 92--101. \textsc{doi}: \href {https://doi.org/10.4204/EPTCS.172.7} {\nolinkurl {10.4204/EPTCS.172.7}}.
%
\bibitem[WB89]{wadler1989type-classes}
Philip Wadler and Stephen Blott. ``How to make ad-hoc polymorphism less ad-hoc''. In: \emph{Conference
Record of the Sixteenth Annual ACM Symposium on Principles of Programming Languages (POPL'89)}
(Austin, Texas, USA, Jan. 11--13, 1989). ACM Press, 1989, pp. 60--76. \textsc{doi}: \href
{https://doi.org/10.1145/75277.75283} {\nolinkurl {10.1145/75277.75283}}.
%
\bibitem[WBA11]{whitfield2011simulation}
James D. Whitfield, Jacob Biamonte, and Alán Aspuru-Guzik. ``Simulation of electronic structure
Hamiltonians using quantum computers''. In: \emph{Molecular Physics} 109.5 (2011), pp. 735--750.
\textsc{doi}: \href {https://doi.org/10.1080/00268976.2011.552441} {\nolinkurl {10.1080/00268976.2011.552441}}.
%
\bibitem[WC20]{wilson2020diagrammatic}
Matt Wilson and Giulio Chiribella. ``A Diagrammatic Approach to Information Transmission in Generalised
Switches''. 2020. \textsc{arXiv}: \href {https://www.arxiv.org/abs/2003.08224} {\nolinkurl {2003.08224}}.
%
\bibitem[WD10]{wille2010effect}
Robert Wille and Rolf Drechsler. ``Effect of BDD optimization on synthesis of reversible and quantum logic''.
In: \emph{Proceedings of the Workshop on Reversible Computation, RC'09} (York, UK). Ed. by
Irek Ulidowski. Vol. 253. Electronic Notes in Theoretical Computer Science 6. 2010, pp. 57--70. \textsc{doi}:
\href {https://doi.org/10.1016/j.entcs.2010.02.006} {\nolinkurl {10.1016/j.entcs.2010.02.006}}.
%
\bibitem[WDAB21]{wechs2021quantum}
Julian Wechs, Hippolyte Dourdent, Alastair A. Abbott, and Cyril Branciard. ``Quantum circuits with classical
versus quantum control of causal order''. In: \emph{PRX Quantum} 2 (3 2021), p. 030335. \textsc{doi}: \href
{https://doi.org/10.1103/PRXQuantum.2.030335} {\nolinkurl {10.1103/PRXQuantum.2.030335}}. \textsc{hal}:
\href {https://hal.archives-ouvertes.fr/hal-03124176} {\nolinkurl {hal-03124176}}. \textsc{arXiv}: \href
{https://www.arxiv.org/abs/2101.08796} {\nolinkurl {2101.08796}}.
%
\bibitem[Wes16]{westerbaan2016quantum}
Abraham Westerbaan. ``Quantum programs as kleisli maps''. In: \emph{Proceedings of the 13th International
Conference on Quantum Physics and Logic, QPL 2016} (Glasgow, Scotland). Ed. by Ross Duncan and
Chris Heunen. Vol. 236. Electronic Proceedings in Theoretical Computer Science. 2016, pp. 215--228.
\textsc{doi}: \href {https://doi.org/10.4204/EPTCS.236.14} {\nolinkurl {10.4204/EPTCS.236.14}}.
%
\bibitem[Wes19]{westerbaan2019category}
Abraham Anton Westerbaan. ``The Category of Von Neumann Algebras''. PhD thesis. Radboud Universiteit
Nijmegen, 2019. \textsc{url}: \url {https://hdl.handle.net/2066/201611}. \textsc{arXiv}: \href
{https://www.arxiv.org/abs/1804.02203} {\nolinkurl {1804.02203}}.
%
\bibitem[WGTDD08]{wille2008revlib}
Robert Wille, Daniel Große, Lisa Teuber, Gerhard W. Dueck, and Rolf Drechsler. ``RevLib: an online resource
for reversible functions and reversible circuits''. In: \emph{Proceedings of the 38th IEEE International
Symposium on Multiple-Valued Logic, ISMVL 2008} (Dallas, Texas, USA, May 22--23, 2008). IEEE Computer
Society, 2008, pp. 220--225. \textsc{isbn}: 978-0-7695-3155-7. \textsc{doi}: \href
{https://doi.org/10.1109/ISMVL.2008.43} {\nolinkurl {10.1109/ISMVL.2008.43}}.
%
\bibitem[Win80]{winskel1980events}
Glynn Winskel. ``Events in Computation''. PhD thesis. University of Edinburgh, 1980.
%
\bibitem[Win87]{winskel1987event}
Glynn Winskel. ``Event structures''. In: \emph{Petri Nets: Applications and Relationships to Other Models of
Concurrency}. Ed. by W. Brauer, W. Reisig, and G. Rozenberg. Springer Berlin Heidelberg, 1987, pp. 325--392.
\textsc{isbn}: 978-3-540-47926-0.
%
\bibitem[WK13]{wiebe2013floating}
Nathan Wiebe and Vadym Kliuchnikov. ``Floating point representations in quantum circuit synthesis''. In:
\emph{New Journal of Physics} 15.9 (2013), p. 093041. \textsc{doi}: \href
{https://doi.org/10.1088/1367-2630/15/9/093041} {\nolinkurl {10.1088/1367-2630/15/9/093041}}. \textsc{arXiv}:
\href {https://www.arxiv.org/abs/1305.5528} {\nolinkurl {1305.5528}}.
%
\bibitem[WOD10]{wille2010syrec}
Robert Wille, Sebastian Offermann, and Rolf Drechsler. ``SyReC: a programming language for synthesis of
reversible circuits''. In: \emph{Proceedings of the 2010 Forum on specification \& Design Languages, FDL
2010, September 14-16, 2010, Southampton, UK}. Ed. by Adam Morawiec and Jinnie Hinderscheit. See also
Journal's version \cite{wille2016syrec}. ECSI, Electronic Chips \& Systems design Initiative, 2010, pp. 184--189.
\textsc{doi}: \href {https://doi.org/10.1049/ic.2010.0150} {\nolinkurl {10.1049/ic.2010.0150}}.
%
\bibitem[WR16]{wiebe2016quantum}
Nathan Wiebe and Martin Roetteler. ``Quantum arithmetic and numerical analysis using
repeat-until-success circuits''. In: \emph{Quantum Information and Computation} 16.1\&2 (2016),
pp. 134--178. \textsc{doi}: \href {https://doi.org/10.26421/QIC16.1-2-9} {\nolinkurl {10.26421/QIC16.1-2-9}}.
\textsc{arXiv}: \href {https://www.arxiv.org/abs/1406.2040} {\nolinkurl {1406.2040}}.
%
\bibitem[WS14]{wecker2014liquid}
Dave Wecker and Krysta M. Svore. ``LIQUi|⟩: A Software Design Architecture and Domain-Specific Language
for Quantum Computing''. 2014. \textsc{arXiv}: \href {https://www.arxiv.org/abs/1402.4467} {\nolinkurl
{1402.4467}}.
%
\bibitem[WSSD16]{wille2016syrec}
Robert Wille, Eleonora Schönborn, Mathias Soeken, and Rolf Drechsler. ``SyReC: a hardware description
language for the specification and synthesis of reversible circuits''. In: \emph{Integration, the VLSI Journal}
53 (2016). See also extended abstract presented at FDL'10 \cite{wille2010syrec}., pp. 39--53. \textsc{doi}: \href
{https://doi.org/10.1016/j.vlsi.2015.10.001} {\nolinkurl {10.1016/j.vlsi.2015.10.001}}.
%
\bibitem[Wüt11]{wuthrich2011genesis}
Adrian Wüthrich. \emph{The Genesis of Feynman Diagrams}. Vol. 26. Archimedes. Springer, 2011.
%
\bibitem[WY22]{van2022building}
John van de Wetering and Lia Yeh. ``Building qutrit diagonal gates from phase gadgets''. In: \cite{qpl2022}.
\textsc{doi}: \href {https://doi.org/10.4204/EPTCS.394.4} {\nolinkurl {10.4204/EPTCS.394.4}}.
%
\bibitem[XVY21]{xu2021reasoning}
Zhaowei Xu, Benoît Valiron, and Mingsheng Ying. ``Reasoning about Recursive Quantum Programs''. Draft,
to appear in ACM TOCL. 2021. \textsc{arXiv}: \href {https://www.arxiv.org/abs/2107.11679} {\nolinkurl
{2107.11679}}.
%
\bibitem[YAG12]{yokoyama2011reversible}
Tetsuo Yokoyama, Holger Bock Axelsen, and Robert Glück. ``Towards a reversible functional language''. In:
\emph{Revised Papers of the Third International Workshop on Reversible Computation, RC'11} (Gent,
Belgium, July 4--5, 2011). Ed. by Alexis De Vos and Robert Wille. Vol. 7165. Lecture Notes in Computer
Science. Springer, 2012, pp. 14--29. \textsc{doi}: \href {https://doi.org/10.1007/978-3-642-29517-1_2} {\nolinkurl
{10.1007/978-3-642-29517-1_2}}.
%
\bibitem[YAG16]{yokoyama2016fundamentals}
Tetsuo Yokoyama, Holger Bock Axelsen, and Robert Glück. ``Fundamentals of reversible flowchart
languages''. In: \emph{Theoretical Computer Science} 611 (2016), pp. 87--115. \textsc{doi}: \href
{https://doi.org/10.1016/j.tcs.2015.07.046} {\nolinkurl {10.1016/j.tcs.2015.07.046}}.
%
\bibitem[Yao93]{yao1993quantum}
A. Chi-Chih Yao. ``Quantum circuit complexity''. In: \emph{Proceedings of the 34th Annual Symposium on
Foundations of Computer Science (FOCS'93)} (Washington, DC, USA). IEEE Computer Society, 1993,
pp. 352--361. \textsc{doi}: \href {https://doi.org/10.1109/SFCS.1993.366852} {\nolinkurl
{10.1109/SFCS.1993.366852}}.
%
\bibitem[YG07]{yokoyama2007reversible}
Tetsuo Yokoyama and Robert Glück. ``A reversible programming language and its invertible self-interpreter''.
In: \emph{Proceedings of the 2007 ACM SIGPLAN Workshop on Partial Evaluation and Semantics-based
Program Manipulation, PEPM 2007, Nice, France, January 15-16, 2007}. Ed. by G. Ramalingam and
Eelco Visser. 2007, pp. 144--153. \textsc{doi}: \href {https://doi.org/10.1145/1244381.1244404} {\nolinkurl
{10.1145/1244381.1244404}}.
%
\bibitem[Yin11]{ying2011floyd-hoare}
Mingsheng Ying. ``Floyd-Hoare logic for quantum programs''. In: \emph{ACM Transactions on Programming
Languages and Systems} 33.6 (2011), 19:1--19:49. \textsc{doi}: \href {https://doi.org/10.1145/2049706.2049708}
{\nolinkurl {10.1145/2049706.2049708}}. \textsc{arXiv}: \href {https://www.arxiv.org/abs/0906.4586} {\nolinkurl
{0906.4586}}.
%
\bibitem[Yin19]{ying2019automatic}
Mingsheng Ying. ``Toward automatic verification of quantum programs''. In: \emph{Formal Aspects of
Computing} 31.1 (2019), pp. 3--25. \textsc{doi}: \href {https://doi.org/10.1007/s00165-018-0465-3} {\nolinkurl
{10.1007/s00165-018-0465-3}}. \textsc{arXiv}: \href {https://www.arxiv.org/abs/1807.11610} {\nolinkurl
{1807.11610}}.
%
\bibitem[YLYF14]{ying2014model-checking}
Mingsheng Ying, Yangjia Li, Nengkun Yu, and Yuan Feng. ``Model-checking linear-time properties of
quantum systems''. In: \emph{ACM Transactions on Computational Logic} 15.3 (2014), 22:1--22:31.
\textsc{doi}: \href {https://doi.org/10.1145/2629680} {\nolinkurl {10.1145/2629680}}.
%
\bibitem[YYFD13]{ying2013verification}
Mingsheng Ying, Nengkun Yu, Yuan Feng, and Runyao Duan. ``Verification of quantum programs''. In:
\emph{Science of Computer Programming} 78.9 (2013), pp. 1679--1700. \textsc{doi}: \href
{https://doi.org/10.1016/j.scico.2013.03.016} {\nolinkurl {10.1016/j.scico.2013.03.016}}.
%
\bibitem[YYW17]{ying2017invariants}
Mingsheng Ying, Shenggang Ying, and Xiaodi Wu. ``Invariants of quantum programs: characterisations and
generation''. In: \cite{popl2017}, pp. 818--832. \textsc{doi}: \href {https://doi.org/10.1145/3009837.3009840}
{\nolinkurl {10.1145/3009837.3009840}}.
%
\bibitem[YZLF22]{ying2022proof}
Mingsheng Ying, Li Zhou, Yangjia Li, and Yuan Feng. ``A proof system for disjoint parallel quantum
programs''. In: \emph{Theoretical Computer Science} 897 (2022), pp. 164--184. \textsc{doi}: \href
{https://doi.org/10.1016/j.tcs.2021.10.025} {\nolinkurl {10.1016/j.tcs.2021.10.025}}.
%
\bibitem[ZF10]{zhao2010dcpo-completion}
Dongsheng Zhao and Taihe Fan. ``Dcpo-completion of posets''. In: \emph{Theoretical Computer Science}
411.22-24 (2010), pp. 2167--2173. \textsc{doi}: \href {https://doi.org/10.1016/j.tcs.2010.02.020} {\nolinkurl
{10.1016/j.tcs.2010.02.020}}.
%
\bibitem[ZW17]{zulehner2017improving}
Alwin Zulehner and Robert Wille. ``Improving synthesis of reversible circuits: exploiting redundancies in
paths and nodes of QMDDs''. In: \cite{rc2017}, pp. 232--247. \textsc{doi}: \href
{https://doi.org/10.1007/978-3-319-59936-6_18} {\nolinkurl {10.1007/978-3-319-59936-6_18}}.
%
\bibitem[ZYC20]{zhao2020quantum}
Xiaobin Zhao, Yuxiang Yang, and Giulio Chiribella. ``Quantum metrology with indefinite causal order''. In:
\emph{Physical Review Letters} 124 (19 2020), p. 190503. \textsc{doi}: \href
{https://doi.org/10.1103/PhysRevLett.124.190503} {\nolinkurl {10.1103/PhysRevLett.124.190503}}.
%
\bibitem[ZYY19]{zhou2019applied}
Li Zhou, Nengkun Yu, and Mingsheng Ying. ``An applied quantum Hoare logic''. In: \emph{Proceedings of
the 40th ACM SIGPLAN Conference on Programming Language Design and Implementation, PLDI 2019}
(Phoenix, AZ, USA, June 22--26, 2019). Ed. by Kathryn S. McKinley and Kathleen Fisher. ACM, 2019,
pp. 1149--1162. \textsc{isbn}: 978-1-4503-6712-7. \textsc{doi}: \href {https://doi.org/10.1145/3314221.3314584}
{\nolinkurl {10.1145/3314221.3314584}}.
\end{thebibliography}

}
\end{document}